\documentclass[sigconf, nonacm]{acmart}
\usepackage{adjustbox}
\usepackage{subcaption}
\usepackage{multirow}
\usepackage[dvipsnames]{xcolor}

\begin{document}
\title{VADER: Filtered Vector Search with Declarative Recall\texorpdfstring{\\{}}{ }[Extended Version]}

\author{\texorpdfstring{
  Manos Chatzakis$^1$,
  Duo Lu$^2$,
  Helena Caminal$^3$,
  Yannis Chronis$^{3,4}$,
  Fatma Özcan$^3$,
  Yannis Papakonstantinou$^3$,
  Timofey Asyrkin$^3$,
  Sebastian Infante Murguia$^3$,
  Itai Rosenblatt$^3$,\newline
  Themis Palpanas$^1$
}{Manos Chatzakis, Duo Lu, Helena Caminal, Yannis Chronis, Fatma Özcan,
  Yannis Papakonstantinou, Timofey Asyrkin, Sebastian Infante Murguia,
  Itai Rosenblatt, Themis Palpanas}}
\affiliation{
  \institution{
    $^1$Université Paris Cité, LIPADE \quad
    $^2$Brown University \quad
    $^3$Google \quad
    $^4$ETH Zürich
  }
}
\email{
  manos.chatzaki@gmail.com, duo\_lu@brown.edu,
  {hcaminal, fozcan, yannispap, tasyrkin, sebinf, itairos}@google.com
}
\email{
  chronis@ethz.ch,
  themis@mi.parisdescartes.fr
}

\begin{abstract}
Approximate filtered vector search (FVS), a core operation in many data management tasks that combine structured data with vector embeddings, exhibits increased complexity due to the characteristics of filtering predicates.
Each predicate is defined by {selectivity} (i.e., the fraction of vectors that satisfy the predicate) and {correlation} (i.e., the relationship between the filter and the vector space), which can significantly affect search difficulty even for the same query vector.
This poses a key challenge for users aiming to integrate vector search with structured data, as efficient execution often requires extensive manual tuning of algorithm parameters.
In this paper, we present VADER, the first approach that eliminates hyperparameter tuning by introducing declarative recall for approximate filtered vector search.
With declarative recall, users specify a desired recall target, and VADER executes FVS queries to meet this target without requiring manual configuration.
VADER achieves this by employing a filter-aware recall predictor that generalizes across varying selectivities and correlations without explicit tuning, and by performing early termination once the predicted recall reaches the user-defined target.
Through extensive experimental evaluation, we show that VADER achieves near-optimal early termination, while providing significant speedups of up to 53\% faster and improved result quality of 28\% compared to the best-performing baseline.
This paper appeared in VLDB 2027.
\end{abstract}

\maketitle

\pagestyle{plain}

\section{Introduction}
\label{sec:introduction}

Approximate nearest neighbor vector search is a core operation in many data science tasks~\cite{lewis2020rag,sanca2023context,chatzakis2021rdfsim,ferhatosmanoglu2001ANNS-multimedia-db,huang2020embedding,chen2022recommendation}.
Recently, 
\emph{filtered approximate vector search (FVS)} has gained traction, as combining structured attributes with vector embeddings is common in retrieval applications~\cite{chronis2025filtered,li2025sieve,lu2026fvsbenchmark}.
In this type of vector search, vector data are combined with structured attribute metadata, and the goal is to retrieve the top-$k$ most similar vectors for a given query vector while ensuring that the returned results satisfy a specified \emph{predicate}.

FVS can be performed by \emph{pre-filtering} (applying the filter before a brute-force search), \emph{post-filtering} (filtering the results of a vector search), iterative scans (repeated post-filtering until enough results are found), or \emph{inline filtering}, which blends the two.
Inline filtering stands out because it retains the efficiency of vector search indexes while incorporating filtering into the search rather than decoupling the two.
Among inline methods, \emph{filter-agnostic} ones make minimal assumptions about predicate characteristics and require only limited modifications to the underlying index, thereby supporting a wide range of filtering operations and enabling practical adoption in large-scale systems~\cite{weaviate_filtering, pinecone_filtering}.

\noindent \textbf{The Complexity of FVS.}
Since queries process both vectors with predicates in structured columns, each query is characterized by a specific \emph{selectivity} (i.e., the fraction of rows/vectors that satisfy the predicate) and \emph{correlation} between the filter selectivity and the distance from the query vector.
These predicate characteristics significantly increase the difficulty of achieving the desired result quality (measured as \emph{Recall}, the fraction of ground-truth vectors retrieved by the search) and latency.

\begin{figure}[tb]
     \centering
     \begin{subfigure}[b]{0.47\textwidth}
         \centering
         \includegraphics[page=6,trim=2 3.5cm 0 0,clip,width=0.93\textwidth]{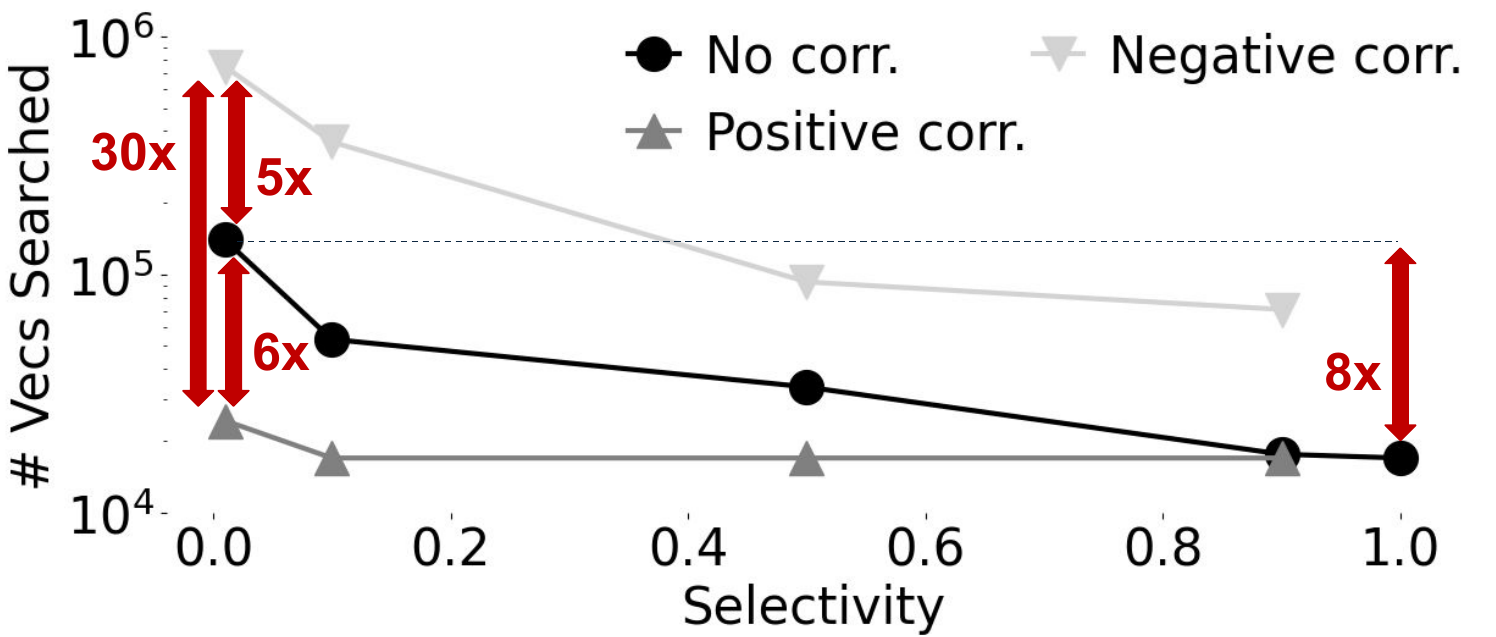}
     \end{subfigure}
    \vspace{-0.3cm}
    \caption{Search effort of a query with different filter correlations in the GLOVE1M dataset for the same recall target.}
    \vspace{-0.8cm}
    \label{fig:filter-motivation}
\end{figure}

As a motivating example, Figure~\ref{fig:filter-motivation} shows the number of vectors we need to search for a query to reach recall 0.9 on GLOVE1M~\cite{pennington2014glove, aumuller2020annbenchmarks}, under varying selectivity and correlation.
The required search effort varies by orders of magnitude: fixing selectivity at 0.01 by only varying correlation we need to search up to $30\times$ more vectors to achieve the same recall.
Similarly for fixed no correlation, varying selectivity from 0.01 to 1.0 we need to search up to $8\times$ more vectors.

\noindent \textbf{Practical Implications.} 
The task of tuning FVS hyperparameters to achieve a desired recall without excessive resource consumption is very challenging for end users. 
One option could be to tune parameters (e.g., $ef\_search$ in $k$-NN graphs) per selectivity and correlation combination, which is impractical: correlation is not known in advance, and selectivity estimates, when available, may be inaccurate~\cite{leis2015howgoodoptimizersare}.
The alternative, reusing values that generalize across selectivities (e.g., tuned on unfiltered workloads), depends heavily on the generalization of the underlying algorithm and typically yields configurations that are either too conservative or too aggressive for the given filter.

Moreover, the tuning procedures underlying both approaches introduce additional challenges~\cite{chatzakis2025darth, zhang2025adaef}. 
Existing methods apply the same hyperparameter values 
across all queries in a workload. 
As a result, easy queries that require minimal effort to reach the target recall 
oversearch the index, leading to wasted computation and increased latency.
Conversely, 
difficult queries may undersearch, producing results that fail to meet the desired recall target. 

Iterative scan~\cite{pgvector} does not provide a universal solution either. 
Iterative scan resumes the search if fewer than $k$ results remain after filtering, by using a resumable post-filtering strategy. 
However, this approach does not fully address the complexity of achieving a target recall in filtered vector search, as there may exist other vectors that satisfy the predicate and are closer to the query, but the hard-stop policy prevents the search from discovering them.
We discuss iterative scan in details in Section~\ref{sec:relatedwork}. 

\noindent \textbf{Declarative Recall for Unfiltered Vector Search.}
Recent work addresses the same tuning burden for unfiltered vector search through \emph{declarative recall} with \emph{early termination}~\cite{chatzakis2025darth, mohoney2025quake, zhang2025adaef, wang2026annie}: users specify a target recall and the system terminates each query once that target is reached, eliminating manual tuning.
The increased complexity of FVS, however, has so far prevented the development of declarative recall methods tailored to this setting.

\noindent \textbf{Contributions.}
We summarize our contributions as follows.

\noindent $\bullet$
We present VADER, the first declarative recall solution for filtered vector search based on early termination.
VADER eliminates the need for hyperparameter tuning, and generalizes seamlessly across different filtering configurations.

\noindent $\bullet$
We introduce a filter-aware recall predictor along with a novel set of input features that enable highly accurate recall prediction with no prior knowledge of predicate selectivity or correlation.

\noindent $\bullet$
We introduce a novel asymmetric loss function for the recall predictor, specifically designed for declarative recall, which automatically penalizes more cases of overpredictions of recall, and ensures that VADER minimizes the number of queries that fall below the target recall, without significantly affecting prediction quality or requiring additional tuning.

\noindent $\bullet$
We implement VADER on top of popular filter-agnostic FVS algorithms of different families, including (without loss of generality) $k$-NN graphs (on top of Sweeping~\cite{weaviate_filtering} and ACORN~\cite{patel2024acorn}) and IVF (on top of Filtered-ScaNN~\cite{guo2020scann, alloydb_filtering}), demonstrating its versatility.

\noindent $\bullet$
We evaluate VADER through an extensive experimental study spanning multiple datasets and filtering configurations.
VADER can be trained within a few minutes and delivers accurate performance independent of selectivity, correlation, and target recall.
VADER terminates search within 4\% of the optimal termination point, meets the desired recall targets across selectivities and correlations without introducing overhead, and achieves up to $25\times$ speedup over search without early termination.
Against the best-performing baseline, which (unlike VADER) is tuned for each selectivity and recall target, VADER is simultaneously up to 28\% better in quality and 53\% faster, establishing the state of the art for FVS declarative recall.

\vspace{-0.1cm}
\section{Preliminaries}
\label{sec:preliminaries}

\subsection{Background}
\label{sec:background}

\subsubsection{Filtered Approximate Nearest Neighbor Vector Search (FVS)}
Given a collection of vectors $V$, a query vector $q$, a distance measure $D$, a number $k$, and a filter predicate $F$, $k$-nearest neighbor vector search refers to the task of retrieving the $k$ most similar vectors (nearest neighbors) to $q$ in $V$ that satisfy $F$, according to $D$.
This task can be either exact~\cite{palpanas2019evolution,del2026daisy,chatzakis2023odyssey,chatzakis2026odysseyplus,echihabi2020hydra1}, where all retrieved nearest neighbors are correct, or approximate~\cite{echihabi2020hydra2,azizi2025graphbenchmark}, where some errors in the retrieved results are tolerated in exchange for significantly faster query processing.
In FVS, the approximation only affects the proximity of vectors in the high dimensional space, and the filtering constraint is satisfied for all returned results.

In this work, we focus on approximate search and remain agnostic to the type of predicate $F$, which may represent any filtering operation over structured data (e.g., joins, selections).
When using approximate search, the quality of the results is evaluated using two main measures.
The first measure is \emph{search quality}, typically quantified using recall (the fraction of true filtered nearest neighbors that are correctly retrieved).
The second measure is \emph{latency}, defined as the time required to process the query.

\subsubsection{Vector Search Index.}
Approximate vector search is typically performed by constructing an index over the vector collection; the most widely used index types are the $k$-NN graph and inverted files (IVF).
$k$-NN graph methods build an approximate proximity graph in which vectors are nodes and edges connect similar vectors.
Variants differ in construction (e.g., HNSW~\cite{malkov2018hnsw} builds a hierarchical multilayer graph, DiskANN~\cite{jayaram2019diskann} relaxes the edge-creation policy) but share the search procedure: starting from a predefined entry point, the search proceeds greedily toward nodes progressively closer to the query.
Its extent is controlled by $efSearch$, which governs the trade-off between efficiency and accuracy.
IVF methods (such as ScaNN~\cite{guo2020scann}) cluster the vectors into $nList$ partitions using k-means and then retrieve the nearest neighbors of a given query by examining the vectors that belong to the partition of the closest $nProbe$ cluster centroids.
Similar to $efSearch$, the $nProbe$ parameter controls the trade-off between efficiency and accuracy at query time.

\subsubsection{Filter-Agnostic FVS}
Filter-agnostic FVS methods assume no prior knowledge of the structured filter predicates, and so generalize to settings where the structured data may evolve.
They construct a standard vector index over the vector collection (e.g., a $k$-NN graph). 
During search, the algorithm evaluates the structured information of each vector using a simple Boolean test to check if it satisfies the given predicate.
This design allows these methods to generalize across different types of filtering mechanisms and facilitates their adoption in large-scale vector databases~\cite{pinecone_filtering,weaviate_filtering,milvus_filtering,alloydb_filtering,pgvector}.

\subsubsection{Filtering Strategies for Vector Search}
There are $3$ main FVS approaches: pre-, inline-, and post-filtering~\cite{lu2026fvsbenchmark}.
Pre-filtering applies the predicate before vector search, removing vectors that do not satisfy the predicate and then performing exhaustive search on the remaining vectors. 
Post-filtering first performs an unfiltered vector search and then removes the returned vectors that do not satisfy the predicate. 
Inline-filtering integrates search and filtering by applying the predicate during the search process, rather than separating the two steps. 
Inline filtering methods are the focus of this work, as they offer greater opportunities for adaptive optimization.

\subsubsection{Declarative Recall for Filtered Vector Search}
Current FVS algorithms are typically invoked in the form $FVS(q, I, k, F, sp)$, where $q$ is the query, $I$ is the index built over the vector collection, $k$ is the number of neighbors to retrieve, $F$ is the predicate over the structured data, and $sp$ denotes the search parameter values (e.g., $ef\_search$ for $k$-NN graphs).
However, selecting $sp$ for a given recall target $R_t$ is challenging, as the behavior of each filter predicate $F$ (in terms of selectivity and correlation) is not known in advance and varies across queries. 
VADER is the first approach to provide declarative recall for FVS, supporting an interface of the form $FVS(q, I, k, F, R_t)$ and eliminating the need for hyperparameter tuning by requiring only the desired target recall $R_t$.

\subsection{Related Work}
\label{sec:relatedwork}

\subsubsection{Filter-Agnostic FVS Algorithms}
Several filter-agnostic approaches for FVS have been proposed, with a comprehensive overview provided in~\cite{lu2026fvsbenchmark}.
ACORN~\cite{patel2024acorn} builds on a $k$-NN graph index, specifically HNSW, and traverses at query time only the subgraph of nodes that satisfy the predicate.
Since such subgraphs can be sparse or disconnected, ACORN maintains connectivity through run-time 2-hop expansion: at each step it collects 1-hop and 2-hop neighbors, applies the filter, and computes distances only for nodes that satisfy the predicate.
ACORN-1 relies solely on this search-time expansion, while ACORN-$\gamma$ additionally densifies the graph during construction, using compression to control index size.
NaviX~\cite{sehgal2025navix} extends ACORN by incorporating additional heuristics into the traversal process.
Sweeping~\cite{weaviate_filtering} constructs the candidate set in the HNSW graph based on proximity, similar to an unfiltered traversal, while retaining in the result set only vectors that satisfy the predicate.
This ensures that the search operates over a well-connected graph.
Iterative scan~\cite{pgvector} decouples HNSW traversal from predicate evaluation, resuming the index search whenever fewer than $k$ results remain after filtering.
It cannot fully address the complexity of achieving a target recall, however: its hard-stop policy prevents the search from discovering other vectors that satisfy the predicate and are closer to the query.
Filtered-ScaNN~\cite{guo2020scann} operates similarly to a standard unfiltered IVF-based vector index.
To retrieve the top-$k$ results for a query, Filtered-ScaNN processes the nearest neighbors of the partitions of the $nProbe$ closest centroids, and only vectors that satisfy the filter are retrieved and added to the final result set.

\subsubsection{Early Termination for Unfiltered Vector Search}
Early termination approaches, i.e., vector search methods that terminate query processing once a predefined target recall has been met~\cite{chatzakis2026vectorsearchquest}, have been proposed only for unfiltered vector search.
Most such methods operate on a single index type, such as IVF~\cite{chen2021spann,zhang2023auncel,mohoney2025quake,horchidan2025conann,busolin2024pee}, $k$-NN graphs~\cite{teofili2025pip,zhang2025adaef,wang2026annie}, or iSAX~\cite{echihabi2023pros,gogolou2019progressive} (an index for exact vector search~\cite{camerra2010isax}), and often rely on assumptions about the underlying distance measure~\cite{zhang2025adaef,echihabi2023pros,gogolou2019progressive}.
Among recent methods that support arbitrary index types (including both $k$-NN-graph and IVF approaches~\cite{li2020laet}), DARTH and DARTH+~\cite{chatzakis2025darth,chatzakis2026darthplus} stand out as the only methods that natively support declarative recall without making assumptions about the underlying distance measure.
However, to the best of our knowledge, there is still a lack of early termination approaches able to provide declarative recall specifically for FVS.

\section{The VADER Approach}
\label{sec:vader}

\begin{figure*}[tb]
     \centering
     \begin{subfigure}[b]{0.99\textwidth}
         \centering
         \includegraphics[page=1,trim=0 7.3cm 0 0,clip,width=\textwidth]{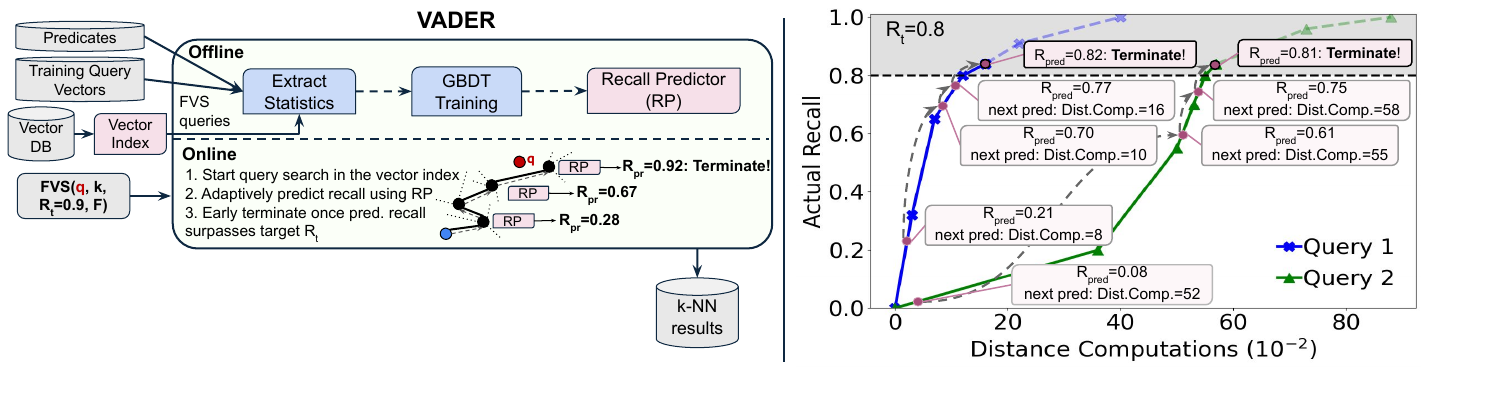}
        \end{subfigure}
    \vspace{-0.8cm}
    \caption{Overview of VADER (left) and example of recall predictor usage in VADER for two FVS queries (right).}
    \vspace{-0.4cm}
    \label{fig:vader-overview}
\end{figure*}

In this section, we present the details of the VADER approach, with an overview provided in Figure~\ref{fig:vader-overview}.
VADER operates on top of a vector index (e.g., a $k$-NN graph or IVF).
In an offline phase, VADER combines training query vectors with predicates generated across different selectivities and correlations.
It then extracts query search progression training data by executing the FVS training queries on the vector index. 
This training data is used to train a recall predictor model.
During the online phase, users submit FVS queries with a declarative recall target $R_t$.
Upon query submission, VADER executes the query using the vector index, invoking the recall predictor at appropriate points and performing early termination once the predicted recall meets or exceeds the target.

We now provide details on both phases.
First, we describe how the recall predictor is used within VADER (Section~\ref{sec:vader:early-term-recall-prediction}), then we describe the architecture of the model (Section~\ref{sec:vader:recall-predictor-architecture}), followed by the procedure for constructing and training the predictor (Section~\ref{sec:vader:recall-predictor-training}).
Finally, we present implementation-specific details for integrating VADER into filter-agnostic inline filtering algorithms (Section~\ref{sec:vader:integration-to-fvs}).

\subsection{Early Termination with Recall Prediction}
\label{sec:vader:early-term-recall-prediction}

VADER is a successor of DARTH~\cite{chatzakis2025darth} and DARTH+~\cite{chatzakis2026darthplus}, which introduce a recall predictor that estimates the recall of a query at any point during the search, enabling a declarative recall interface without per-query parameter tuning.
An important problem they address is the invocation strategy: when to call the predictor during query execution. 
Since vector search operates at millisecond-scale latency, too many invocations introduce overhead that offsets the benefit of early termination, while too few cause the system to miss optimal termination points. 
The adaptive strategy of DARTH/DARTH+ invokes the predictor more frequently as the predicted recall approaches the target, and less frequently when it is far from it, enabling fine-grained invocation near the optimal termination point and avoiding overhead during the early stages.

To implement this adaptive strategy, DARTH/DARTH+ use the number of distance computations as the invocation interval, evaluating at each call
$pi = mpi + (ipi - mpi) \cdot (R_t - R_p),$
where $R_t$ is the target recall, $R_p$ the predicted recall, $pi$ the number of distance computations after which the next prediction occurs, and $mpi$ and $ipi$ the minimum and maximum (initial) prediction intervals.
Although $ipi$ and $mpi$ are hyperparameters, they are set automatically based on the training data: the average number of distance computations required to reach each recall target, $dist_{R_t}$, is obtained at no additional cost during training data collection, and the strategy sets $ipi = \frac{dist_{R_t}}{2}$ and $mpi = \frac{dist_{R_t}}{10}$.
Note that we experimentally verify and discuss the performance of this heuristic for FVS queries in Section~\ref{sec:experiments:main-results}.
Since DARTH/DARTH+ do not support FVS queries, VADER builds upon them to provide an efficient declarative recall interface for filtered vector search, but adopts the recall prediction and adaptive invocation mechanisms from them.

Figure~\ref{fig:vader-overview} (right) shows this for $2$ FVS GLOVE1M queries with $R_t$=$0.8$, marking the points at which the predictor is invoked; dashed lines denote search without early termination.
Note that the predictor is invoked more frequently as the predicted recall approaches the target, and the search terminates once the target is exceeded.

\subsection{Model Architecture}
\label{sec:vader:recall-predictor-architecture}

We now describe our design choices for the architectural details of VADER's recall predictor.

\subsubsection{Recall Predictor Model}
The model must achieve high prediction accuracy while keeping inference latency low enough not to introduce overhead into the usually millisecond-scale vector search.
Complex deep learning models are therefore impractical, while simpler regression models (e.g., linear regression) lack sufficient accuracy~\cite{chatzakis2025darth,li2020laet}.
These constraints motivate Gradient Boosted Decision Trees (GBDT)~\cite{friedman2002stochasticgradientboosting}, an ensemble of decision trees in which each tree corrects the errors of the previous ones, capturing complex, non-linear relationships at low computational cost. 
Thus, GBDT achieves a favorable balance between prediction accuracy and inference efficiency.
Such models have been successfully applied to similar tasks~\cite{li2020laet,chatzakis2025darth,chatzakis2026darthplus}, particularly using optimized libraries such as LightGBM~\cite{ke2017lightgbm}, which offer fast training and near-constant-time, sub-millisecond inference.
VADER therefore adopts GBDT as the architecture of its recall predictor.

\subsubsection{Asymmetric Loss Function}
Classic GBDT training uses a standard loss function such as the Mean Squared Error (MSE):
$\mathcal{L}_{MSE} = \frac{1}{2} \sum_{i=1}^{n} (R_{pred,i} - R_{act,i})^2$, where $R_{pred,i}, R_{act,i} \in [0,1]$ denote the predicted and the actual recall, and the (optional) factor $\frac{1}{2}$ makes the gradient with respect to $R_{pred,i}$ equal to the residual, avoiding an additional multiplicative constant. 
This penalizes recall overpredictions (predicting a recall higher than the actual value) and underpredictions equally.
However, in recall-critical applications, such as those in the legal or medical domain or High-Recall-Retrieval (HRR) settings~\cite{kim2022legaldocretrieval,yang2024splademedicaldocretrieval,song2019effectivehrr}, ensuring that queries meet the target recall is essential, making overpredictions more harmful than underpredictions~\footnote{Note that overpredictions lead to undersearching for hard queries since it triggers the early termination signal earlier than it should}. 
This is because in declarative recall the objective is to reach the target recall (there is an implicit bias in the definition).
To address this, VADER introduces an optional asymmetric loss function that penalizes overpredictions more heavily than underpredictions, yielding more conservative recall estimates and reducing the likelihood of failing to meet the target.
Defining a penalty parameter \(\lambda > 1\) on overprediction, the resulting loss for each observation is

\[
\mathcal{L}_{asym}(R_{act,i}, R_{pred,i})
=
\begin{cases}
\dfrac{1}{2}(R_{\text{pred},i} - R_{\text{act},i})^2, & \text{if } R_{\text{pred},i} \le R_{\text{act},i}, \\[6pt]
\dfrac{\lambda}{2}(R_{\text{pred},i} - R_{\text{act},i})^2, & \text{if } R_{\text{pred},i} > R_{\text{act},i}.
\end{cases}
\]

The total empirical loss is \(\mathcal{L} = \sum_{i=1}^{n} \mathcal{L}(R_{\text{act},i}, R_{\text{pred},i})\), which preserves the quadratic structure of the MSE objective while assigning a higher weight \(\lambda\) to positive residuals (i.e., \(R_{\text{pred},i} > R_{act,i}\)), reducing overprediction of recall by making the predictor more conservative.

A key challenge in this design is determining the appropriate penalization factor $\lambda$, as excessive penalization may degrade prediction accuracy. 
Further, hand-tuning such a parameter would contradict VADER’s goal of eliminating hyperparameter tuning.
VADER sets $\lambda$ in a cost free way based on the distribution of the recall values of the training data as $\lambda = 1 + \sqrt{1 - \bar{R}_{act}}$, where $\bar{R}_{act}$ is the average training recall. 
This formula is used to calculate the penalty across different settings and datasets.
Note that we experimentally study the penalty behavior in Section~\ref{sec:experiments:recall-prediction-analysis}.

\subsection{Recall Predictor Training}
\label{sec:vader:recall-predictor-training}

We now focus on the details of constructing an accurate recall predictor, discussing the input features and the training procedure.

\subsubsection{Recall Prediction Features}
\label{sec:vader:recall-prediction-features}

A core design challenge is to derive an effective set of features that capture different characteristics of the filtered search process.
Recall that since our predictor is called multiple times for each query, the derived input features can capture dynamic characteristics of the search progress of each query.
Prior work~\cite{chatzakis2025darth} demonstrated that query recall during unfiltered search can be predicted using features that track vector search progression. 
We argue (and experimentally verify in Section~\ref{sec:experiments}) that this approach is insufficient for accurate recall prediction under the presence of filters.
This limitation arises from relying only on vector search progression features that fail to capture the dynamics of the filtering process. 
Recall that for the same query vector, the characteristics of the filtering predicate can significantly alter the search effort required to achieve the same recall (Figure~\ref{fig:filter-motivation}).

To address this, VADER employs a novel feature set, combining dynamic vector search progression features and nearest neighbor statistics with features that capture dynamic predicate-related characteristics.
Importantly, VADER does not explicitly use predicate selectivity or correlation as input features, as these quantities cannot be assumed to be known in advance.
Moreover, the input features are not tied to a specific algorithm and can be applied across a wide range of vector search indexing and FVS algorithms.
Furthermore, the feature set can be extended with additional algorithm-specific features, depending on the underlying FVS method, as subsequently discussed in Section~\ref{sec:vader:integration-to-fvs}.
We summarize the full set of VADER features in Table~\ref{table:input-features} and describe below the key filtering-related features.

VADER is designed to operate as a filter-agnostic technique. Specifically, we do not assume access to predicate-specific data distribution information before the start of the filtered vector search execution. 
Instead, we rely on features derivable from the filter-agnostic search process itself.
The first three are $vectors\_checked$, $vectors\_passed$, and $vectors\_failed$, which represent the number of vectors for which the predicate has been evaluated, the number that satisfy the predicate, and the number that do not, respectively.
In addition, we include $observed\_selectivity$, defined as the fraction of examined vectors that satisfy the predicate (i.e., $vectors\_passed / vectors\_checked$). 
This feature has proven the most informative of the four and one whose importance is also highlighted in prior work~\cite{sehgal2025navix}.
Section~\ref{sec:experiments:recall-prediction-analysis} studies the contribution of each feature in more depth.

The non-filtering related features of VADER are inherited directly from DARTH and DARTH+~\cite{chatzakis2025darth,chatzakis2026darthplus} that experimentally verified that these features accurately capture vector-specific search characteristics. 
Specifically, the features capture the search progression within the index (\emph{Search-Progression}), statistics of intermediate nearest neighbor distances (\emph{Intermediate-NN-Stats}), and query dimensional distribution properties (\emph{Query-Dim-Stats}).

Note that the features in Table~\ref{table:input-features} capture the core state of an individual query search \emph{and} are common across ANNS methods, so the same base set applies to both $k$-NN graph and IVF algorithms; 
Section~\ref{sec:vader:integration-to-fvs} gives the per-algorithm specifics.

\begin{table}[tb]
\centering
{
\scriptsize
\begin{adjustbox}{max width=\columnwidth}
\begin{tabular}{|| c | c | c ||} 
 \hline
 Type & Features & \parbox{0.5\columnwidth}{\centering Description} \\ 
 \hline\hline

 \multirow{3}{*}{\centering Search Progression} 
 & $nstep$ & Search Step (e.g. $k$-NN graph hops)  \\ 
 & $ndis$ & No. distance calculations \\ 
 & $ninserts$ & No. updates in the NN result set \\ 
 \hline
 
 \multirow{9}{*}{\centering Intermediate-NN-Stats} 
 & $firstNN$ & Distance of the first NN evaluated   \\ 
 & $closestNN$ & Distance of current closest NN \\
 & $furthestNN$ & Distance of current furthest ($k$-th) NN \\
 & $avg$ & Average of distances of the NN \\
 & $var$ & Variance of distances of NN \\
 & $med$ & Median of distances of NN \\
 & $perc25$ & 25th percentile of distances of NN \\
 & $perc75$ & 75th percentile of distances of NN \\
 \hline

 \multirow{8}{*}{\centering Query-Dim-Stats} 
 & $q\_avg$ & Average of query dimensions \\
 & $q\_med$ & Median of query dimensions \\
 & $q\_std$ & Stan. dev. of query dimensions \\
 & $q\_min$ & Min of query dimensions \\
 & $q\_max$ & Max of query dimensions \\
 & $q\_range$ & Range of query dimensions ($q\_max-q\_min$) \\
 & $q\_L1$ & L1 norm of query dimensions \\
 & $q\_L2$ & L2 norm of query dimensions \\
 \hline

 \multirow{4}{*}{\centering Novel Filtering Stats} 
 & $vectors\_checked$ & Number of predicates vectors evaluated \\
 & $vectors\_passed$ & Number of vectors passing filter \\
 & $vectors\_failed$ & Number of vectors failing filter \\
 & $observed\_selectivity$ & Ratio of passed to checked vectors \\
 \hline

\end{tabular}
\end{adjustbox}
}
\caption{Input features of VADER for recall prediction.}
\vspace{-1.0cm}
\label{table:input-features}
\end{table}

\vspace{-0.2cm}
\subsubsection{Training FVS Query Generation}
\label{sec:vader:train-query-generation}

To train our filter-aware recall predictor, we need to generate training data with various combinations of filtering selectivities and value-vector correlations. 
The query vectors follow a distribution similar to that of the index vectors in our datasets. 
As we argue in our experimental evaluation in Section~\ref{sec:experiments:recall-prediction-analysis}, as few as 100 query vectors suffice for accurate results; predictor performance stabilizes at 500 query vectors.
To generate predicates, we use the filtered workload generation procedure introduced in~\cite{lu2026fvsbenchmark}.

This generation simulates the result of evaluating a filter predicate, represented as a bitmap, without requiring artificially generated structured data: given a vector dataset, a query vector, a target selectivity, and a correlation type, it sorts all dataset vectors by ascending distance to the query and samples from that sorted list using a method determined by the correlation type (e.g., under positive vector-predicate correlation~\cite{patel2024acorn} vectors closest to the query are more likely to satisfy the filter).
Since filter evaluation does not affect VADER's predictor, simulating it does not impact accuracy and makes training data generation more efficient.
We consider positive, negative, and no correlation, using the same 
definitions as prior work~\cite{lu2026fvsbenchmark, patel2024acorn}, and refer the reader to~\cite{lu2026fvsbenchmark} for the sampling details.
We combine our training queries with the generated predicates to construct FVS training queries, which are executed in the index to produce training data for our recall predictor.

A natural question is which predicate characteristics the predictor needs, and at what training cost.
In practice, a small number of representative selectivities suffices for the predictor to generalize 
to unseen selectivity values.
For correlations, a robust setup spans positive, negative, and no correlation predicates, but since negative correlations are relatively rare in practice VADER also supports faster training on a reduced set, with the predictor retaining strong performance and generalization; Section~\ref{sec:experiments:recall-prediction-analysis} details the accuracy and trade-offs of these choices.

\subsubsection{Extracting Training Data from FVS Queries}
\label{sec:vader:train-data-generation}
Once the FVS training queries are generated, VADER extracts training samples by executing these queries on the vector index.
For each training query, VADER periodically computes the feature values listed in Table~\ref{table:input-features}, and the corresponding actual recall at that point in the search.
Since these queries are derived from ground truth data, their exact recall values are known.
VADER collects training samples more frequently in higher recall ranges (e.g., $\geq 0.80$) and less frequently in lower 
ranges, as declarative recall primarily targets high-recall regimes.
Moreover, for queries that reach maximum recall early in the search, VADER terminates sample collection to reduce data generation overhead.
The VADER training data generation process is embarrassingly parallel and 
can be completed within a few minutes even in the worst case (cf. (Section~\ref{sec:experiments:recall-prediction-analysis}).

\subsection{VADER in Filter-Agnostic FVS Algorithms}
\label{sec:vader:integration-to-fvs}
As discussed in Section~\ref{sec:vader:recall-predictor-training}, the standard VADER features of Table~\ref{table:input-features} are compatible with a wide range of filter-agnostic vector search algorithms and indexes.
This design choice enables VADER to be employed on top of various algorithms for FVS.
Indeed, the VADER features can always be expanded based on the details of the underlying algorithm.
Below, we detail how VADER is implemented on top of three popular, widely adopted filter-agnostic vector search algorithms~\cite{lu2026fvsbenchmark}, namely Sweeping~\cite{weaviate_filtering}, ACORN~\cite{patel2024acorn}, and Filtered-ScaNN~\cite{filtered_scann_alloydb_scann_2025}, without loss of generality for other FVS algorithms.

\subsubsection{Sweeping}
Sweeping~\cite{weaviate_filtering} follows the standard $k$-NN graph search procedure, closely mimicking HNSW~\cite{malkov2018hnsw}, maintaining two priority queues: $C$, holding candidate nearest neighbors to be explored in subsequent steps, and $W$, the result set, which contains only vectors that satisfy the predicate.
The contents of $C$ can therefore differ significantly from those of $W$ depending on the selectivity and correlation of the predicate, so we incorporate additional feature statistics describing the distance distribution of vectors in $C$.
Recall from Table~\ref{table:input-features} that distance-based statistics from $W$ are already included in the intermediate nearest neighbor features.
For $C$, we include features such as $avgC$, the average distance of vectors in $C$; $minC$ and $maxC$, the minimum and maximum distances in $C$, $rangeC$, the range between $minC$ and $maxC$; and $firstNNC$, the distance of the first neighbor included in $C$.
We also include the average distance of vectors that pass the predicate and those that do not, denoted as $avgPassDist$ and $avgFailDist$, respectively.

\subsubsection{ACORN}
For ACORN~\cite{patel2024acorn}, we focus on the ACORN-1 variant, which does not modify the underlying $k$-NN graph index, without loss of generality.
The main design principle of ACORN is the two-hop expansion mechanism.
At each step of the graph search, ACORN explores not only the direct neighbors of the current node but also their neighbors (i.e., two-hop neighbors), and includes in the candidate set $C$ only those vectors that satisfy the predicate.
Distances are then computed for these vectors, and they are inserted into the $C$ and $W$ queues following a strategy similar to HNSW.
Since, in ACORN, both $C$ and $W$ contain only vectors that satisfy the predicate, we observe that $C$-related features similar to those used in Sweeping do not improve prediction quality.
Furthermore, ACORN computes distances only for vectors that satisfy the predicate, preventing the use of features such as $avgPassDist$ and $avgFailDist$.

\subsubsection{Filtered-ScaNN}
Although Filtered-ScaNN~\cite{guo2020scann, alloydb_filtering} belongs to a different index family (IVF), every feature of Table~\ref{table:input-features} is computed and used without change, with $nstep$ denoting the currently explored partition rather than a $k$-NN graph hop; we add only $centroidNN$, the distance of the query to the current cluster centroid.
Since Filtered-ScaNN computes the distances of an entire partition with a highly optimized SIMD routine, we perform recall prediction at partition granularity, leaving all other details unchanged, including the frequency of predictor calls provided by the heuristic of Section~\ref{sec:vader:early-term-recall-prediction}.
Achieving declarative recall on a fundamentally different index with this minimal modification showcases the versatility of VADER.

\section{Experiments}
\label{sec:experiments}

\noindent \textbf{Setup.}
We use a server with Intel(R) Xeon(R) CPU E5-2650 v4 @ 2.20GHz (24 cores, 48 hyperthreads) and 380GB of RAM, running Ubuntu 22.04.5 LTS.
All implementations are in C++, compiled with g++ version 11.4.0, using hnswlib~\cite{hnswlib} with 
SIMD\footnote{Single Instruction Multiple Data (SIMD): A parallel computing method where a single instruction operates simultaneously
on multiple data points.} 
enabled.

\noindent \textbf{Datasets.}
We consider 6 datasets that are widely used in the literature, with a diverse range of sizes, dimensionalities, and structural characteristics.
Their details are summarized in Table~\ref{table:datasets}.

\begin{table}[tb]
{\scriptsize
\centering
\begin{adjustbox}{max width=\columnwidth}
\begin{tabular}{|| c | c | c | c | c ||} 
 \hline
 Dataset & Dim & Vectors & Description & Dist. Meas. \\ 
 \hline\hline
 SIFT1M~\cite{jegou2011SIFT} & 128 & 1M & Image Descriptors & L2 \\ 
 GIST1M~\cite{jegou2010GIST} & 960 & 1M & Spatial Image Descriptors & L2 \\ 
 GLOVE1M~\cite{pennington2014glove,aumuller2020annbenchmarks} & 100 & 1.1M & Word Embeddings & IP \\ 
 DEEP10M~\cite{babenko2016deep} & 96 & 10M & Image Embeddings & L2 \\
 T2I10M~\cite{simhadri2022yantti} & 200 & 10M & Image and Text Embeddings (OOD) & IP \\ 
 CASELAW7M~\cite{annbench2025caselaw} & 1532 & 7.4M & Caselaw text and metadata & L2 \\ 
 \hline
 \hline
\end{tabular}
\end{adjustbox}
\caption{Datasets used in our evaluation.}
\vspace{-1.0cm}
\label{table:datasets}
}
\end{table}

\noindent \textbf{Query Workloads.}
Our testing query workloads consist of 100 queries randomly sampled from the official repositories of the datasets listed in Table~\ref{table:datasets}.
For training the recall predictor, we use 500 queries randomly sampled from the official learning vector workloads provided in the same repositories.
For tasks that require tuning (e.g., tuning competing approaches), we also use the same 500 queries.
To generate predicates for each vector, we construct workloads with varying selectivities (0.01--1.0) and correlations (positive, negative, and no correlation), following the procedure described in~\cite{lu2026fvsbenchmark}.
This setup extends beyond standard FVS benchmarks, combining datasets of diverse sizes, dimensionalities, distance measures, and structures with a wide range of filtering characteristics.
Every query is executed on a single thread, with 48 queries running in parallel, in both training and inference.
The exception is CASELAW7M, which ships with real structured metadata and its own FVS queries rather than generated ones; we therefore examine it separately in Section~\ref{sec:experiments:caselaw}.

\noindent \textbf{Indexing Parameters.}
For each dataset, we construct an HNSW vector index with $M=32$ and $efConstruction=200$ for both Sweeping and ACORN.
The index construction time using 48 threads is approximately 10 minutes for SIFT1M and GLOVE1M, 30 minutes for GIST1M, 20 minutes for DEEP10M, 40 minutes for T2I10M, and 4 hours for CASELAW7M.
The elevated indexing time for CASELAW7M is attributed to the higher dimensionality of the vectors it contains.
When VADER is used, both ACORN and Sweeping are configured with large $efSearch$ values (5K for ACORN and 1K for Sweeping).
These settings allow plain search (i.e., search without early termination) to achieve approximately 0.97 recall for Sweeping and approximately 0.95 recall for ACORN across all datasets and experimental configurations.
We follow the same approach to build a ScaNN index per dataset; since Filtered-ScaNN is a fundamentally different FVS algorithm, we give its details and results separately in Section~\ref{sec:experiments:filtered-scann}.

\noindent \textbf{Values of k, selectivity, correlation, and recall target.}
All experiments run with $k \in \{10, 100, 250\}$, selectivities $sel \in \{0.01, 0.1, ..., 0.9, 1.0\}$ (including values VADER is not trained on, and unfiltered search at $sel=1.0$), recall targets $R_t \in \{0.8, 0.85, 0.9, 0.95\}$, and positive, negative, and no correlation.
For $k$-NN graph experiments, we omit $R_t = 0.99$, which is unattainable in our workloads even with elevated indexing and search hyperparameter values.

\noindent \textbf{VADER Model.}
We train and deploy the VADER models using LightGBM~\cite{ke2017lightgbm} with its default values (100 estimators, 0.1 learning rate) across all experiments.
A predictor averages approximately 380kB, enabling efficient caching; it is initialized once, shared across all threads, and evaluated on a single thread to avoid interfering with parallel query execution, requiring on average 0.1 ms per prediction including feature computation.
It is invoked multiple times during each query, at the frequency given by the strategy of Section~\ref{sec:vader:early-term-recall-prediction}.

\noindent \textbf{Baselines.}
Since VADER is the first approach to provide declarative recall for FVS, there are no direct baselines.
We therefore compare against its predecessor DARTH/DARTH+~\cite{chatzakis2025darth, chatzakis2026darthplus}, which supports declarative recall for unfiltered search, and against the two tuning-based alternatives of Section~\ref{sec:introduction}, which we call REM (Recall-to-Effort Mapping) as they derive hyperparameter values that achieve the desired recall on average~\cite{yang2024vdtuner,daulton2020differentiable,chatzakis2025darth}.
REM-Filtered tunes separately for each selectivity (under no correlation); REM-Unfiltered tunes on unfiltered workloads and applies the same values across all selectivities.
Neither is a declarative recall approach, as both require explicit tuning per recall target and support only the targets they were tuned for, and REM-Filtered further assumes selectivity is known in advance, which does not generally hold in practice~\cite{leis2015howgoodoptimizersare}.
Both REM variants are significantly stronger baselines than iterative scan (Section~\ref{sec:relatedwork}), whose hard stop at the first $k$ predicate-satisfying vectors prevents the broader exploration that retrieving the true $k$ nearest neighbors usually requires.
REM-Filtered searches more effectively because it is explicitly tuned for each selectivity, and REM-Unfiltered explores more broadly because its $efSearch$ values are always $\geq k$ and in most cases $> k$.
All baselines require either training (DARTH) or extensive tuning (both REM variants), for which we use the same 500 training query vectors as VADER; Section~\ref{sec:experiments:main-results} gives the configuration details.

\noindent \textbf{Quality Measures.}
We evaluate result quality primarily by the recall each approach achieves on a query workload.
We additionally examine the recall distribution over individual queries, the deviation of queries from the recall target, and the Ratio of Queries below the recall Target (RQUT).
We compare speed by average query latency, and assess the recall prediction quality with the Mean Absolute Error (MAE) between predicted and true recall.

\subsection{Recall Prediction Analysis}
\label{sec:experiments:recall-prediction-analysis}

In this section, we examine the quality and characteristics of the VADER recall predictor models.
We focus on their prediction accuracy, required number of training queries, training time, feature importance, and loss function configurations.
In this set of experiments, all reported results are obtained by invoking the predictor at every distance computation during the search, in order to evaluate recall prediction accuracy in an unbiased manner.
Throughout this experimental setup, we study three main VADER predictor variants based on the data used for training.
We denote these variants as no-corr (trained on queries combined with predicates exhibiting no correlation), pos-corr (trained on queries combined with predicates exhibiting positive correlation), and all-corr (trained on queries combined with predicates exhibiting positive, negative, and no correlation).

\subsubsection{Recall Prediction Accuracy.}
\begin{figure*}
    \begin{adjustbox}{max width=0.4\textwidth}
        \includegraphics{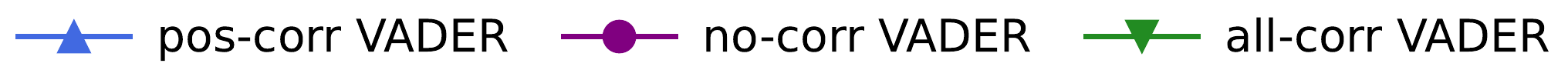}
    \end{adjustbox}
    
    \begin{minipage}[t]{0.99\textwidth}
        \centering
        \begin{subfigure}[t]{0.19\textwidth}
            \centering
            \includegraphics[width=\textwidth]{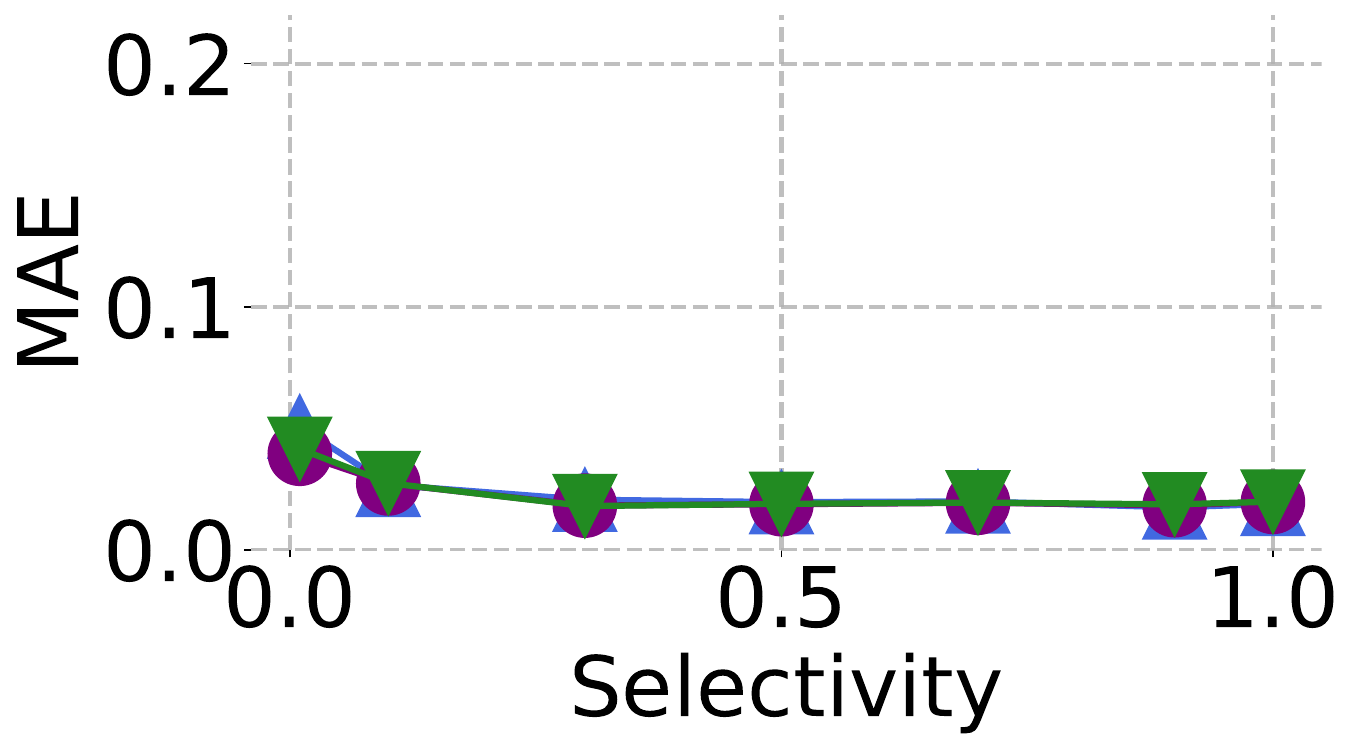}
            \caption{SIFT1M}
        \end{subfigure}
        \hfill
        \begin{subfigure}[t]{0.19\textwidth}
            \centering
            \includegraphics[width=\textwidth]{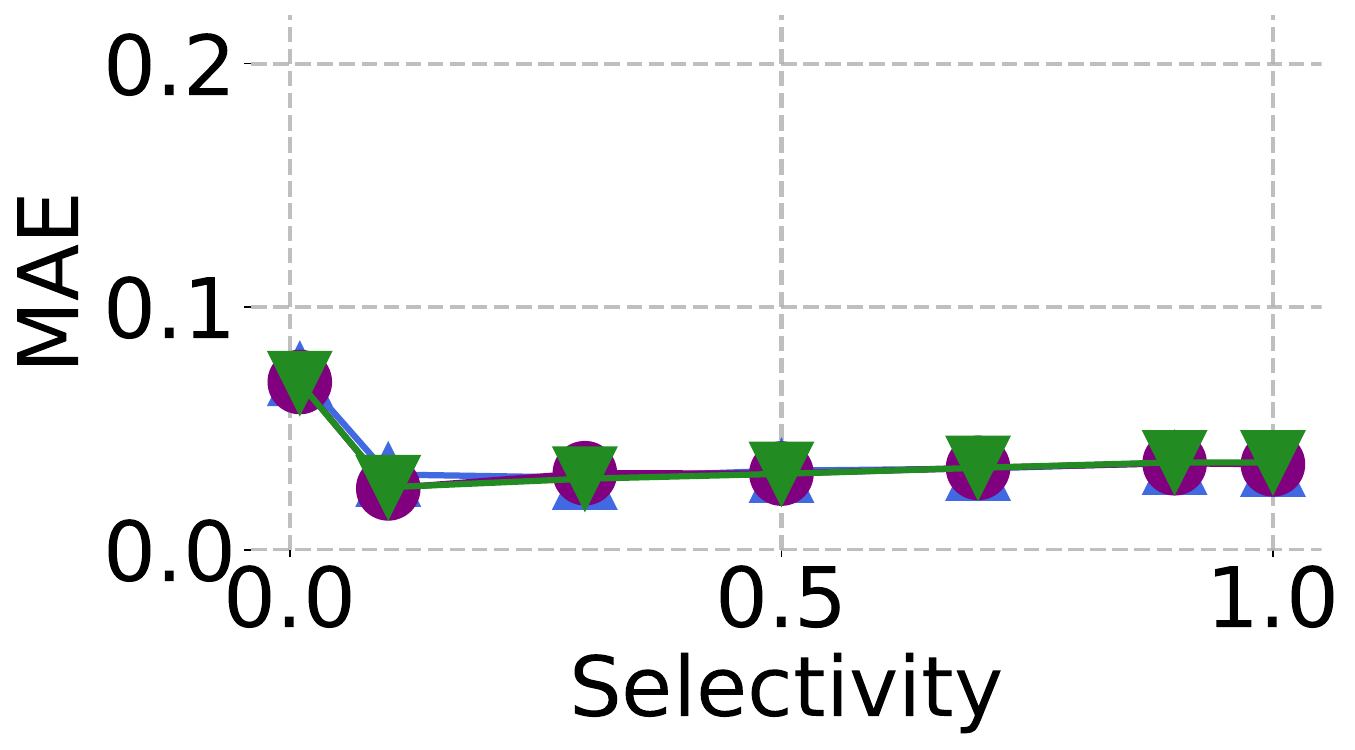}
            \caption{GIST1M}
        \end{subfigure}
        \begin{subfigure}[t]{0.19\textwidth}
            \centering
            \includegraphics[width=\textwidth]{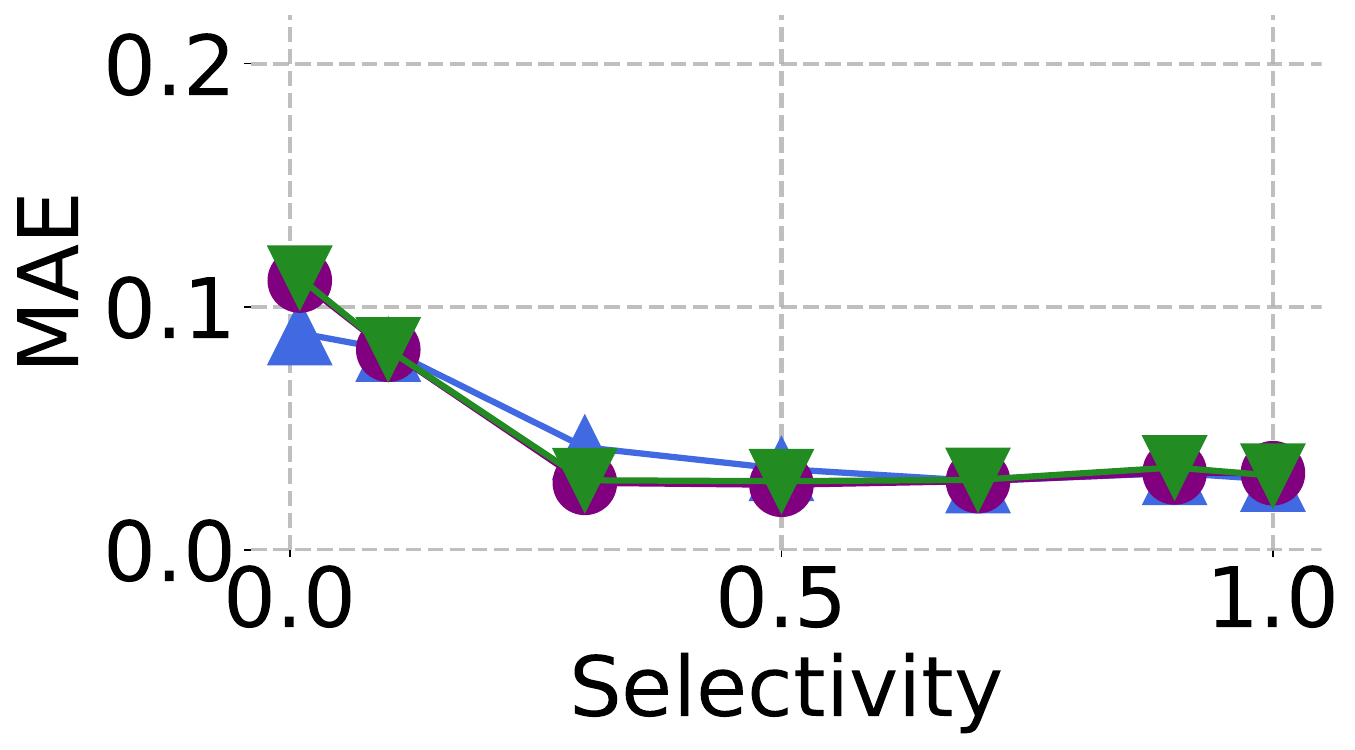}
            \caption{GLOVE1M}
        \end{subfigure}
        \begin{subfigure}[t]{0.19\textwidth}
            \centering
            \includegraphics[width=\textwidth]{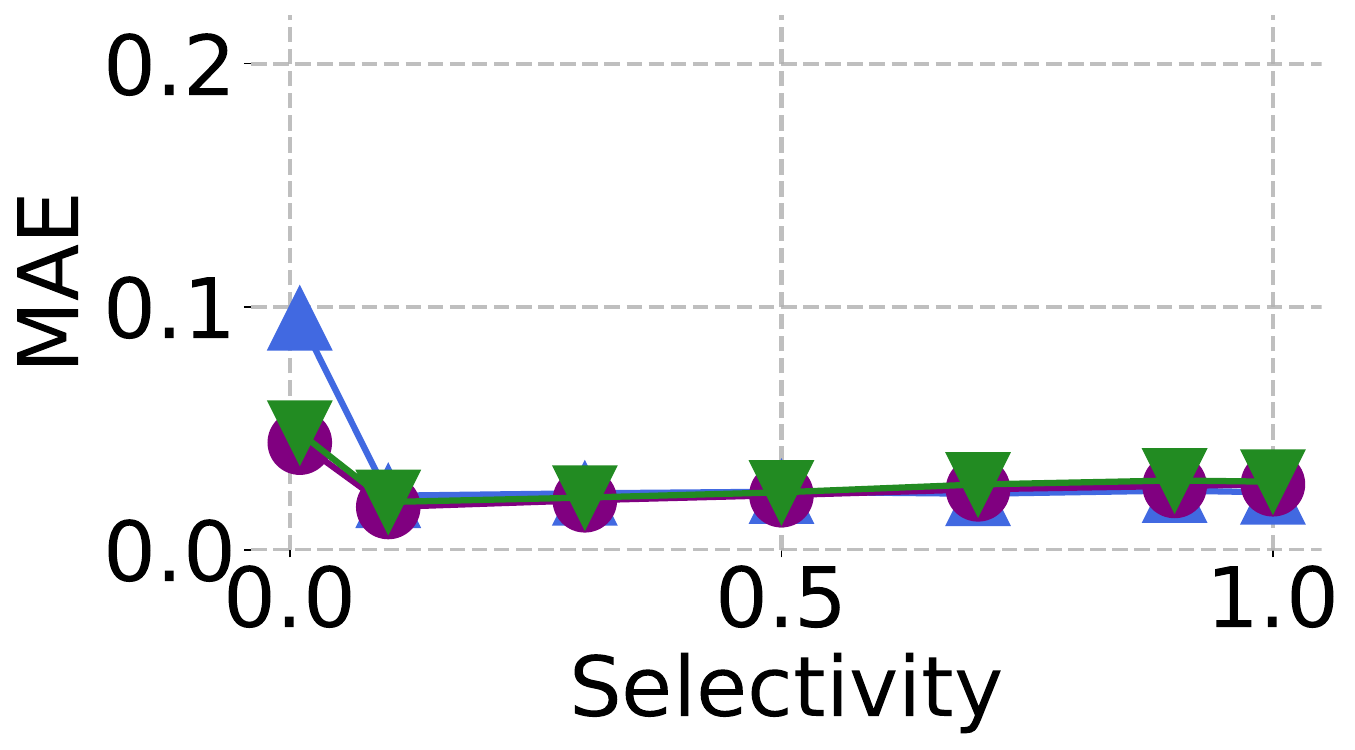}
            \caption{DEEP10M}
        \end{subfigure}
        \begin{subfigure}[t]{0.19\textwidth}
            \centering
            \includegraphics[width=\textwidth]{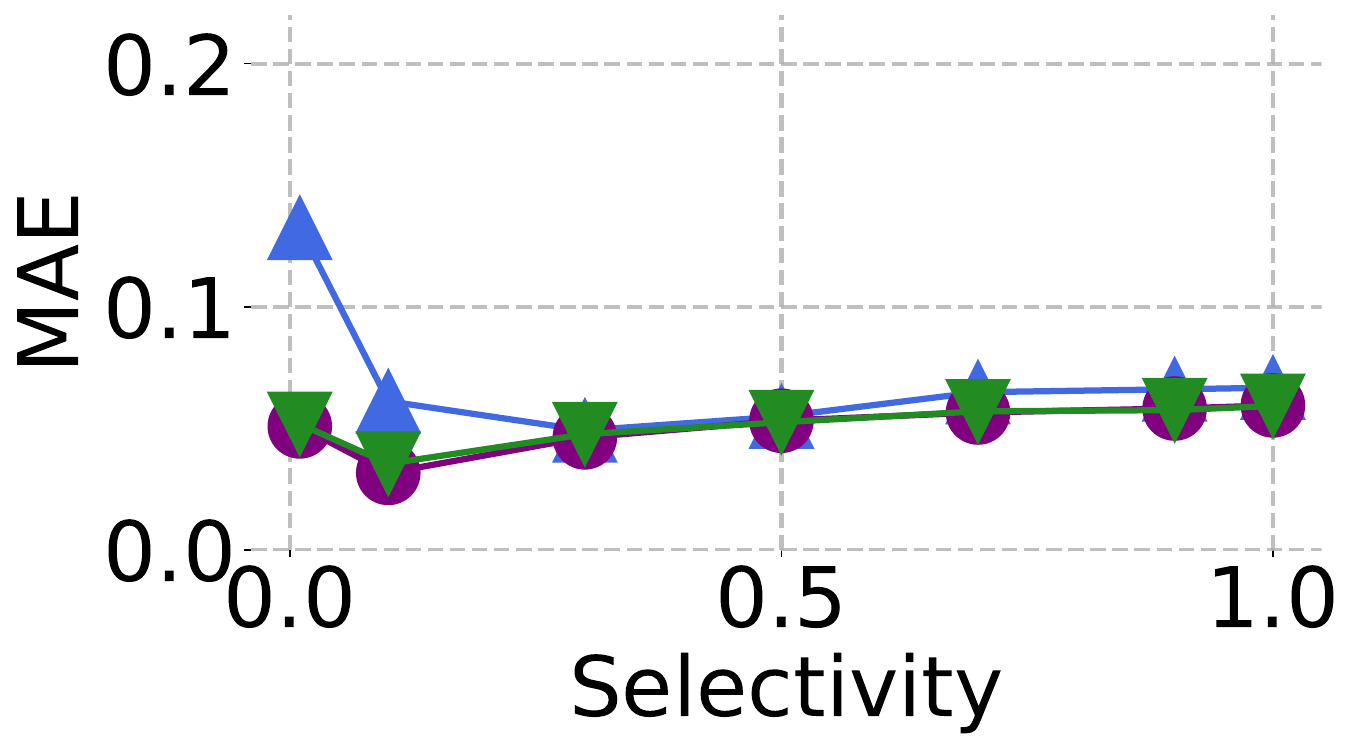}
            \caption{T2I10M}
        \end{subfigure}
        \hfill
    \vspace{-0.3cm}
    \caption{MAE of recall prediction (Sweeping, $k=100$, No Correlation).}
    \label{fig:mse-k100-nocorr}
    \end{minipage}
\end{figure*}

We begin by presenting recall prediction accuracy using the Mean Absolute Error (MAE) between the predicted recall by VADER and the ground-truth recall at each point during the search.
Figure~\ref{fig:mse-k100-nocorr} plots MAE against selectivity for no correlation predicates using Sweeping with $k=100$, with one line per VADER predictor: trained on all correlations (all-corr), only positive correlation (pos-corr), or no correlation (no-corr), across all of our datasets.
Figures for other correlations exhibit very similar trends and are omitted, while the numbers reported below are computed across all configurations.

Across all settings, VADER achieves high prediction accuracy.
The all-corr model consistently delivers the best performance on average, achieving $MAE=0.046$, while the no-corr and pos-corr models achieve $MAE=0.048$ and $MAE=0.054$, respectively.
In particular, the all-corr model maintains strong performance under negative correlation, achieving $MAE=0.070$, compared to $MAE=0.074$ for no-corr and $MAE=0.122$ for pos-corr.
This behavior is expected, as negative correlation cases differ significantly from positive and no correlation cases, and achieving strong performance in such cases requires additional training data.
This performance generalizes across different values of $k$ (specifically $k \in \{10, 100, 250\}$), where the all-corr model achieves an average $MAE=0.052$, compared to $MAE=0.055$ for no-corr and $MAE=0.072$ for pos-corr.
Note that the reported values correspond to averages over the entire search process, including lower recall ranges, to provide an unbiased evaluation of model performance.

When focusing on higher recall ranges (e.g., $\geq 0.80$), the MAE values of all models are further reduced.
For ACORN, we observe similar trends and therefore omit the corresponding figures due to space constraints.
Across all configurations, the all-corr model again achieves the best performance, with $MAE=0.054$, while the no-corr and pos-corr models achieve $MAE=0.086$ and $MAE=0.080$, respectively.

\begin{figure}
    \centering
    \begin{minipage}[t]{\columnwidth}
        \centering
        \begin{subfigure}[t]{0.99\textwidth}
            \centering
            \includegraphics[width=\textwidth]{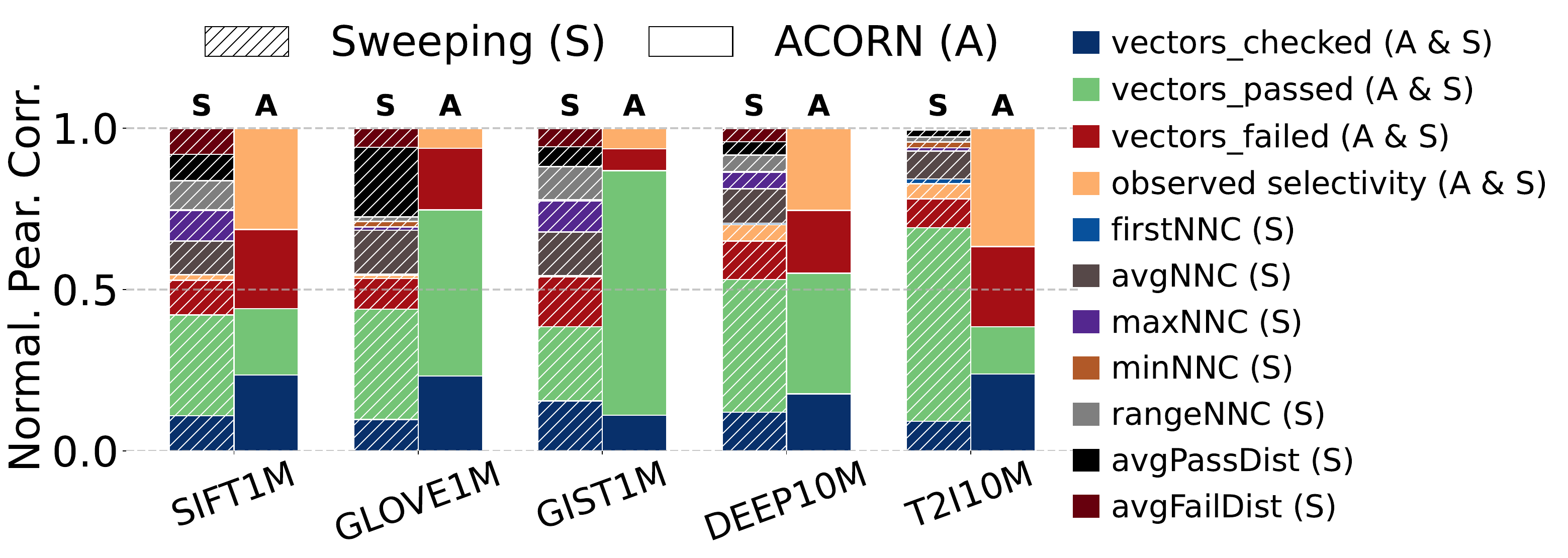}
        \end{subfigure}
    \vspace{-0.4cm}
    \caption{Filtering Feature Correlation to Recall.}
    \label{fig:correlation-to-target}
    \end{minipage}
\end{figure}

\begin{figure*}
    \centering
    \begin{minipage}[t]{0.99\textwidth}
        \centering
        \begin{subfigure}[t]{0.19\textwidth}
            \centering
            \includegraphics[width=\textwidth]{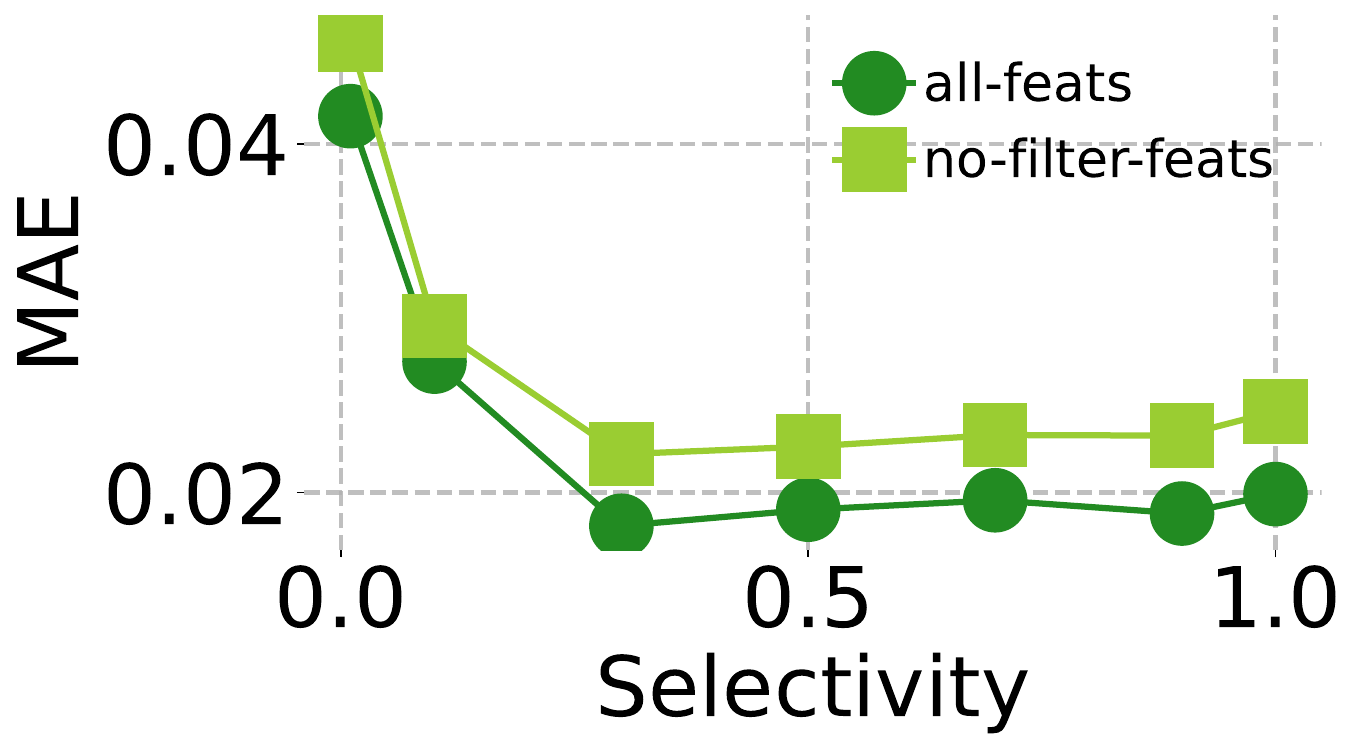}
            \caption{SIFT1M}
        \end{subfigure}
        \hfill
        \begin{subfigure}[t]{0.19\textwidth}
            \centering
            \includegraphics[width=\textwidth]{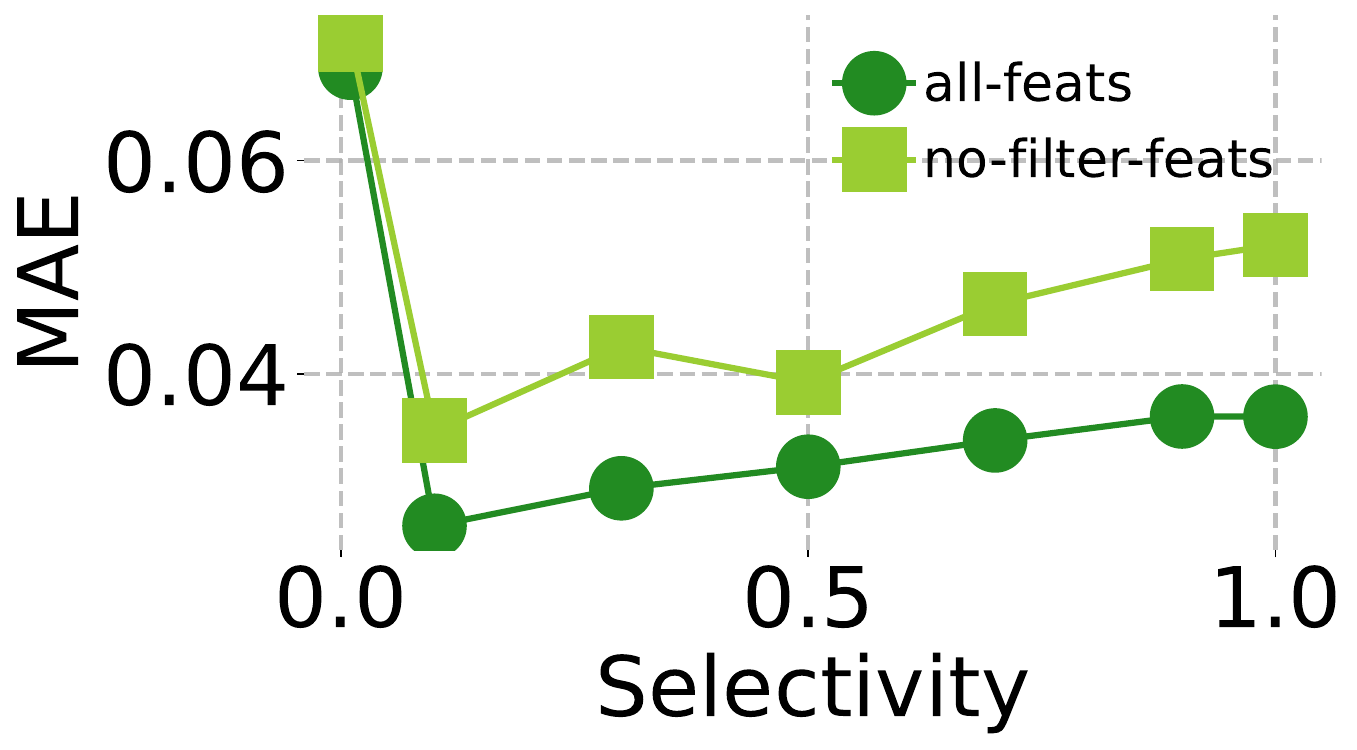}
            \caption{GIST1M}
        \end{subfigure}
        \hfill
        \begin{subfigure}[t]{0.19\textwidth}
            \centering
            \includegraphics[width=\textwidth]{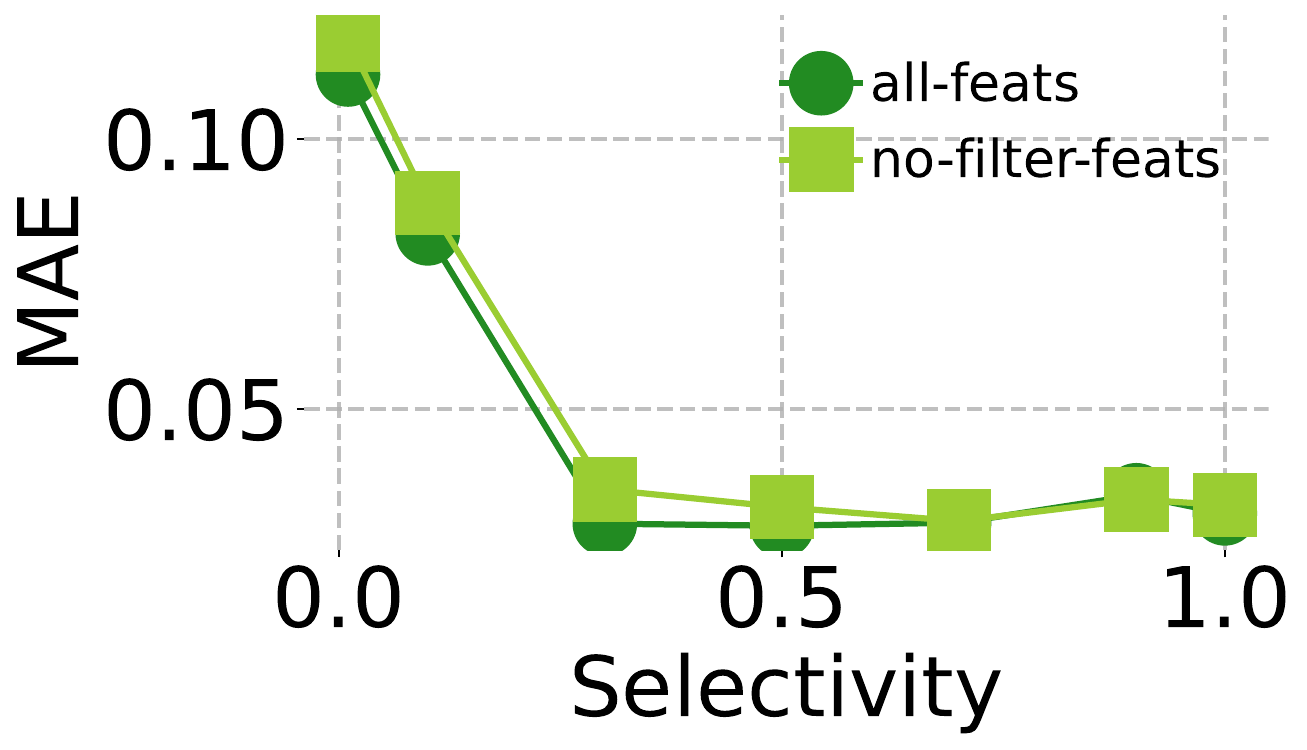}
            \caption{GLOVE1M}
        \end{subfigure}
        \hfill
        \begin{subfigure}[t]{0.19\textwidth}
            \centering
            \includegraphics[width=\textwidth]{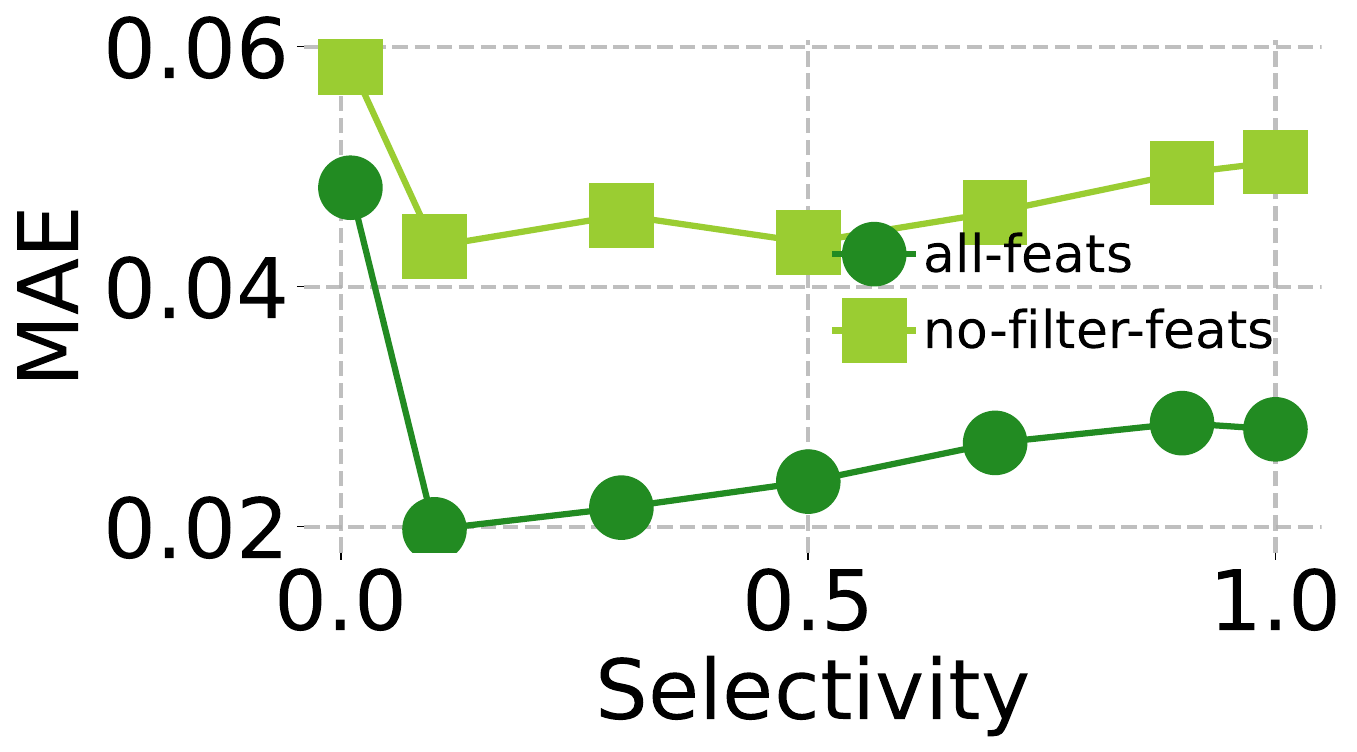}
            \caption{DEEP10M}
        \end{subfigure}
        \hfill
        \begin{subfigure}[t]{0.19\textwidth}
            \centering
            \includegraphics[width=\textwidth]{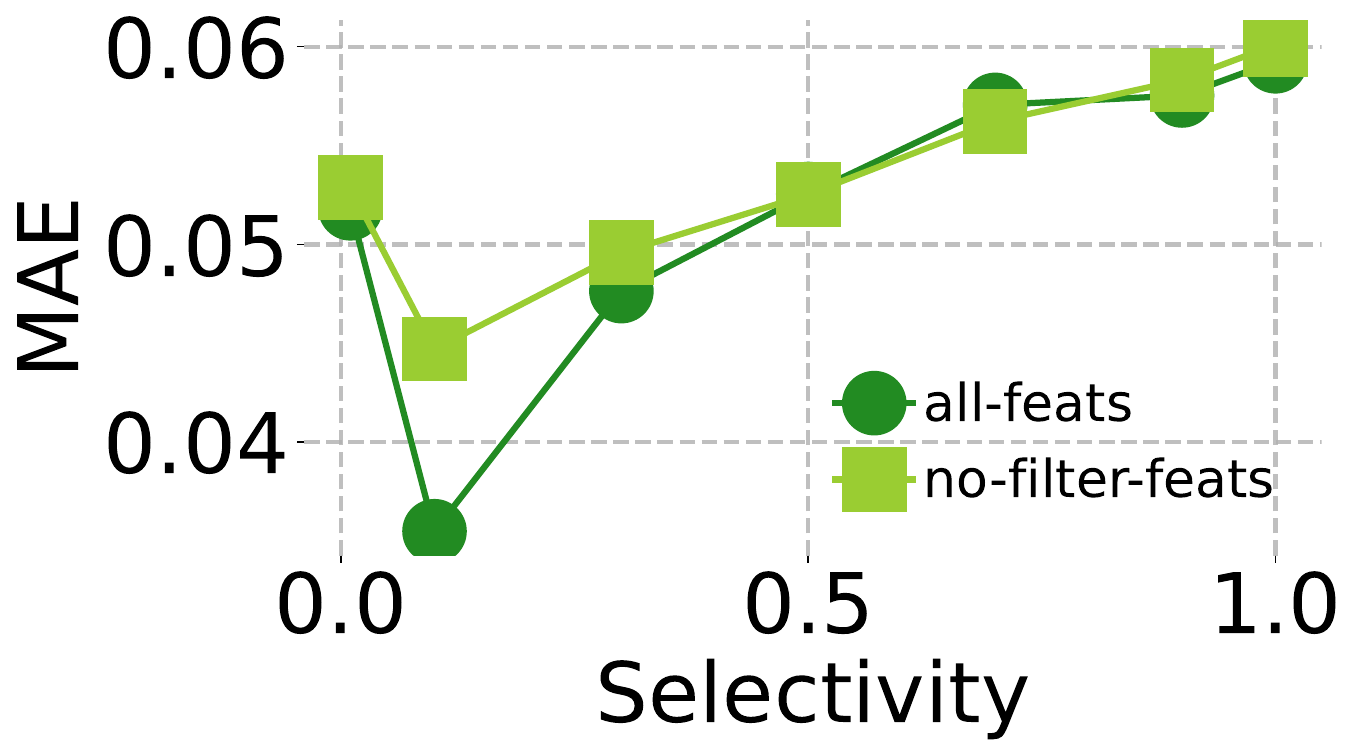}
            \caption{T2I10M}
        \end{subfigure}
    \vspace{-0.3cm}
    \caption{Ablation study for VADER model variants across all datasets (Sweeping, $k=100$, No Correlation).}
    \label{fig:vader-model-variant-ablation-k100-nocorr}
    \end{minipage}
    
    \begin{minipage}[t]{0.99\textwidth}
        \centering
        \begin{adjustbox}{max width=0.4\textwidth}
            \includegraphics{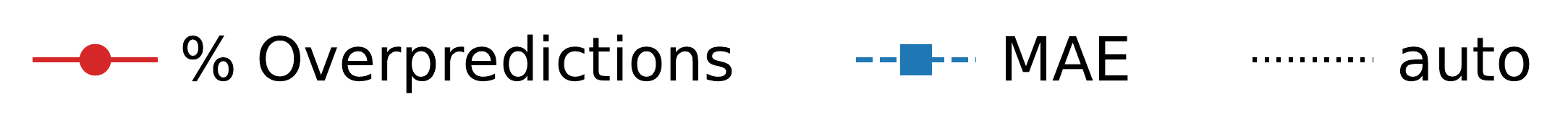}
        \end{adjustbox}

        \begin{subfigure}[t]{0.19\textwidth}
            \centering
            \includegraphics[width=\textwidth]{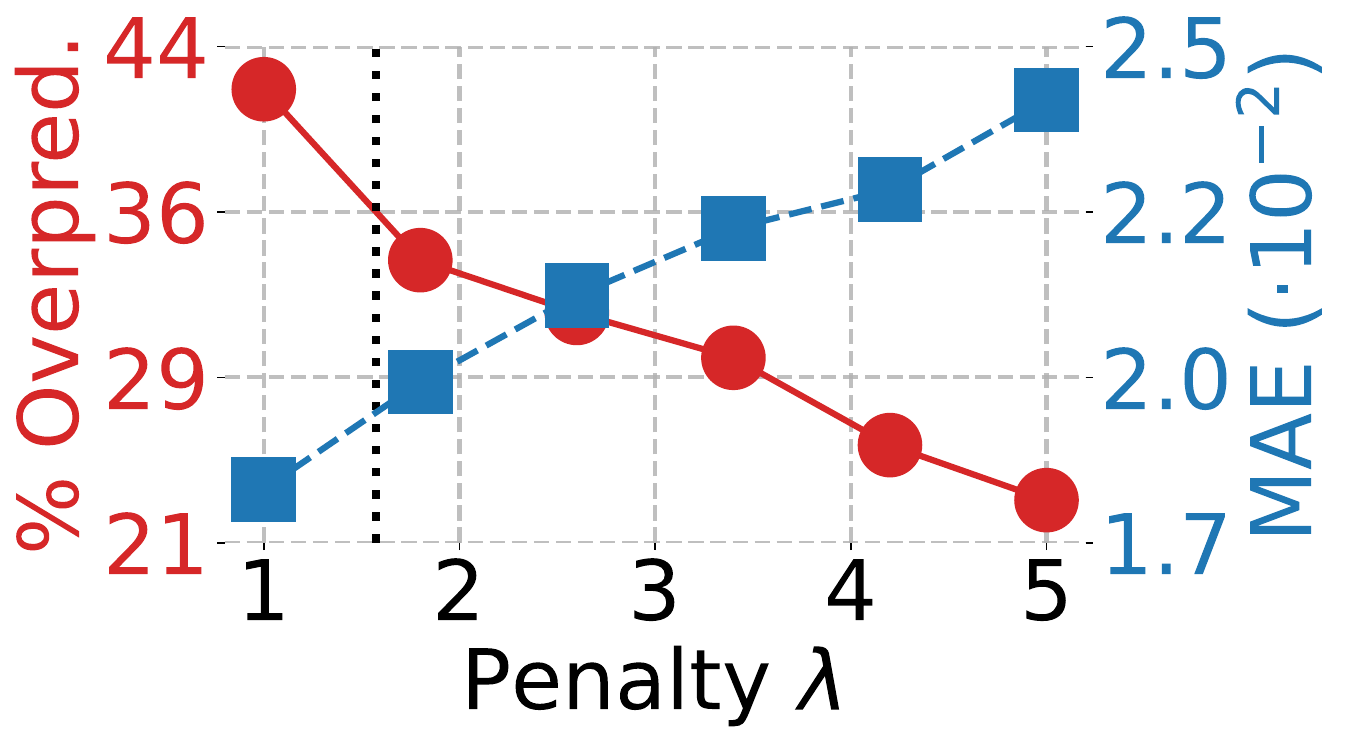}
            \caption{SIFT1M}
        \end{subfigure}
        \hfill
        \begin{subfigure}[t]{0.19\textwidth}
            \centering
            \includegraphics[width=\textwidth]{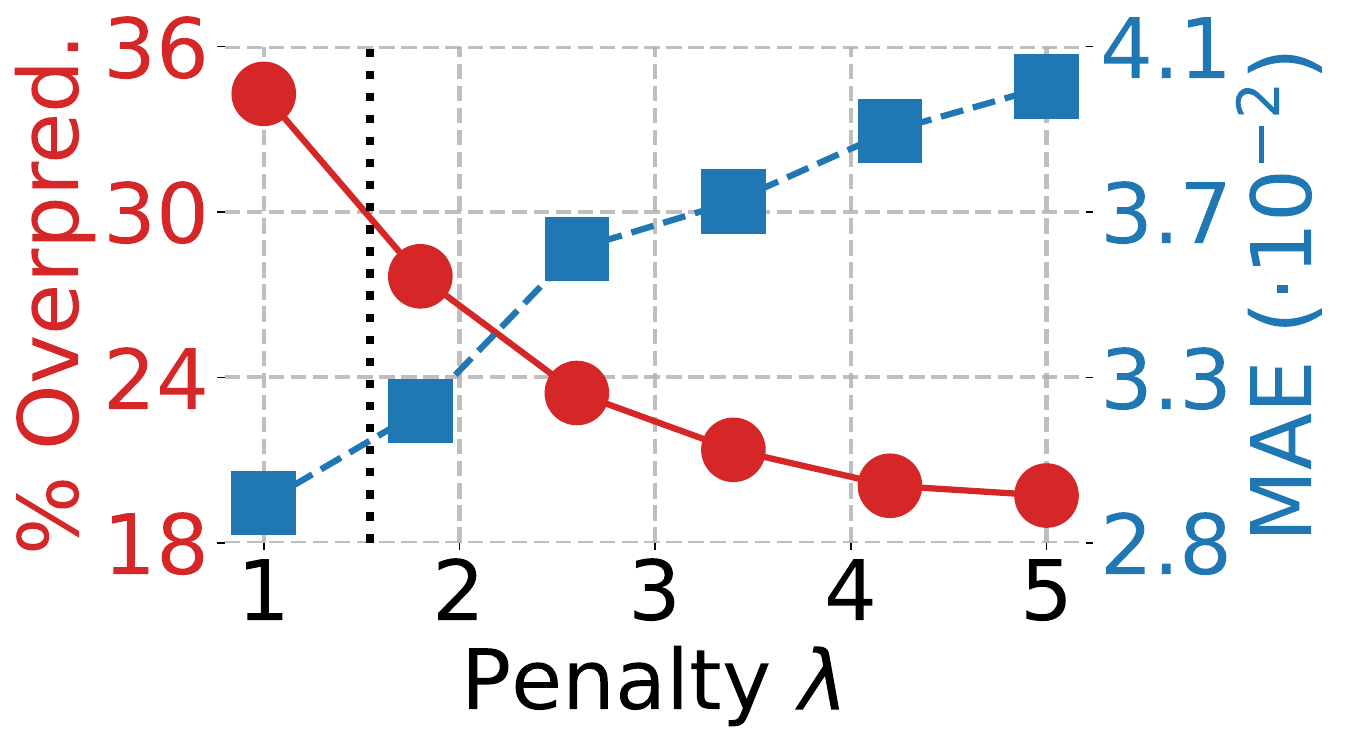}
            \caption{GIST1M}
        \end{subfigure}
        \hfill
        \begin{subfigure}[t]{0.19\textwidth}
            \centering
            \includegraphics[width=\textwidth]{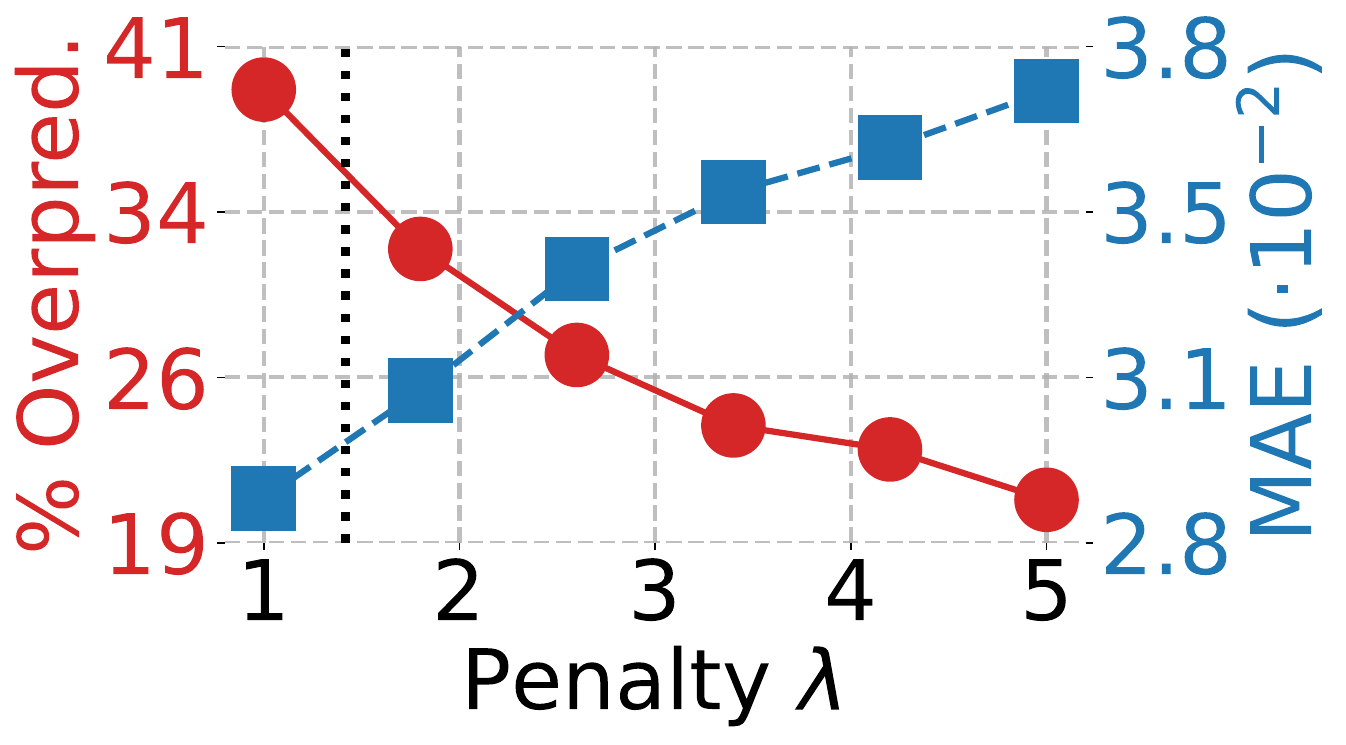}
            \caption{GLOVE1M}
        \end{subfigure}
        \hfill
        \begin{subfigure}[t]{0.19\textwidth}
            \centering
            \includegraphics[width=\textwidth]{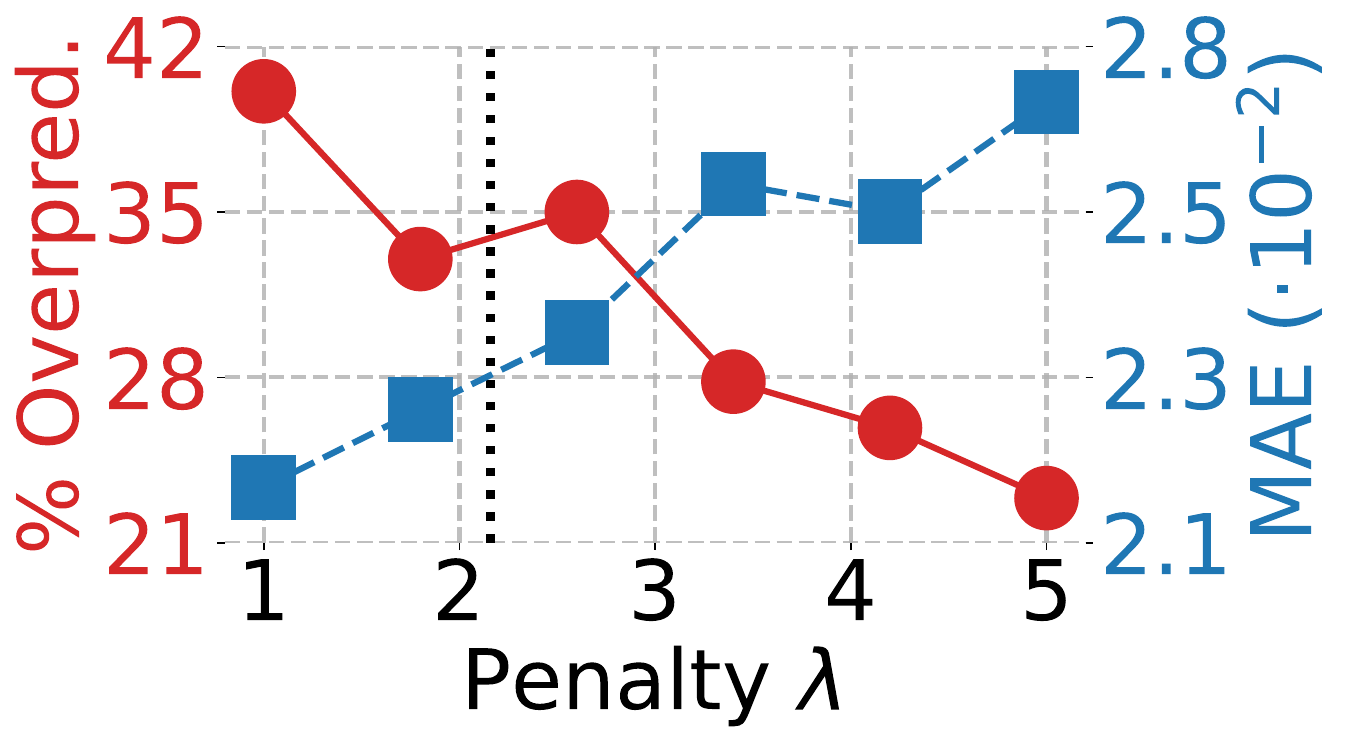}
            \caption{DEEP10M}
        \end{subfigure}
        \hfill
        \begin{subfigure}[t]{0.19\textwidth}
            \centering
            \includegraphics[width=\textwidth]{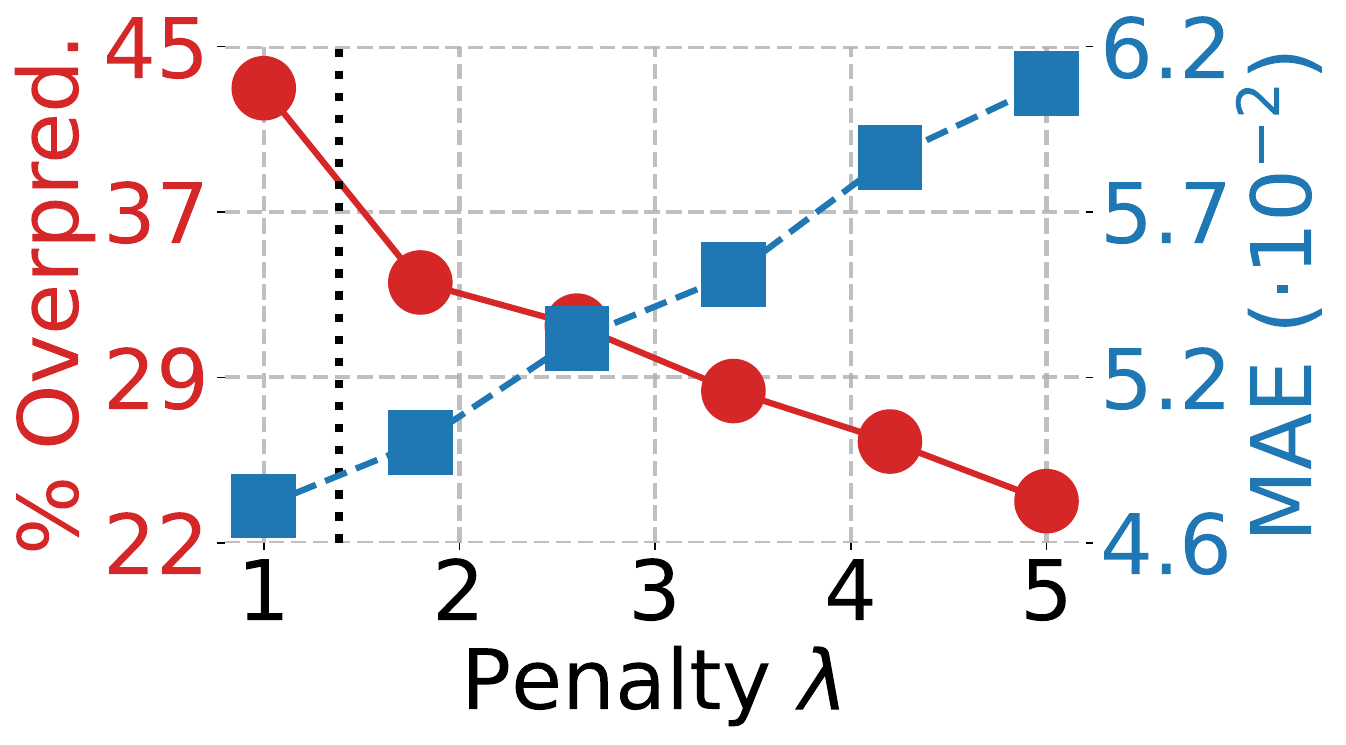}
            \caption{T2I10M}
        \end{subfigure}
    \vspace{-0.3cm}
    \caption{Asymmetric penalty sensitivity study (Sweeping, $k=100$, No Correlation, $sel=0.3$).}
    \label{fig:penalty-ablation-k100-nocorr}
    \end{minipage}
\end{figure*}

\subsubsection{Training Queries.}
We evaluate the recall predictor with 100, 500, and 1000 training query vectors, each combined with predicates of various selectivities and correlations (for the all-corr variant, $sel\in\{0.01,0.1,0.3,0.5,0.7,0.9,1.0\}$ with positive, negative, and no correlation).
Performance stabilizes at approximately 500 queries for the no-corr and all-corr variants, while pos-corr peaks at 1K.
The average MAE across all datasets, selectivities, and correlations for Sweeping is summarized in Table~\ref{tab:training_queries}.
For all-corr, 100/500/1000 queries yield $MAE=0.053/0.052/0.051$, and the corresponding ACORN values are $0.057/0.054/0.054$, with the other variants following very similar trends.
Accuracy is therefore high even with as few as 100 training queries, and we use 500 as the default for the remainder of the evaluation.

\begin{table}[tb]
{\scriptsize
\centering
\begin{adjustbox}{max width=\columnwidth}
\begin{tabular}{|| c | c | c | c ||} 
 \hline
 Training Query Vectors & pos-corr & no-corr & all-corr \\ 
 \hline\hline
 100  & 0.083 & 0.057 & 0.053 \\ 
 500  & 0.072 & 0.055 & 0.052 \\ 
 1000 & 0.064 & 0.054 & 0.051 \\ 
 \hline
\end{tabular}
\end{adjustbox}
\caption{Mean Absolute Error of VADER models across a varying number of training queries.}
\vspace{-0.8cm}
\label{tab:training_queries}
}
\end{table}

\subsubsection{Training Sample Sizes.}
Although we use 500 queries as the default number of training query vectors, the number of training samples extracted from these queries varies across VADER model variants.
For example, the all-corr model is trained on more samples than the no-corr and pos-corr variants for two main reasons.
First, the same set of 500 queries is combined with predicates spanning positive, negative, and no correlation configurations, resulting in a larger training set.
Second, as correlation shifts from positive to negative, filtered vector search becomes more challenging, requiring more extensive search to reach high recall levels, so queries under negative correlation generate more training samples, further increasing the total number of samples used by the all-corr model.
This effect is evident in the number of training samples used for GBDT training across the three VADER variants: for Sweeping with $k=100$ (without loss of generality, with similar trends observed for ACORN), the all-corr model uses approximately 40M samples for SIFT1M, 16M for GIST1M, 50M for GLOVE1M, 17M for DEEP10M, and 20M for T2I10M.
In comparison, the no-corr model uses 10M samples for SIFT1M, 5M for GIST1M, 15M for GLOVE1M, and 3M for T2I10M.
Finally, the pos-corr model, which exhibits the lowest recall prediction accuracy, uses 3M samples for SIFT1M, 3M for GIST1M, 5M for GLOVE1M, and 1M for DEEP10M and T2I10M.

Note that the variation in sample size across datasets is not solely determined by dataset size, as the intrinsic complexity of the vector space plays a more significant role.
For example, GLOVE1M consistently requires the largest number of training samples across all configurations, as it is a word embedding dataset with a highly clustered structure~\cite{pennington2014glove, cha2017embeddingclustering, shi2017weLDAembeddingclustering}; under inner-product similarity, it becomes more challenging to distinguish true nearest neighbors at high recall levels, prolonging the search process, particularly under negative correlation, and generating a larger number of training samples.

\subsubsection{Training Times.}
Training a recall predictor consists of two stages: (i) generating training data from FVS training queries, and (ii) training a GBDT model using this data.
Both complete within a few minutes, including for the all-corr variant on the largest datasets.
For Sweeping, all-corr requires 5 minutes for data generation and 0.85 minutes for GBDT training, under 6 minutes in total, while no-corr requires 1 minute.
For ACORN the corresponding figures are 8 and 0.46 minutes for all-corr, under 9 minutes in total, and approximately 3 minutes for no-corr.

\subsubsection{Feature Selection.}
\label{sec:experiments-feature-selection}
To verify the contribution of our novel filtering features independently from the trained model, we compute the normalized Pearson correlation~\cite{benesty2009pearsoncorr} of each feature to the true recall on our training data, presented in Figure~\ref{fig:correlation-to-target} for Sweeping and ACORN.
We focus exclusively on these features, as the non-filtering ones are established in prior work~\cite{li2020laet, chatzakis2025darth, chatzakis2026darthplus}; recall that Sweeping contributes additional features compared to ACORN (Section~\ref{sec:vader:integration-to-fvs}).
Our analysis shows that all filtering-related features exhibit a non-negligible correlation with recall.
In particular, $vectors\_checked$, $vectors\_passed$, and $vectors\_failed$ demonstrate significant correlation in both algorithms.
However, $observed\_selectivity$ exhibits higher correlation in ACORN and lower correlation in Sweeping.
This difference can be attributed to the distinct traversal strategies of the two algorithms.
In ACORN, search expansion is strictly based on the subset of neighbors that satisfy the predicate, making observed selectivity a more influential factor.
In contrast, Sweeping traverses the graph using unfiltered neighbors, reducing the relative importance of observed selectivity.

We also evaluate the contribution of VADER's filter-aware features through an ablation study.
We train two variants of the VADER recall predictor in a controlled setting where all configurations, including the model architecture and the training data, are identical, differing only in the features provided to the predictor: all\_features uses the complete feature set presented in Table~1 and corresponds to the VADER model used throughout the paper, while no\_filter\_features excludes the filter-aware features, retaining all remaining ones.
The results, presented in Figure~\ref{fig:vader-model-variant-ablation-k100-nocorr}, demonstrate the importance of the proposed filter-aware features: despite introducing only a small number of additional features (discussed in Sections~\ref{sec:vader:recall-prediction-features} and ~\ref{sec:vader:integration-to-fvs}), the complete VADER model achieves 12\% lower MAE on average for Sweeping, and 16\% lower MAE for ACORN, across all values of $k$ and correlation settings.

Finally, we examine the importance the GBDT model assigns to each feature category after VADER training~\cite{ke2017lightgbm}, computed as the sum of the gain-based importance scores of the individual features across all configurations, for both Sweeping and ACORN.
For Sweeping, the importance is distributed as 21\% for search progression features, 30\% for intermediate nearest neighbor statistics, 17\% for query-dimensional statistics, and 31\% for filtering-related features. 
For ACORN, the corresponding values are 11\%, 46\%, 25\%, and 17\%, respectively.
These results highlight an important aspect of the recall prediction task: no single feature category dominates, indicating that all selected feature categories contribute meaningfully to prediction performance. 
This also means that accurate recall prediction requires combining diverse signals, namely search progression dynamics, query characteristics, intermediate results, and filtering behavior.

\subsubsection{Asymmetric Loss Sensitivity Study.}
We finally examine the effect of the asymmetric loss function using the all-corr model variant, reporting results averaged across different correlations, selectivities, and values of $k$ for Sweeping (results for ACORN follow very similar trends).

We perform a sensitivity study over the $\lambda$ penalty value (cf. Section~\ref{sec:vader:recall-predictor-architecture}) to study its effect on VADER's recall prediction quality and the percentage of recall overpredictions among all model predictions.
We present the results in Figure~\ref{fig:penalty-ablation-k100-nocorr} for all of our datasets, using no correlation, $k=100$, $sel=0.3$, and Sweeping (results are similar for other configurations and are omitted for brevity).
As expected, increasing the penalty value results in a much smaller number of recall overpredictions, since the recall predictor becomes more conservative, but the recall prediction error (expressed by MAE) increases as well, revealing a tradeoff between reducing the number of overpredictions and maintaining recall prediction quality in general.
This tradeoff is also reflected in the end-to-end performance, with higher penalty values resulting in fewer queries falling below the target, at the cost of increased latency.

Striking a balance for the asymmetric loss is a task-specific objective, and the penalty can be adjusted in VADER based on the application requirements. 
In our implementation, however, we use an automatic method to set it that requires no tuning, as discussed in Section~\ref{sec:vader:recall-predictor-architecture}.
The penalty values obtained using this automatic method, depicted in Figure~\ref{fig:penalty-ablation-k100-nocorr} as black dotted vertical lines, reduce the percentage of recall overpredictions to around 33\% across all configurations, compared with 40\% achieved by the symmetric loss function, without noticeable degradation in prediction quality ($\mathrm{MAE}=0.053$, compared with $\mathrm{MAE}=0.052$ for the symmetric loss).
Therefore, we use this automatically configured asymmetric loss function throughout our evaluation.

\subsection{End-to-end Performance Results}
\label{sec:experiments:main-results}

We now focus on the end-to-end performance of VADER.
Unless stated otherwise, we use the all-corr variation with the asymmetric loss function as our VADER recall predictor.

\subsubsection{Summary of Achieved Recalls and Speedups over Plain Search}
\begin{figure*}
    \begin{adjustbox}{max width=0.4\textwidth}
        \includegraphics{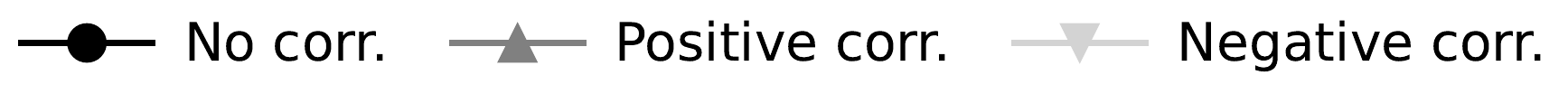}
    \end{adjustbox}
    
    \begin{minipage}[t]{0.99\textwidth}
        \centering
        
        \begin{subfigure}[t]{0.19\textwidth}
            \centering
            \includegraphics[width=\textwidth]{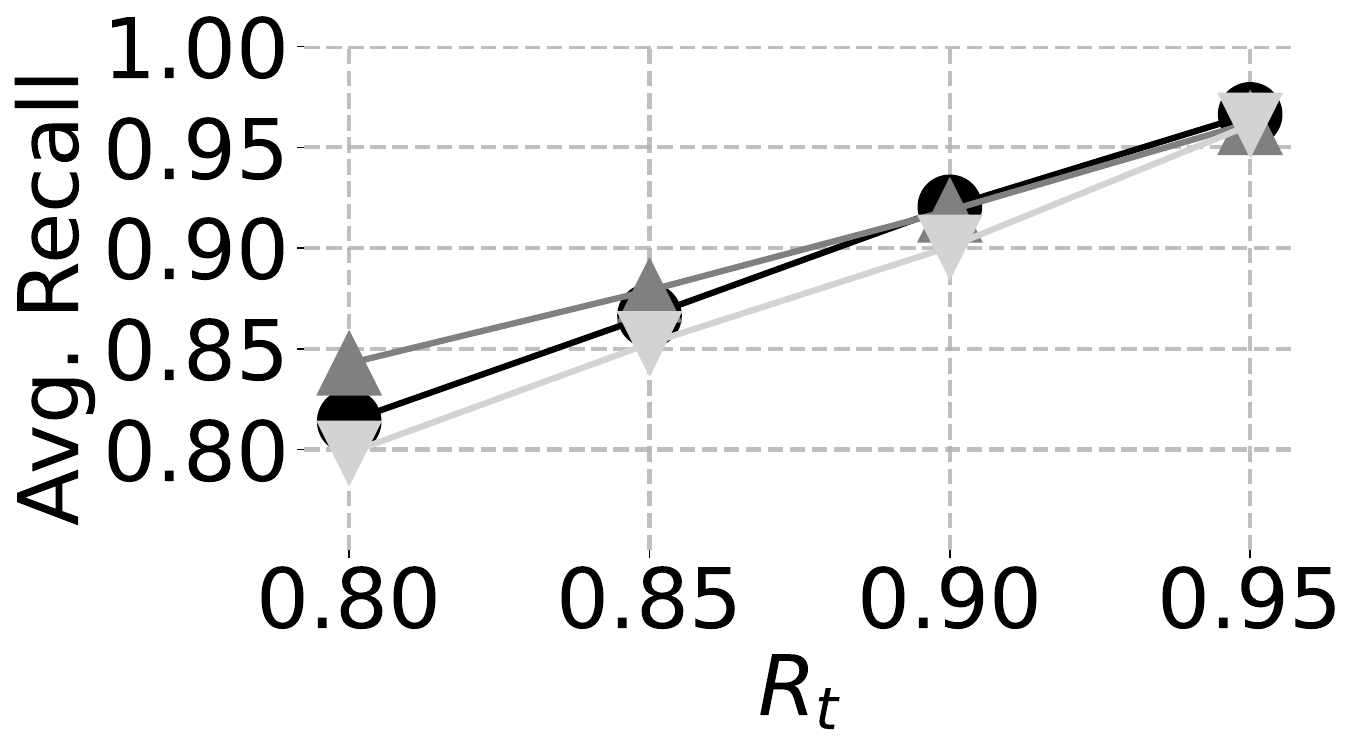}
            \caption{SIFT1M}
        \end{subfigure}
        \hfill
        \begin{subfigure}[t]{0.19\textwidth}
            \centering
            \includegraphics[width=\textwidth]{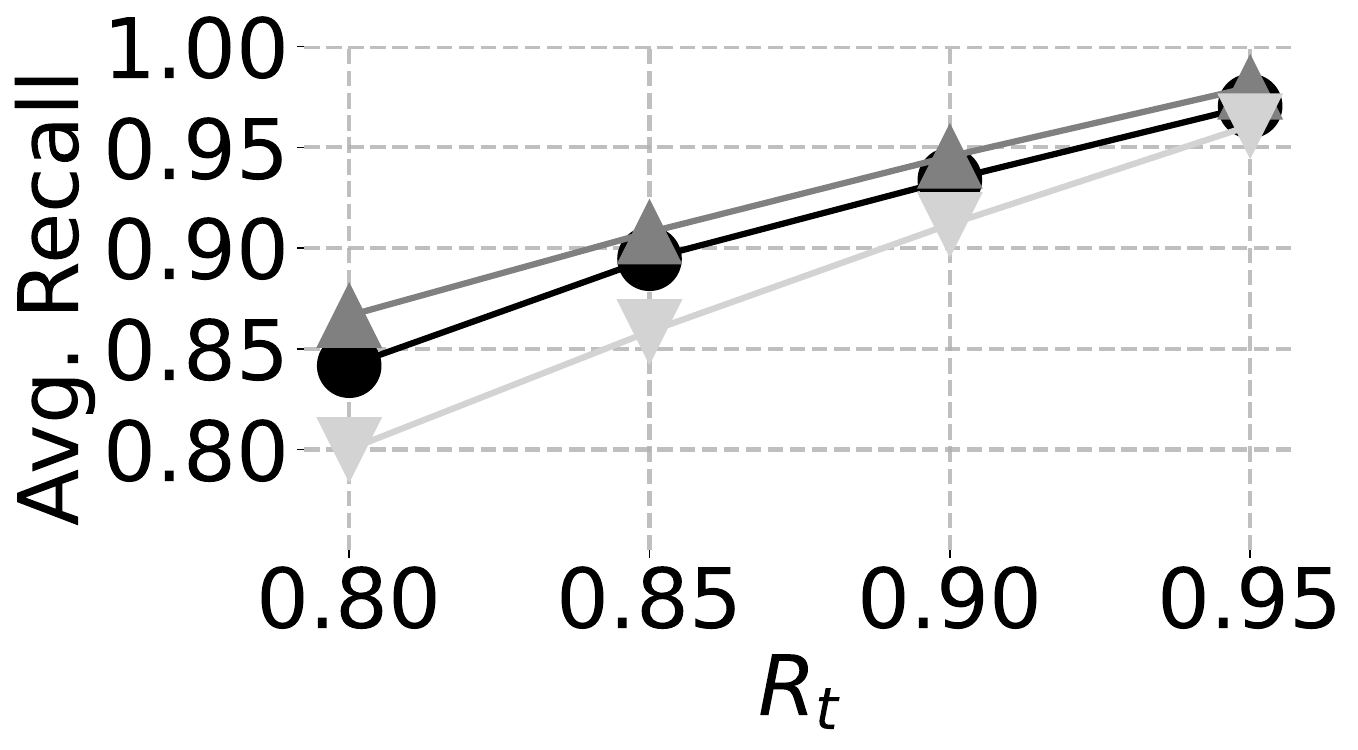}
            \caption{GIST1M}
        \end{subfigure}
        \hfill
        \begin{subfigure}[t]{0.19\textwidth}
            \centering
            \includegraphics[width=\textwidth]{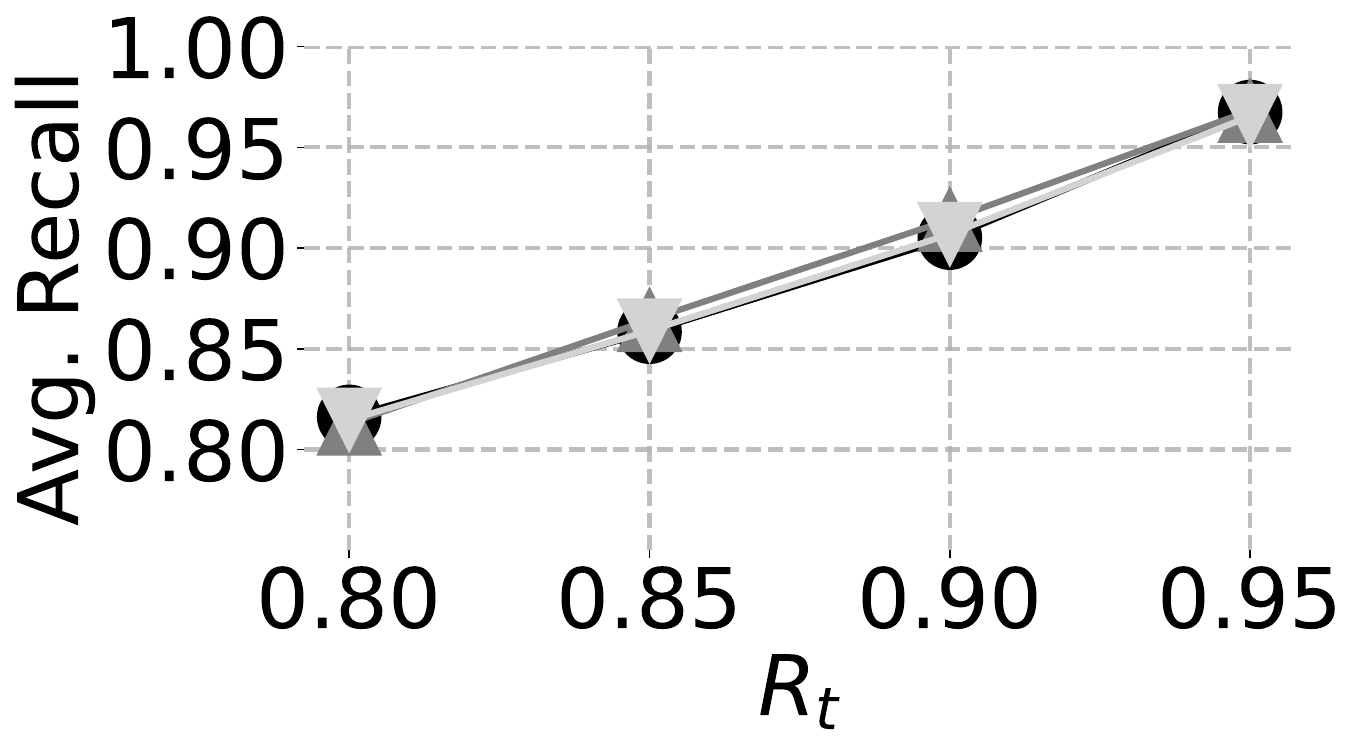}
            \caption{GLOVE1M}
        \end{subfigure}
        \hfill
        \begin{subfigure}[t]{0.19\textwidth}
            \centering
            \includegraphics[width=\textwidth]{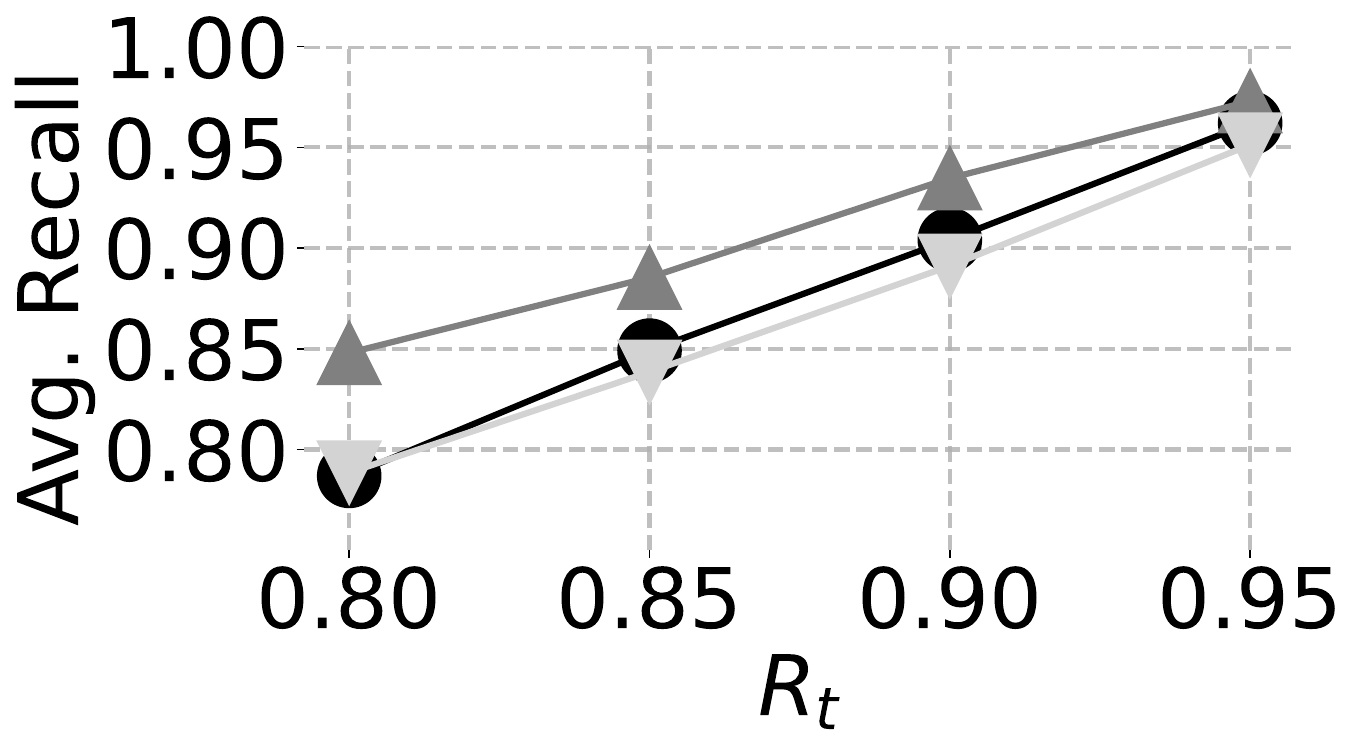}
            \caption{DEEP10M}
        \end{subfigure}
        \hfill
        \begin{subfigure}[t]{0.19\textwidth}
            \centering
            \includegraphics[width=\textwidth]{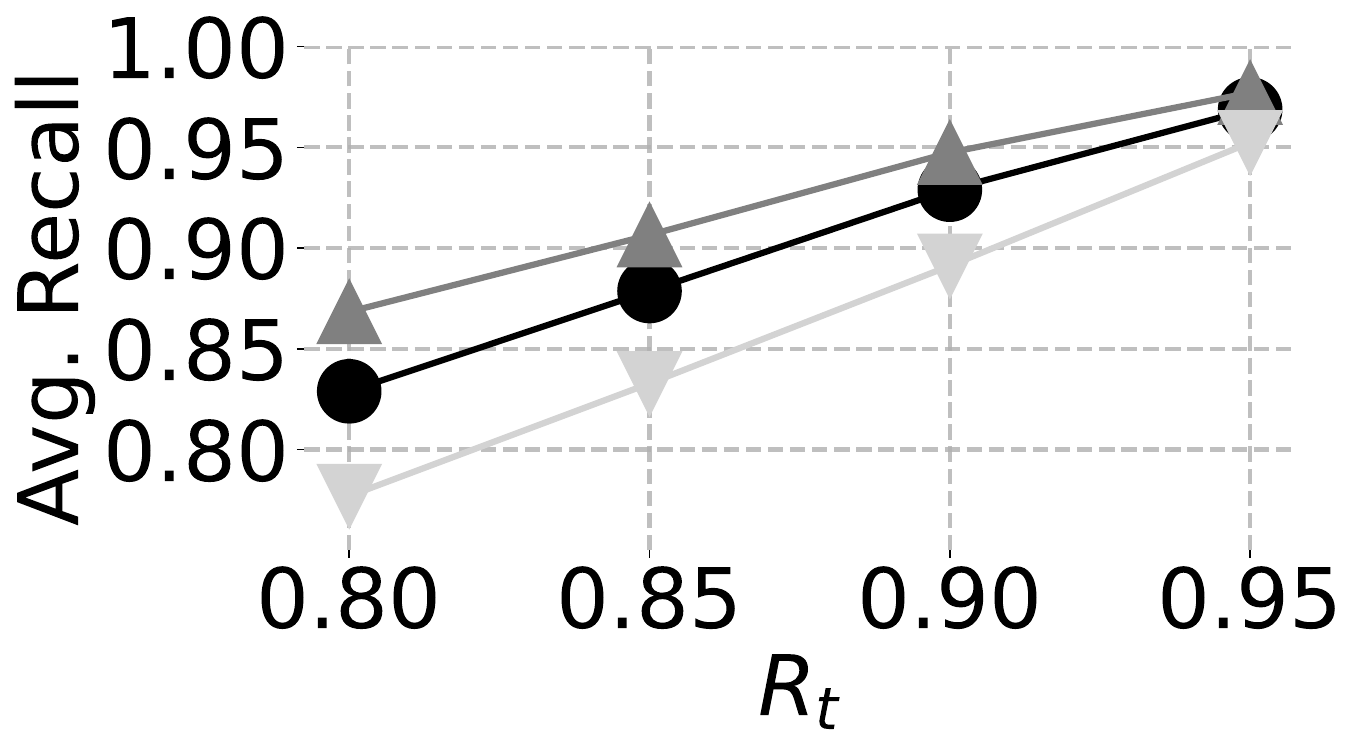}
            \caption{T2I10M}
        \end{subfigure}
    \vspace{-0.3cm}
    \caption{Achieved Recalls of VADER across FVS query workloads with different correlations, $k=100$, $sel=0.3$, Sweeping}
    \label{fig:vader-overview-recalls}
    \end{minipage}

    \begin{minipage}[t]{0.99\textwidth}
        \centering
        
        \begin{subfigure}[t]{0.19\textwidth}
            \centering
            \includegraphics[width=\textwidth]{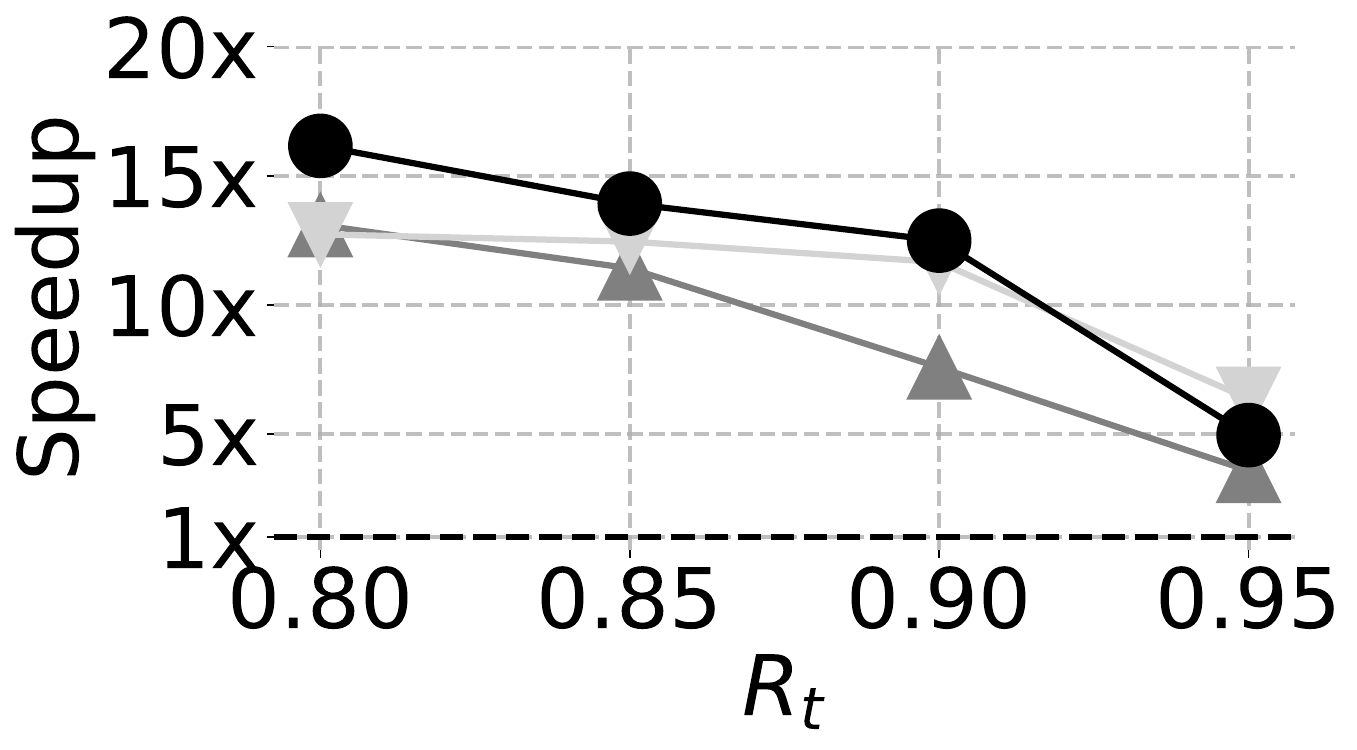}
            \caption{SIFT1M}
        \end{subfigure}
        \hfill
        \begin{subfigure}[t]{0.19\textwidth}
            \centering
            \includegraphics[width=\textwidth]{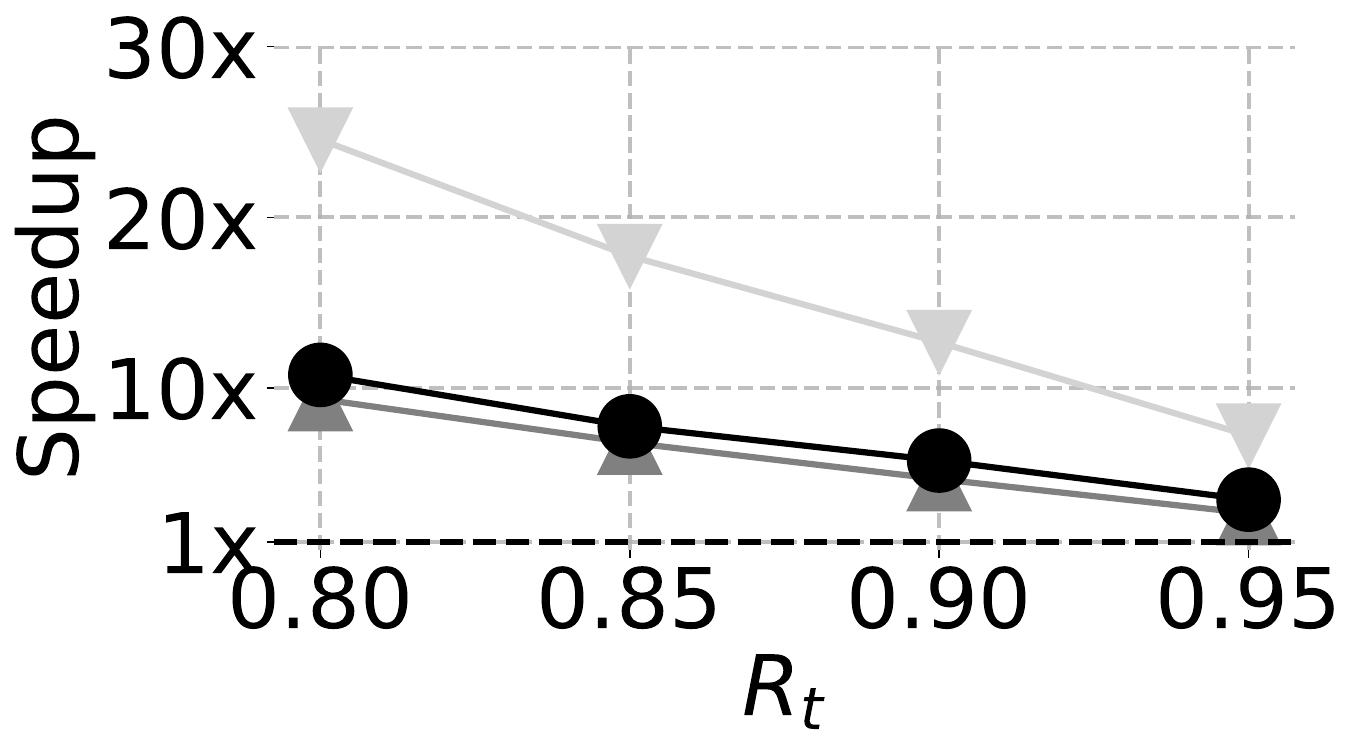}
            \caption{GIST1M}
        \end{subfigure}
        \hfill
        \begin{subfigure}[t]{0.19\textwidth}
            \centering
            \includegraphics[width=\textwidth]{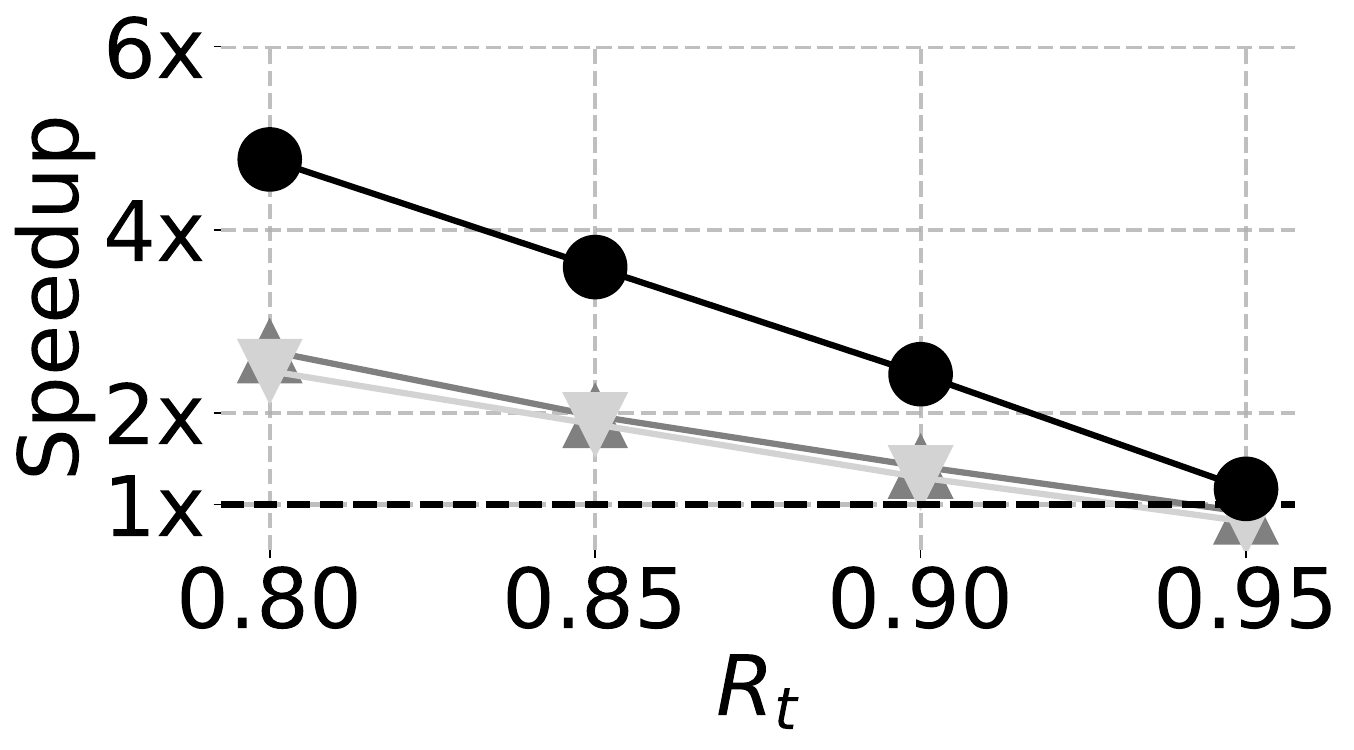}
            \caption{GLOVE1M}
        \end{subfigure}
        \hfill
        \begin{subfigure}[t]{0.19\textwidth}
            \centering
            \includegraphics[width=\textwidth]{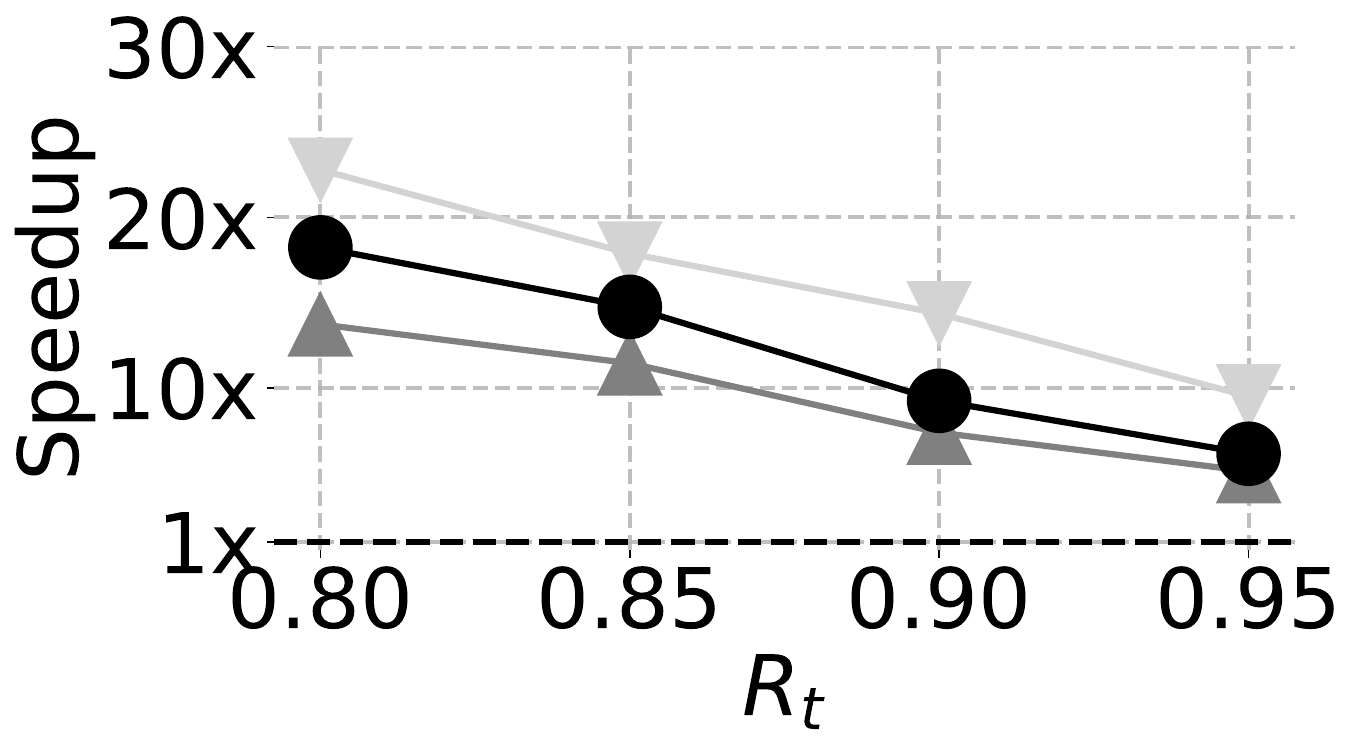}
            \caption{DEEP10M}
        \end{subfigure}
        \hfill
        \begin{subfigure}[t]{0.19\textwidth}
            \centering
            \includegraphics[width=\textwidth]{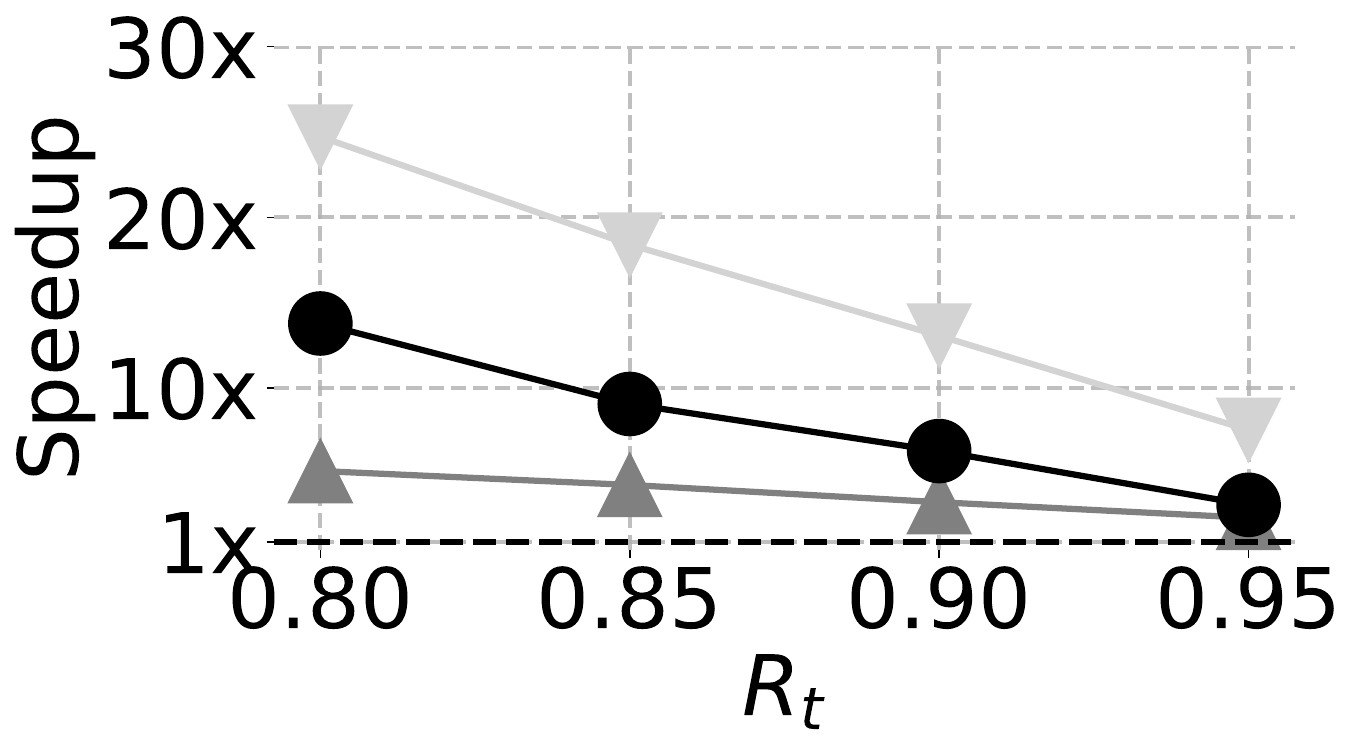}
            \caption{T2I10M}
        \end{subfigure}
    \vspace{-0.3cm}
    \caption{Speedups of VADER over plain search across different correlations (Sweeping, $k=100$, $sel=0.3$).}
    \label{fig:vader-overview-speedup}
    \end{minipage}
\end{figure*}

We evaluate the recall VADER achieves and its speedup over plain search (i.e., search without early termination, at fixed $ef\_search$) for both Sweeping and ACORN, reporting speedup relative to plain search to highlight that the recall predictor not only avoids significant overhead but enables substantial gains.
Figures~\ref{fig:vader-overview-recalls} and~\ref{fig:vader-overview-speedup} give the achieved recall and the speedup for varying recall targets across all datasets and correlations, at $sel=0.3$, $k=100$, using Sweeping; other selectivities and values of $k$ are similar.
The results demonstrate that VADER consistently meets the specified recall targets across different correlation settings.
The achieved recall across all queries of the workload closely matches the specified target, while achieving an average $RQUT=0.27$, with the average recall deviation of queries from the target recall, both for queries above and below the target, is only 0.04.
At the same time, VADER is achieving significant speedups of up to $24.8\times$ and $8.8\times$ on average.
For ACORN, we observe similar trends.
VADER still achieves recalls that closely match the desired recall target, ($RQUT=0.35$ and average recall deviation of 0.05), while providing substantial speedups over plain search, reaching up to $15.1\times$ and $3.4\times$ on average.

\vspace{-0.1cm}
\subsubsection{Recall Distributions}
\begin{figure*}
    \begin{minipage}[t]{0.99\textwidth}
        \centering
        
        \begin{subfigure}[t]{0.19\textwidth}
            \centering
            \includegraphics[width=\textwidth]{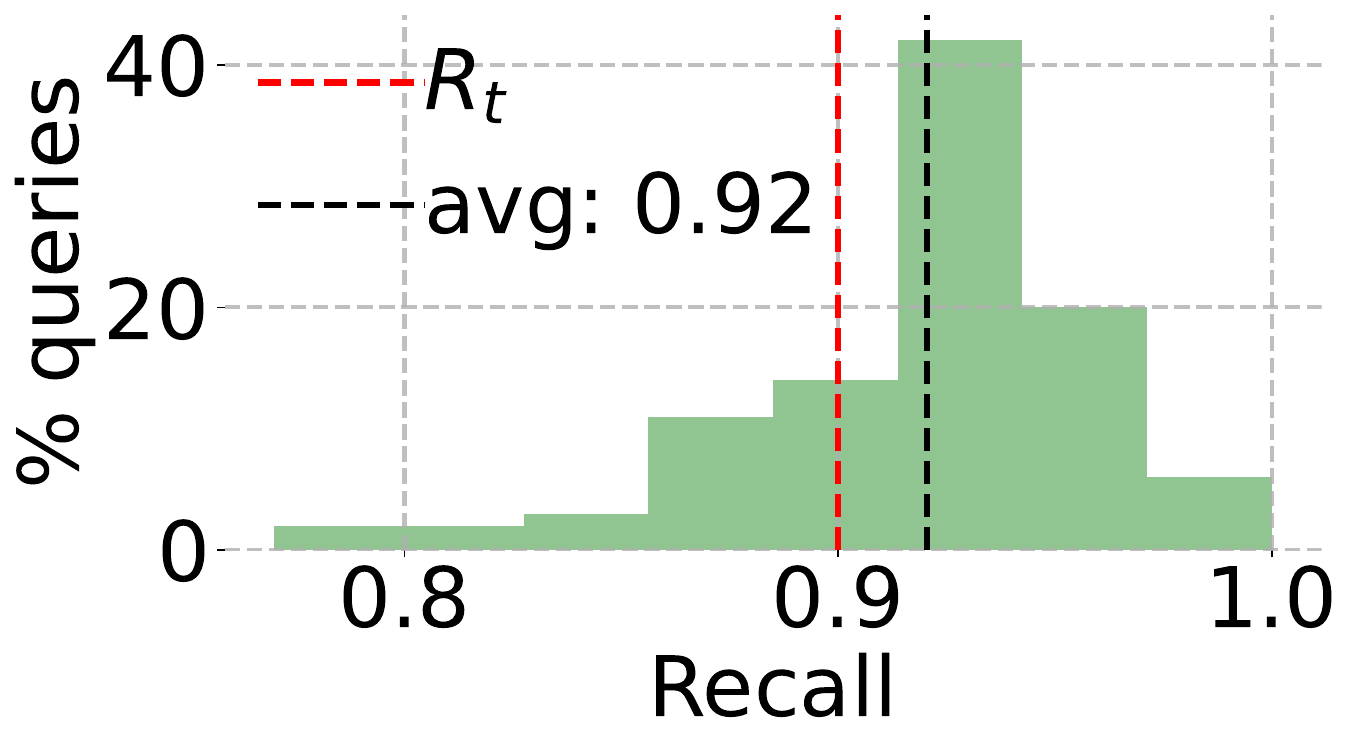}
            \caption{SIFT1M}
        \end{subfigure}
        \hfill
        \begin{subfigure}[t]{0.19\textwidth}
            \centering
            \includegraphics[width=\textwidth]{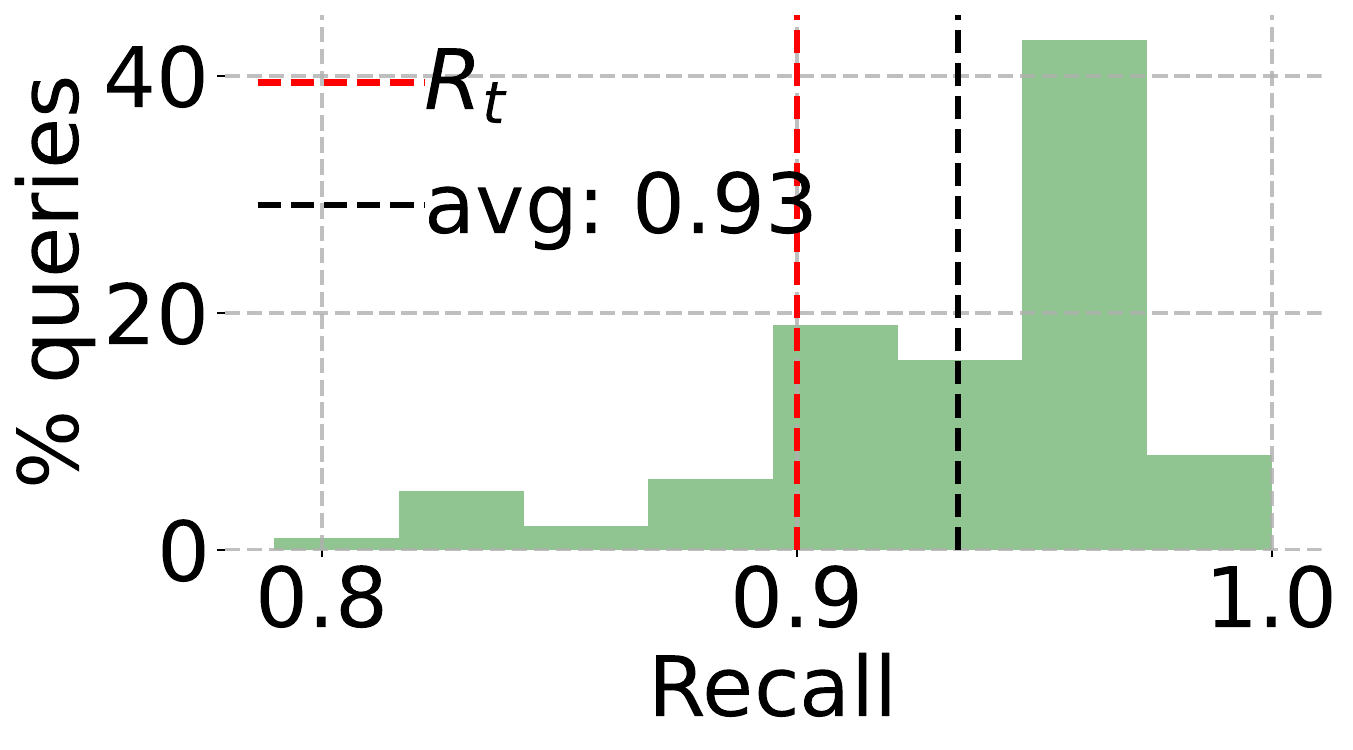}
            \caption{GIST1M}
        \end{subfigure}
        \hfill
        \begin{subfigure}[t]{0.19\textwidth}
            \centering
            \includegraphics[width=\textwidth]{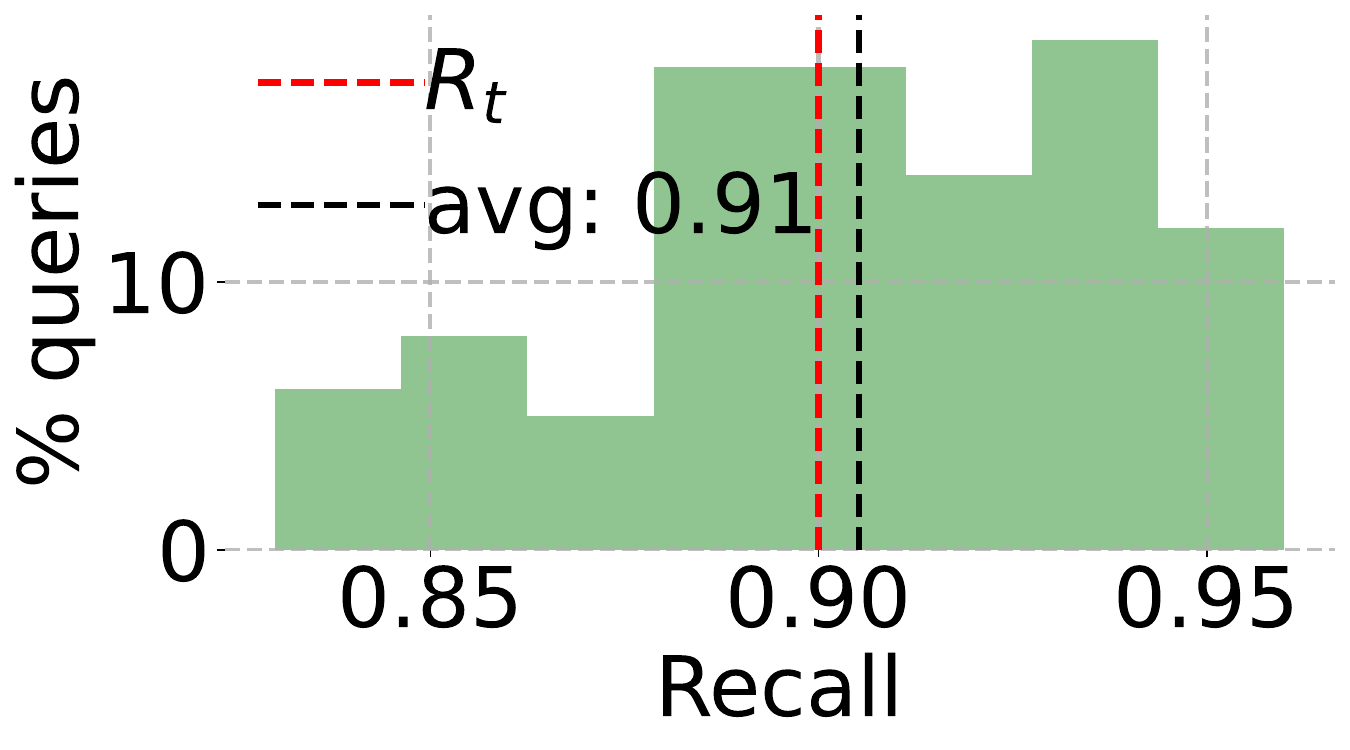}
            \caption{GLOVE1M}
        \end{subfigure}
        \hfill
        \begin{subfigure}[t]{0.19\textwidth}
            \centering
            \includegraphics[width=\textwidth]{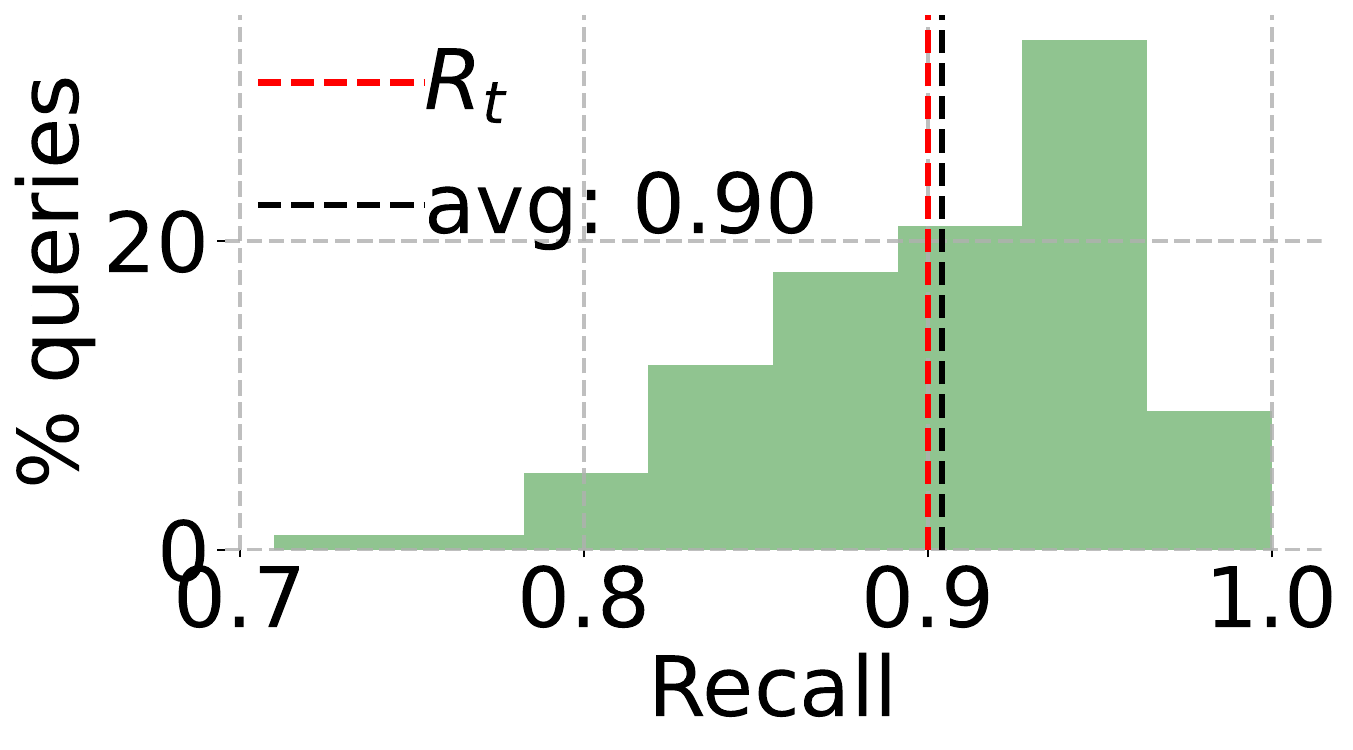}
            \caption{DEEP10M}
        \end{subfigure}
        \hfill
        \begin{subfigure}[t]{0.19\textwidth}
            \centering
            \includegraphics[width=\textwidth]{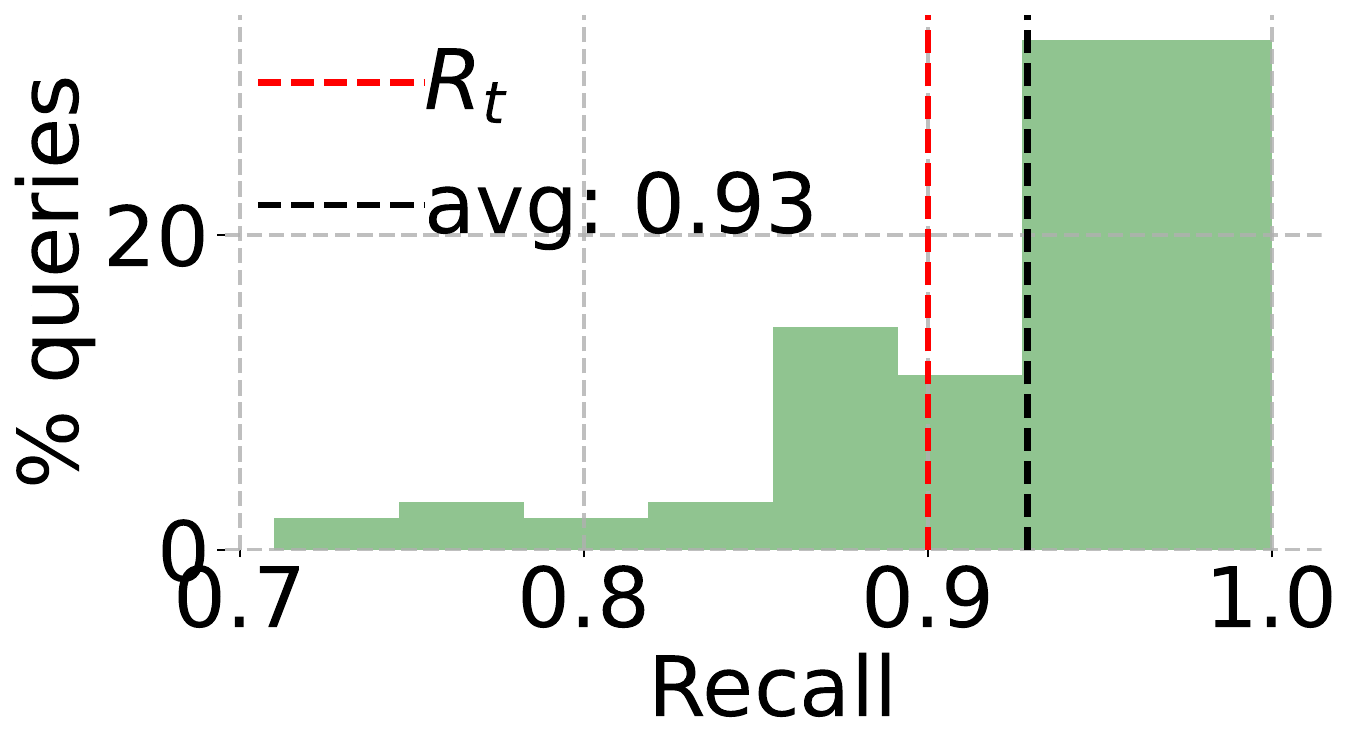}
            \caption{T2I10M}
        \end{subfigure}
        \hfill
    \vspace{-0.3cm}
    \caption{Recall distr. (Sweeping, $k=100$, $R_t=0.90$, No correlation, $sel=0.3$).}
    \label{fig:vader-recall-distr-rt0.9-nocorr}
    \end{minipage}
    
    \begin{minipage}[t]{0.99\textwidth}
        \centering
         \begin{adjustbox}{max width=0.4\textwidth}
            \includegraphics{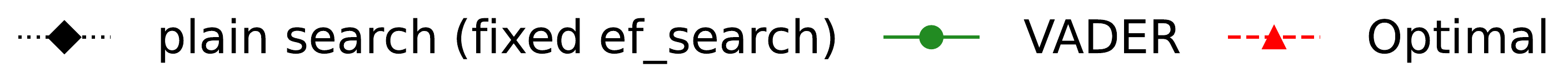}
        \end{adjustbox}
        
        \begin{subfigure}[t]{0.19\textwidth}
            \centering
            \includegraphics[width=\textwidth]{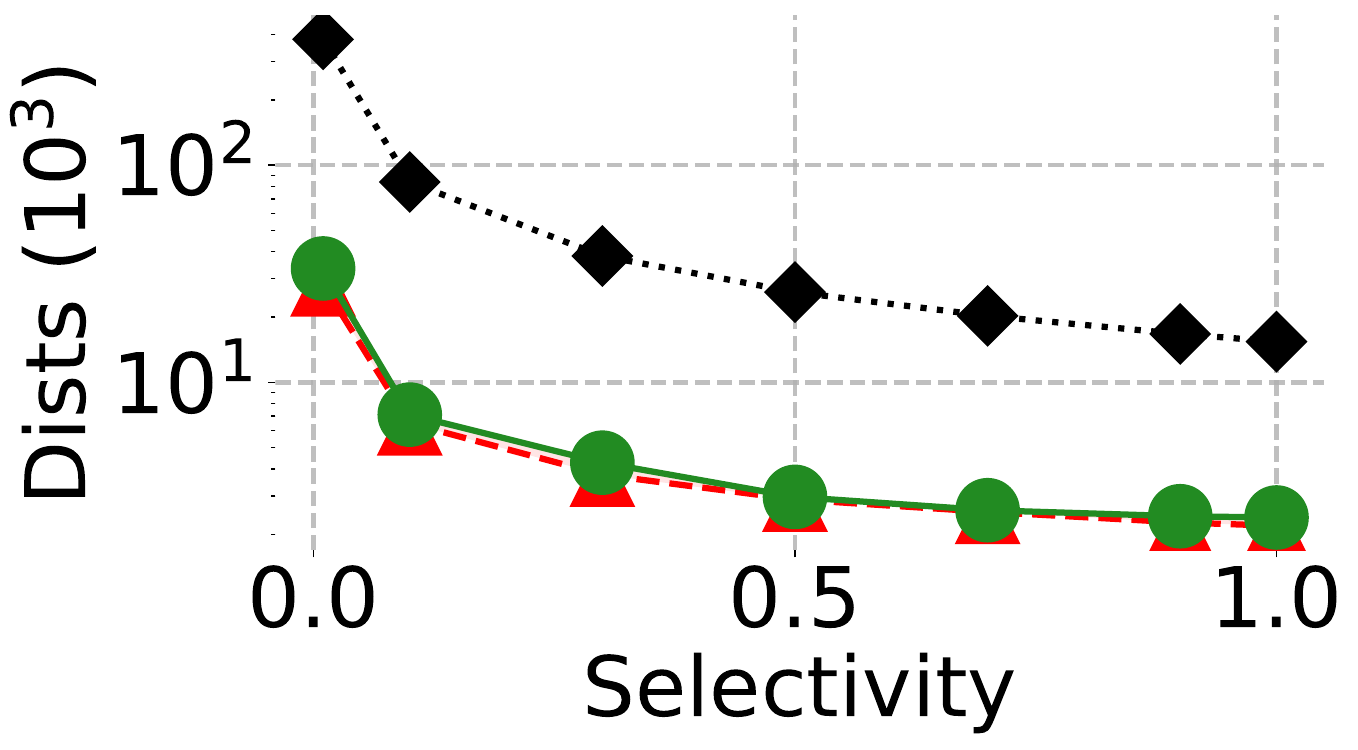}
            \caption{SIFT1M}
        \end{subfigure}
        \hfill
        \begin{subfigure}[t]{0.19\textwidth}
            \centering
            \includegraphics[width=\textwidth]{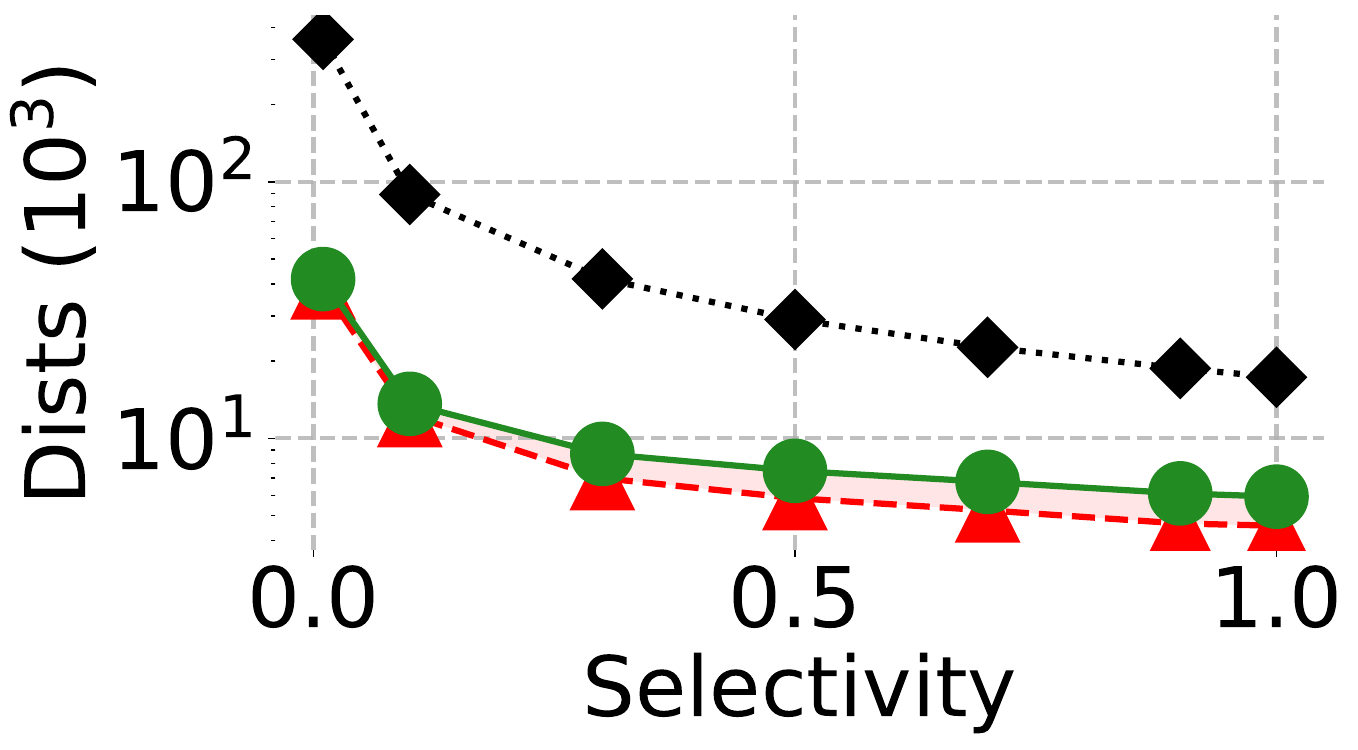}
            \caption{GIST1M}
        \end{subfigure}
        \hfill
         \begin{subfigure}[t]{0.19\textwidth}
             \centering
             \includegraphics[width=\textwidth]{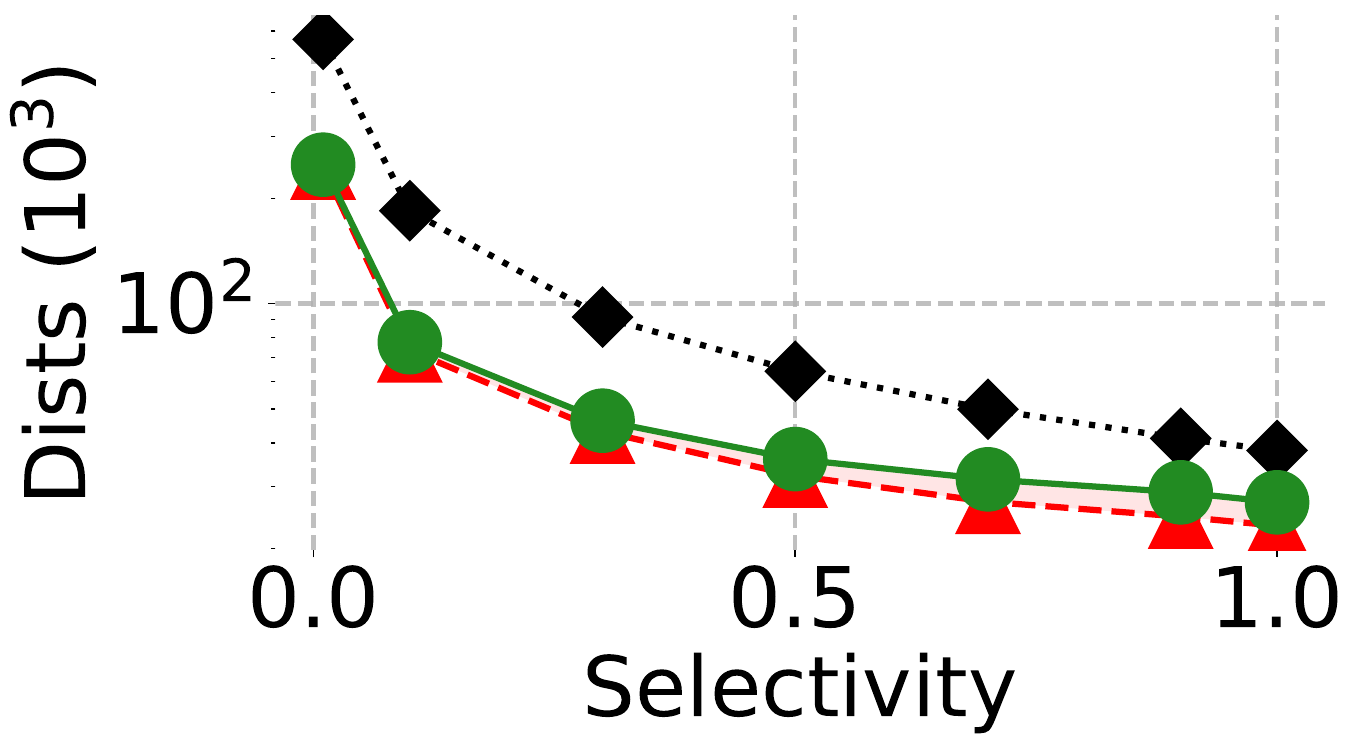}
             \caption{GLOVE1M}
         \end{subfigure}
         \hfill
         \begin{subfigure}[t]{0.19\textwidth}
             \centering
             \includegraphics[width=\textwidth]{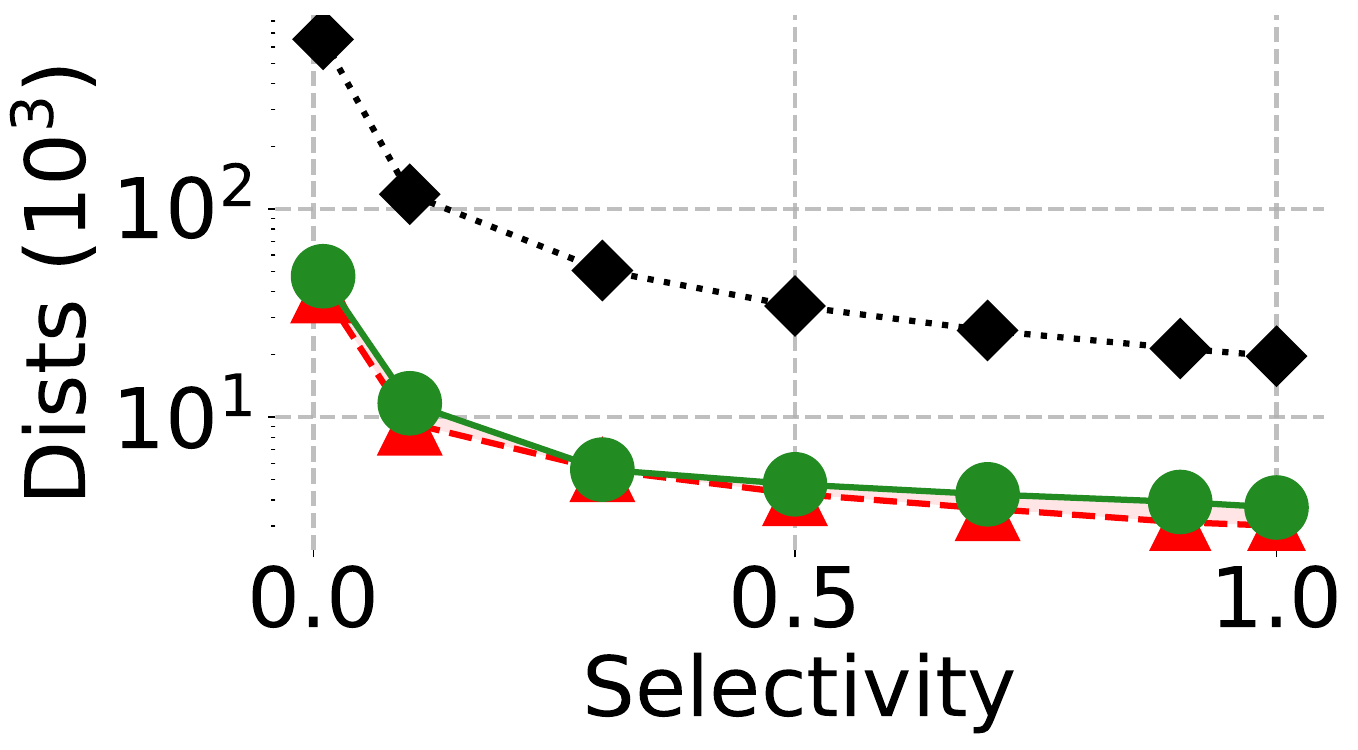}
             \caption{DEEP10M}
         \end{subfigure}
         \hfill
         \begin{subfigure}[t]{0.19\textwidth}
             \centering
             \includegraphics[width=\textwidth]{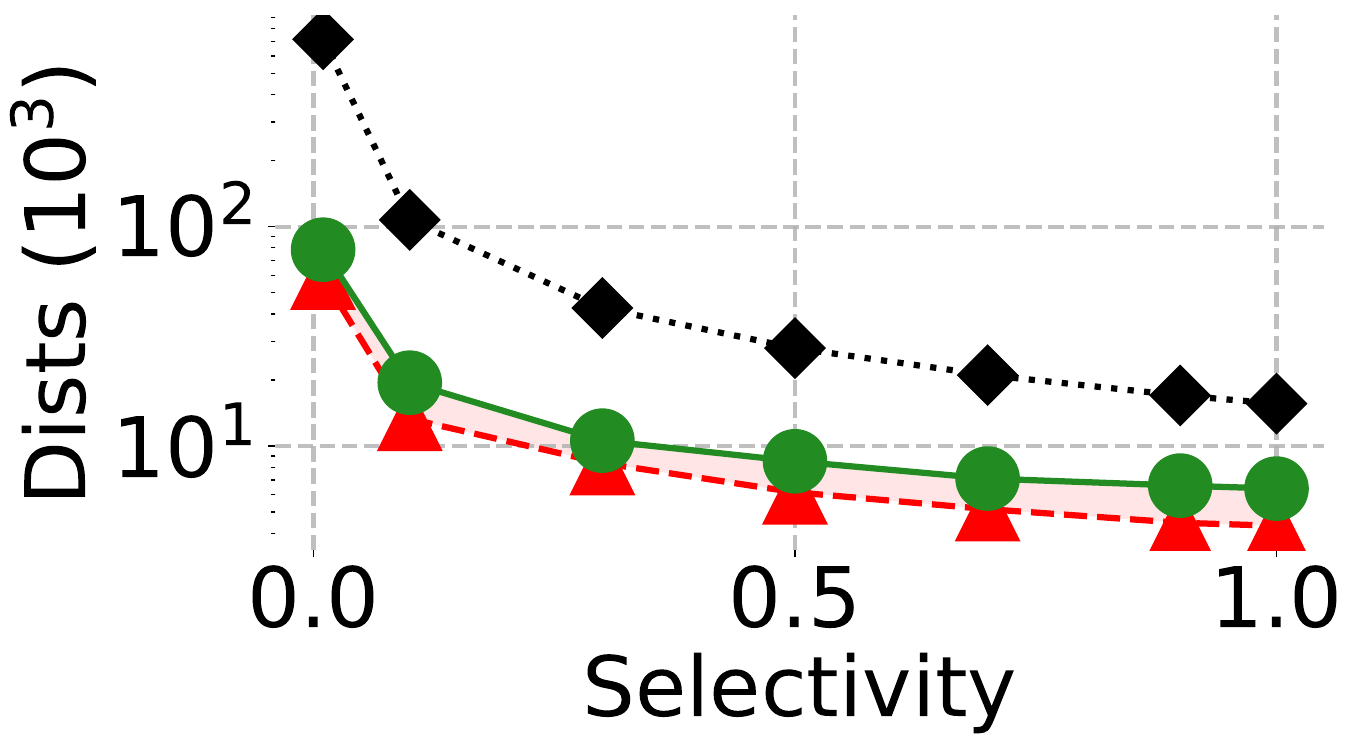}
             \caption{T2I10M}
         \end{subfigure}
    \vspace{-0.3cm}
    \caption{Early termination quality of VADER against the Optimal (Sweeping, $k=100$, $R_t=0.9$, No Correlation).}
    \label{fig:optimal-dists-vader-k100-rt0.9-no-corr}
    \end{minipage}

    \begin{minipage}[t]{0.99\textwidth}
        \centering 
        \begin{subfigure}[t]{0.24\textwidth}
            \centering
            \includegraphics[width=\textwidth]{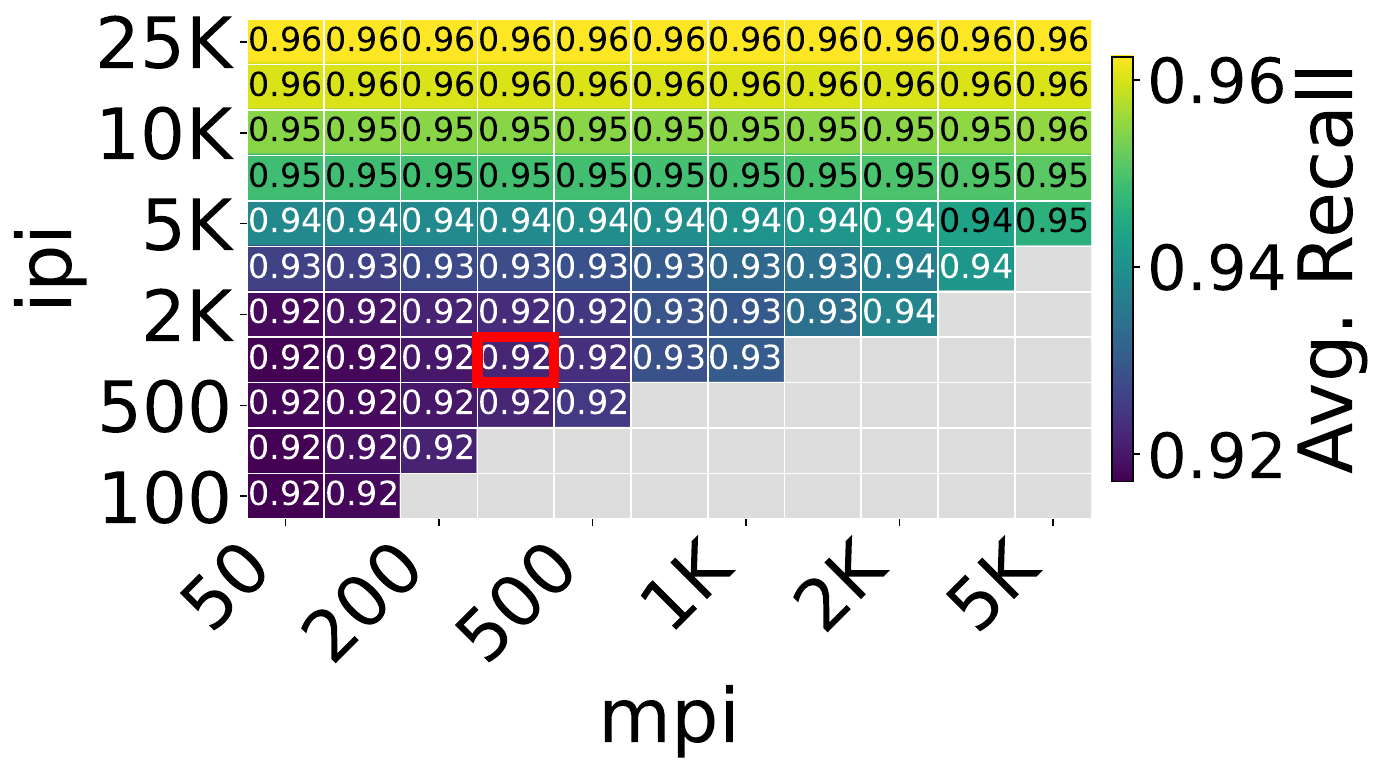}
            \caption{Avg. Recall}
        \end{subfigure}
        \hfill
        \begin{subfigure}[t]{0.24\textwidth}
            \centering
            \includegraphics[width=\textwidth]{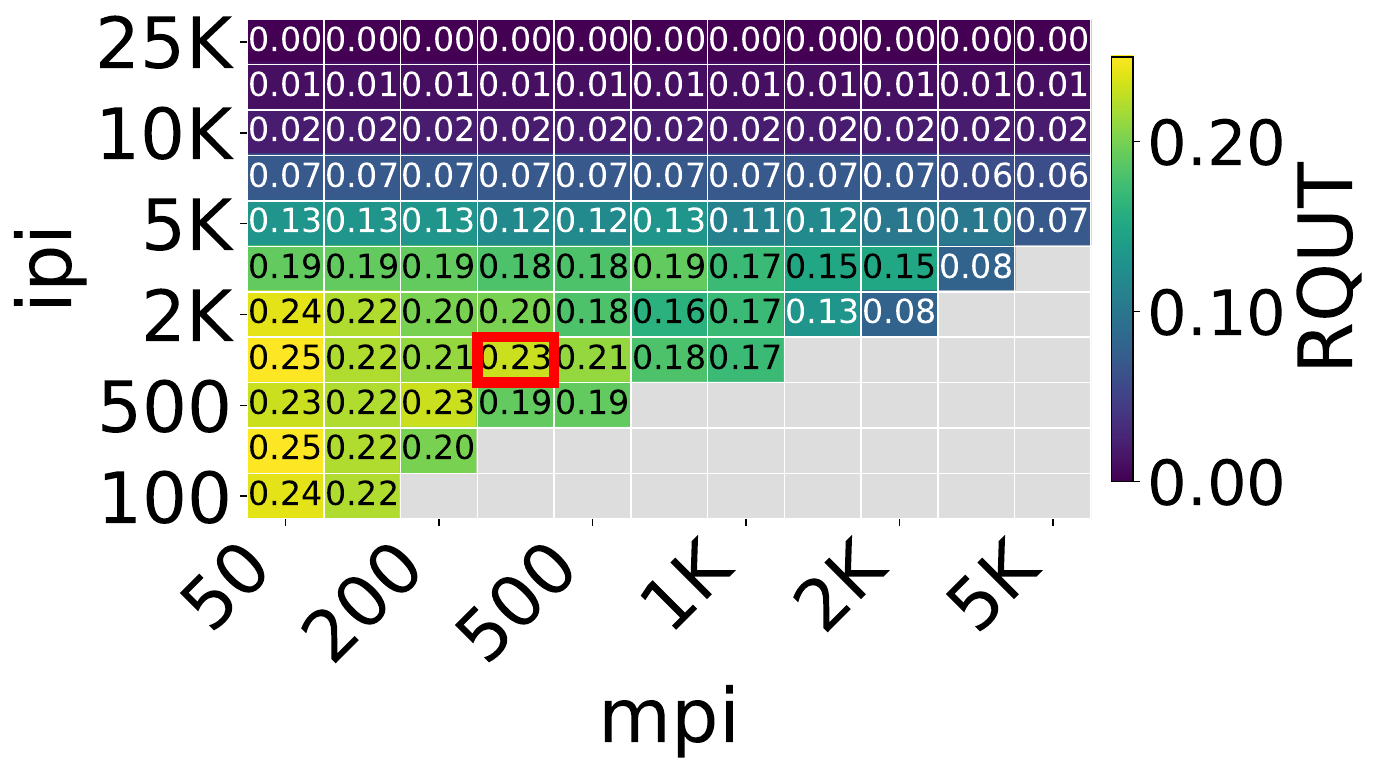}
            \caption{RQUT}
        \end{subfigure}
        \begin{subfigure}[t]{0.24\textwidth}
            \centering
            \includegraphics[width=\textwidth]{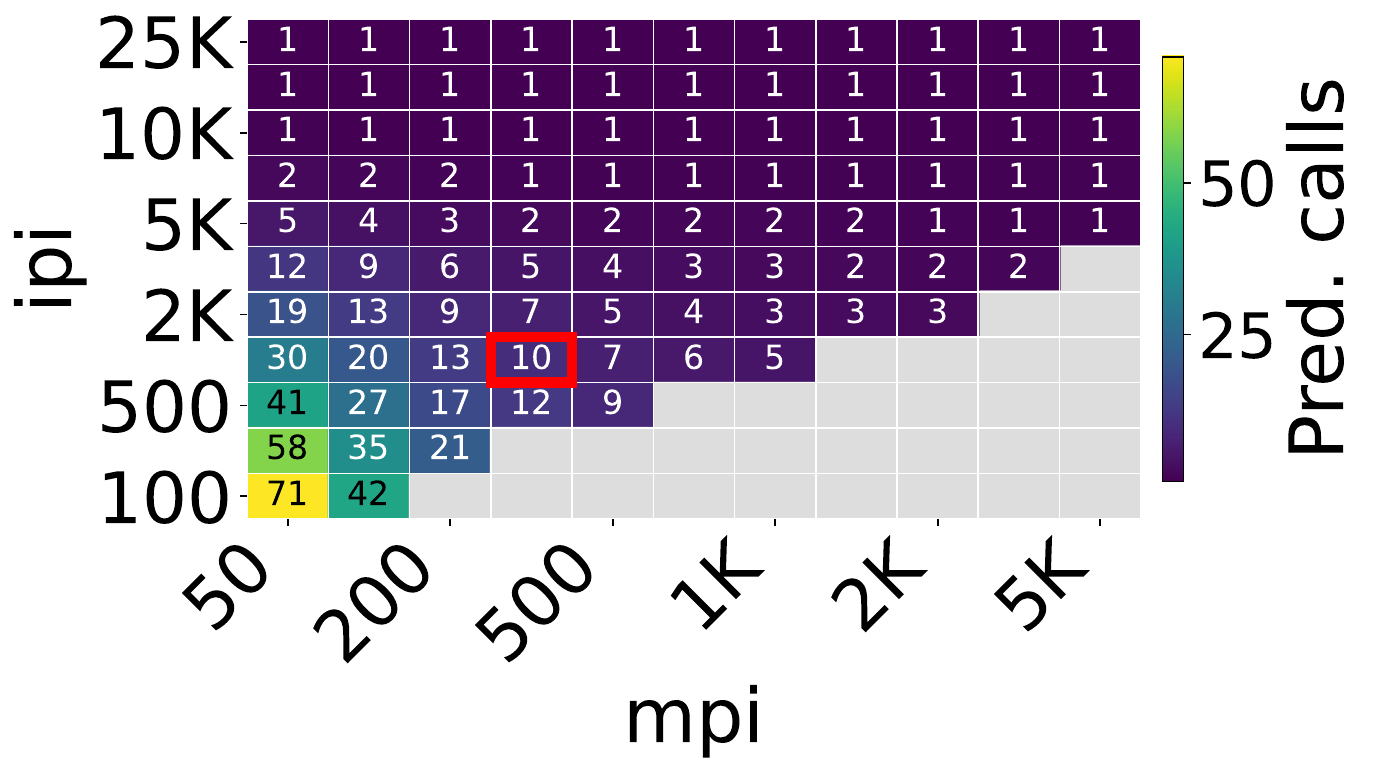}
            \caption{Predictor Calls}
        \end{subfigure}
        \hfill
        \begin{subfigure}[t]{0.24\textwidth}
            \centering
            \includegraphics[width=\textwidth]{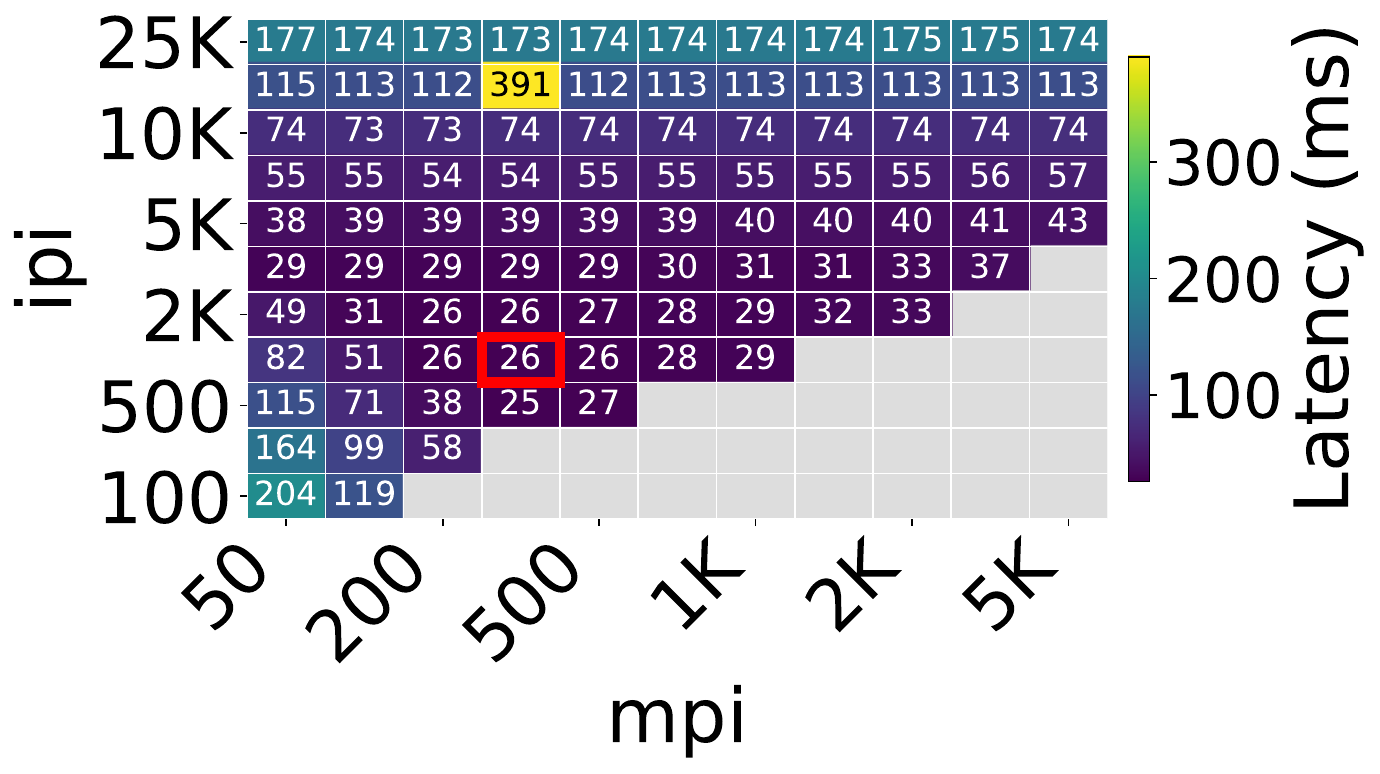}
            \caption{Latency (ms)}
        \end{subfigure}
    \vspace{-0.3cm}
    \caption{Predictor invocation sensitivity study (Sweeping, $k=100$, No correlation, $R_t=0.9$, $sel=0.3$, SIFT1M).}
    \label{fig:pi-sensitivity-k100-rt0.9}
    \end{minipage}

    \begin{minipage}[t]{0.99\textwidth}
        \centering
        \begin{adjustbox}{max width=0.4\textwidth}
            \includegraphics{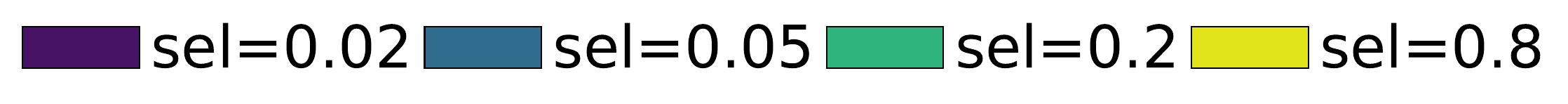}
        \end{adjustbox}
        
        \begin{subfigure}[t]{0.19\textwidth}
            \centering
            \includegraphics[width=\textwidth]{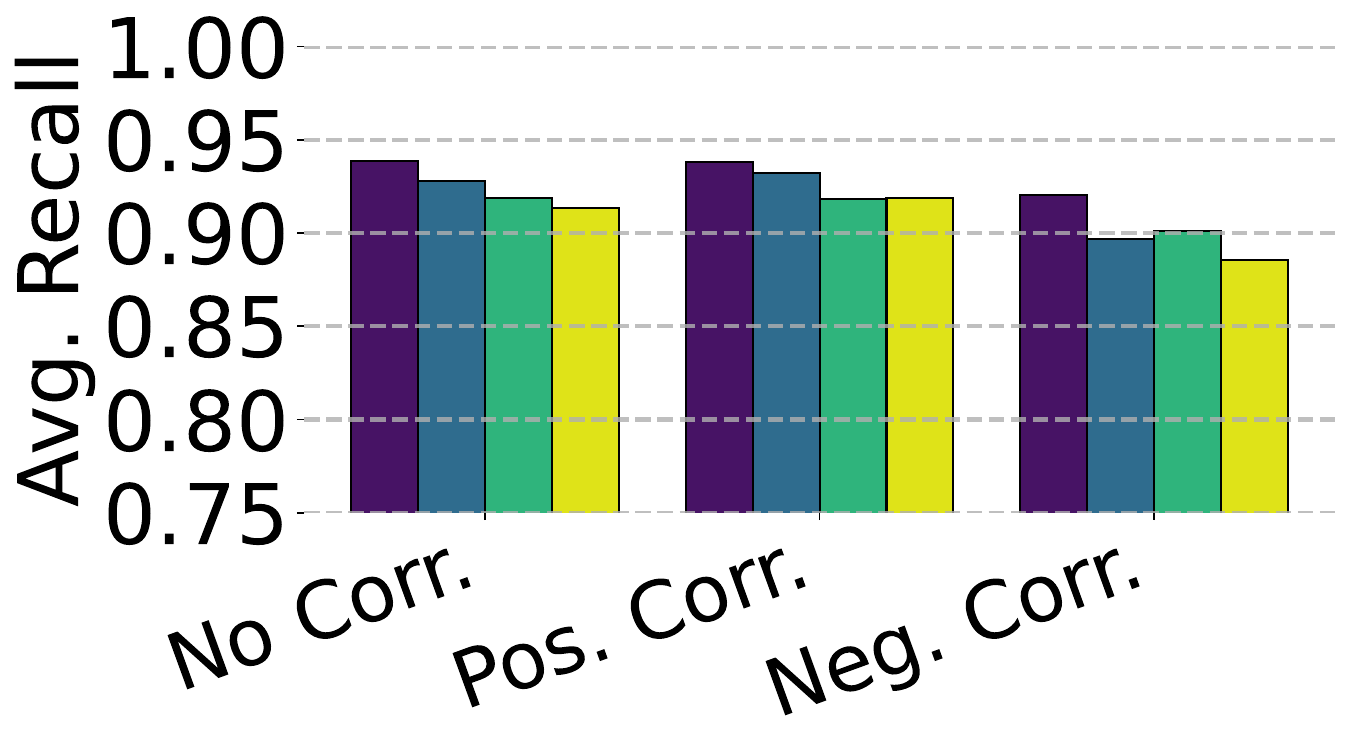}
            \caption{SIFT1M}
        \end{subfigure}
        \hfill
        \begin{subfigure}[t]{0.19\textwidth}
            \centering
            \includegraphics[width=\textwidth]{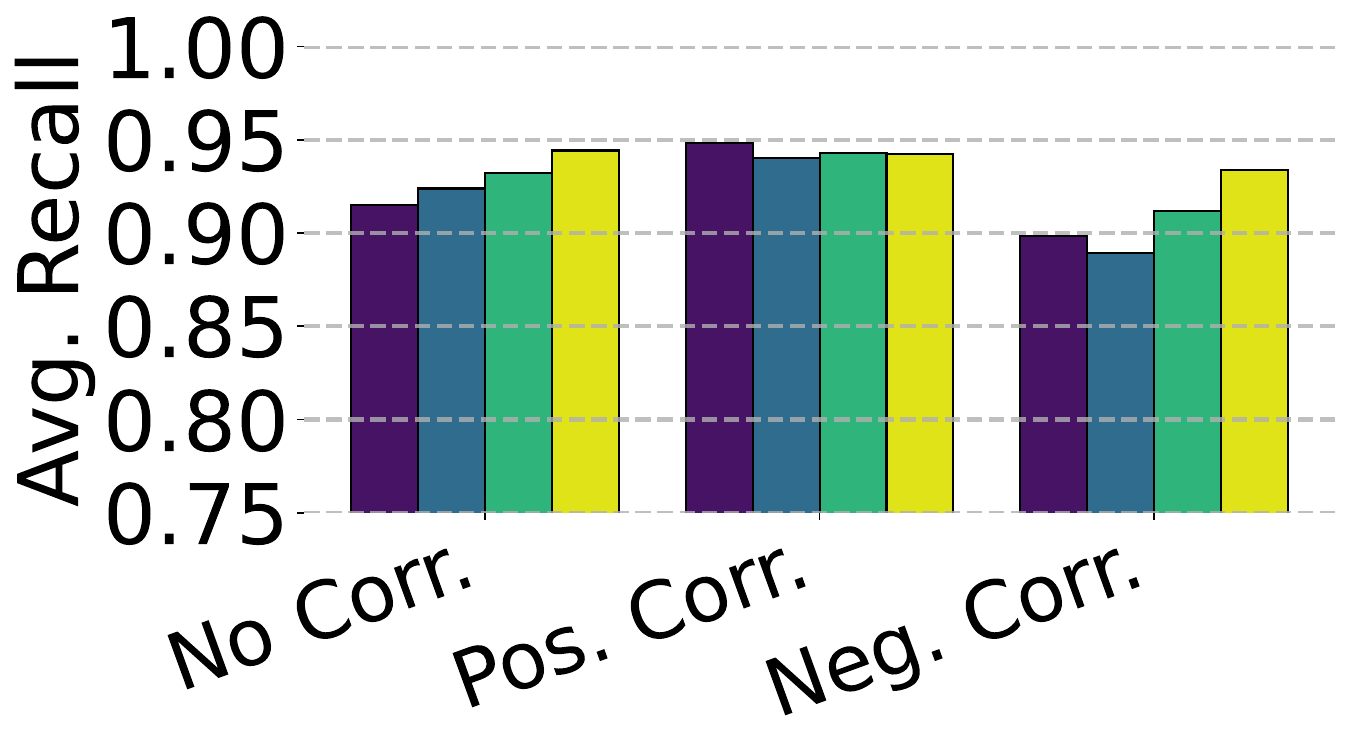}
            \caption{GIST1M}
        \end{subfigure}
        \hfill
        \begin{subfigure}[t]{0.19\textwidth}
            \centering
            \includegraphics[width=\textwidth]{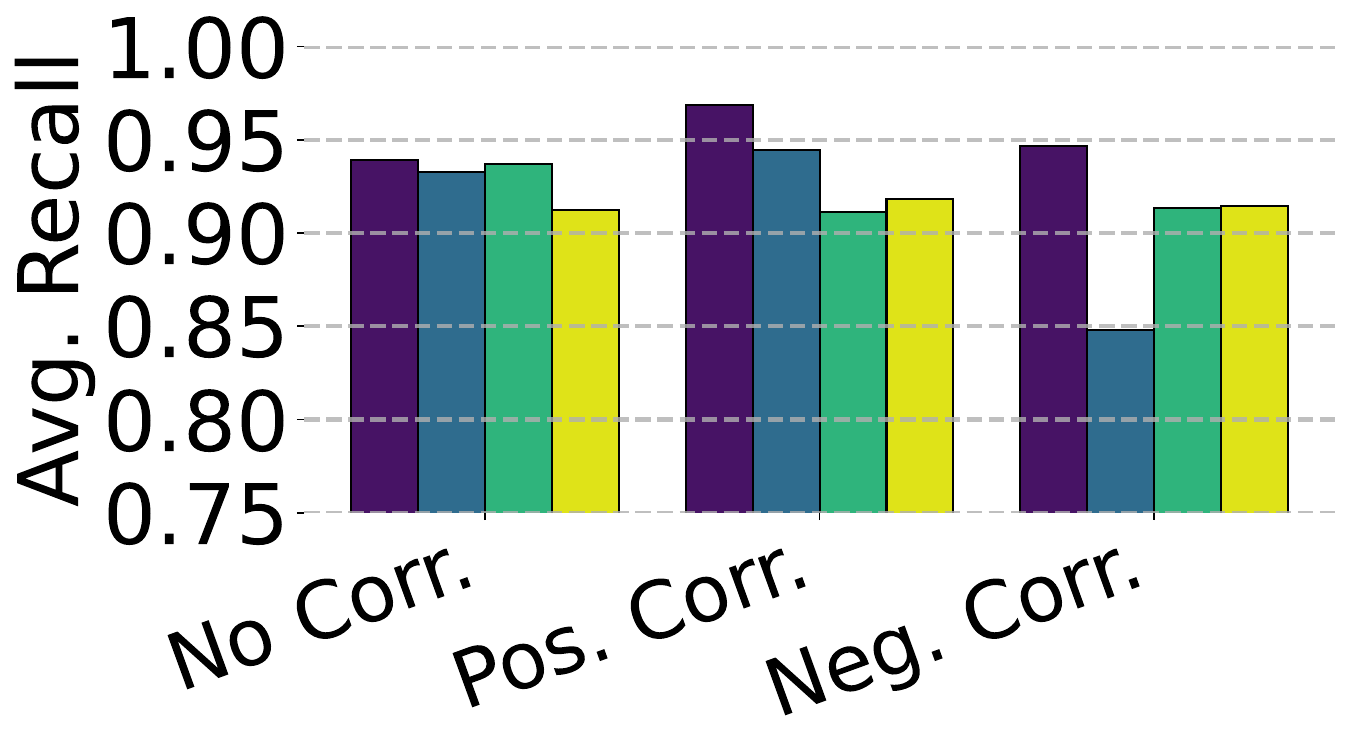}
            \caption{GLOVE1M}
        \end{subfigure}
        \hfill
        \begin{subfigure}[t]{0.19\textwidth}
            \centering
            \includegraphics[width=\textwidth]{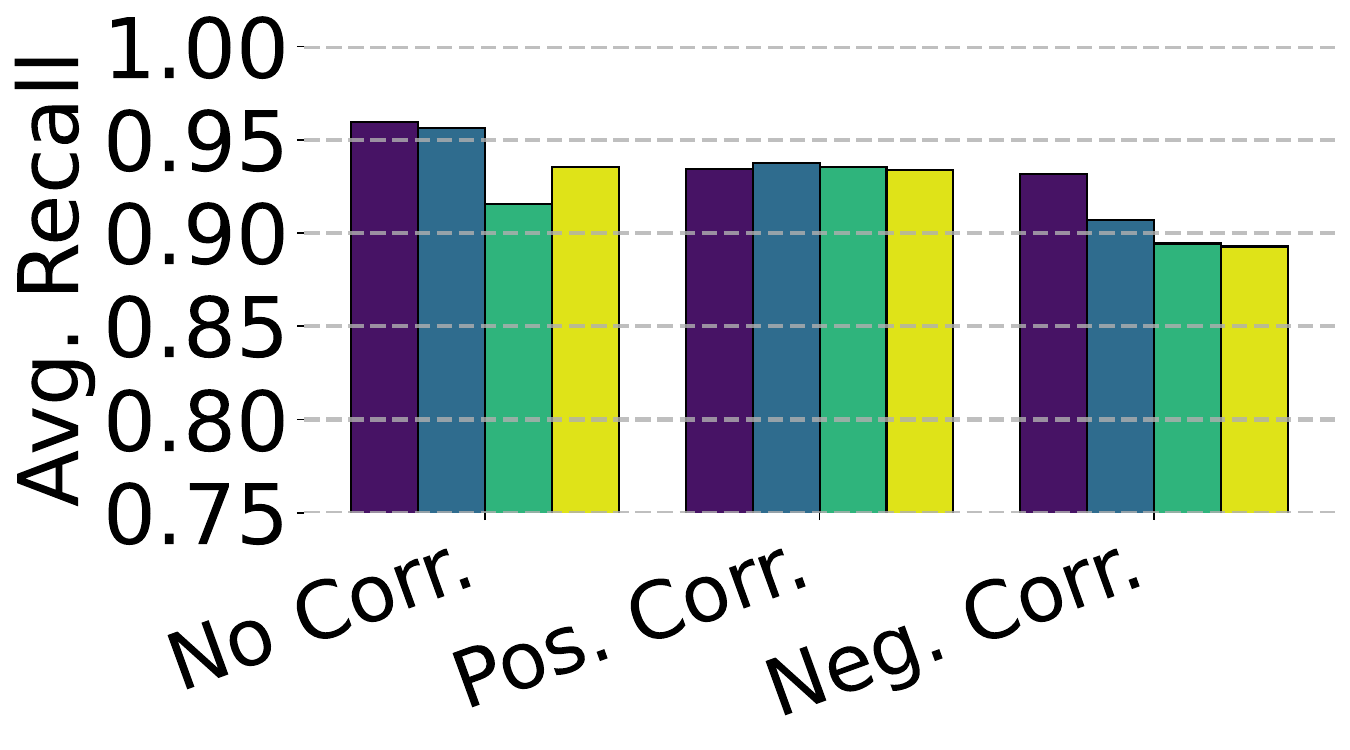}
            \caption{DEEP10M}
        \end{subfigure}
        \hfill
        \begin{subfigure}[t]{0.19\textwidth}
            \centering
            \includegraphics[width=\textwidth]{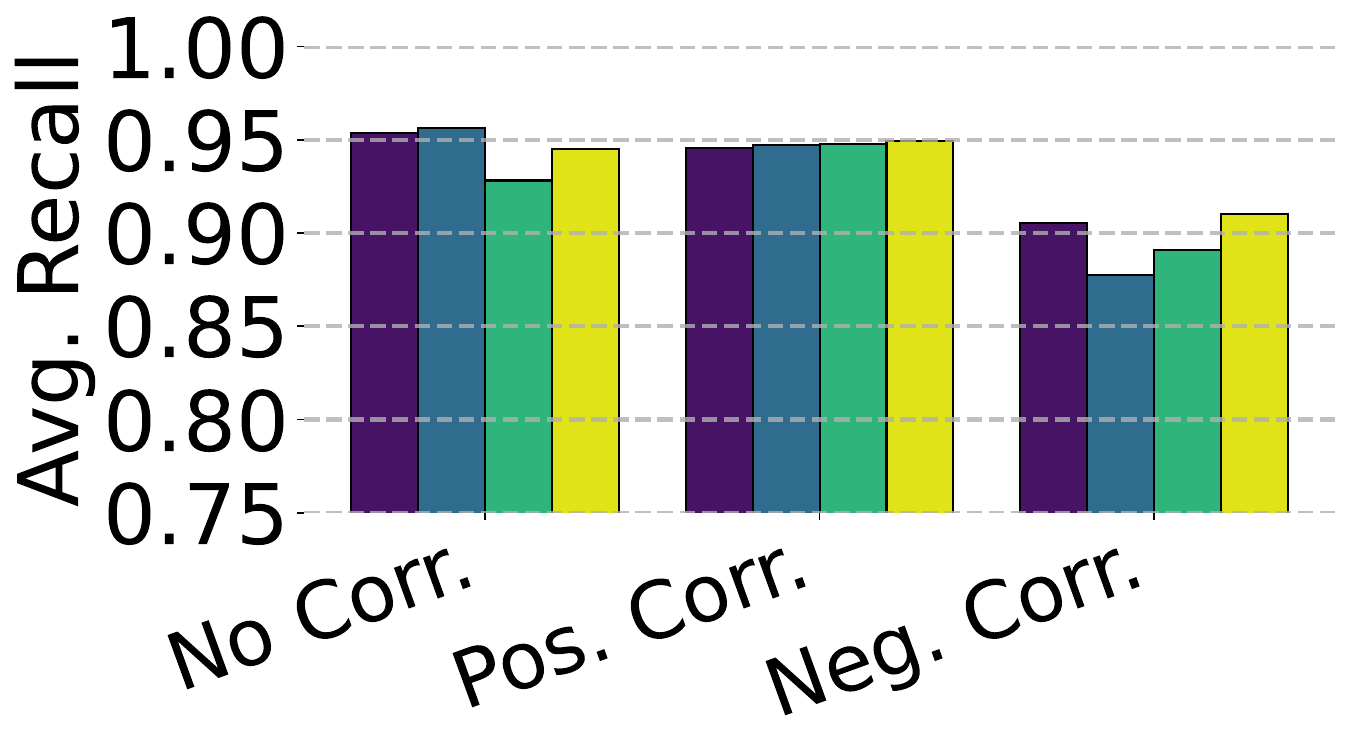}
            \caption{T2I10M}
        \end{subfigure}
    \vspace{-0.3cm}
    \caption{Achieved recalls of VADER on selectivities is was not trained on (Sweeping, $k=100$, $R_t=0.9$).}
    \label{fig:unseen-sweeping-k100-rt0.9}
    \end{minipage}
\end{figure*}

In this set of experiments, we report the recall distribution of all queries for each dataset and correlation setting, using Sweeping with $k=100$, $R_t=0.9$, and $sel=0.3$ (results for other configurations and algorithms exhibit similar trends and are therefore omitted).
Recall distributions for all datasets in no correlation are shown in Figure~\ref{fig:vader-recall-distr-rt0.9-nocorr} (results for other correlations are similar).
The results indicate that VADER consistently achieves recall values very close to the target for individual queries across all datasets and correlation settings.
Across all configurations, the average deviation from the target recall is 0.04 (with an $RQUT=0.27$), demonstrating that even queries that do not exactly meet the target still achieve recall values very close to it.
Examining the worst-case results, the lowest observed recall across all workloads is 0.76, which remains relatively close to the target recall.

\subsubsection{Optimality of Early Termination Points}
We examine the quality of VADER's termination points against an oracle that computes, for each query, selectivity, and correlation, the minimum number of distance computations required to reach $R_t$; this reference is infeasible in practice but serves as a useful benchmark.
For additional context we also report the distance computations of plain search without early termination.
Figure~\ref{fig:optimal-dists-vader-k100-rt0.9-no-corr} shows the results for no correlation with Sweeping at $k=100$ and $R_t=0.9$; positive and negative correlation, and ACORN, exhibit very similar trends and are included in the numbers below.
Across all datasets, selectivities, and correlations, VADER terminates near-optimally, deviating from the oracle by only 4.1\% with respect to plain search.


\subsubsection{Predictor Invocations Heuristics}
Recall that the frequency at which the VADER recall predictor is invoked during query execution is governed by the heuristic described in Section~\ref{sec:vader:early-term-recall-prediction}.
We perform a sensitivity analysis to examine how different values of $mpi$ and $ipi$ affect VADER's performance and to evaluate the effectiveness of the heuristically selected values.
Specifically, we explore a grid of $mpi$ and $ipi$ values, with $mpi\in\{50,\allowbreak 75,\allowbreak 100,\allowbreak 150,\allowbreak ...,\allowbreak 5K,\allowbreak 7.5K,\allowbreak 10K\}$ and $ipi\in\{100,\allowbreak 200,\allowbreak 300,\allowbreak ...,\allowbreak 50K,\allowbreak 75K,\allowbreak 100K\}$, and evaluate VADER in terms of achieved recall, RQUT, query latency, and number of predictor invocations.
Figure~\ref{fig:pi-sensitivity-k100-rt0.9} presents the results for SIFT1M with $k=100$, no correlation, $sel=0.3$, and $R_t=0.9$, showing heatmaps of the achieved recall (a) and RQUT (b) for varying $ipi$ and $mpi$, along with the corresponding number of predictor invocations (c) and query latency (d); similar trends are observed across other datasets and configurations.
In general, large values of $mpi$ and $ipi$ result in slightly higher recall ($\sim$0.95--0.96) and very low RQUT ($\sim$0.00--0.02), but the resulting coarse-grained invocation schedule leads to very few predictor calls (typically 1--2), causing VADER to miss near-optimal termination points and perform redundant distance computations, and thus substantially increasing query latency, in some cases exceeding 100\,ms.
Conversely, small values of $mpi$ and $ipi$ result in recall values closer to the target ($\sim$0.92) and higher RQUT ($\sim$0.24), as frequent predictor invocations allow VADER to terminate more precisely near the target recall. 
However, these configurations invoke the predictor substantially more often than necessary (e.g., 31--71 times per query), and the resulting inference overhead again leads to high query latency, in some cases exceeding 100ms.
These results reveal an important trade-off in selecting $mpi$ and $ipi$: both excessively small and excessively large values can result in high latency, albeit for different reasons, with intermediate values providing the best balance across recall, RQUT, predictor invocations, and query latency.
This observation motivates the heuristic introduced in Section~\ref{sec:vader:early-term-recall-prediction}, which automatically identifies a near-optimal operating point without requiring explicit tuning.

The configuration selected by our heuristic is highlighted by the red square in each heatmap.
Despite the large and complex parameter space, the heuristically selected $mpi$ and $ipi$ values achieve a recall of $0.92$ with an RQUT of $0.23$, while requiring only approximately 10 predictor invocations per query and achieving an average query latency of approximately 26\,ms.
Finally, we observe that small variations of the heuristic, such as modifying the scaling factors or using the median instead of the mean, produce similar results.
This indicates that the underlying heuristic is robust and consistently provides an effective balance between result quality and search latency without requiring parameter tuning.

\subsubsection{Generalization to Unseen Selectivities}
We now examine the generalization ability of VADER to selectivities that are not observed during training.
Thus, we examined the recall quality of VADER with $k=100$, $R_t=0.9$, and selectivities with $sel \in \{0.02,0.05,0.2,0.8\}$, that were not included in the training of the model.
The results, illustrated in Figure~\ref{fig:unseen-sweeping-k100-rt0.9} across all datasets, show that, even for unseen selectivities, VADER consistently achieves high-quality performance across both highly selective cases and less selective cases.
Specifically, VADER deviates by 0.045 on average from the target recall across all configurations, compared to 0.041 for selectivities observed during training.
The slight increase in deviation for unseen selectivities is expected, as the model is not explicitly trained on these configurations; nevertheless, the results remain close to the target, demonstrating the strong generalization capability of VADER.

\subsubsection{Setting Up Baselines}
For REM-Unfiltered we manually tune the $ef\_search$ value that achieves the desired recall target on average at $sel=1.0$, and use it across all experiments.
For REM-Filtered we tune extensively, per selectivity and recall target under no correlation, applying each tuned value across all correlation settings since correlation is not known in advance.
Both variants are tuned twice, once for ACORN and once for Sweeping.
This places VADER at a clear disadvantage: the per-selectivity, per-target tuning of REM-Filtered is impractical in deployment but sets a high bar, and exceeding it implies VADER would also exceed the more realistic REM configurations.
We present our comparisons across varying selectivities, recall targets and values of $k$.
We start by presenting a full comparison of VADER across the above configurations for no correlation, and then we present the figures for positive and negative correlation.

\subsubsection{Performance for Varying Selectivities.}
\begin{figure*}
    \centering
    \vspace{-0.1cm}
    \begin{adjustbox}{max width=0.6\textwidth}
        \includegraphics{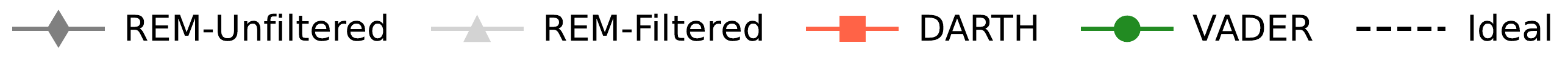}
    \end{adjustbox}

    \vspace{-0.1cm}
    \makebox[0.19\textwidth][c]{\textbf{SIFT1M}} \hfill
    \makebox[0.19\textwidth][c]{\textbf{GIST1M}} \hfill
    \makebox[0.19\textwidth][c]{\textbf{GLOVE1M}} \hfill
    \makebox[0.19\textwidth][c]{\textbf{DEEP10M}} \hfill
    \makebox[0.19\textwidth][c]{\textbf{T2I10M}}
    
    
    \begin{minipage}[t]{\textwidth}
        \includegraphics[width=0.19\textwidth]{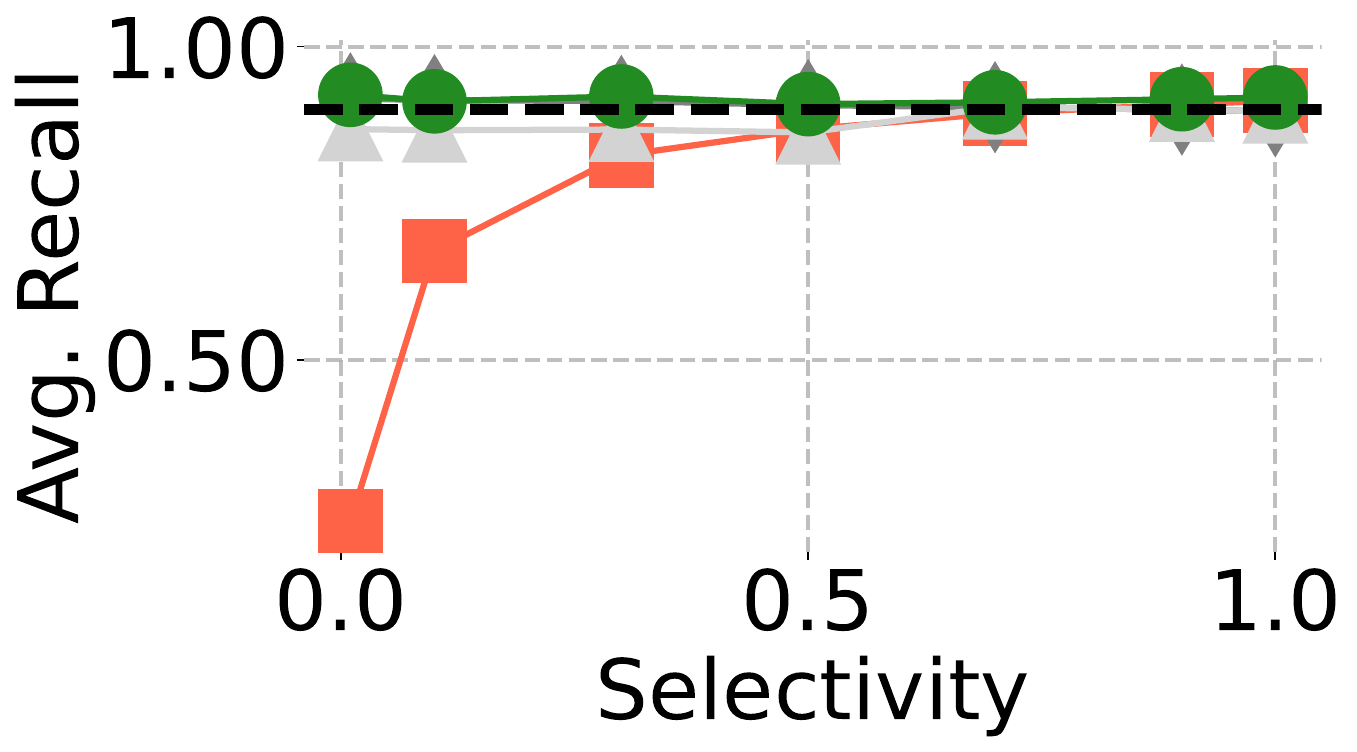}\hfill
        \includegraphics[width=0.19\textwidth]{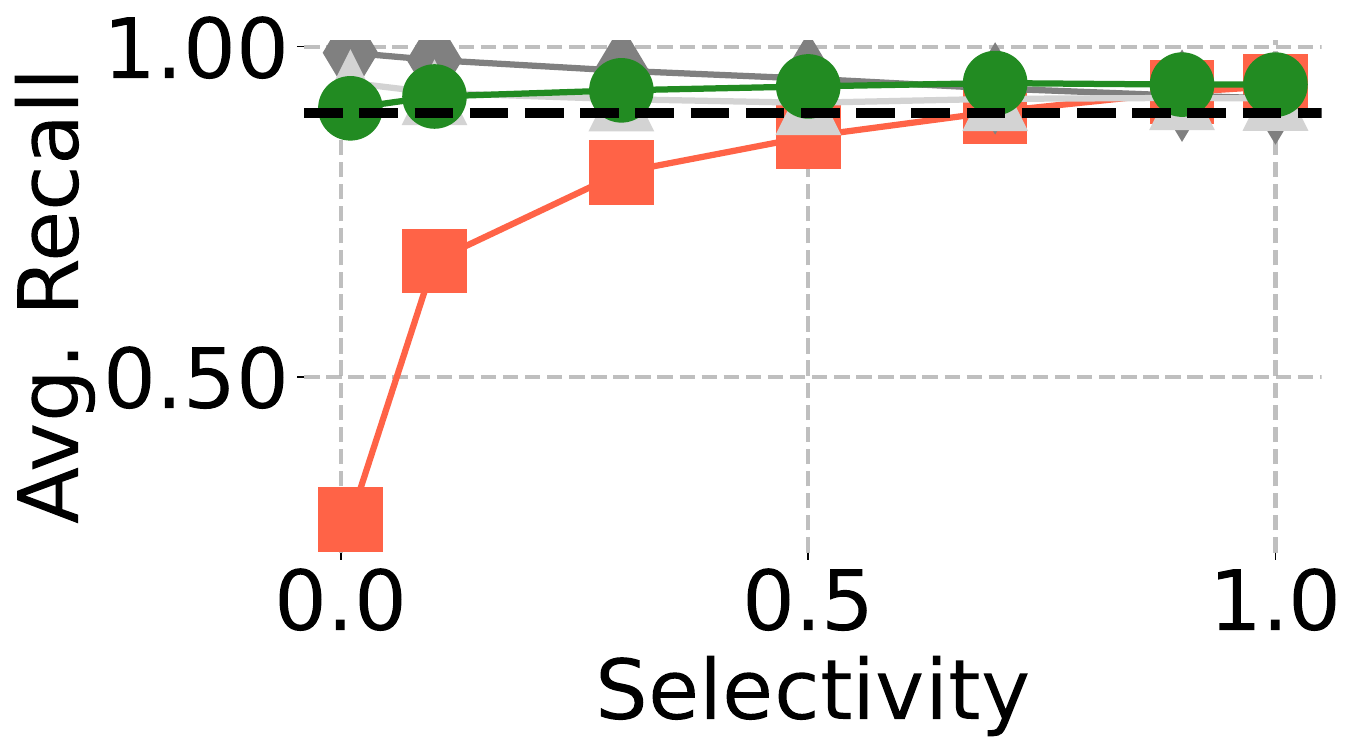}\hfill
        \includegraphics[width=0.19\textwidth]{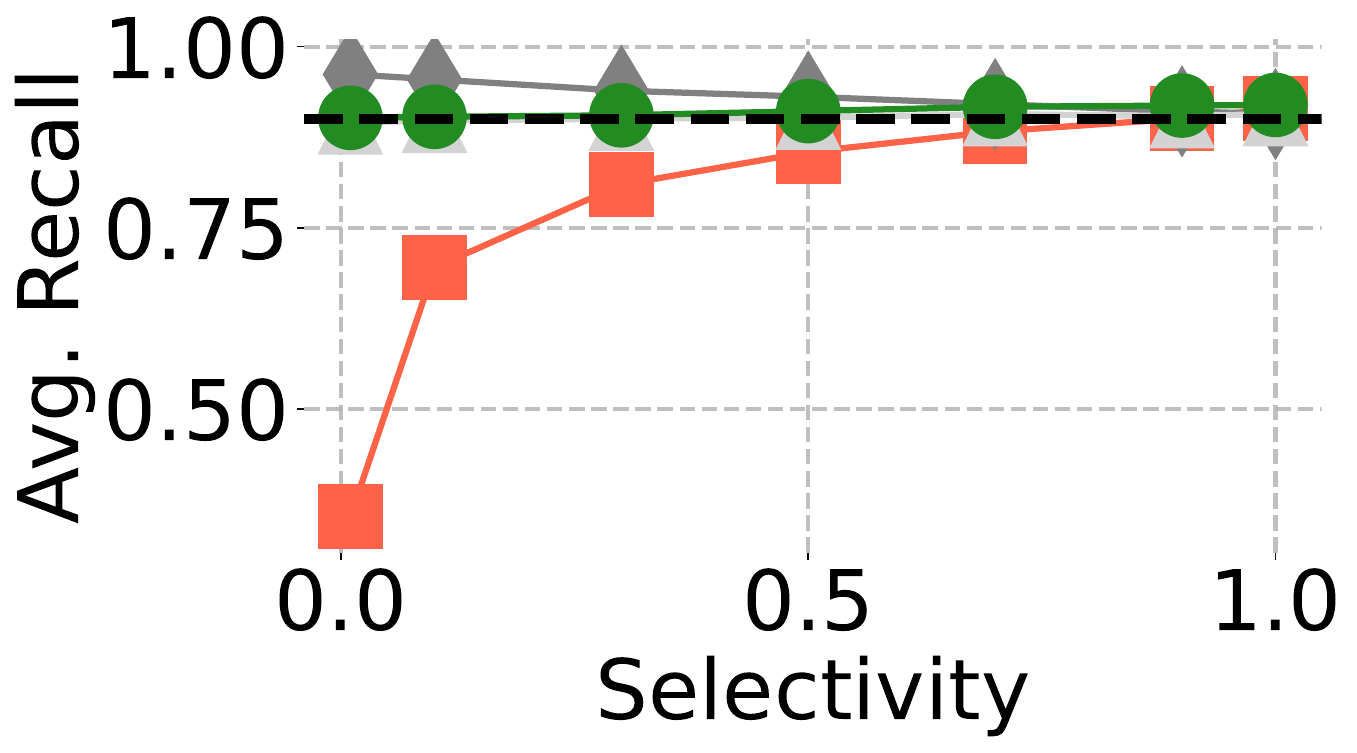}\hfill
        \includegraphics[width=0.19\textwidth]{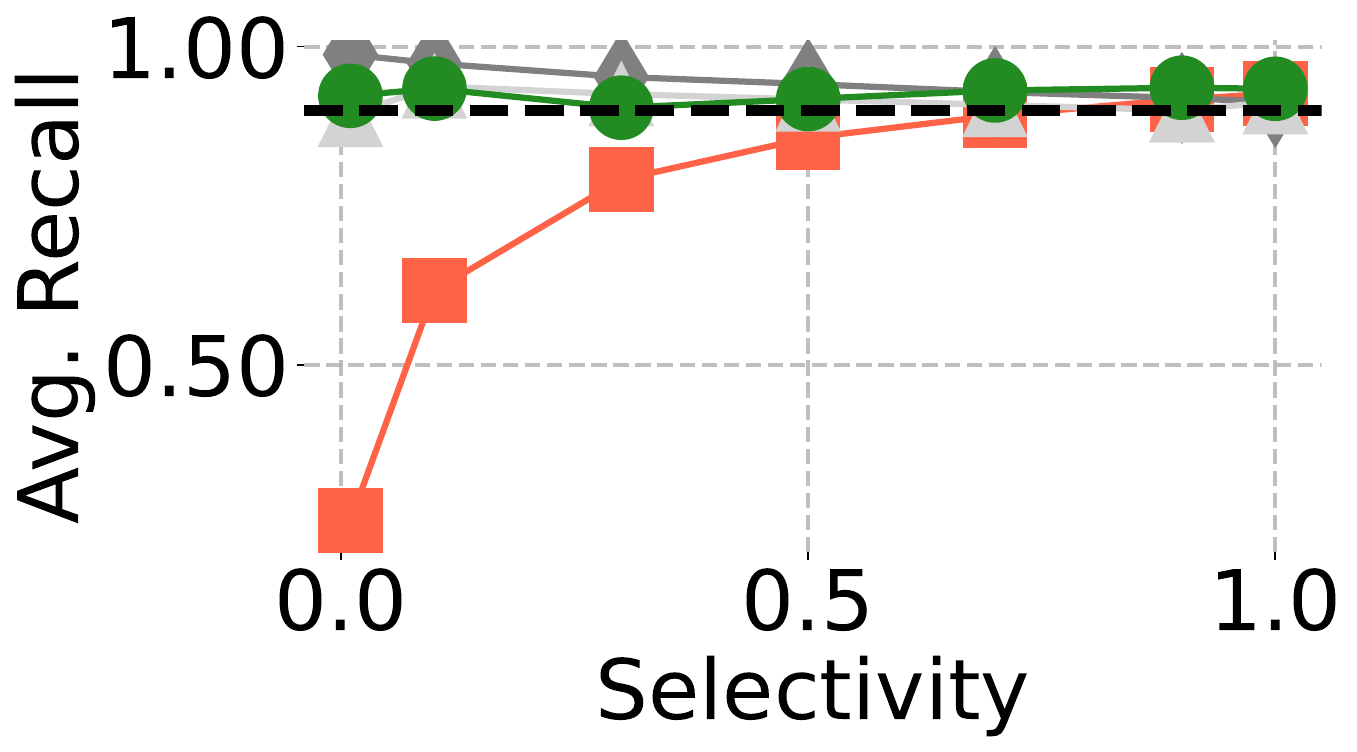}\hfill
        \includegraphics[width=0.19\textwidth]{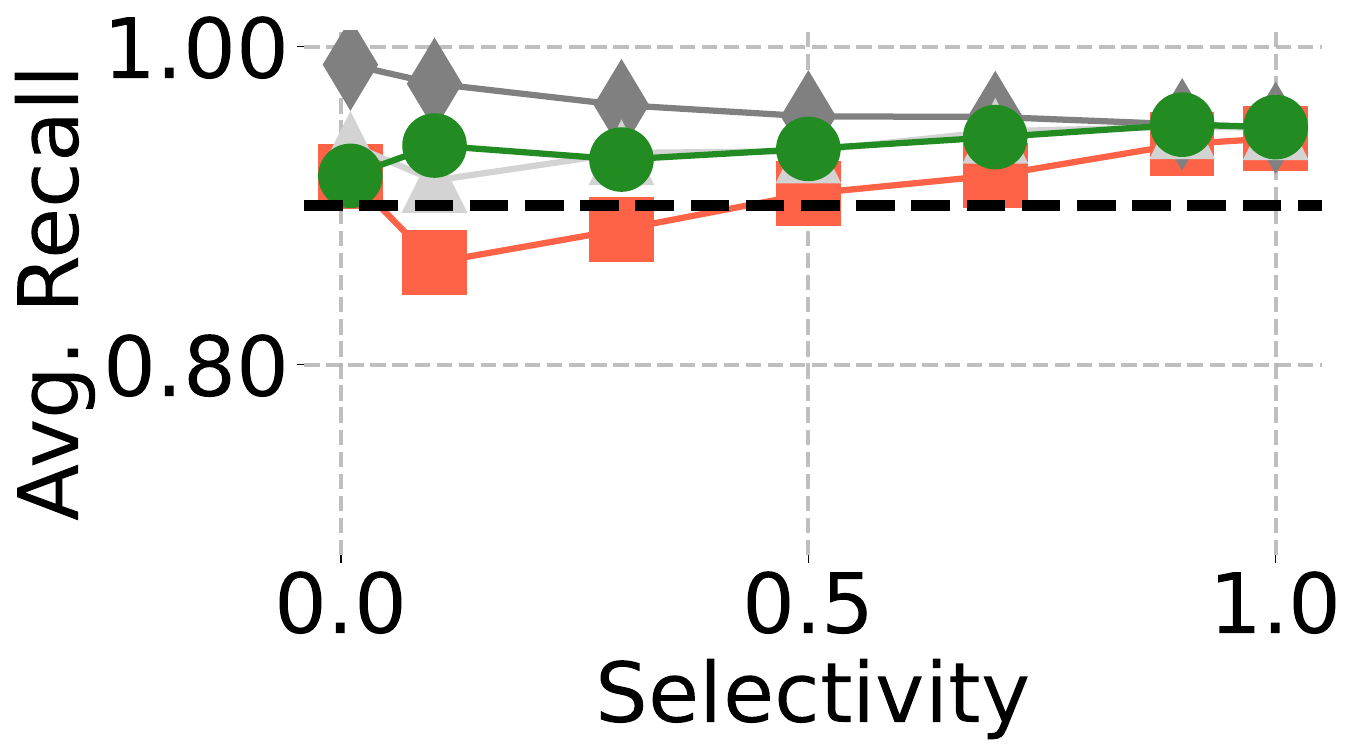}
        \vspace{-0.4cm}
        \caption{Sweeping: Achieved recalls for varying selectivity (No Correlation, $k=100$, $R_t=0.9$).}
        \label{fig:recall-vs-selectivity-rt0.9-k100-sweeping-no-corr}
    \end{minipage}

    \begin{minipage}[t]{\textwidth}
        \includegraphics[width=0.19\textwidth]{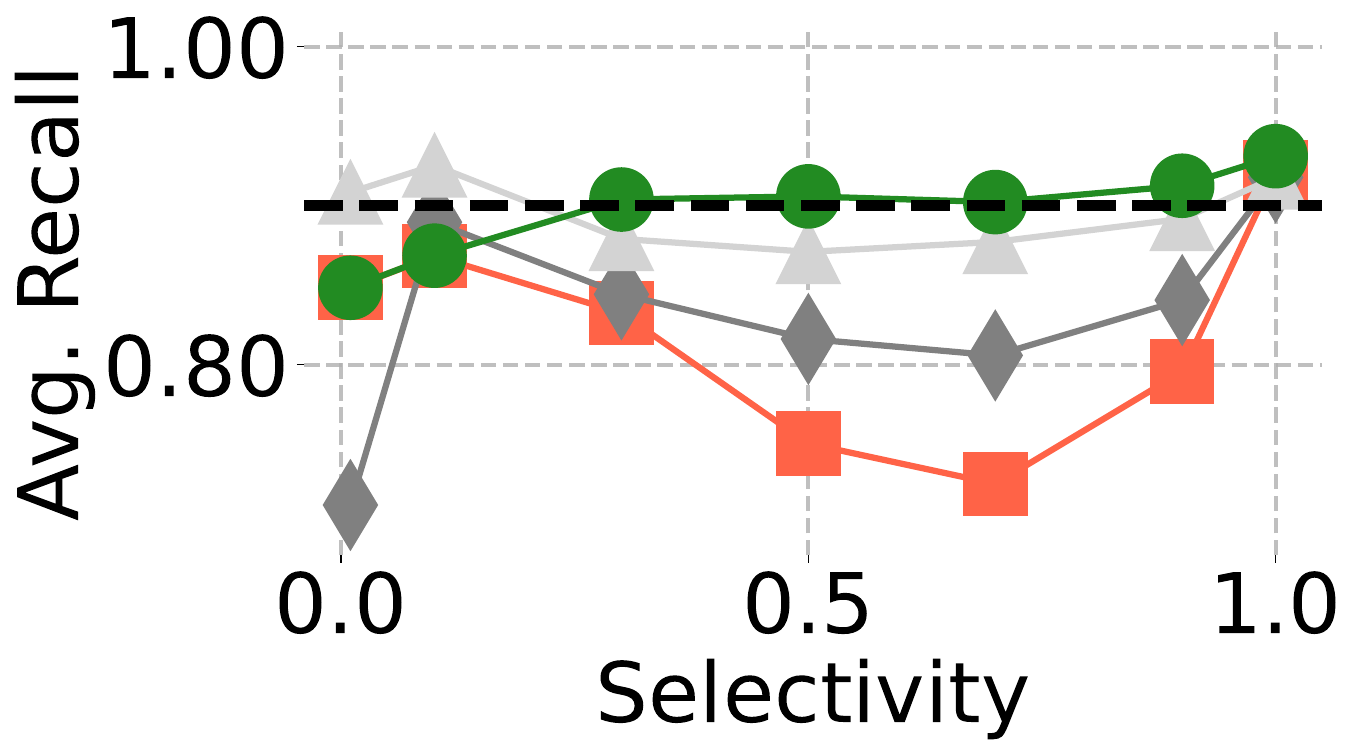}\hfill
        \includegraphics[width=0.19\textwidth]{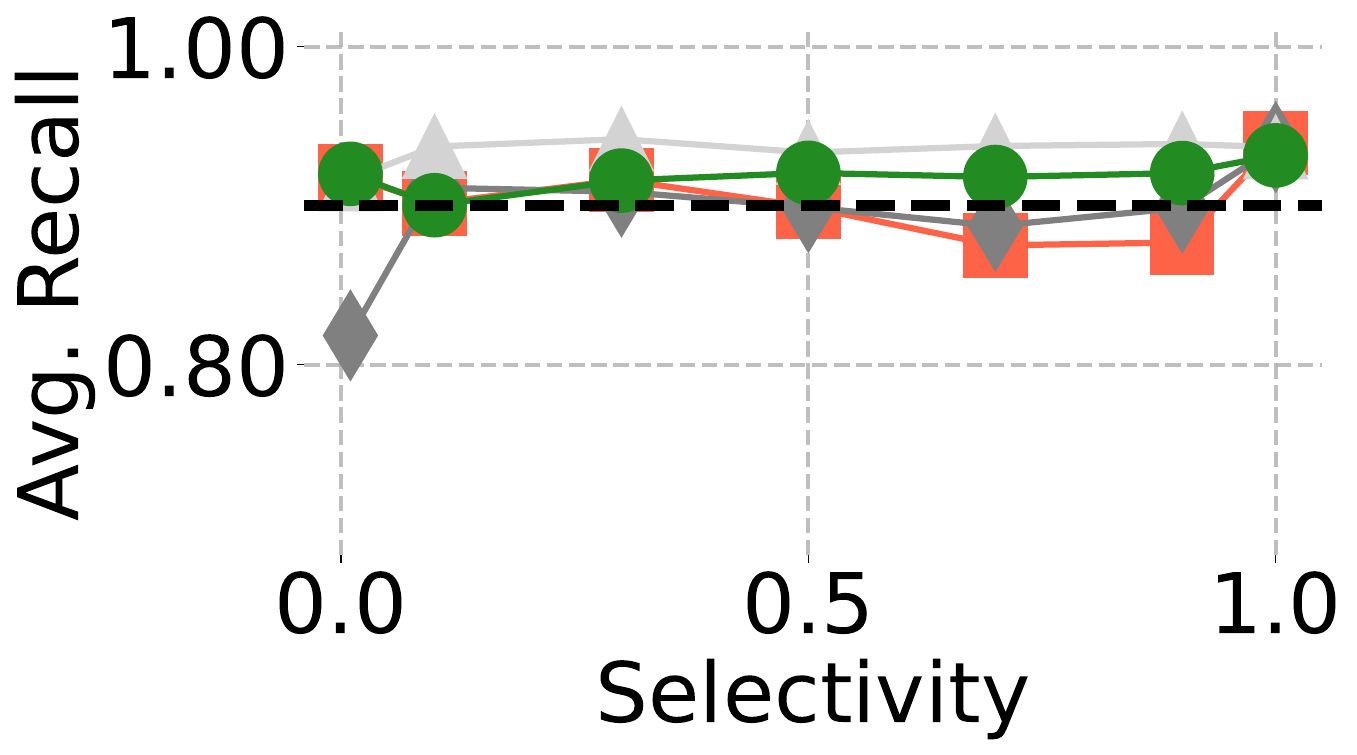}\hfill
        \includegraphics[width=0.19\textwidth]{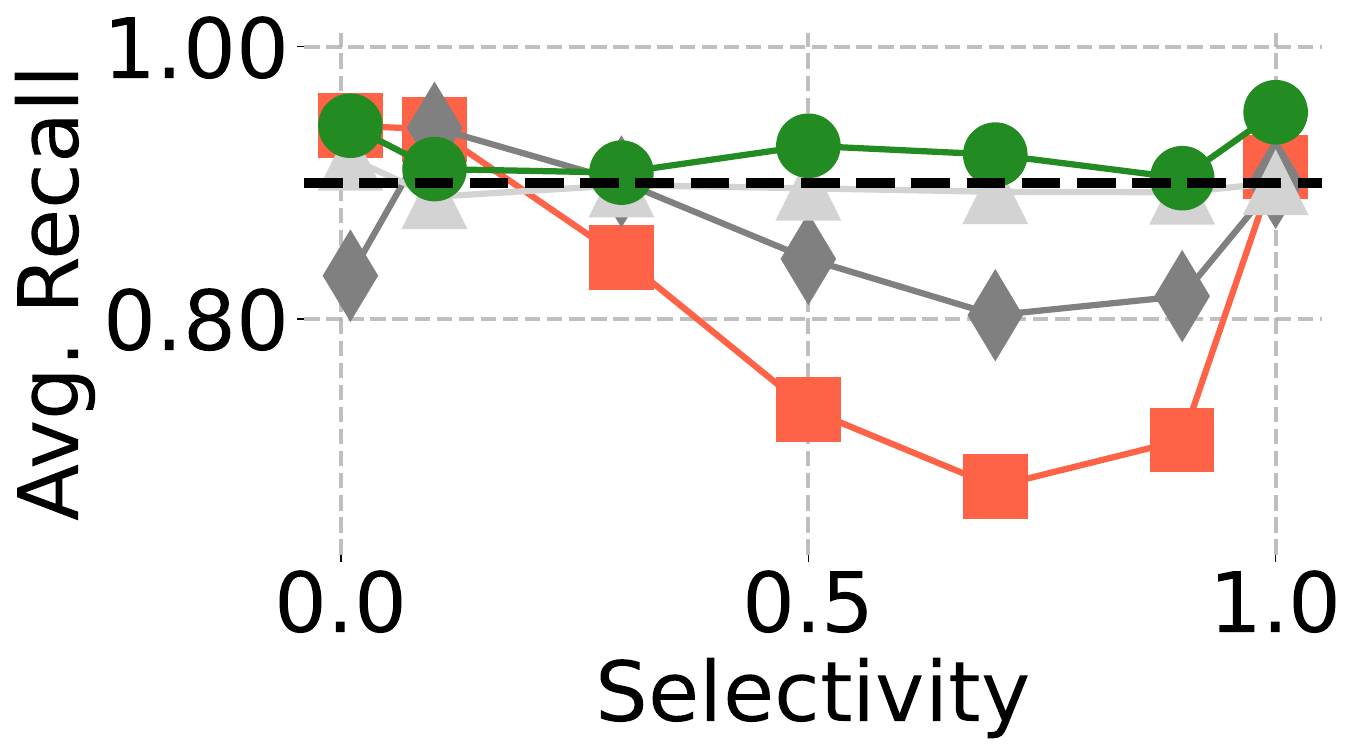}\hfill
        \includegraphics[width=0.19\textwidth]{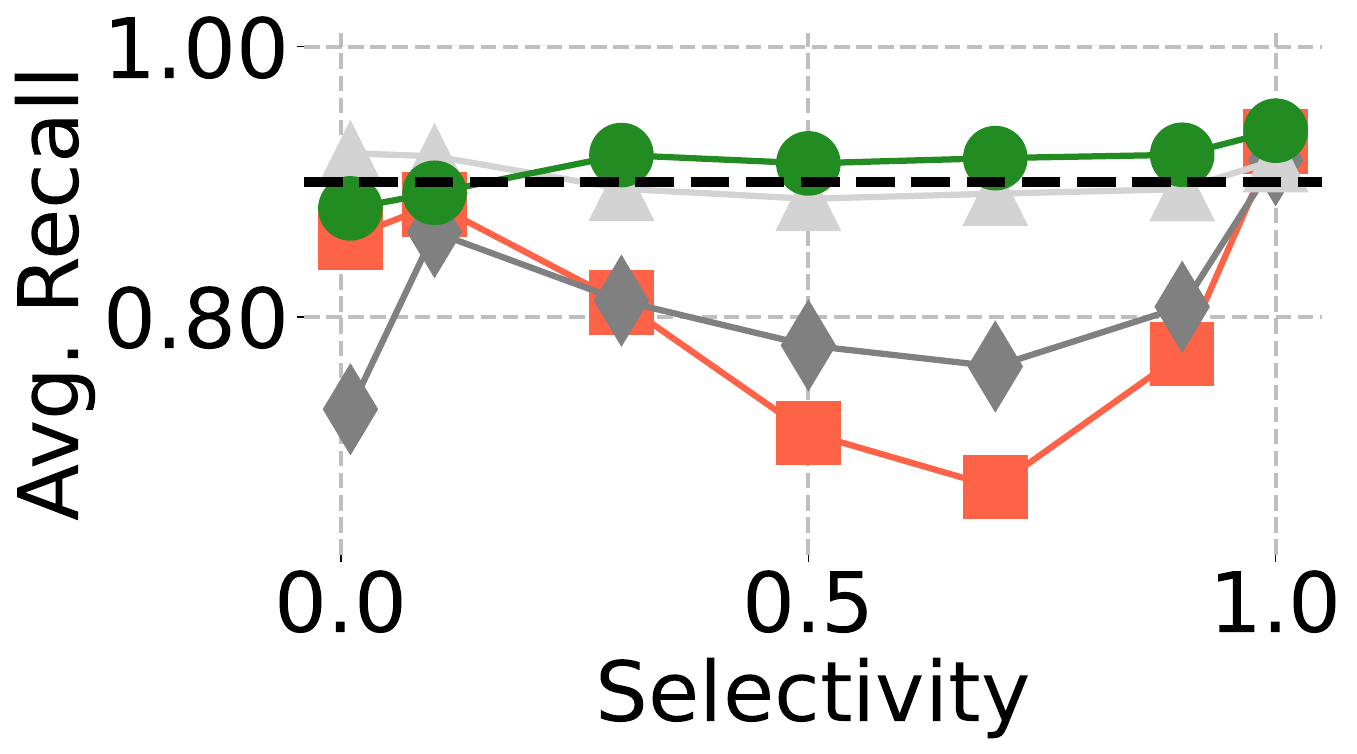}\hfill
        \includegraphics[width=0.19\textwidth]{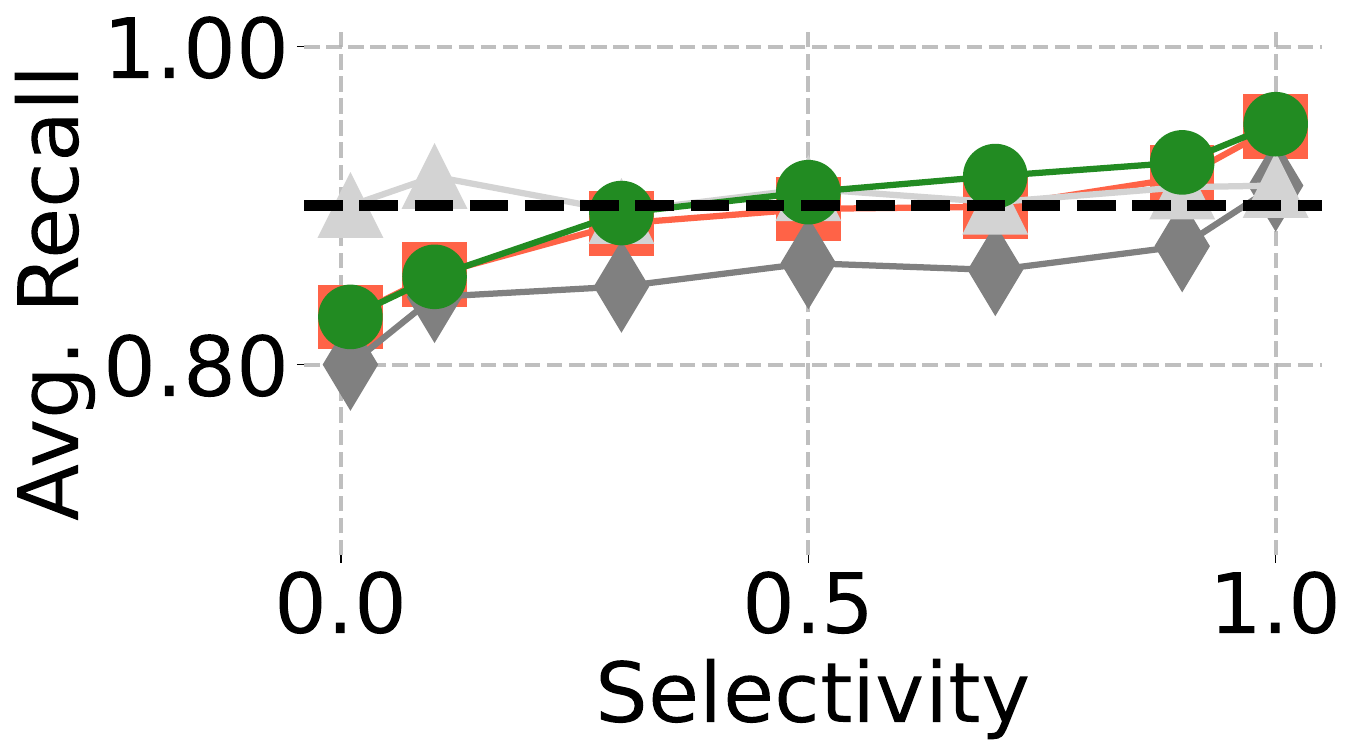}
        \vspace{-0.4cm}
        \caption{ACORN: Achieved recalls for varying selectivity (No Correlation, $k=100$, $R_t=0.9$).}
        \label{fig:recall-vs-selectivity-rt0.9-k100-acorn-no-corr}
    \end{minipage}

    
    \begin{minipage}[t]{\textwidth}
        \includegraphics[width=0.19\textwidth]{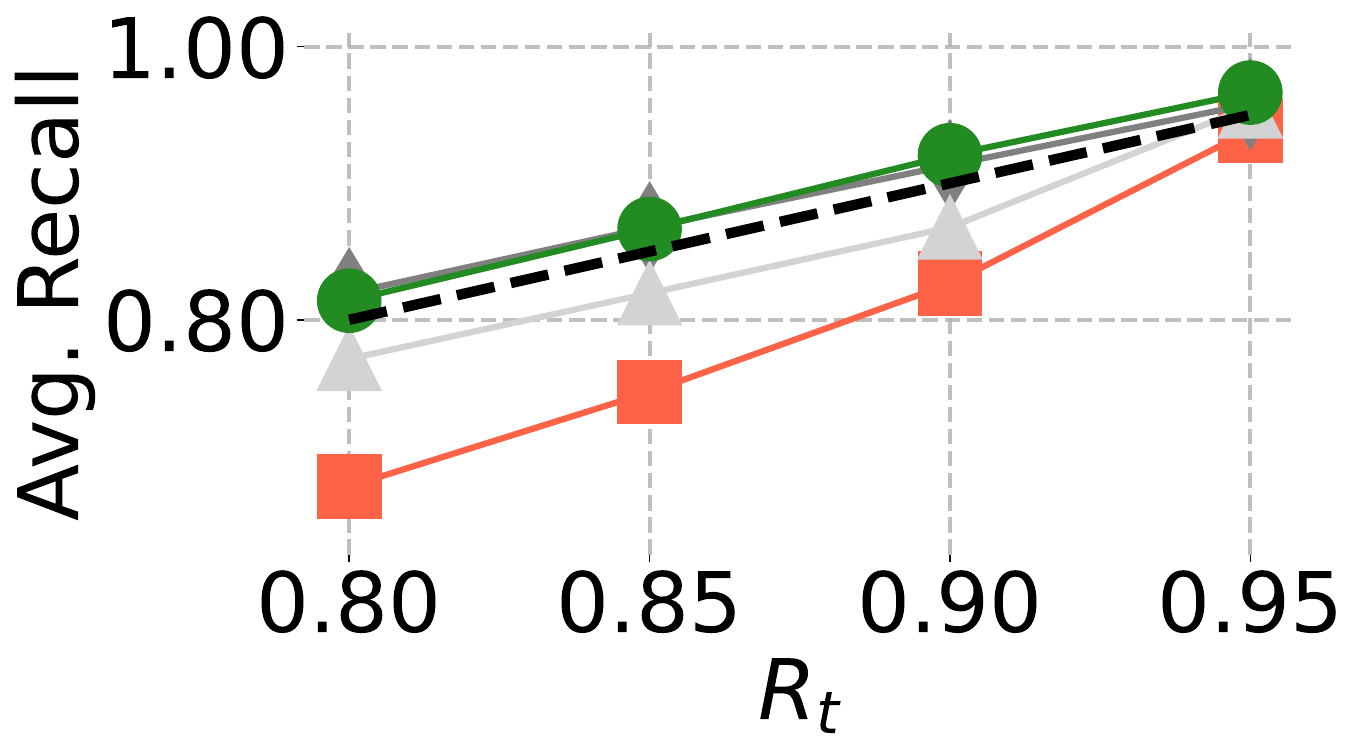}\hfill
        \includegraphics[width=0.19\textwidth]{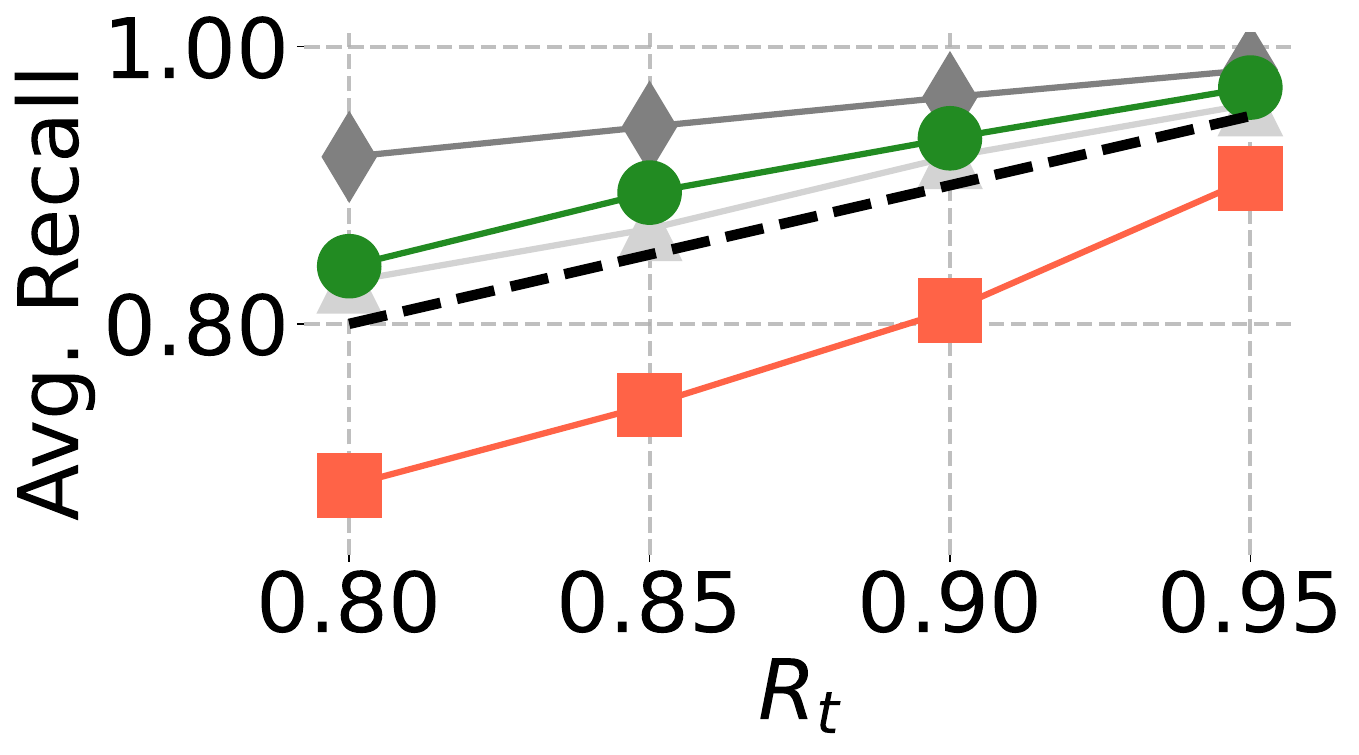}\hfill
        \includegraphics[width=0.19\textwidth]{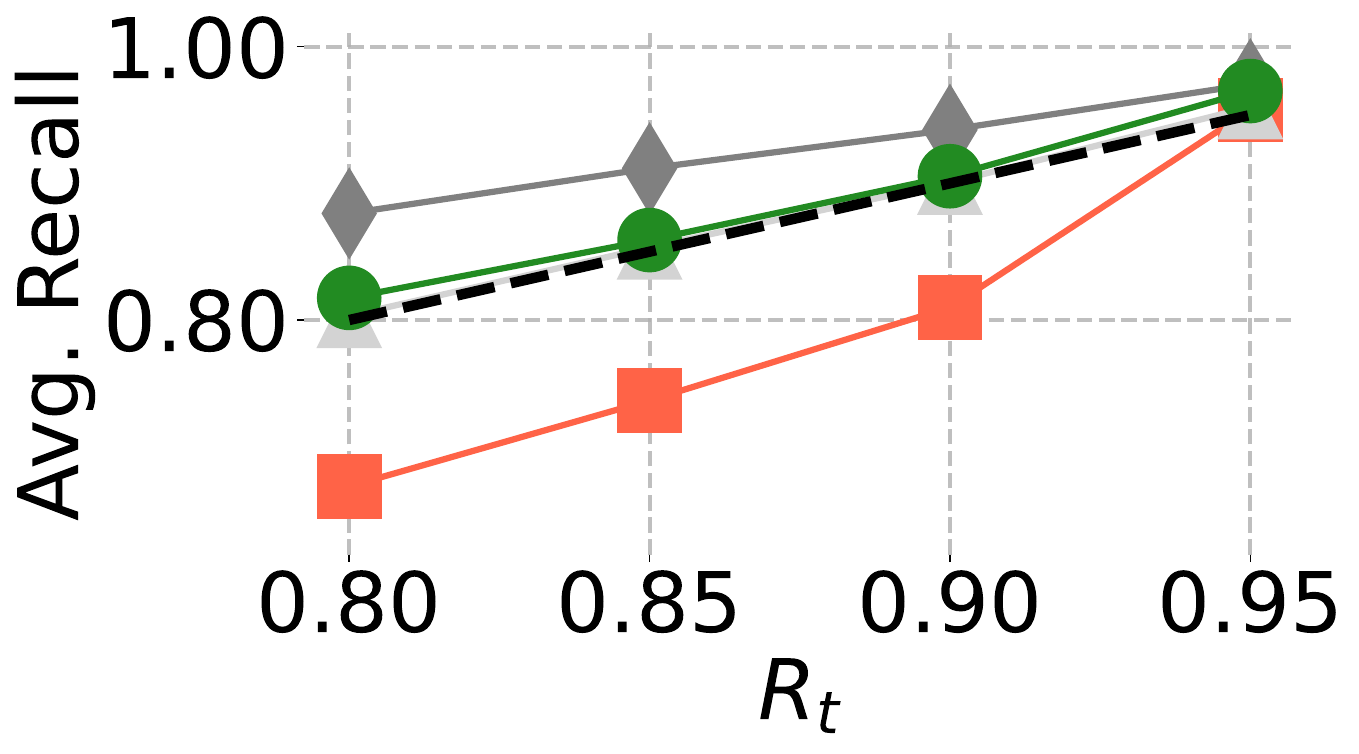}\hfill
        \includegraphics[width=0.19\textwidth]{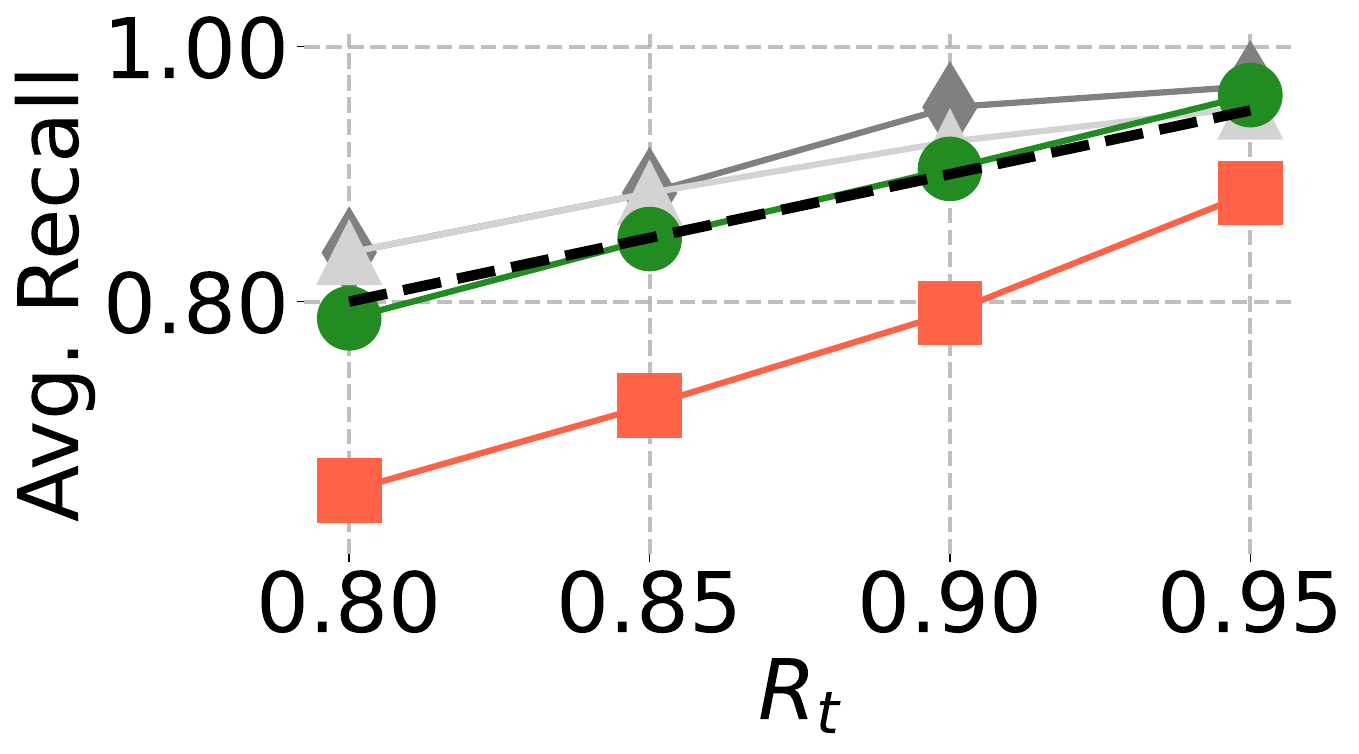}\hfill
        \includegraphics[width=0.19\textwidth]{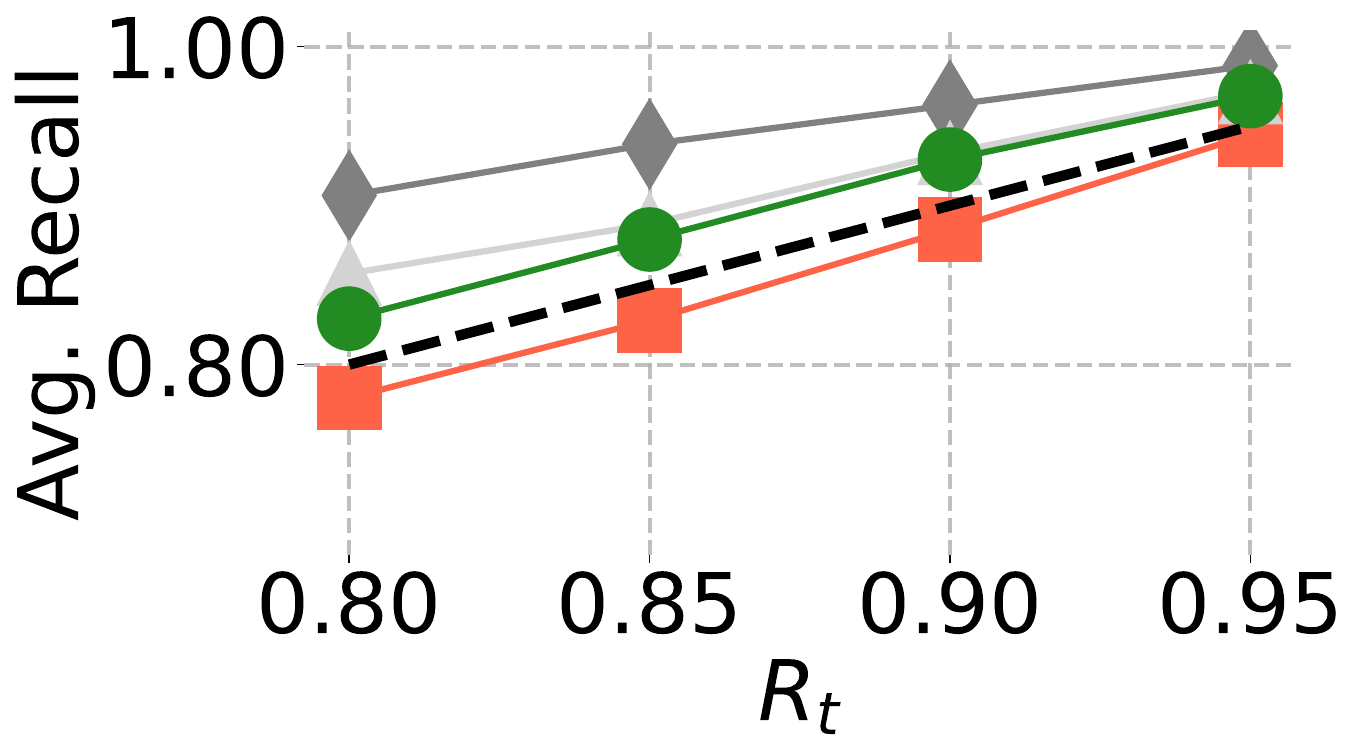}
        \vspace{-0.4cm}
        \caption{Sweeping: Achieved recalls for varying recall target (No Correlation, $k=100$, $sel=0.3$).}
        \label{fig:recall-vs-rt-k100-sel0.3-sweeping-no-corr}
    \end{minipage}

    \begin{minipage}[t]{\textwidth}
        \includegraphics[width=0.19\textwidth]{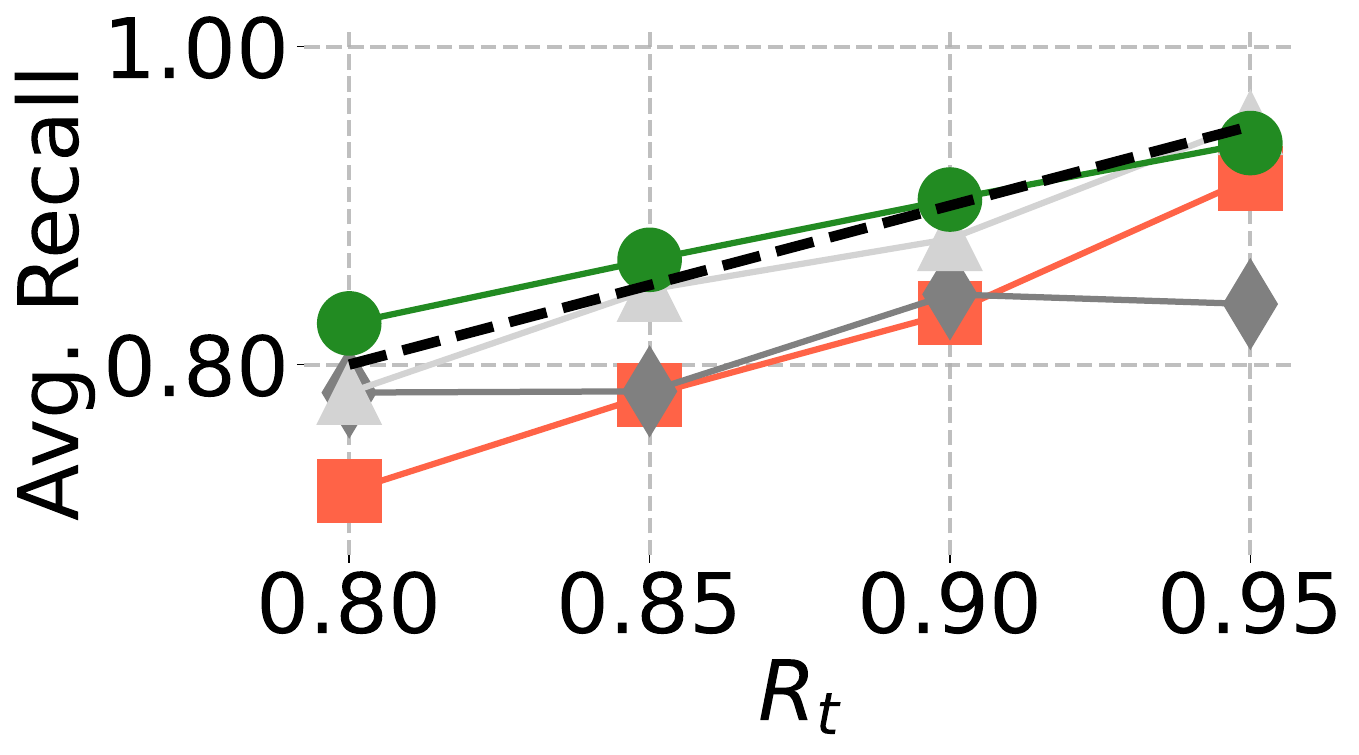}\hfill
        \includegraphics[width=0.19\textwidth]{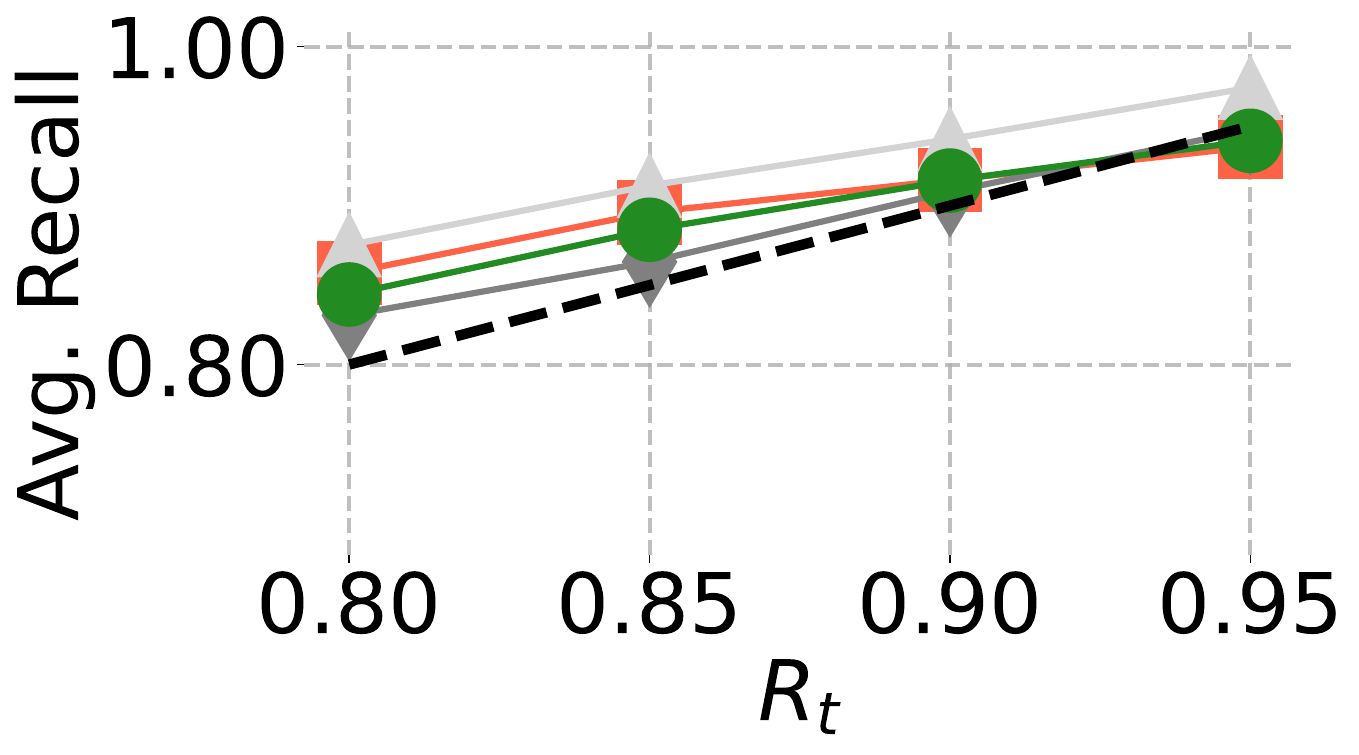}\hfill
        \includegraphics[width=0.19\textwidth]{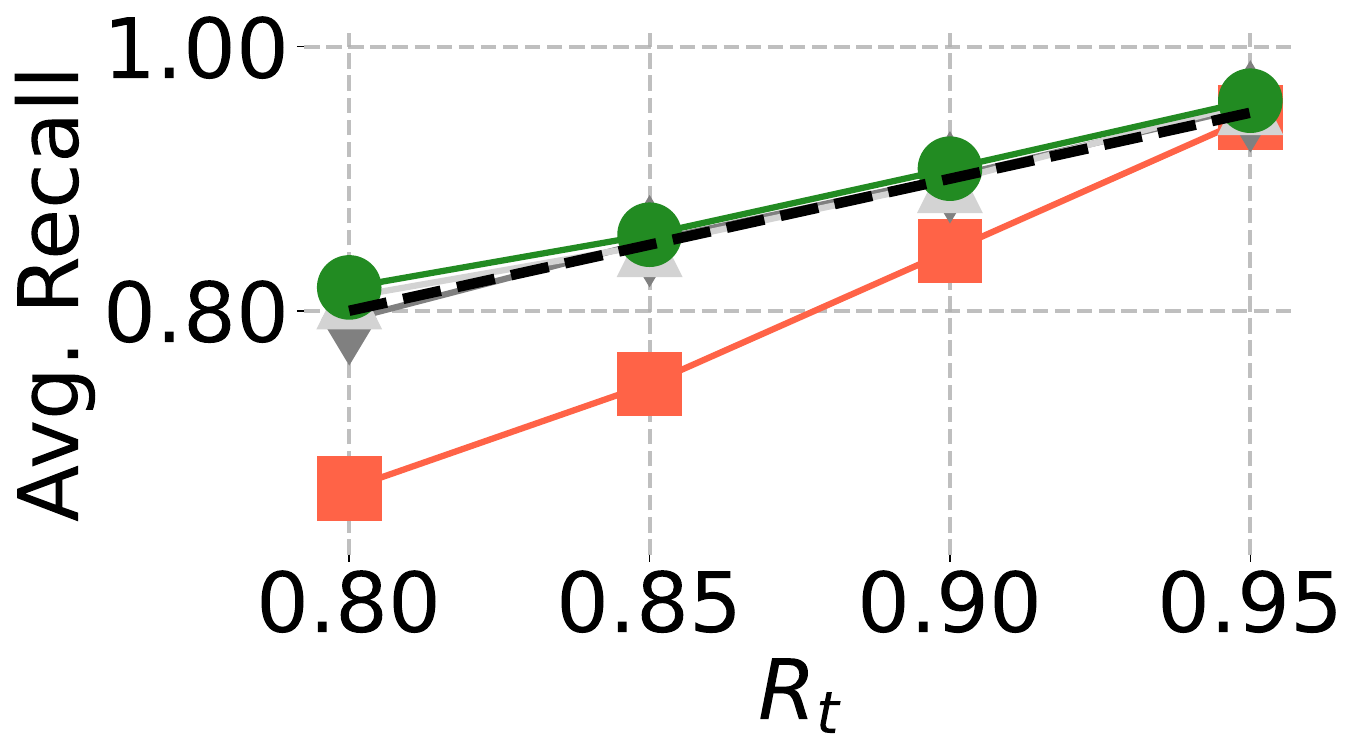}\hfill
        \includegraphics[width=0.19\textwidth]{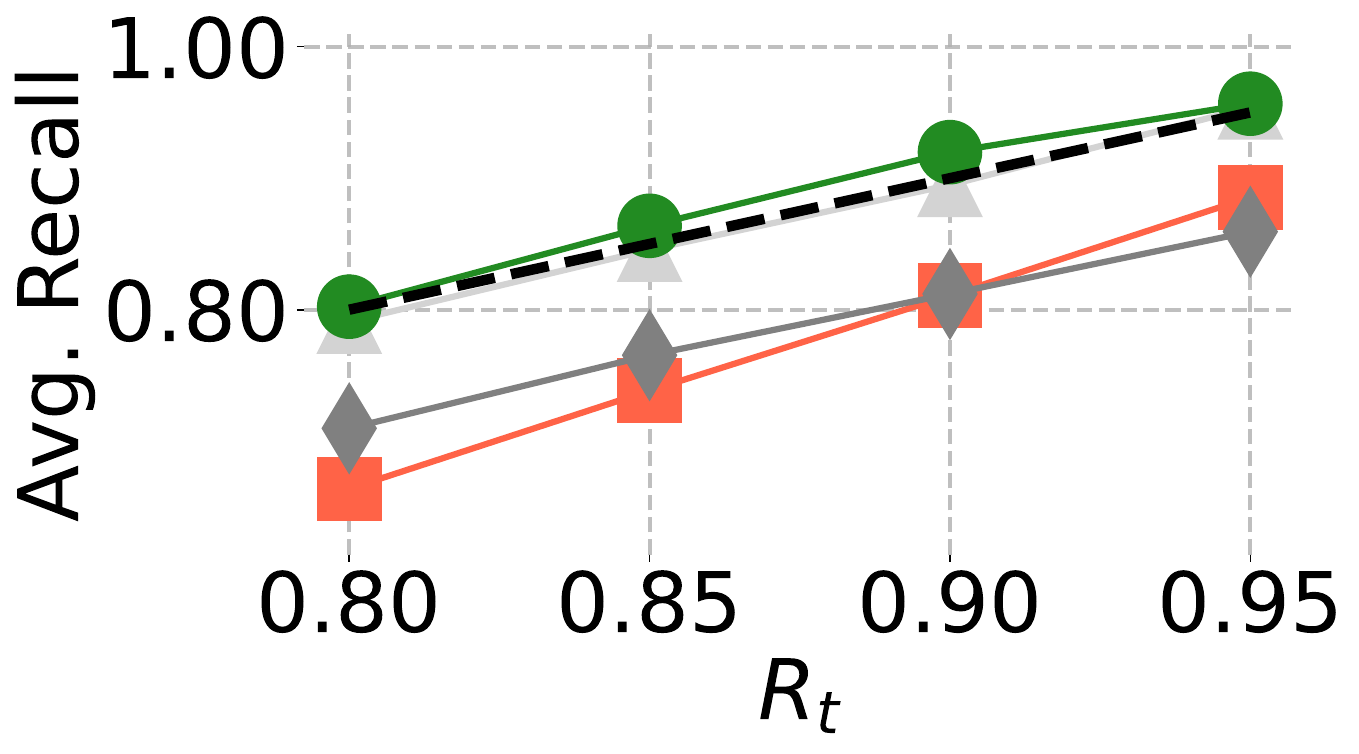}\hfill
        \includegraphics[width=0.19\textwidth]{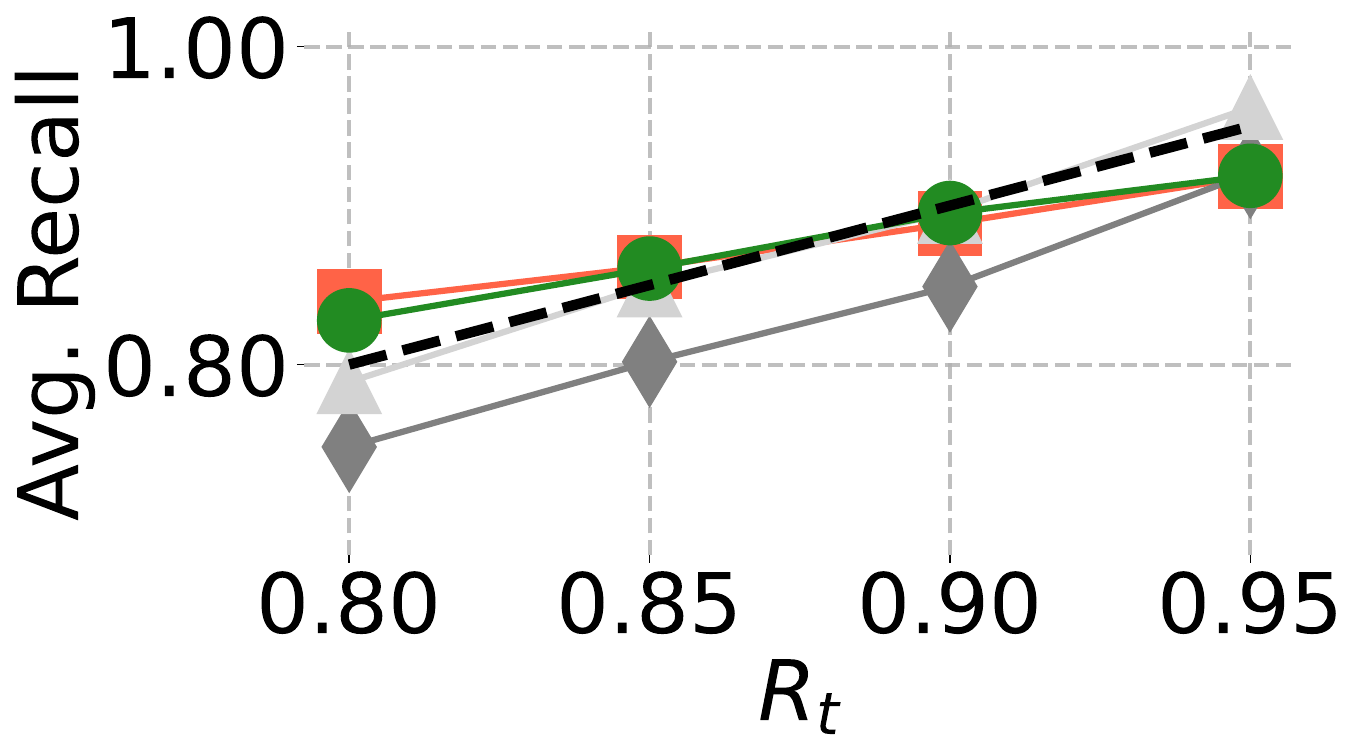}
        \vspace{-0.4cm}
        \caption{ACORN: Achieved recalls for varying recall target (No Correlation, $k=100$, $sel=0.3$).}
        \label{fig:recall-vs-rt-k100-sel0.3-acorn-no-corr}
    \end{minipage}

    
    \begin{minipage}[t]{\textwidth}
        \includegraphics[width=0.19\textwidth]{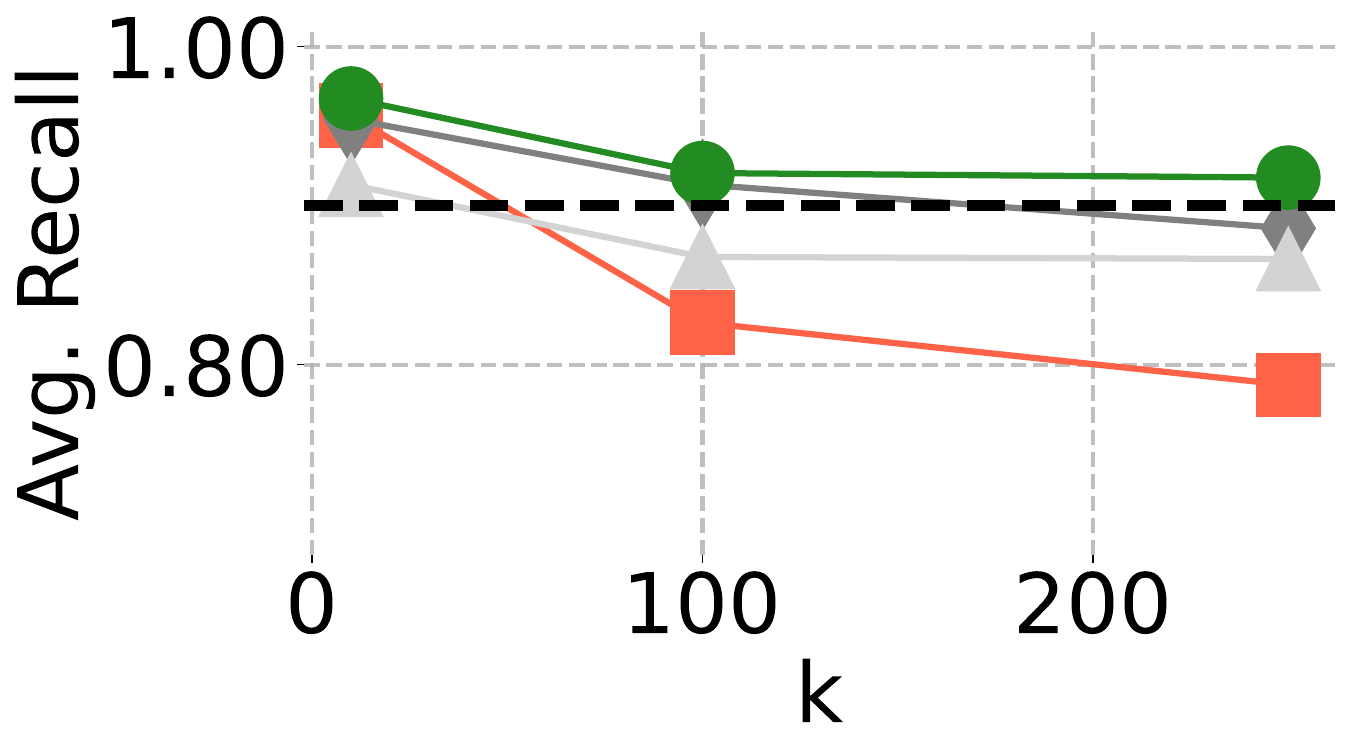}\hfill
        \includegraphics[width=0.19\textwidth]{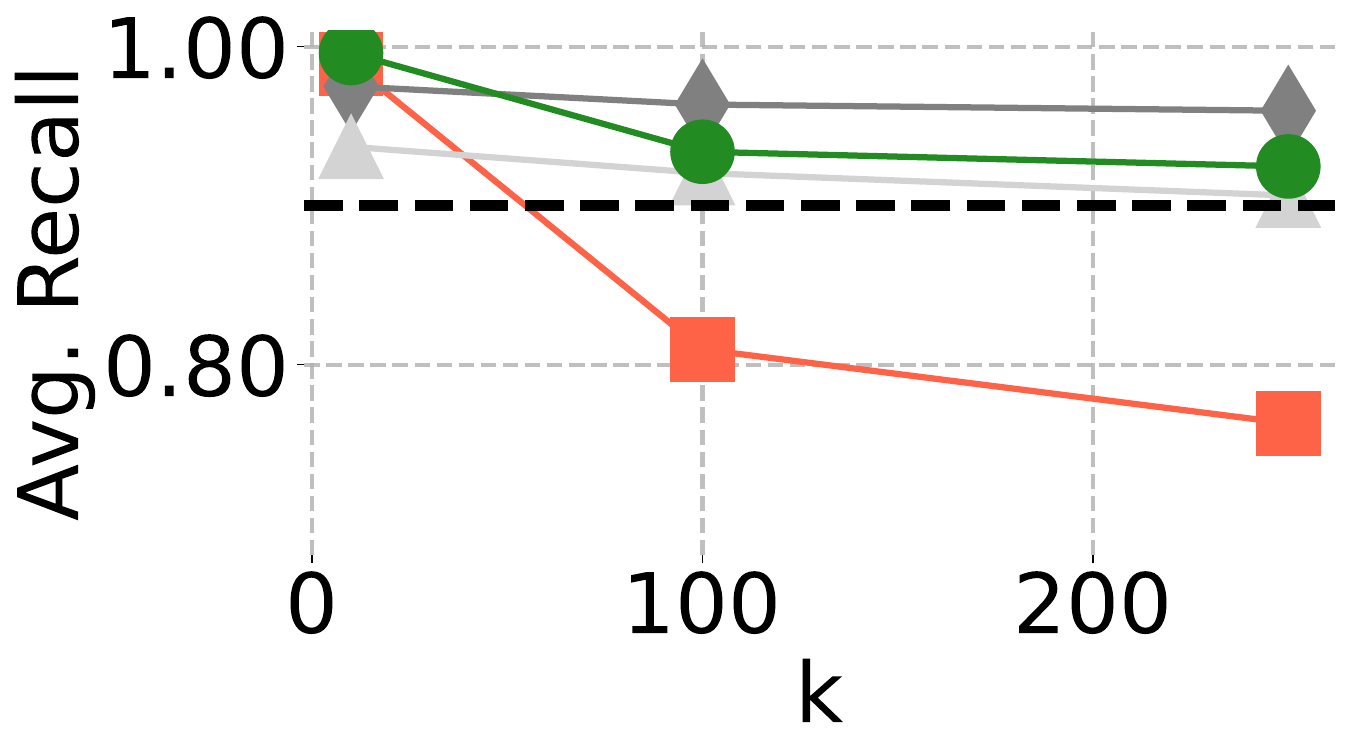}\hfill
        \includegraphics[width=0.19\textwidth]{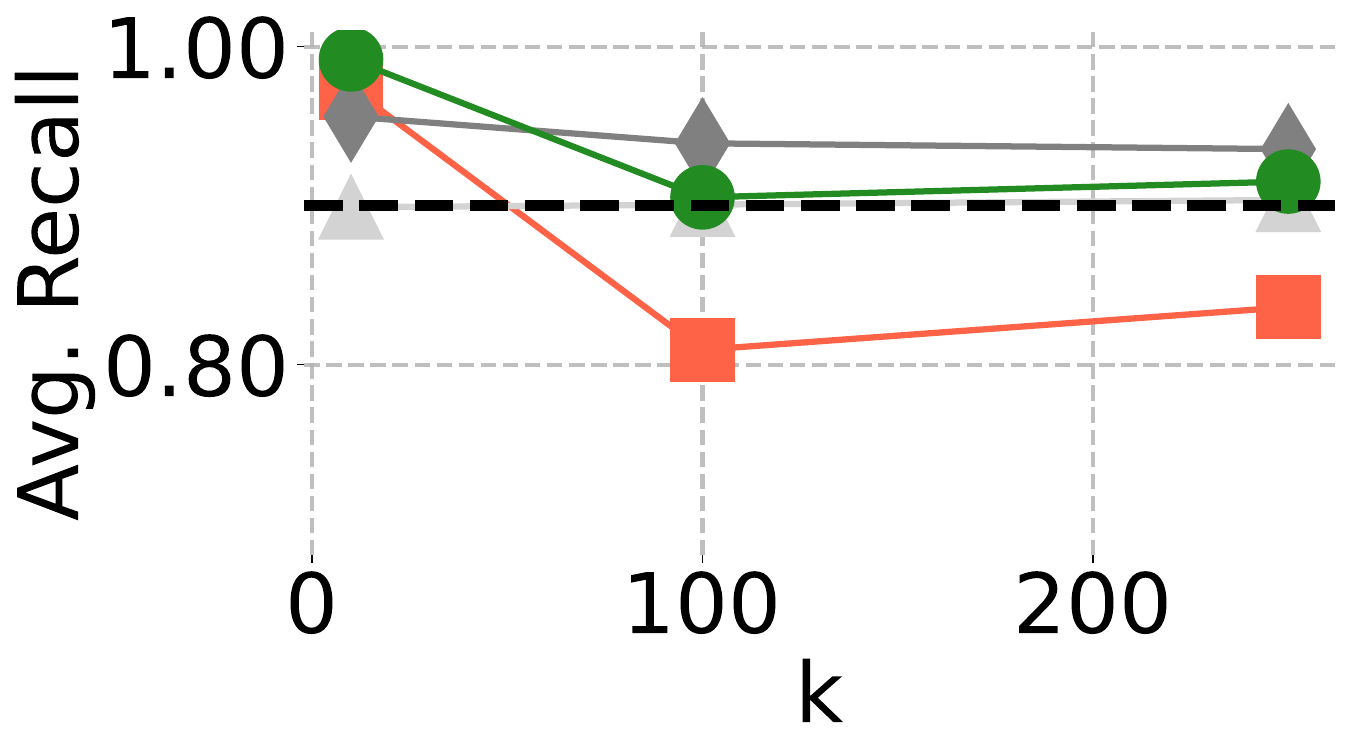}\hfill
        \includegraphics[width=0.19\textwidth]{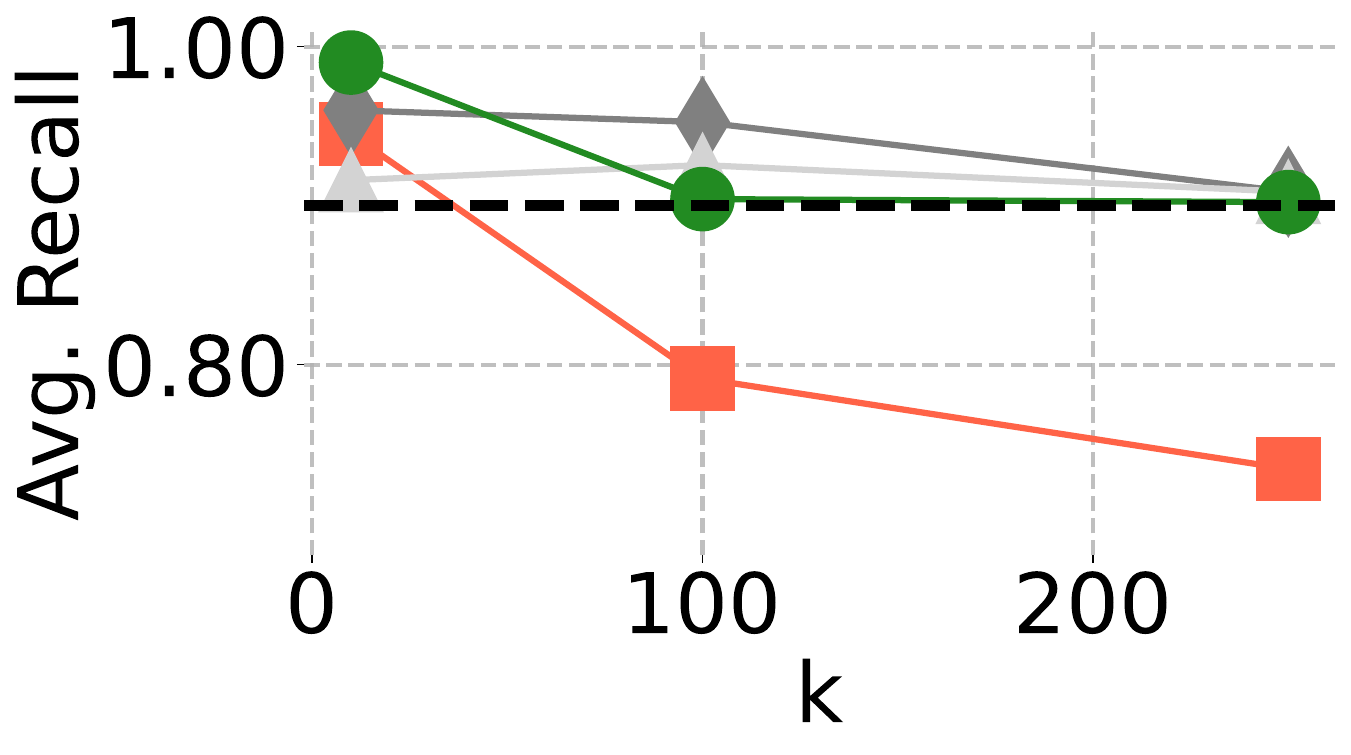}\hfill
        \includegraphics[width=0.19\textwidth]{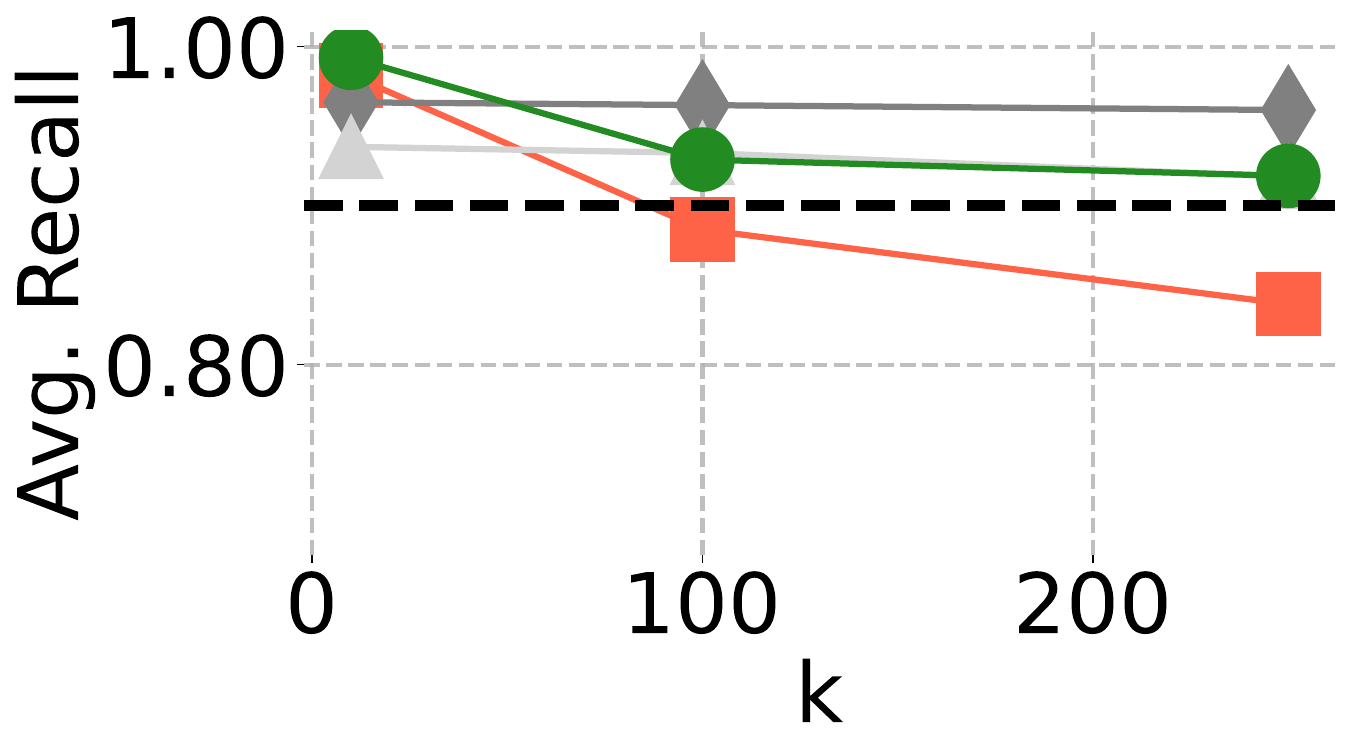}
        \vspace{-0.4cm}
        \caption{Sweeping: Achieved recalls for varying value of $k$ (No correlation, $R_t=0.9$, $sel=0.3$).}
        \label{fig:recall-vs-k-rt0.9-sel0.3-sweeping-no-corr}
    \end{minipage}
    
    \begin{minipage}[t]{\textwidth}
        \includegraphics[width=0.19\textwidth]{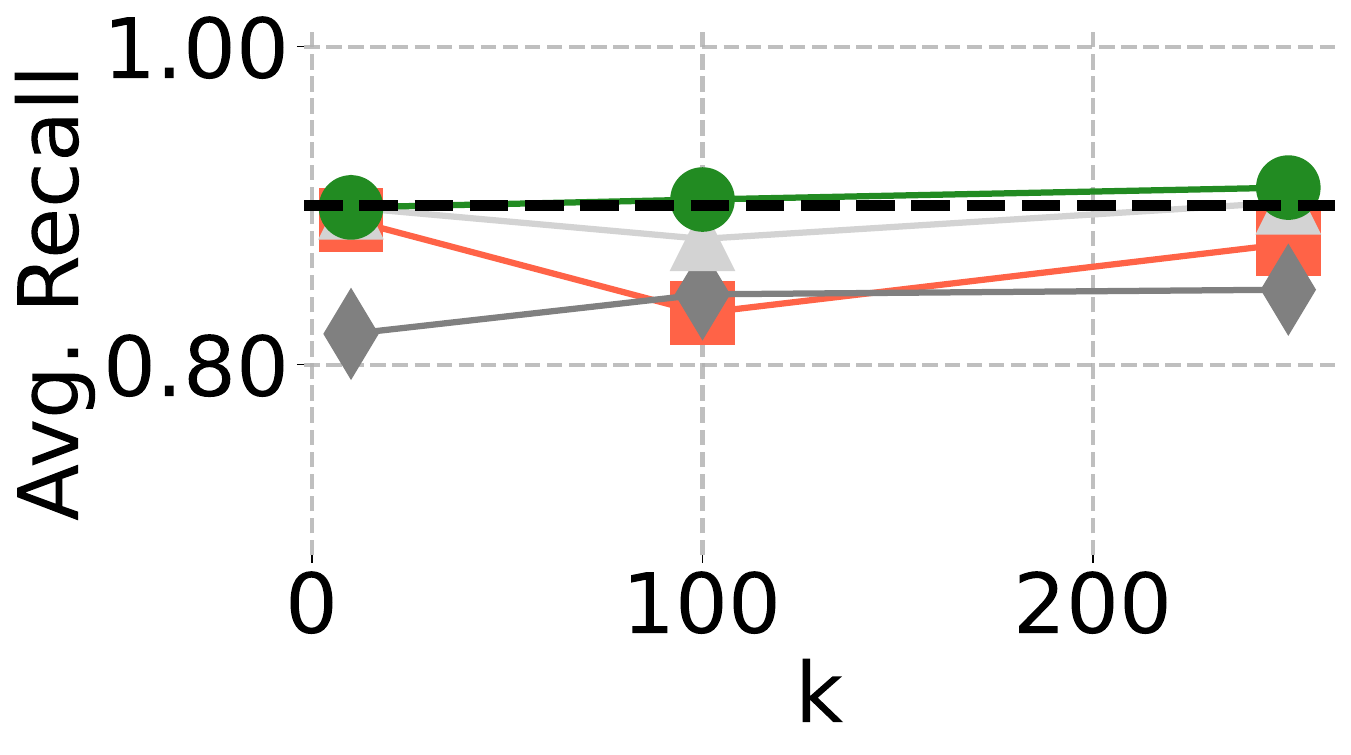}\hfill
        \includegraphics[width=0.19\textwidth]{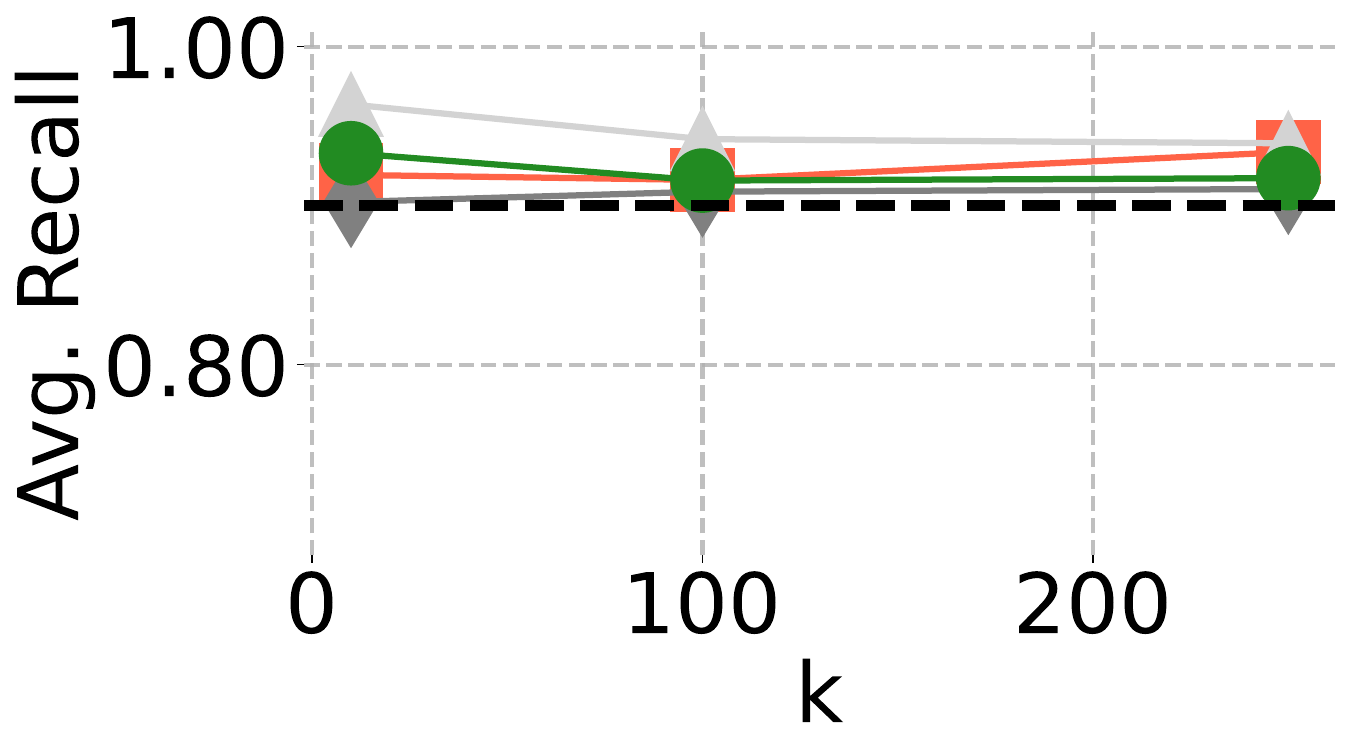}\hfill
        \includegraphics[width=0.19\textwidth]{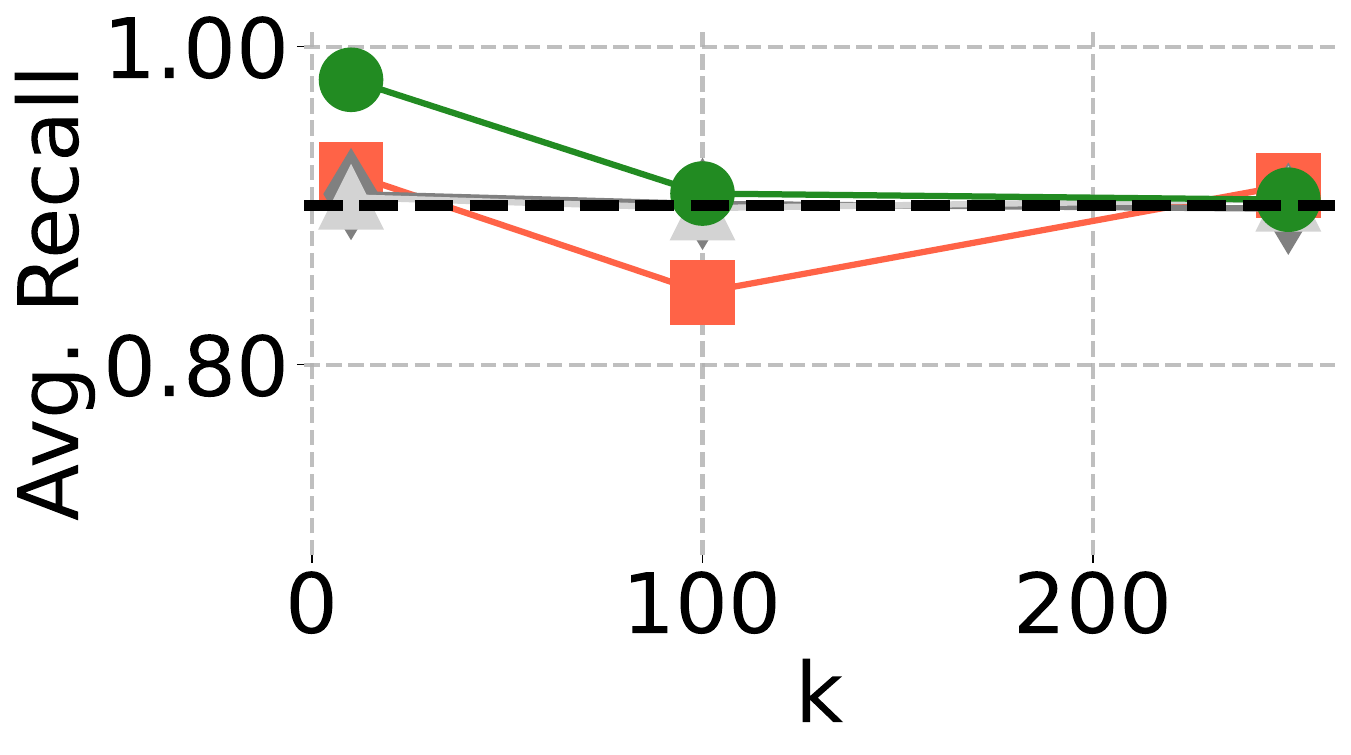}\hfill
        \includegraphics[width=0.19\textwidth]{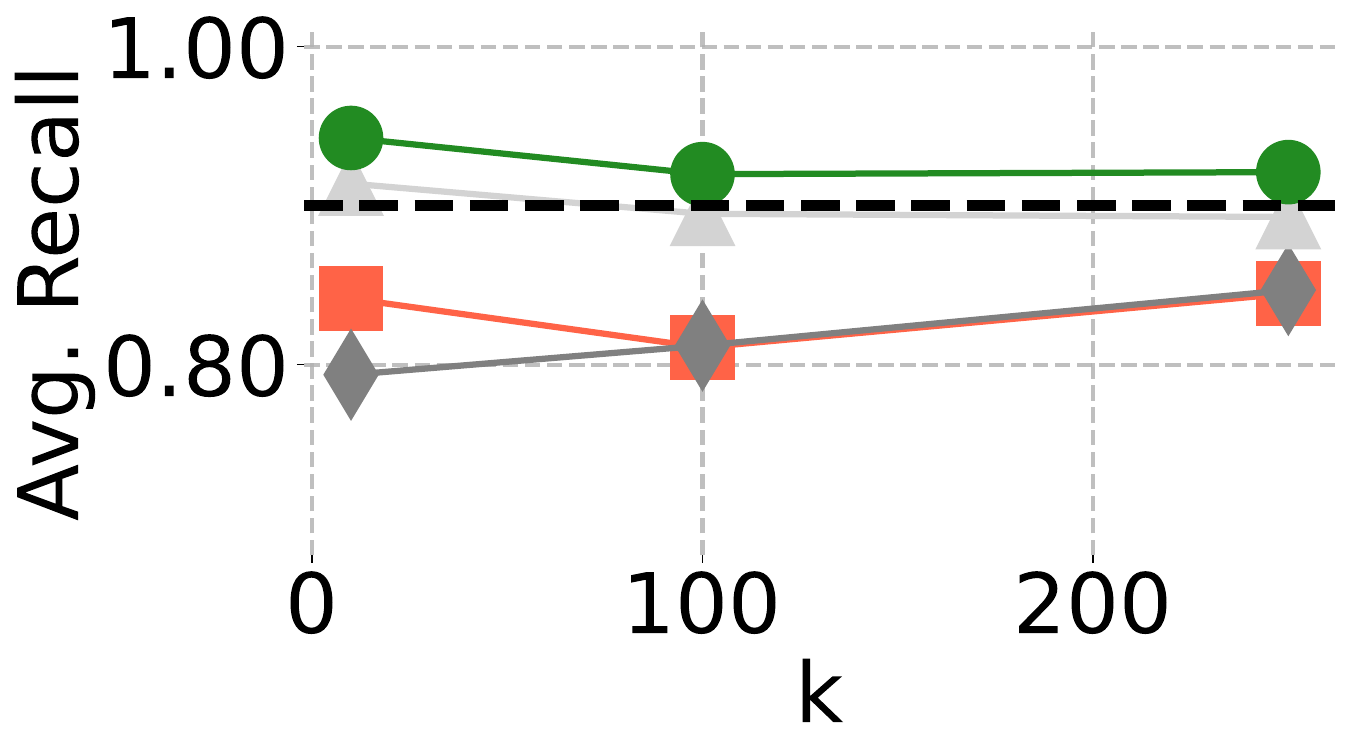}\hfill
        \includegraphics[width=0.19\textwidth]{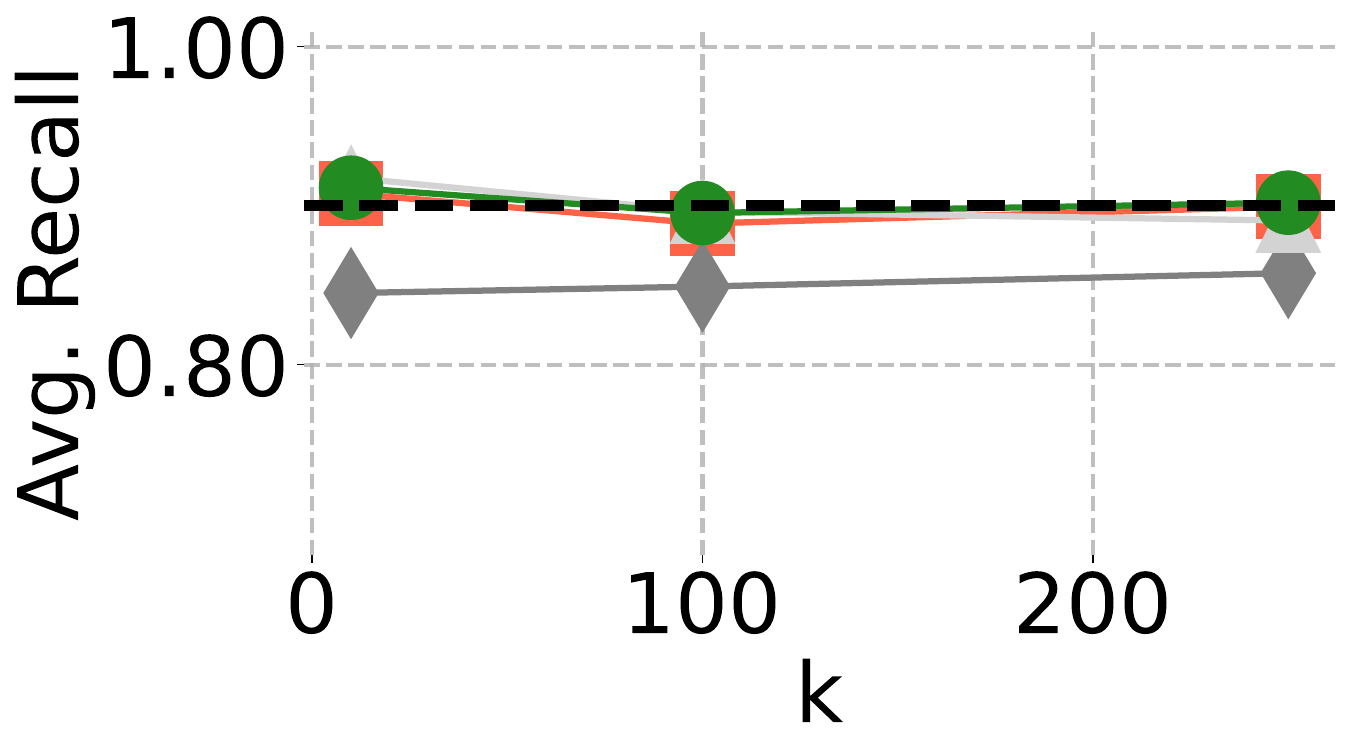}
        \vspace{-0.4cm}
        \caption{ACORN: Achieved recalls for varying value of $k$ (No correlation, $R_t=0.9$, $sel=0.3$).}
        \label{fig:recall-vs-k-rt0.9-sel0.3-acorn-no-corr}
    \end{minipage}
\end{figure*}

\begin{figure*}
    \centering
    \vspace{-0.1cm}
    \begin{adjustbox}{max width=0.6\textwidth}
        \includegraphics{figs/legend_lines.pdf}
    \end{adjustbox}

    \vspace{-0.1cm}
    \makebox[0.19\textwidth][c]{\textbf{SIFT1M}} \hfill
    \makebox[0.19\textwidth][c]{\textbf{GIST1M}} \hfill
    \makebox[0.19\textwidth][c]{\textbf{GLOVE1M}} \hfill
    \makebox[0.19\textwidth][c]{\textbf{DEEP10M}} \hfill
    \makebox[0.19\textwidth][c]{\textbf{T2I10M}}
    
    
    \begin{minipage}[t]{\textwidth}
        \includegraphics[width=0.19\textwidth]{figs/avg_recall_vs_selectivity_SIFT1M_sweeping_corr_uar_full_positive_k100_selvar_rt0.9.pdf}\hfill
        \includegraphics[width=0.19\textwidth]{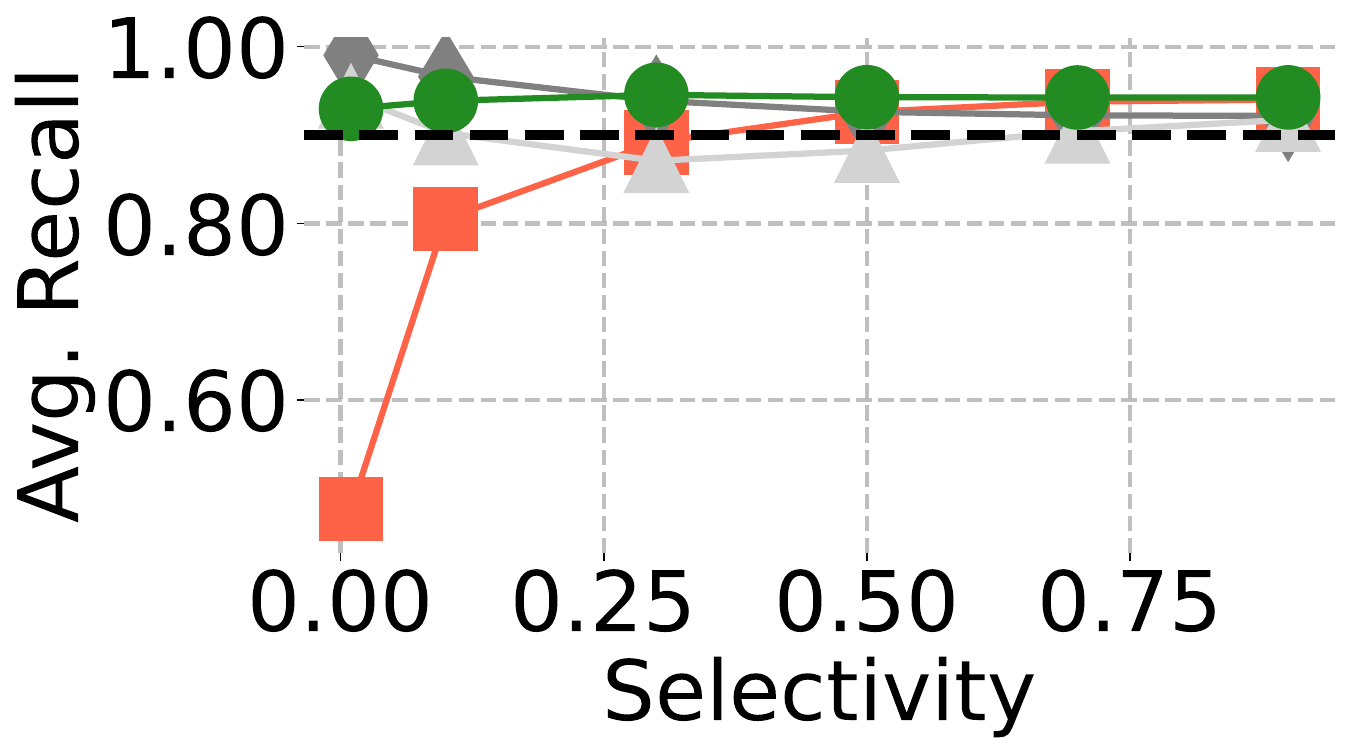}\hfill
        \includegraphics[width=0.19\textwidth]{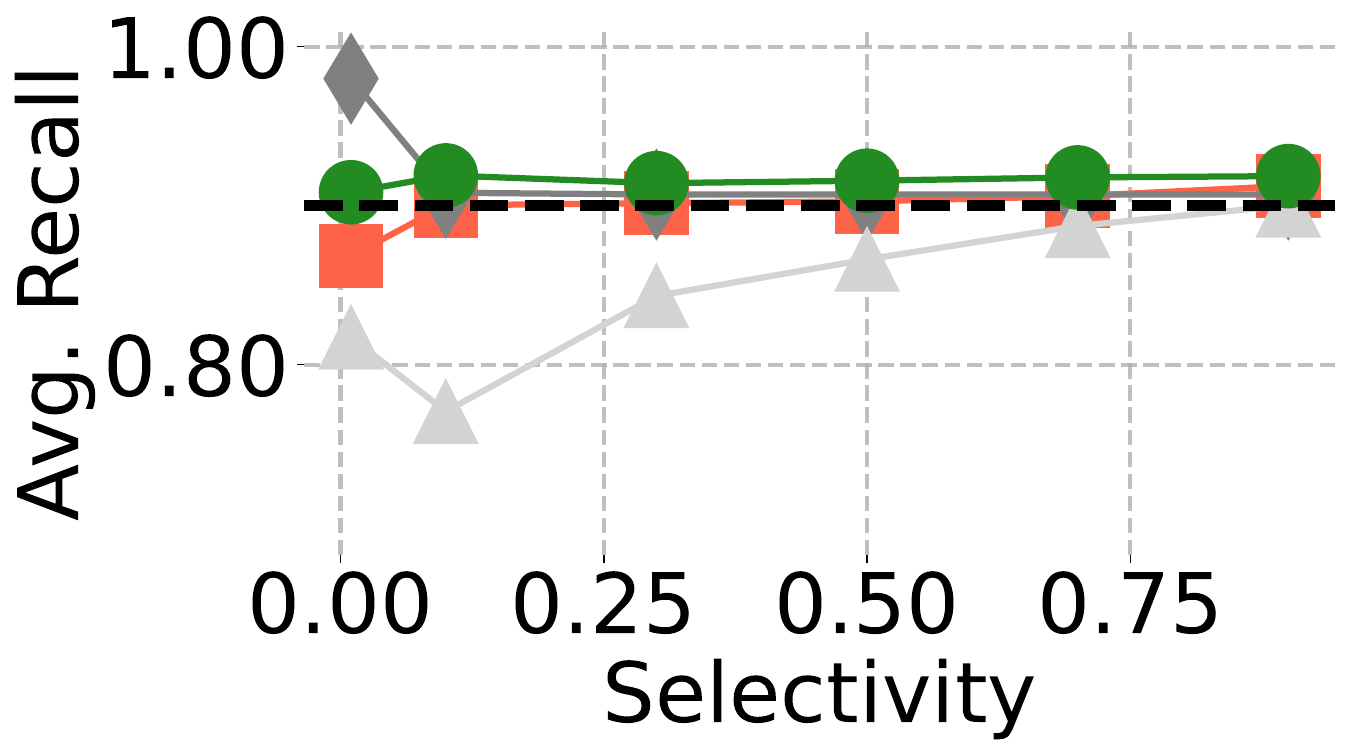}\hfill
        \includegraphics[width=0.19\textwidth]{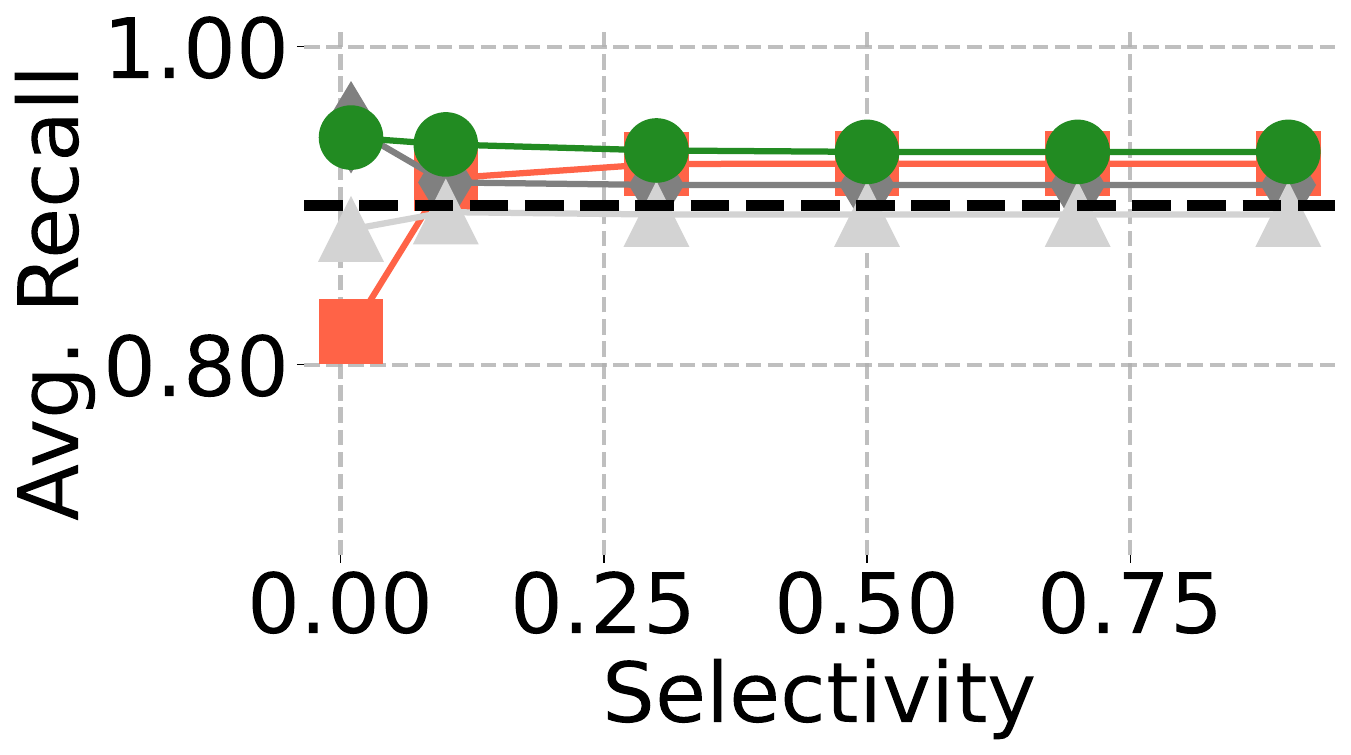}\hfill
        \includegraphics[width=0.19\textwidth]{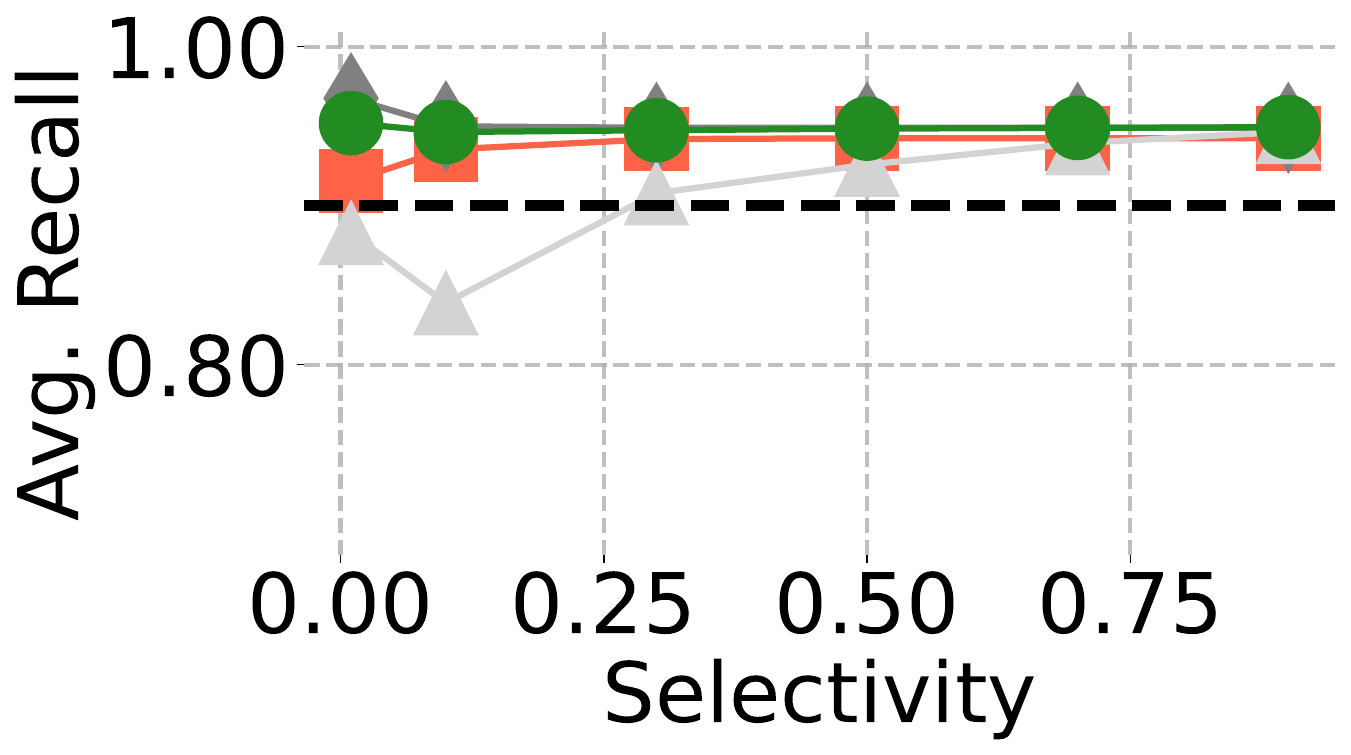}
        \vspace{-0.4cm}
        \caption{Sweeping: Achieved recalls for varying selectivity (Positive Correlation, $k=100$, $R_t=0.9$).}
        \label{fig:recall-vs-selectivity-rt0.9-k100-sweeping-pos-corr}
    \end{minipage}

    \begin{minipage}[t]{\textwidth}
        \includegraphics[width=0.19\textwidth]{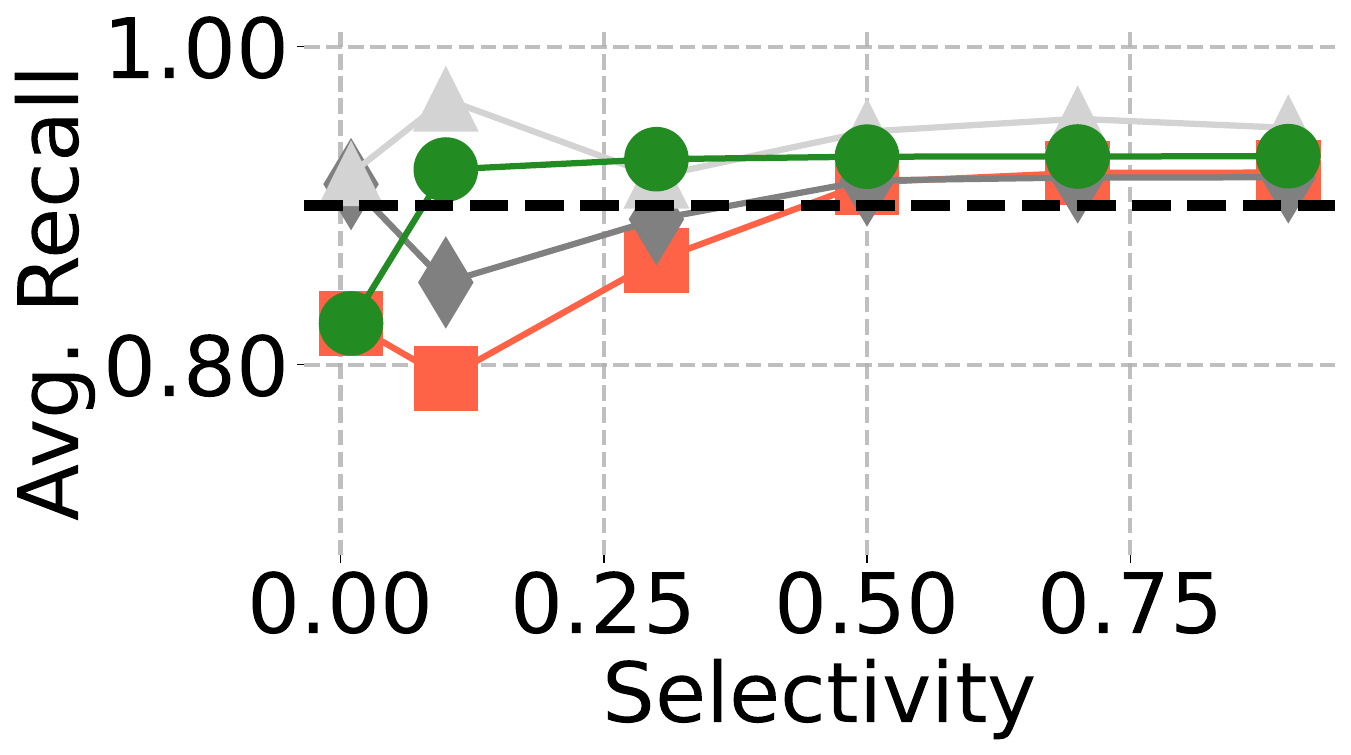}\hfill
        \includegraphics[width=0.19\textwidth]{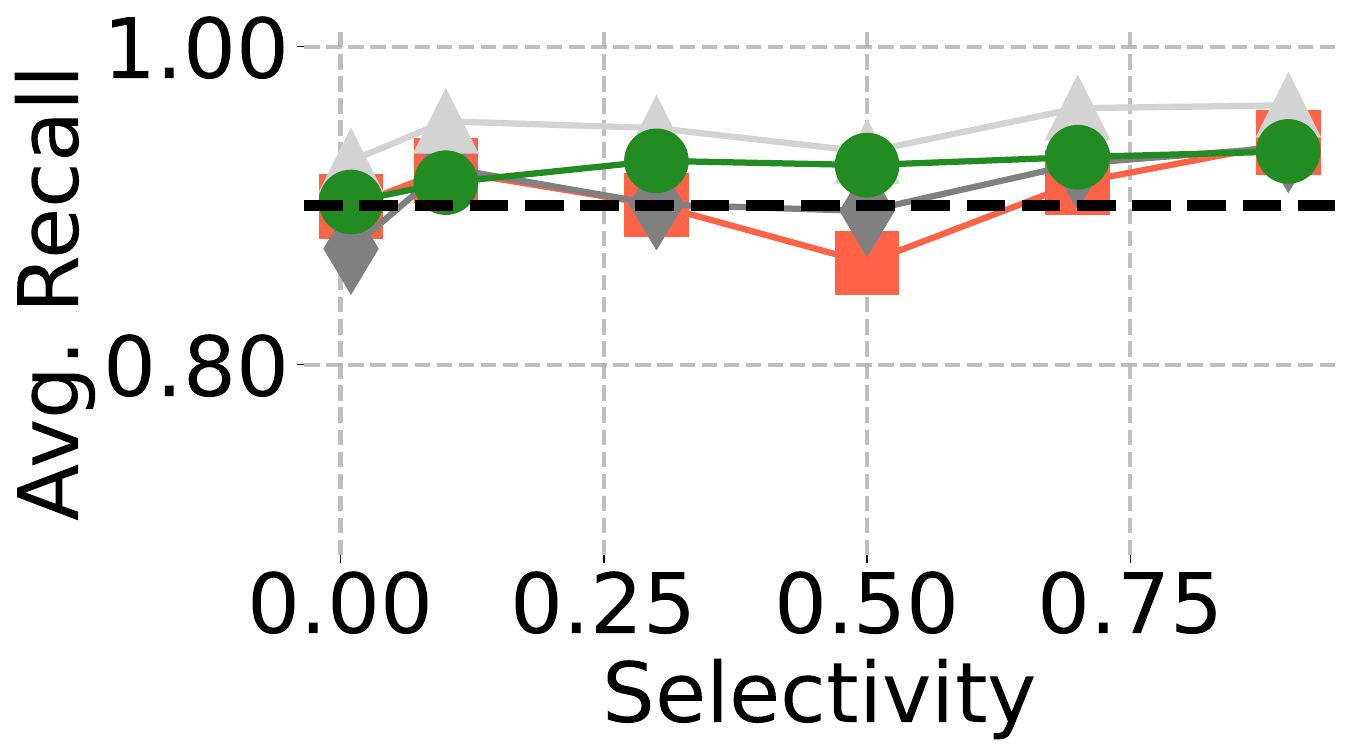}\hfill
        \includegraphics[width=0.19\textwidth]{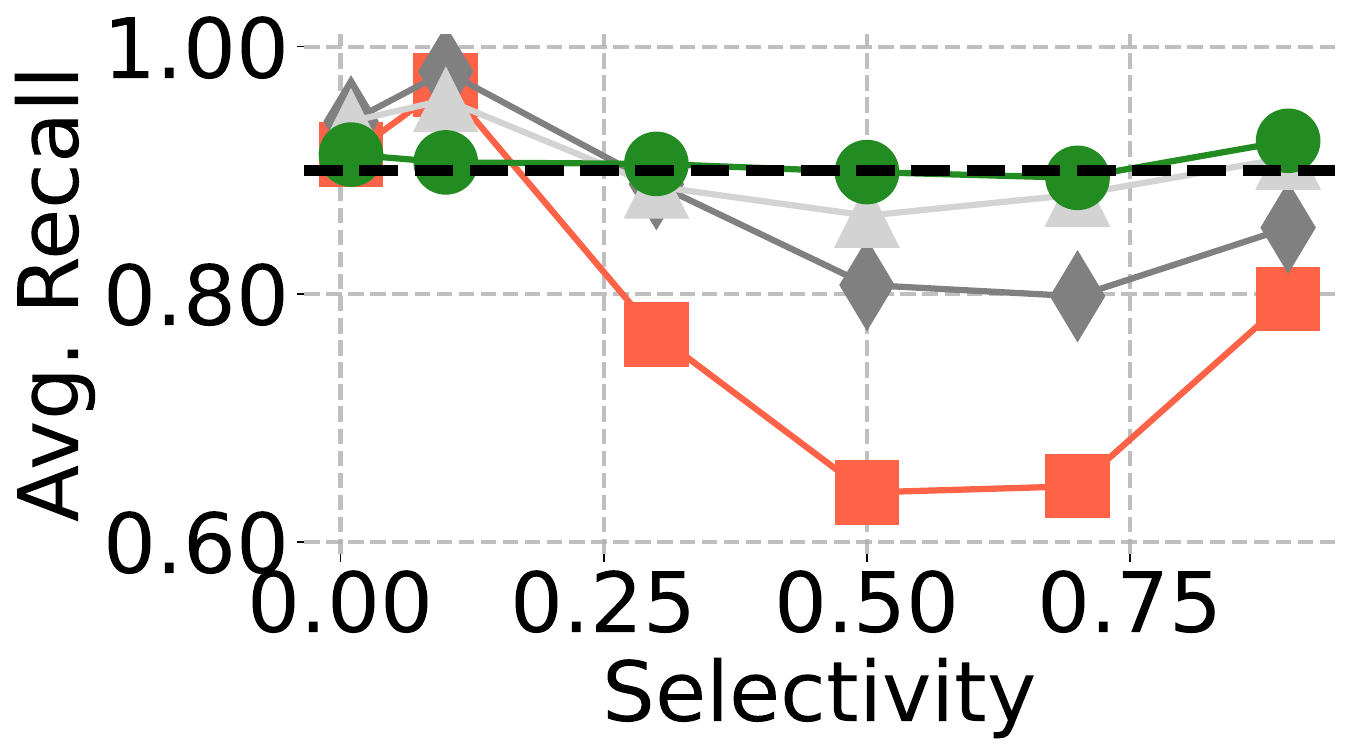}\hfill
        \includegraphics[width=0.19\textwidth]{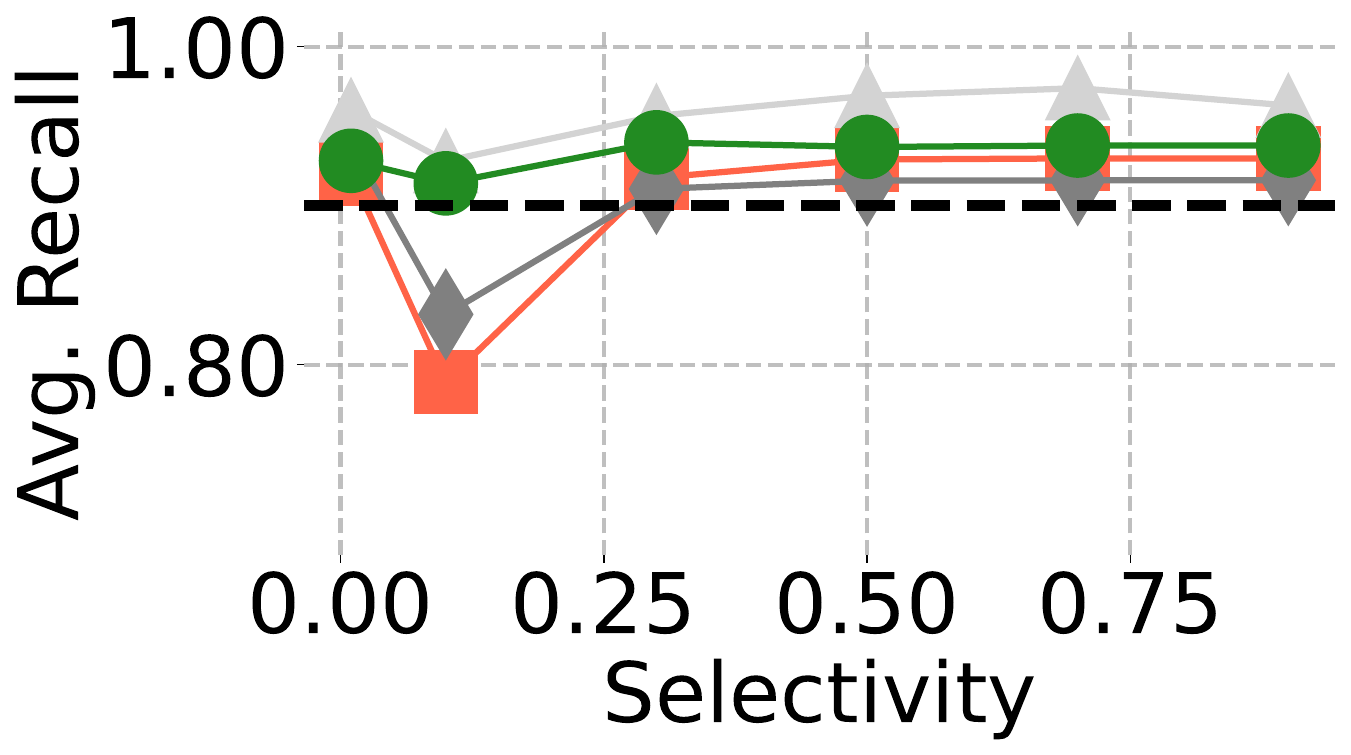}\hfill
        \includegraphics[width=0.19\textwidth]{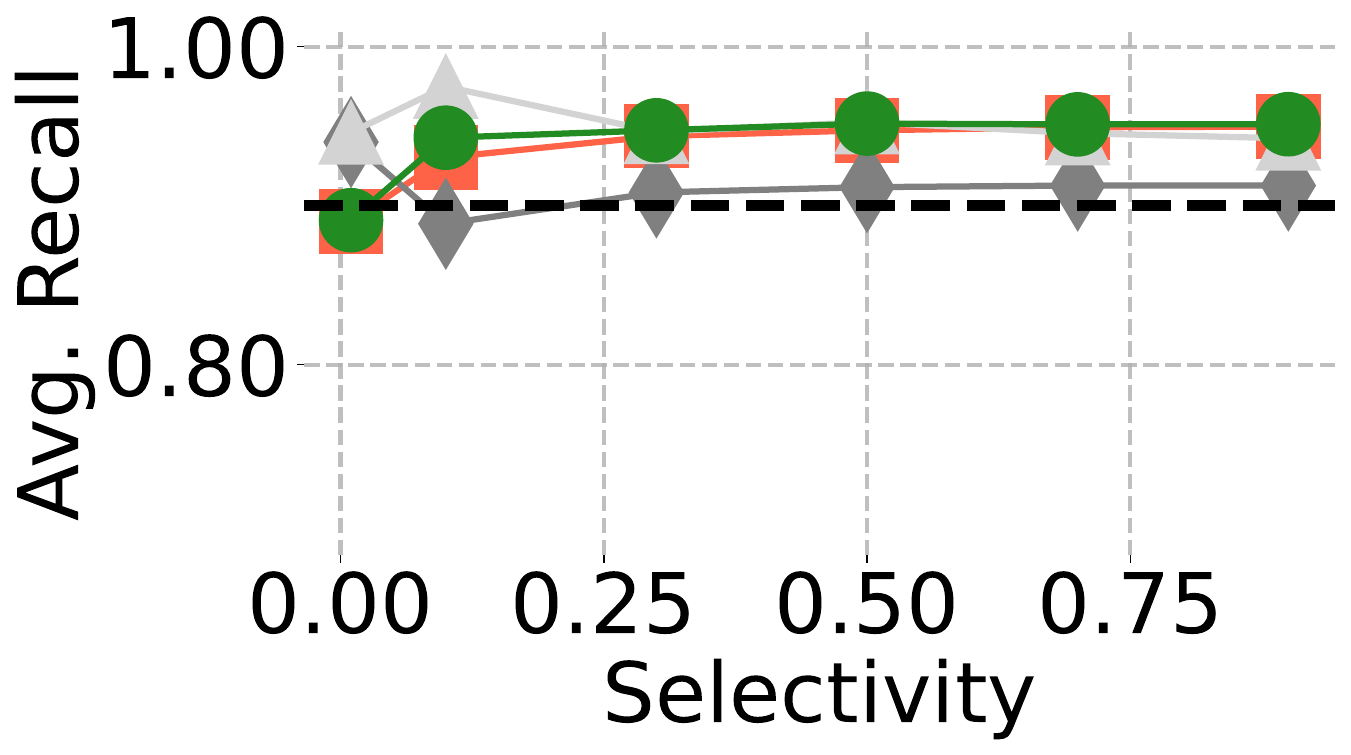}
        \vspace{-0.4cm}
        \caption{ACORN: Achieved recalls for varying selectivity (Positive Correlation, $k=100$, $R_t=0.9$).}
        \label{fig:recall-vs-selectivity-rt0.9-k100-acorn-pos-corr}
    \end{minipage}

    
    \begin{minipage}[t]{\textwidth}
        \includegraphics[width=0.19\textwidth]{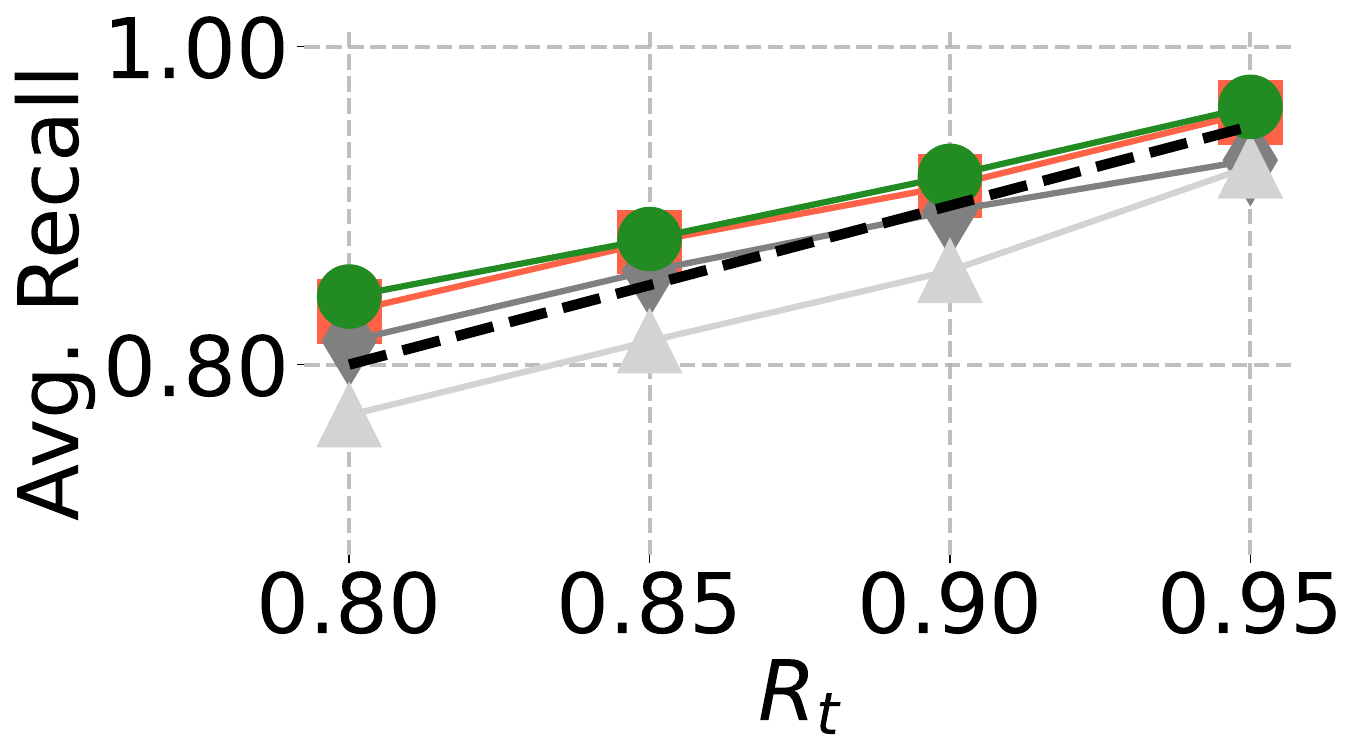}\hfill
        \includegraphics[width=0.19\textwidth]{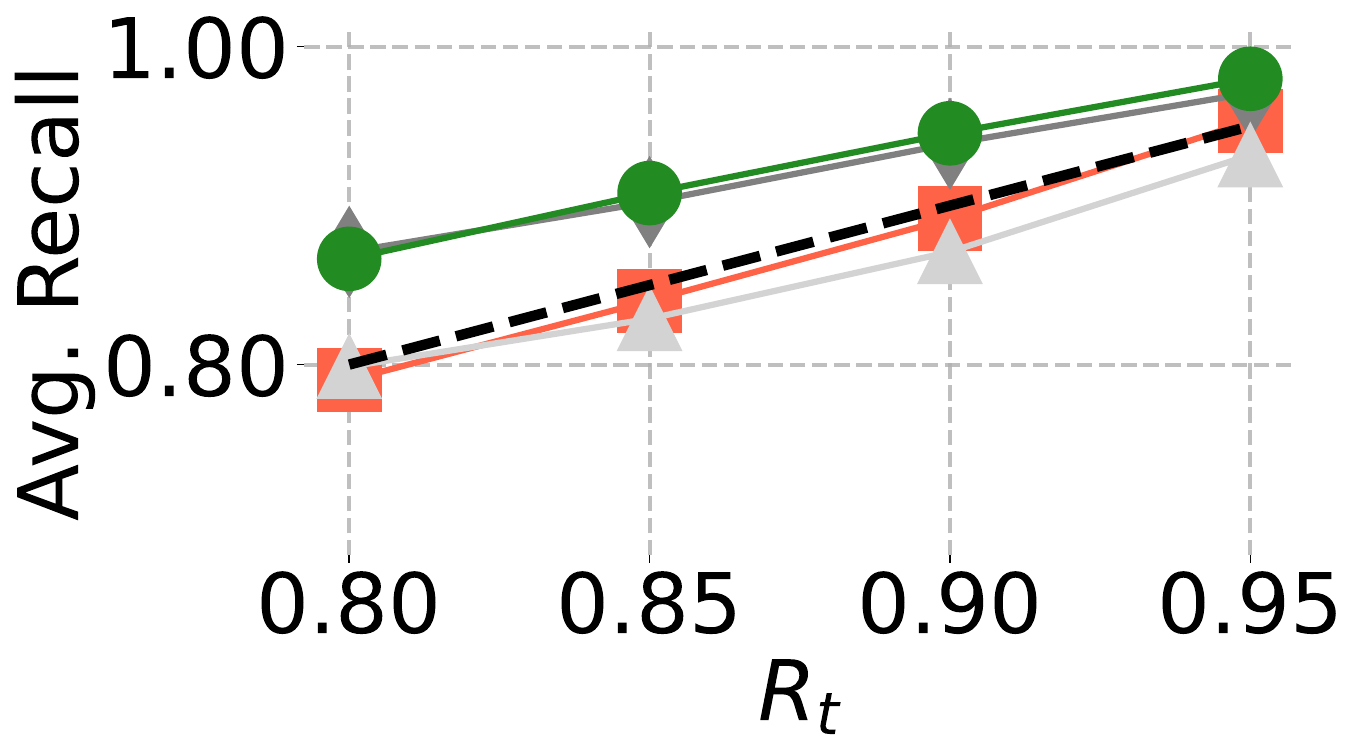}\hfill
        \includegraphics[width=0.19\textwidth]{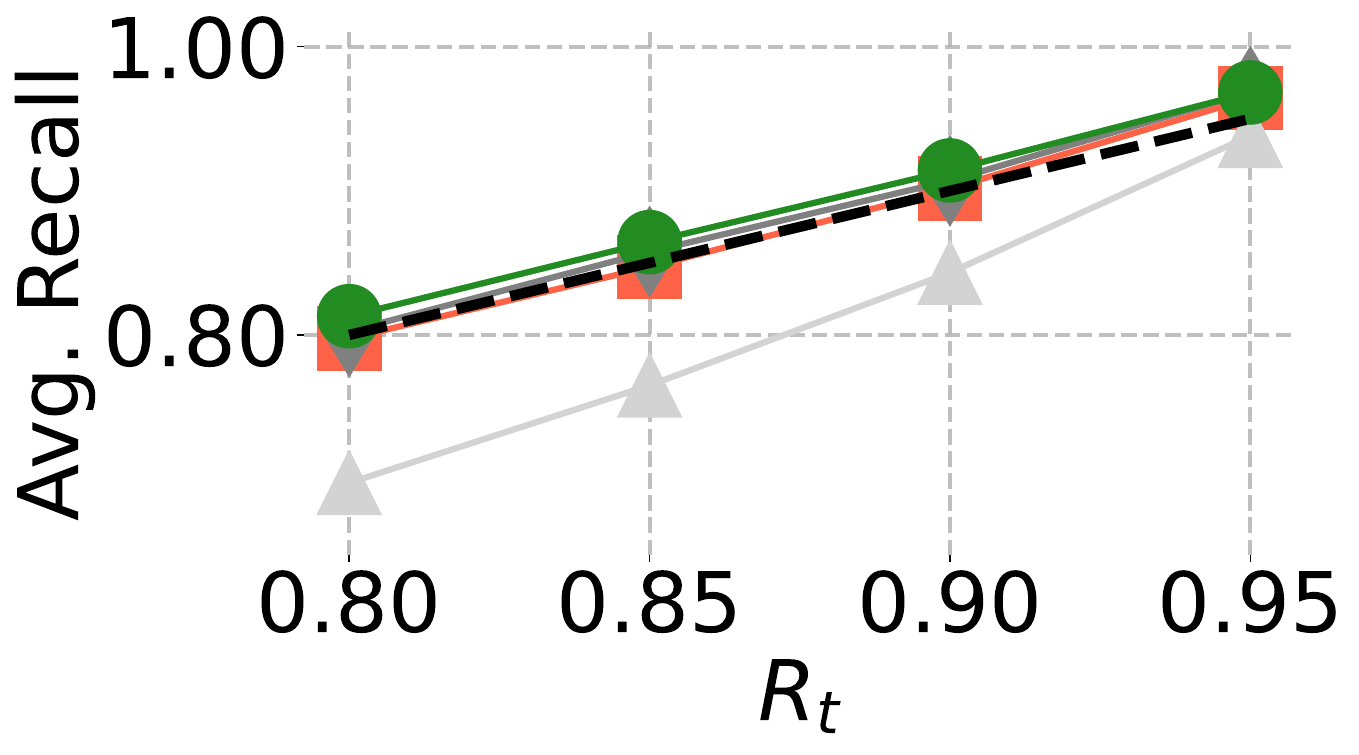}\hfill
        \includegraphics[width=0.19\textwidth]{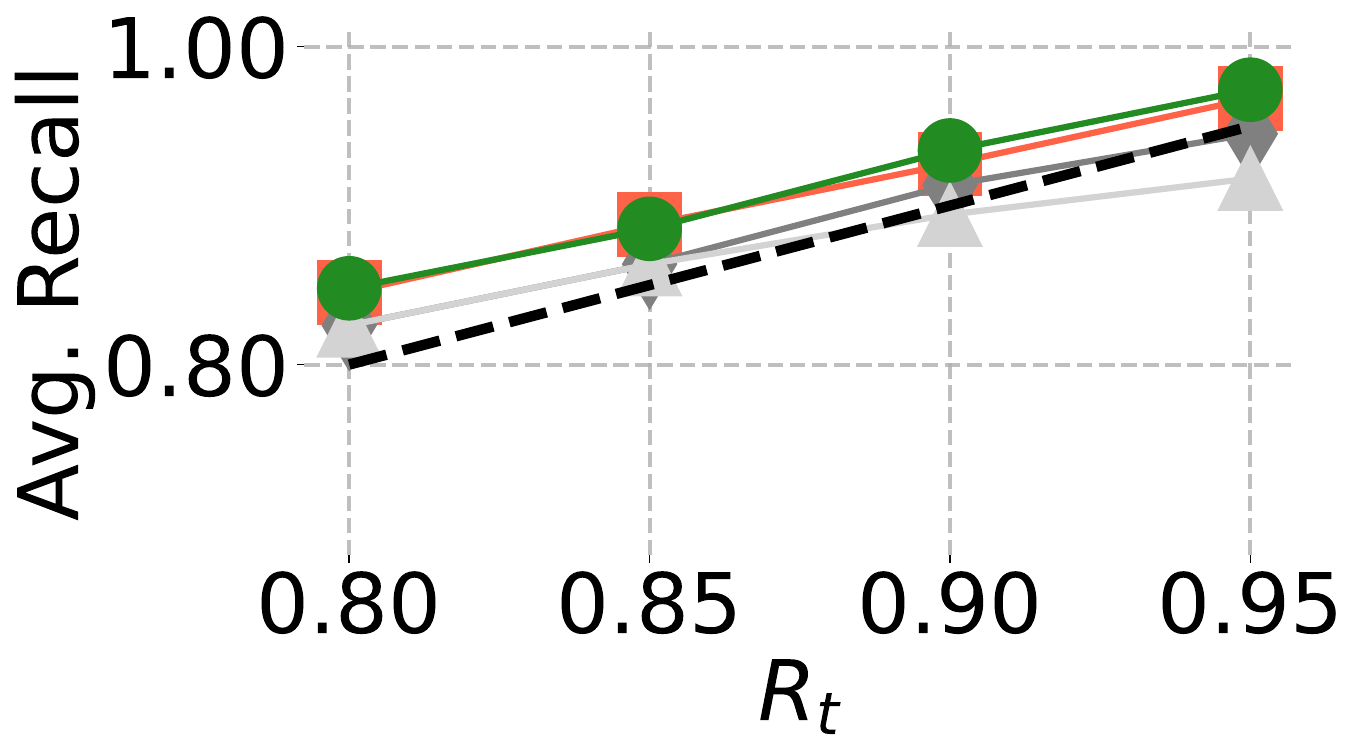}\hfill
        \includegraphics[width=0.19\textwidth]{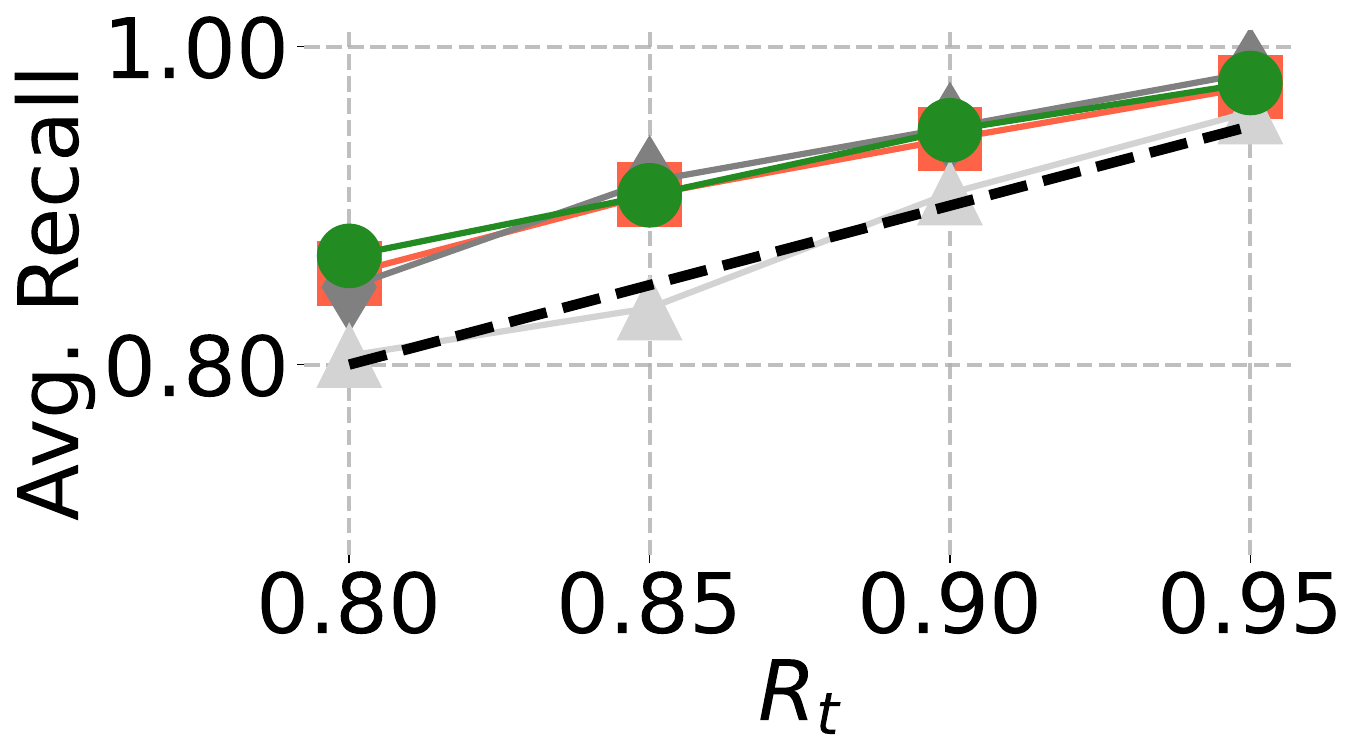}
        \vspace{-0.4cm}
        \caption{Sweeping: Achieved recalls for varying recall target (Positive Correlation, $k=100$, $sel=0.3$).}
        \label{fig:recall-vs-rt-k100-sel0.3-sweeping-pos-corr}
    \end{minipage}

    \begin{minipage}[t]{\textwidth}
        \includegraphics[width=0.19\textwidth]{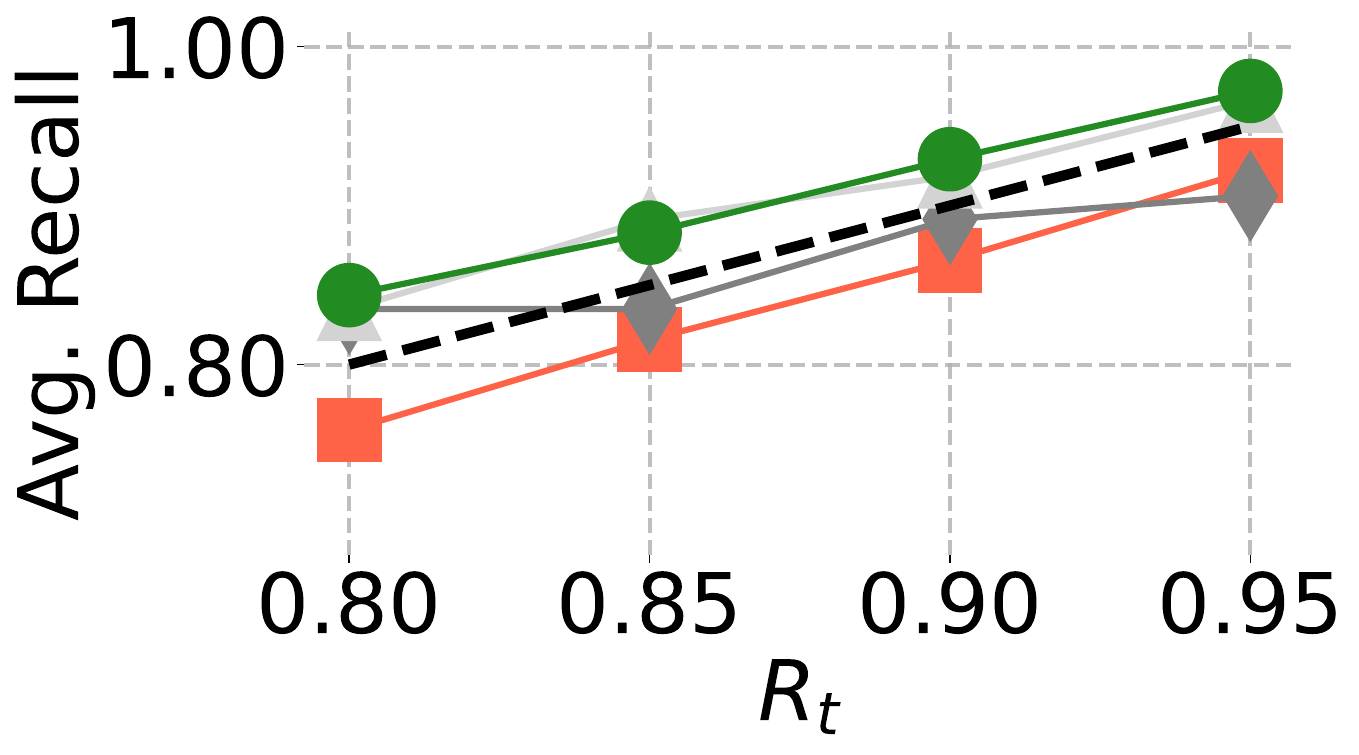}\hfill
        \includegraphics[width=0.19\textwidth]{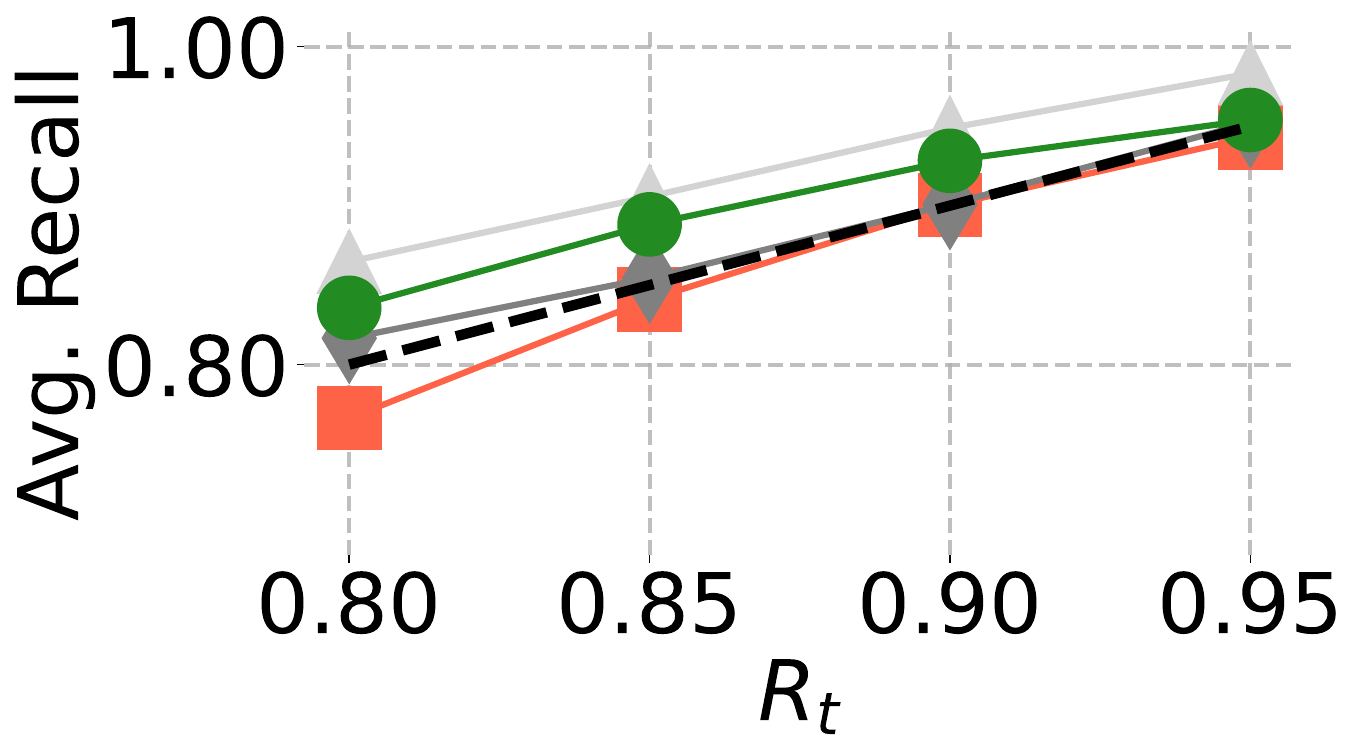}\hfill
        \includegraphics[width=0.19\textwidth]{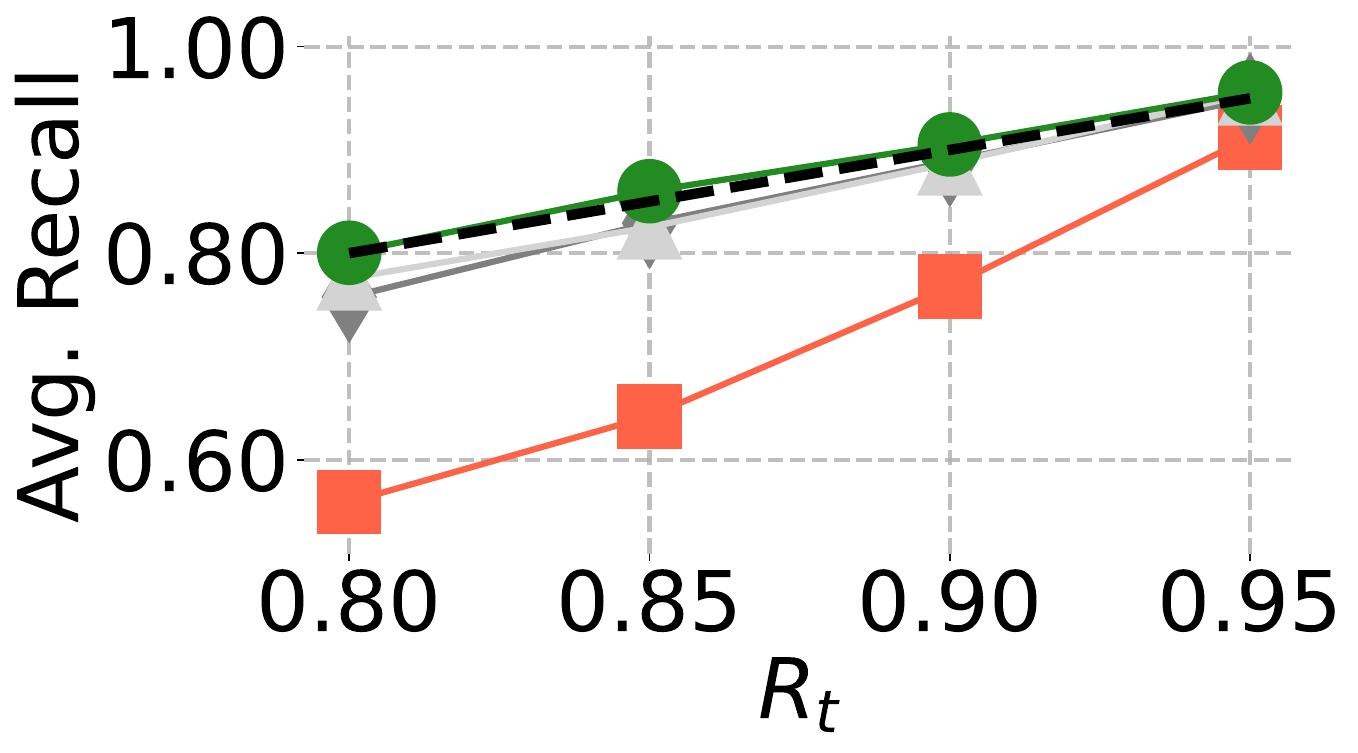}\hfill
        \includegraphics[width=0.19\textwidth]{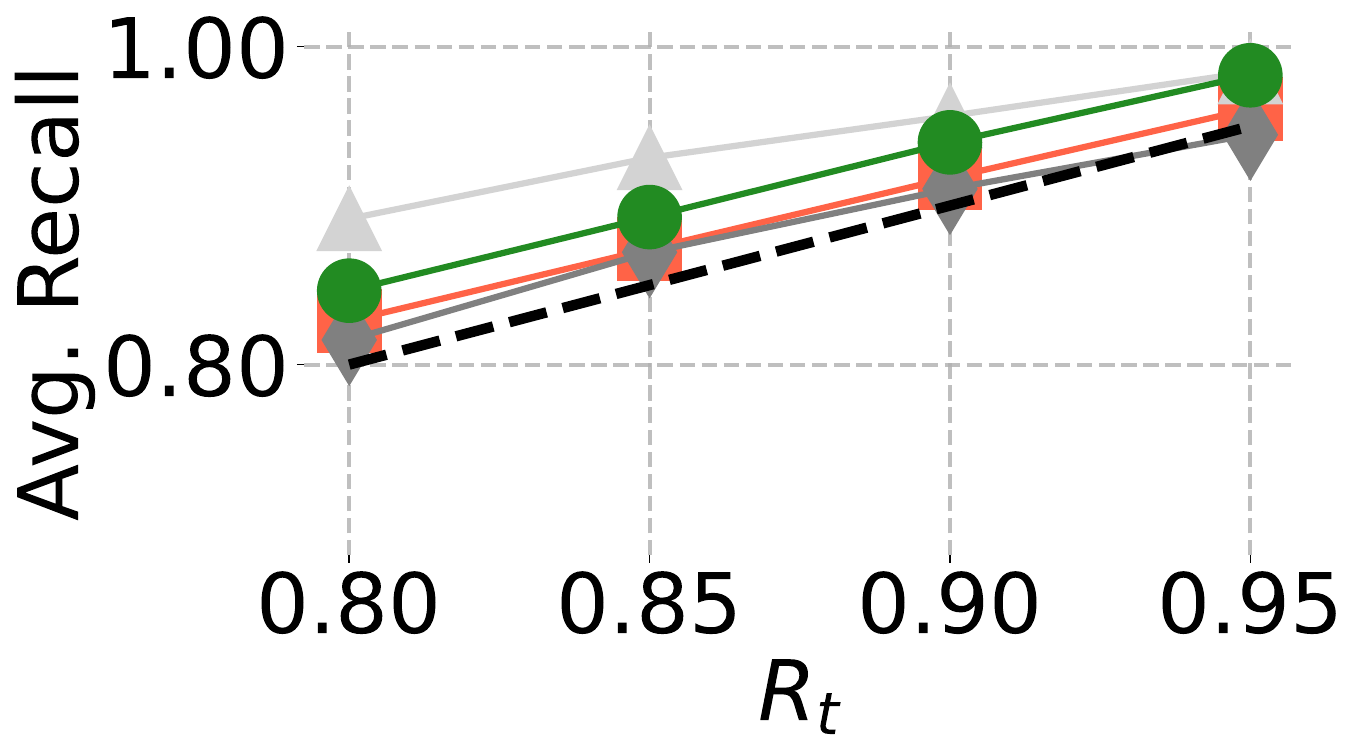}\hfill
        \includegraphics[width=0.19\textwidth]{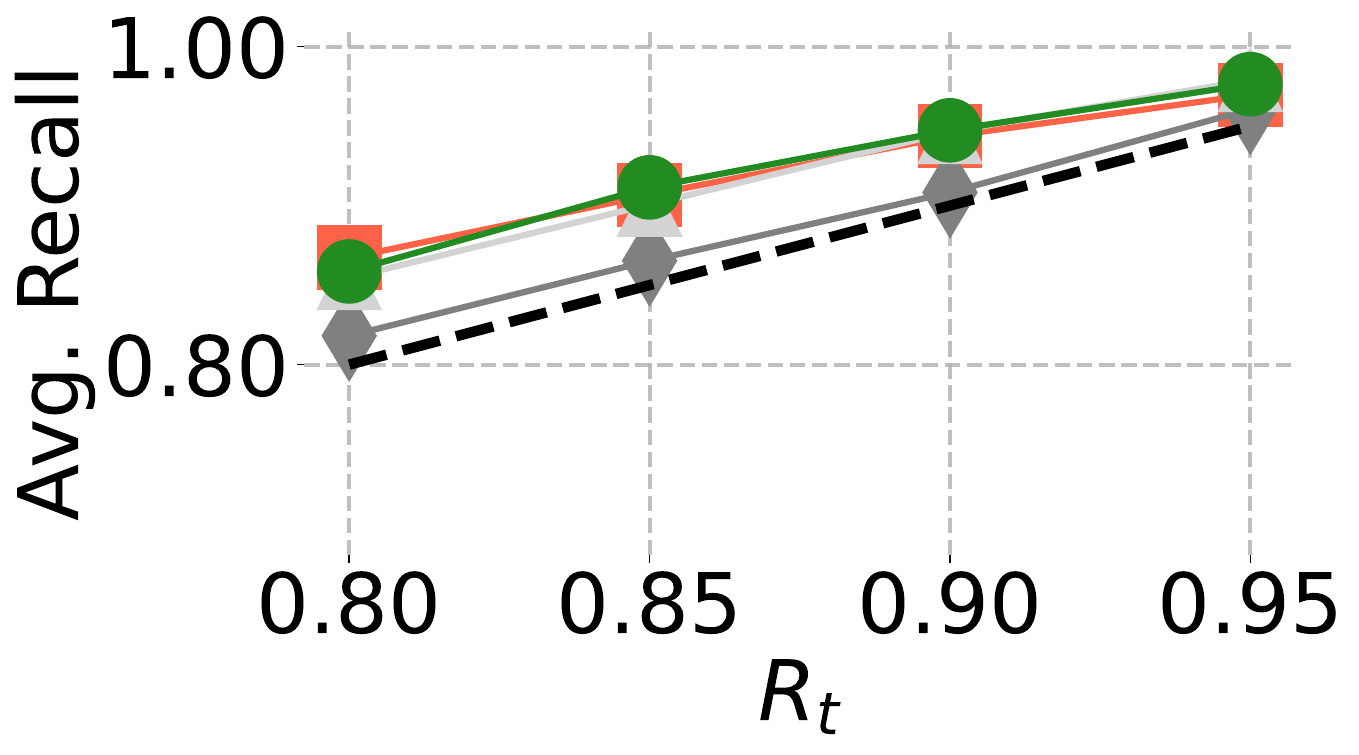}
        \vspace{-0.4cm}
        \caption{ACORN: Achieved recalls for varying recall target (Positive Correlation, $k=100$, $sel=0.3$).}
        \label{fig:recall-vs-rt-k100-sel0.3-acorn-pos-corr}
    \end{minipage}

    
    \begin{minipage}[t]{\textwidth}
        \includegraphics[width=0.19\textwidth]{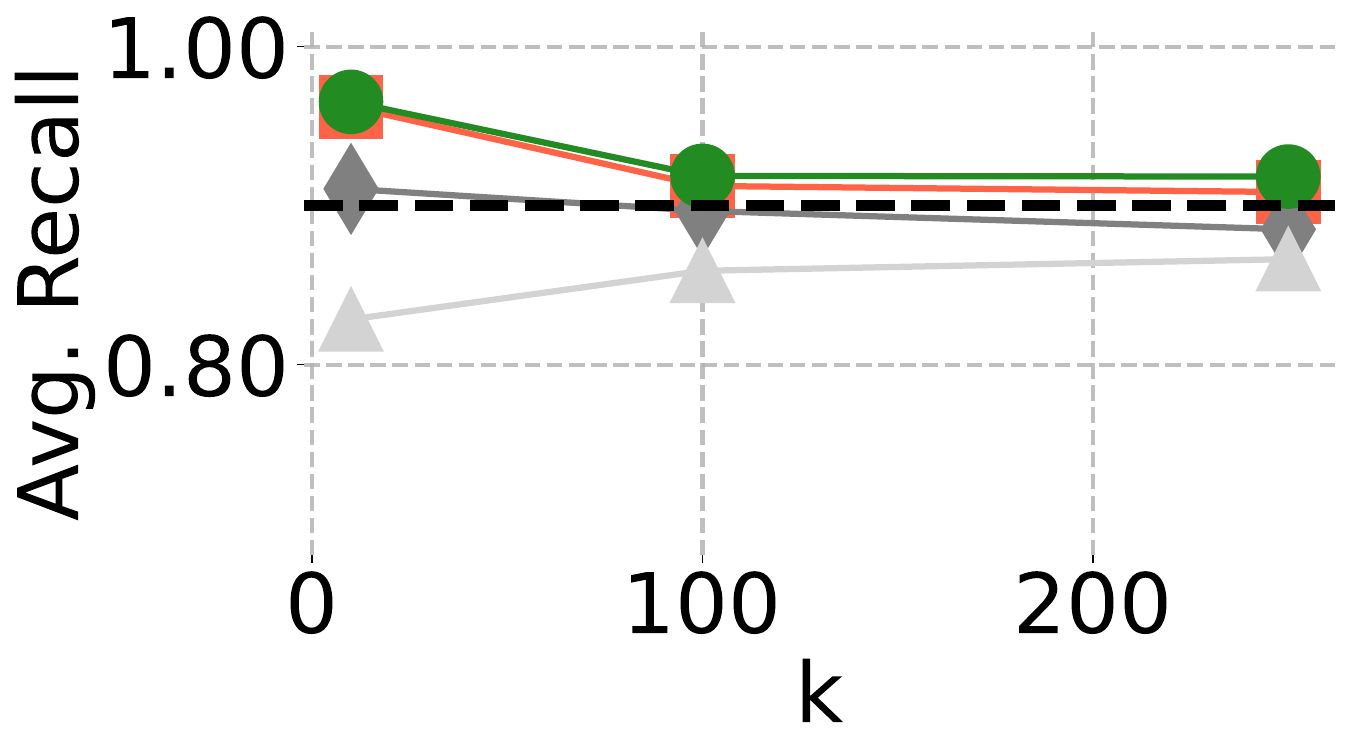}\hfill
        \includegraphics[width=0.19\textwidth]{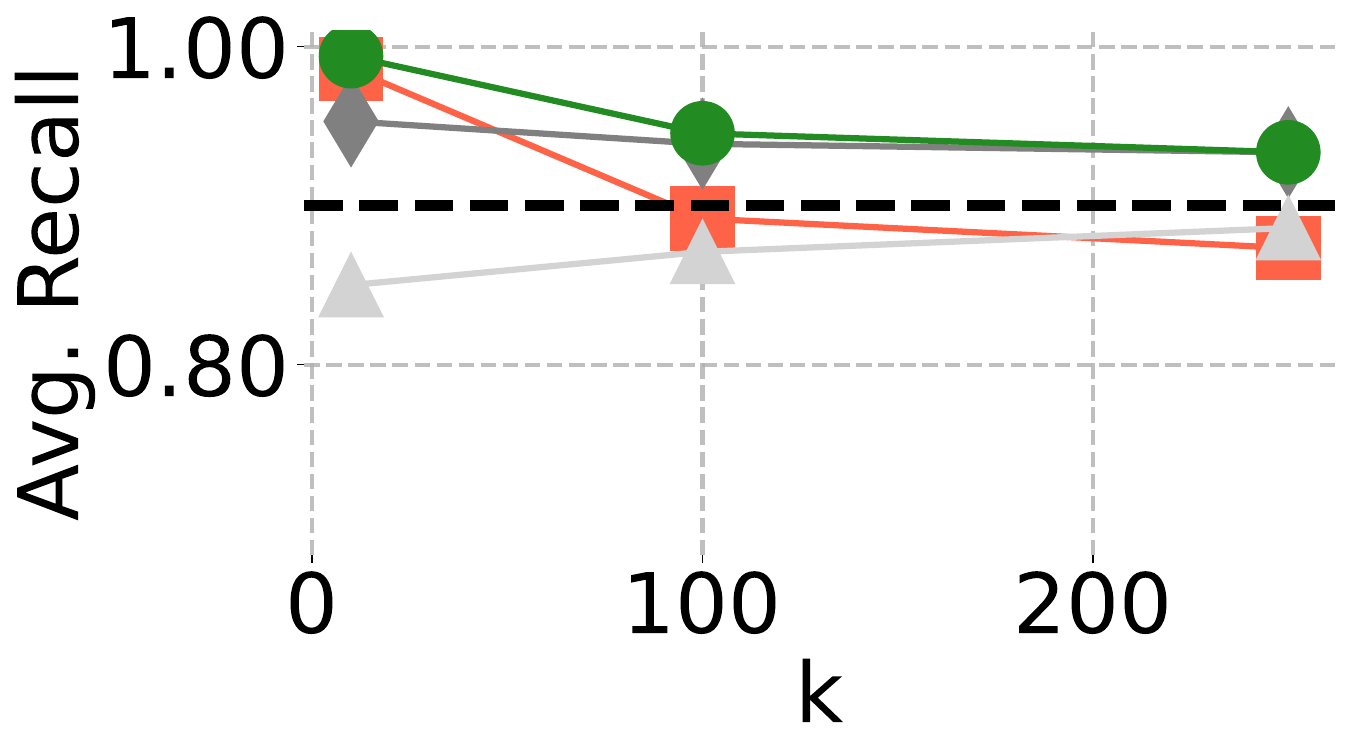}\hfill
        \includegraphics[width=0.19\textwidth]{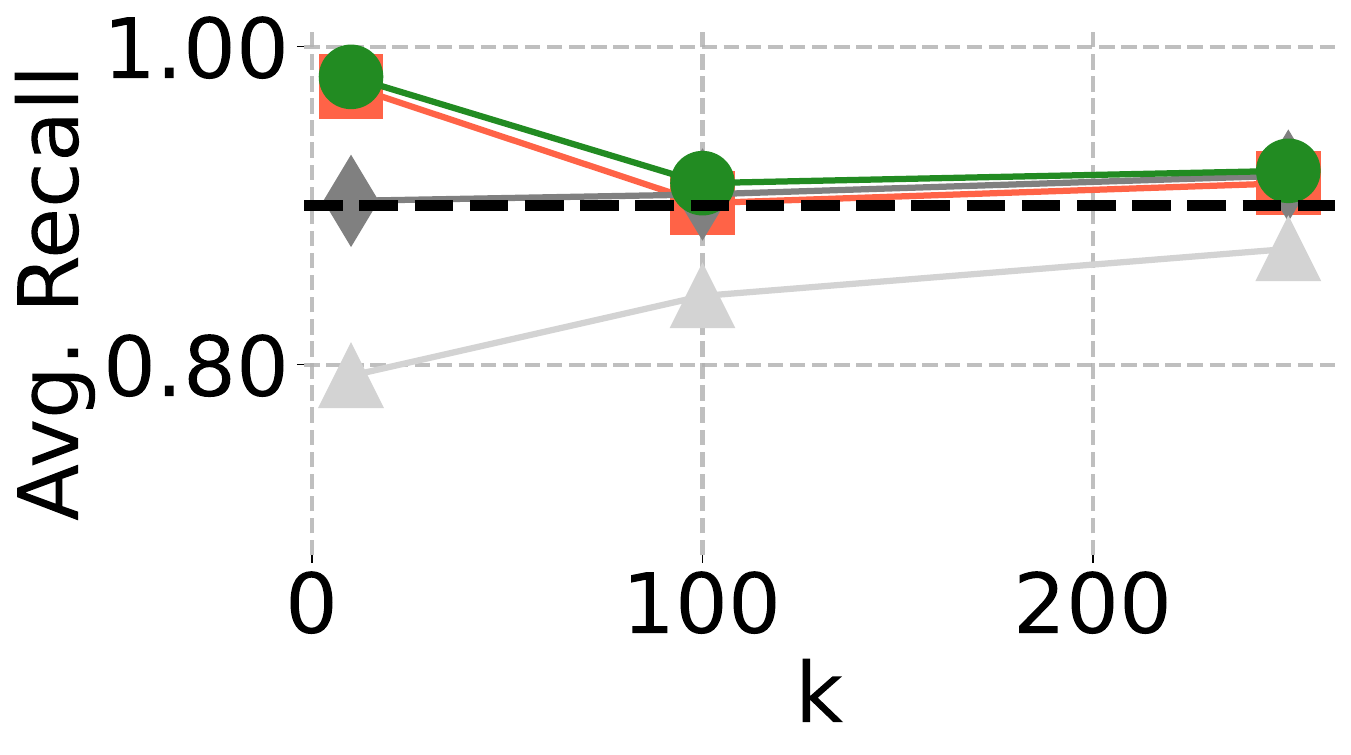}\hfill
        \includegraphics[width=0.19\textwidth]{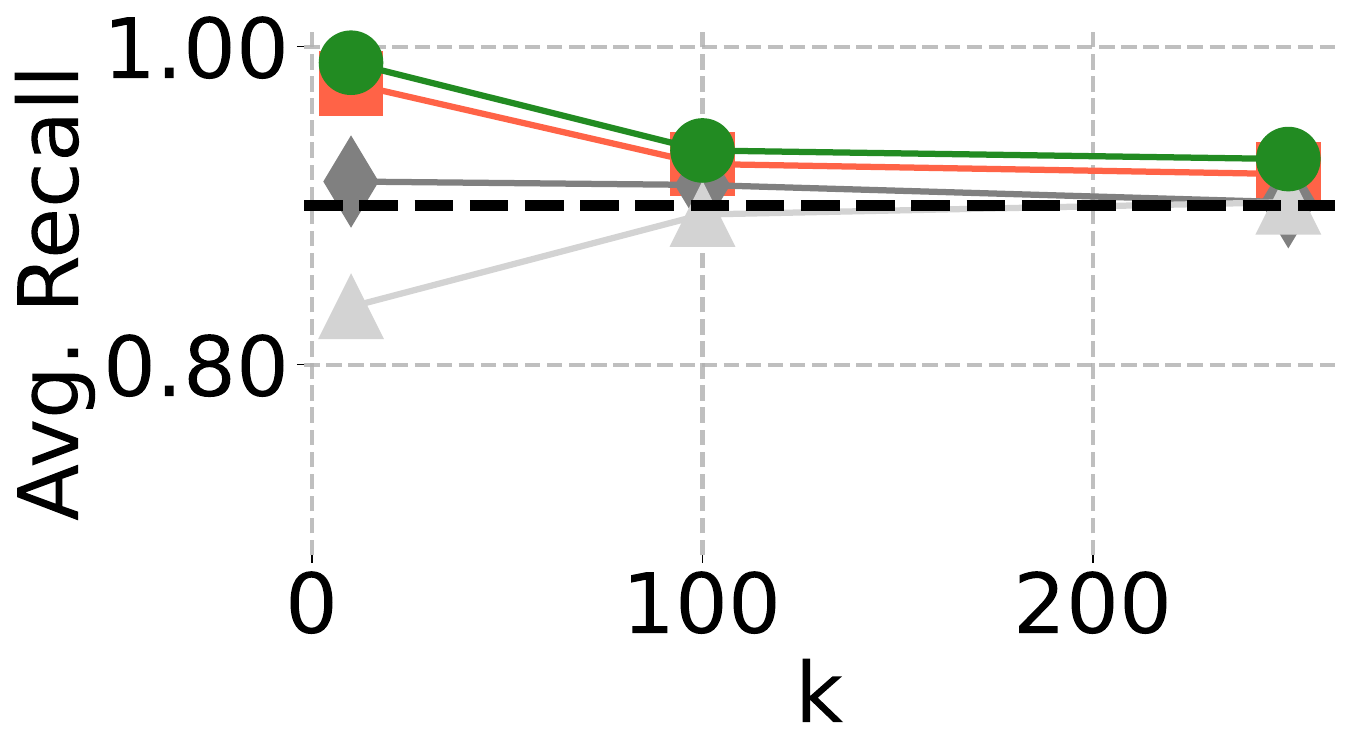}\hfill
        \includegraphics[width=0.19\textwidth]{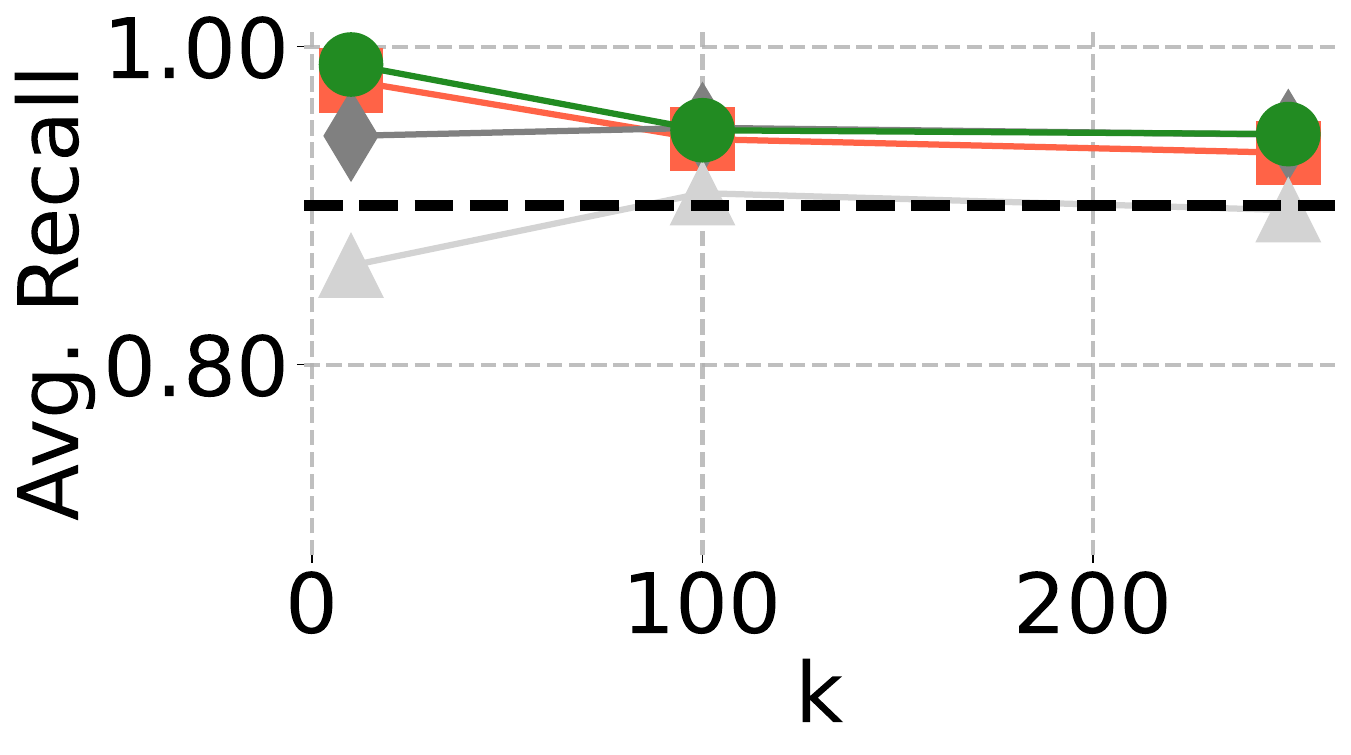}
        \vspace{-0.4cm}
        \caption{Sweeping: Achieved recalls for varying value of $k$ (Positive Correlation, $R_t=0.9$, $sel=0.3$).}
        \label{fig:recall-vs-k-rt0.9-sel0.3-sweeping-pos-corr}
    \end{minipage}
    
    \begin{minipage}[t]{\textwidth}
        \includegraphics[width=0.19\textwidth]{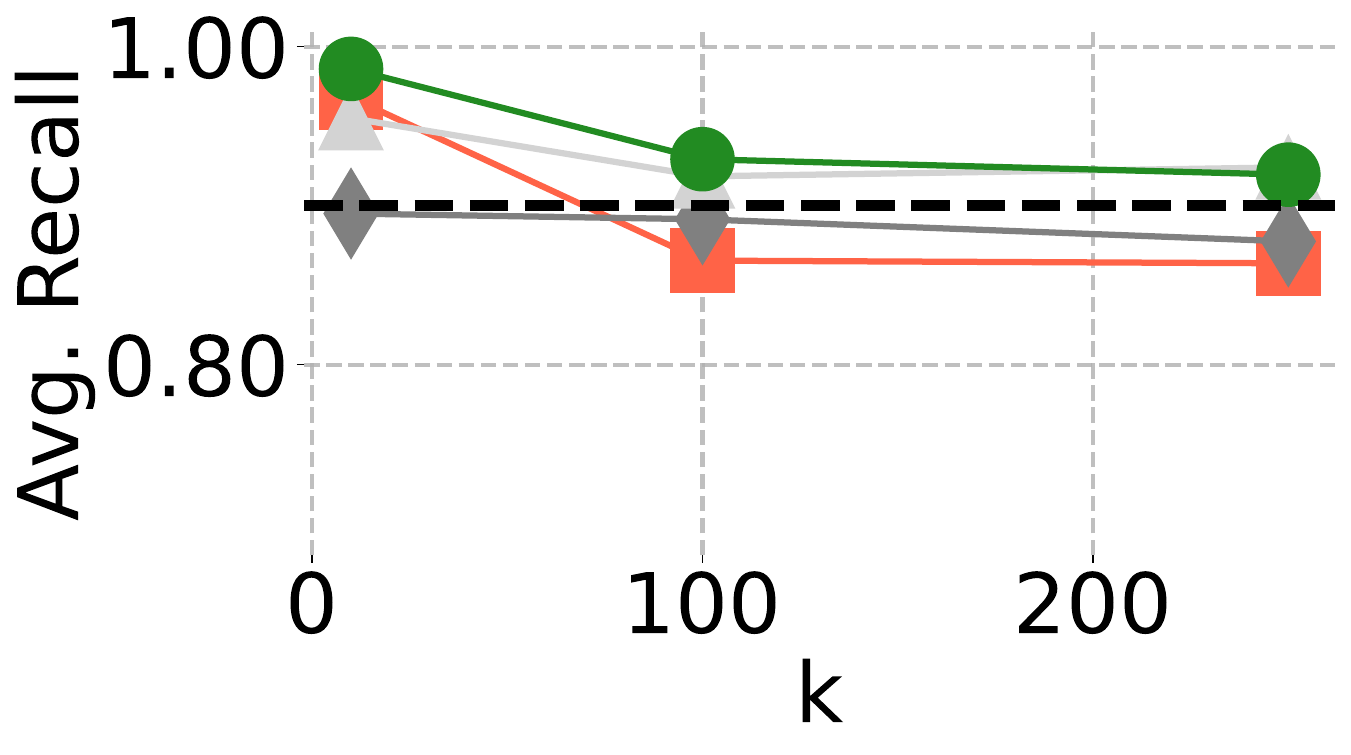}\hfill
        \includegraphics[width=0.19\textwidth]{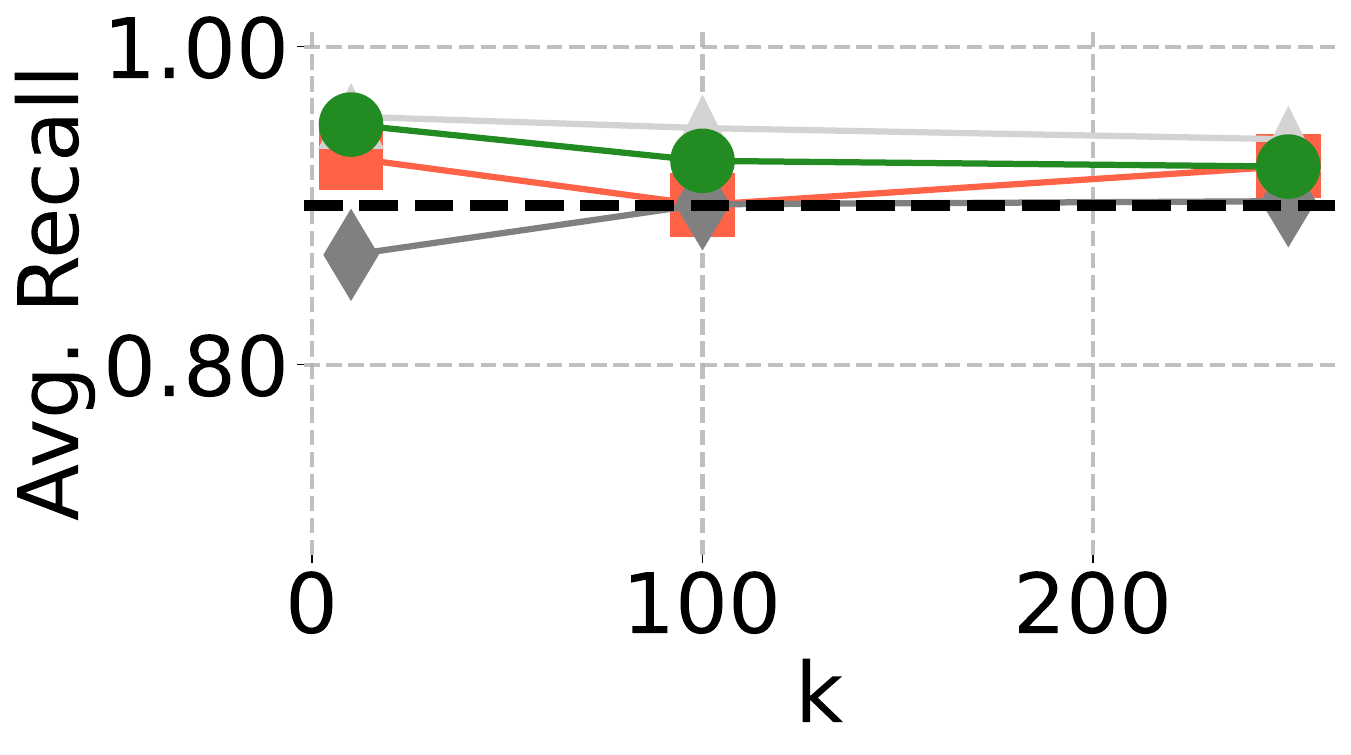}\hfill
        \includegraphics[width=0.19\textwidth]{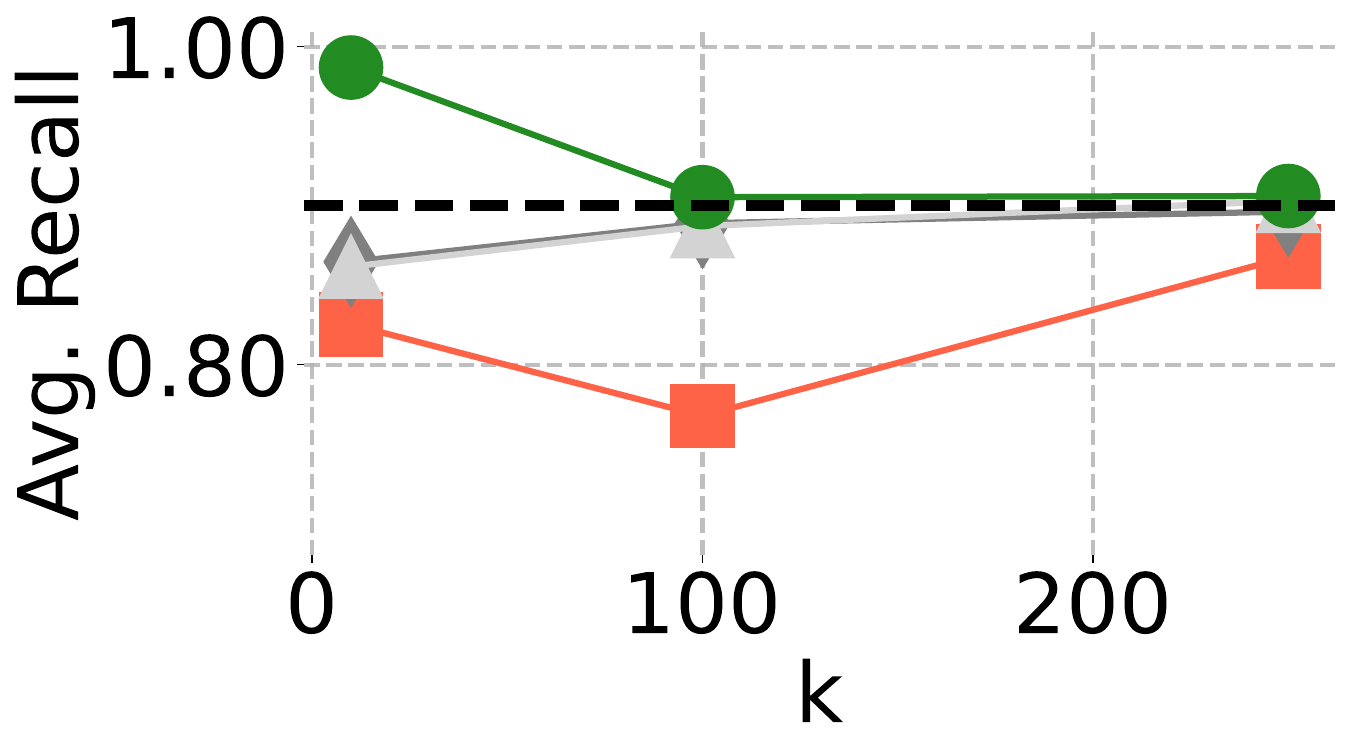}\hfill
        \includegraphics[width=0.19\textwidth]{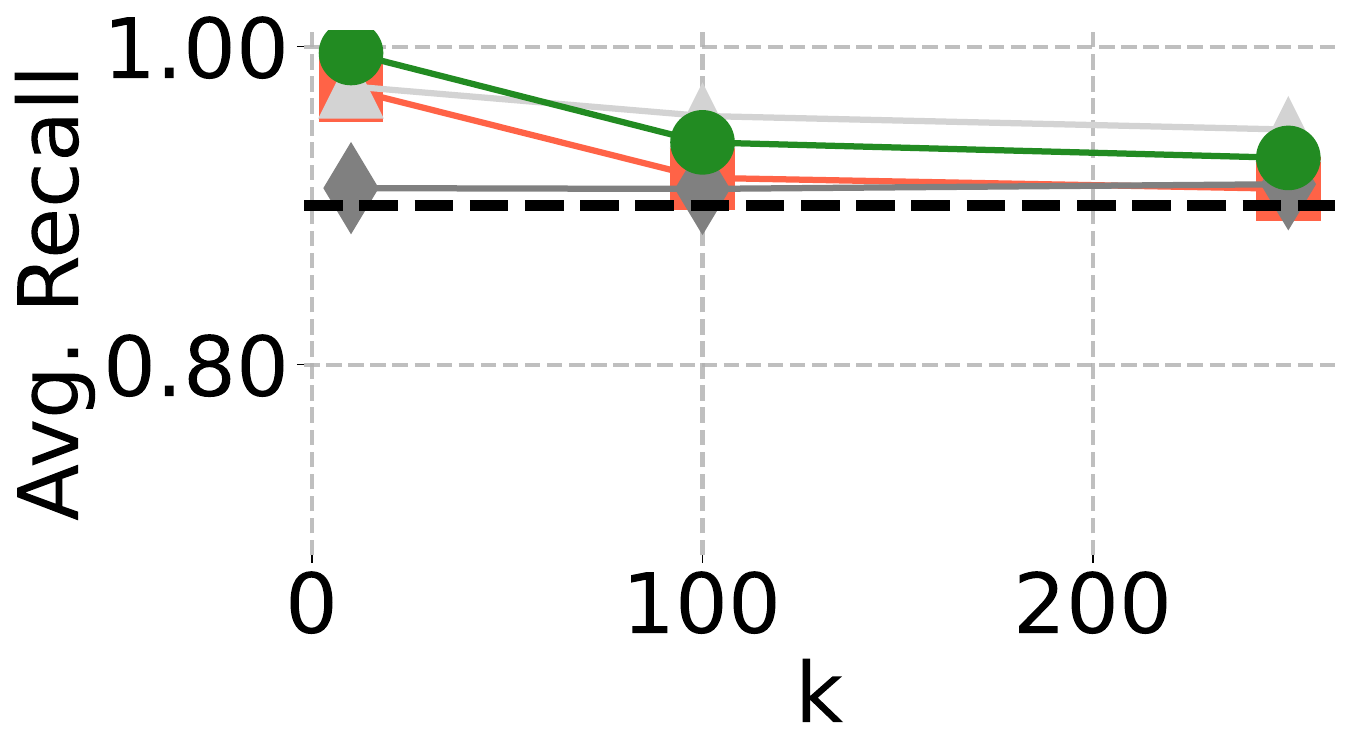}\hfill
        \includegraphics[width=0.19\textwidth]{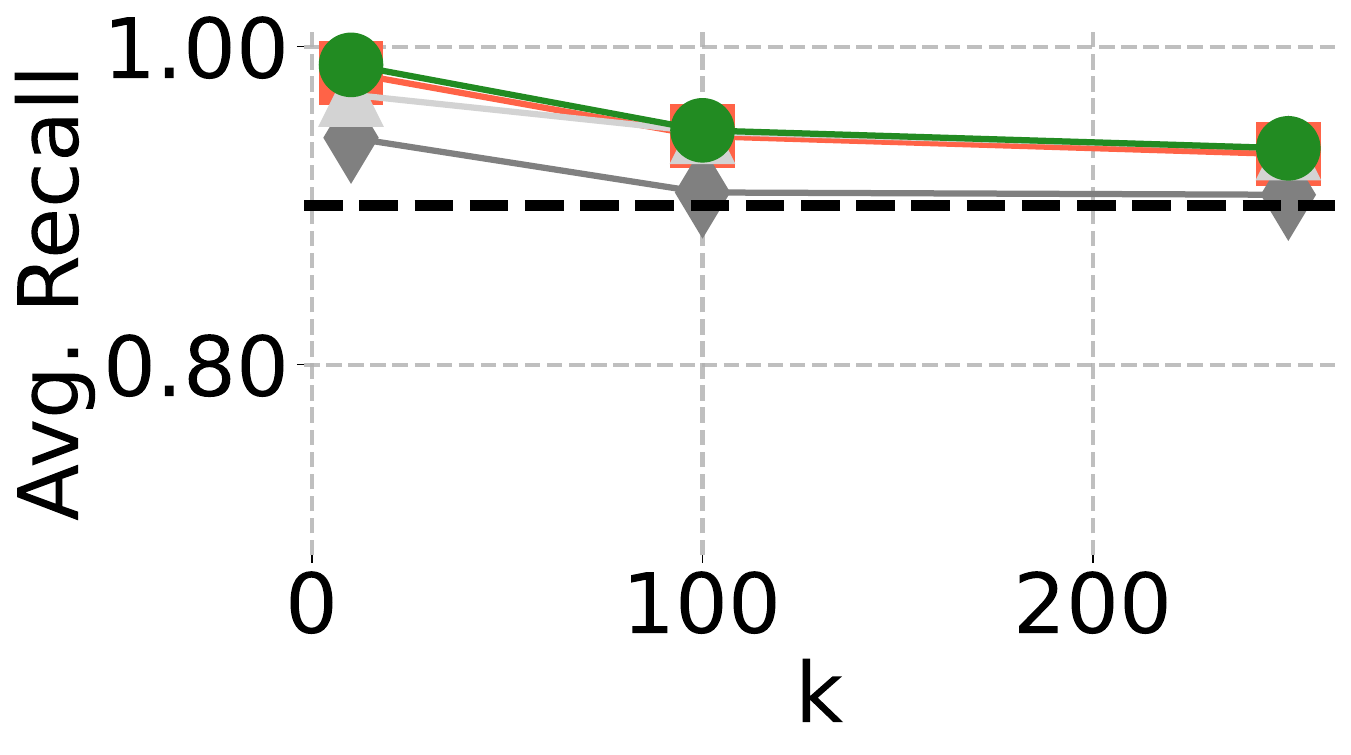}
        \vspace{-0.4cm}
        \caption{ACORN: Achieved recalls for varying value of $k$ (Positive Correlation, $R_t=0.9$, $sel=0.3$).}
        \label{fig:recall-vs-k-rt0.9-sel0.3-acorn-pos-corr}
    \end{minipage}
\end{figure*}

\begin{figure*}
    \centering
    \vspace{-0.1cm}
    \begin{adjustbox}{max width=0.6\textwidth}
        \includegraphics{figs/legend_lines.pdf}
    \end{adjustbox}

    \vspace{-0.1cm}
    \makebox[0.19\textwidth][c]{\textbf{SIFT1M}} \hfill
    \makebox[0.19\textwidth][c]{\textbf{GIST1M}} \hfill
    \makebox[0.19\textwidth][c]{\textbf{GLOVE1M}} \hfill
    \makebox[0.19\textwidth][c]{\textbf{DEEP10M}} \hfill
    \makebox[0.19\textwidth][c]{\textbf{T2I10M}}
    
    
    \begin{minipage}[t]{\textwidth}
        \includegraphics[width=0.19\textwidth]{figs/avg_recall_vs_selectivity_SIFT1M_sweeping_corr_uar_full_positive_k100_selvar_rt0.9.pdf}\hfill
        \includegraphics[width=0.19\textwidth]{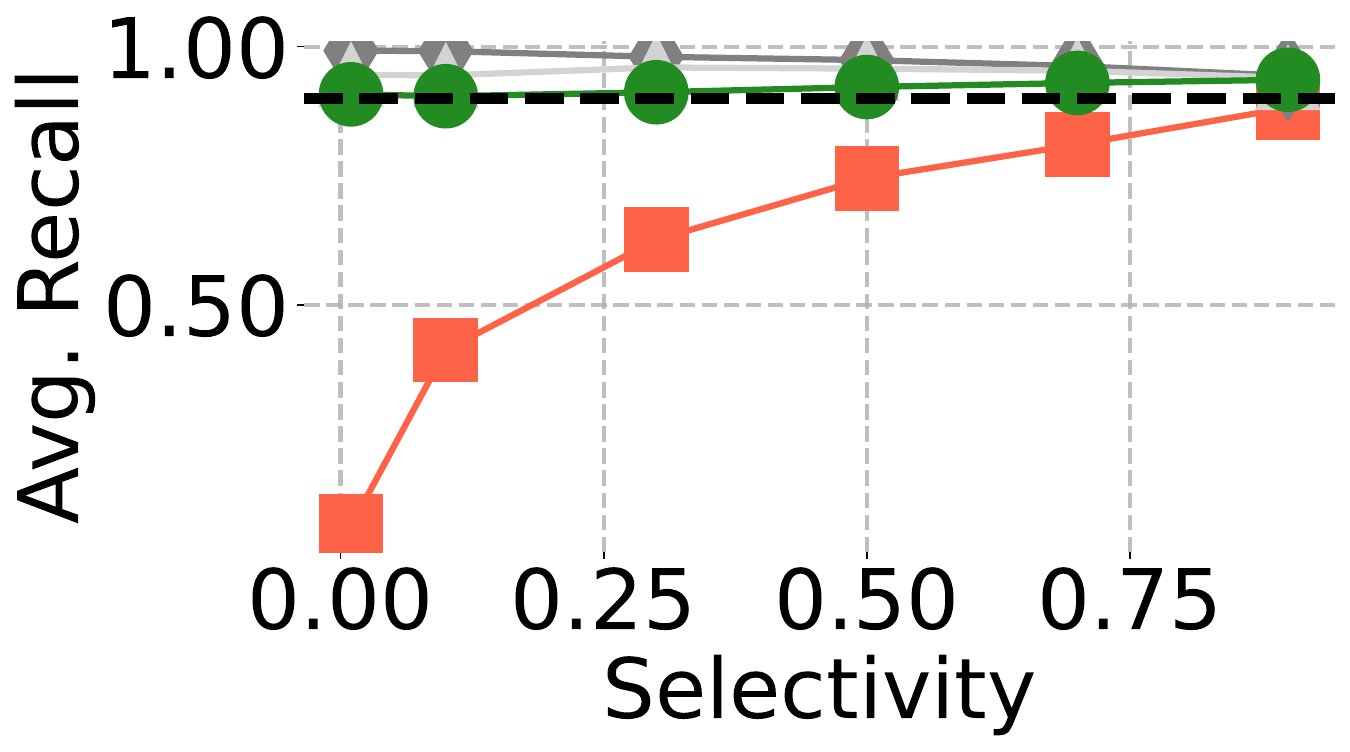}\hfill
        \includegraphics[width=0.19\textwidth]{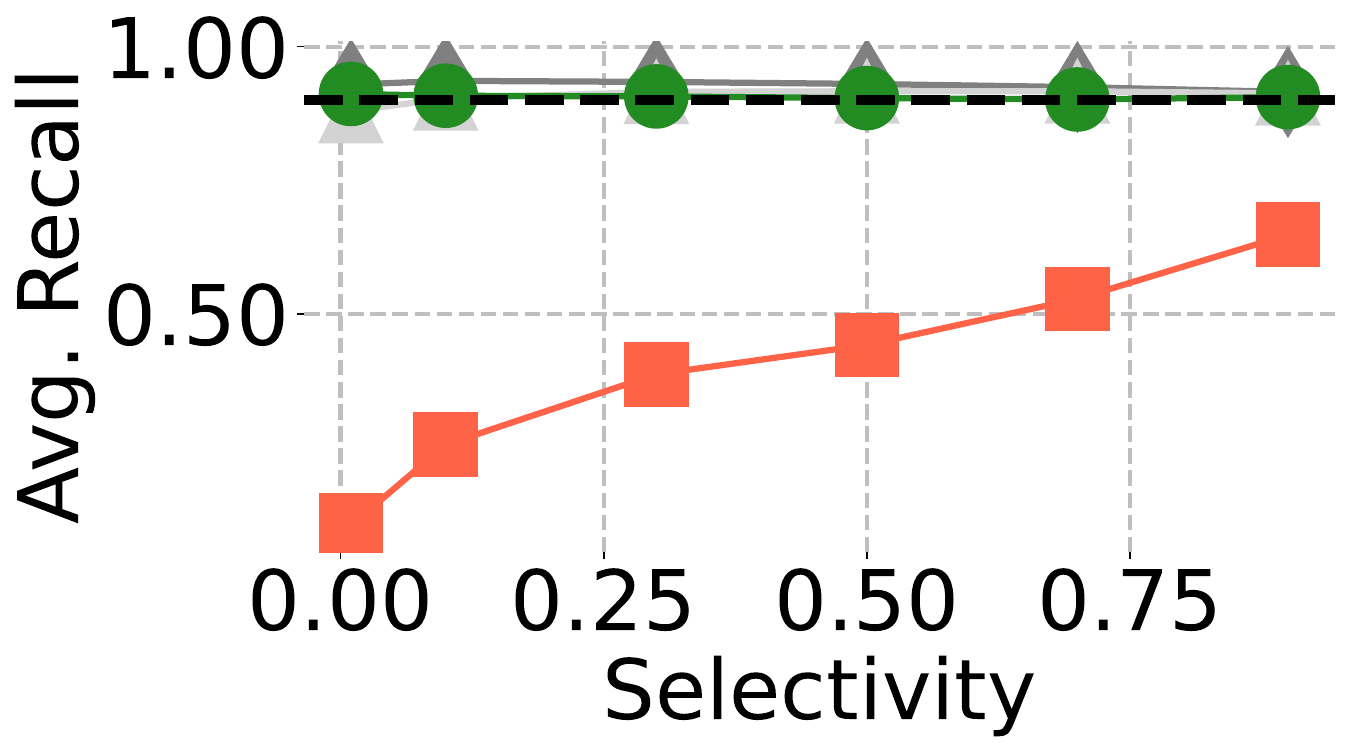}\hfill
        \includegraphics[width=0.19\textwidth]{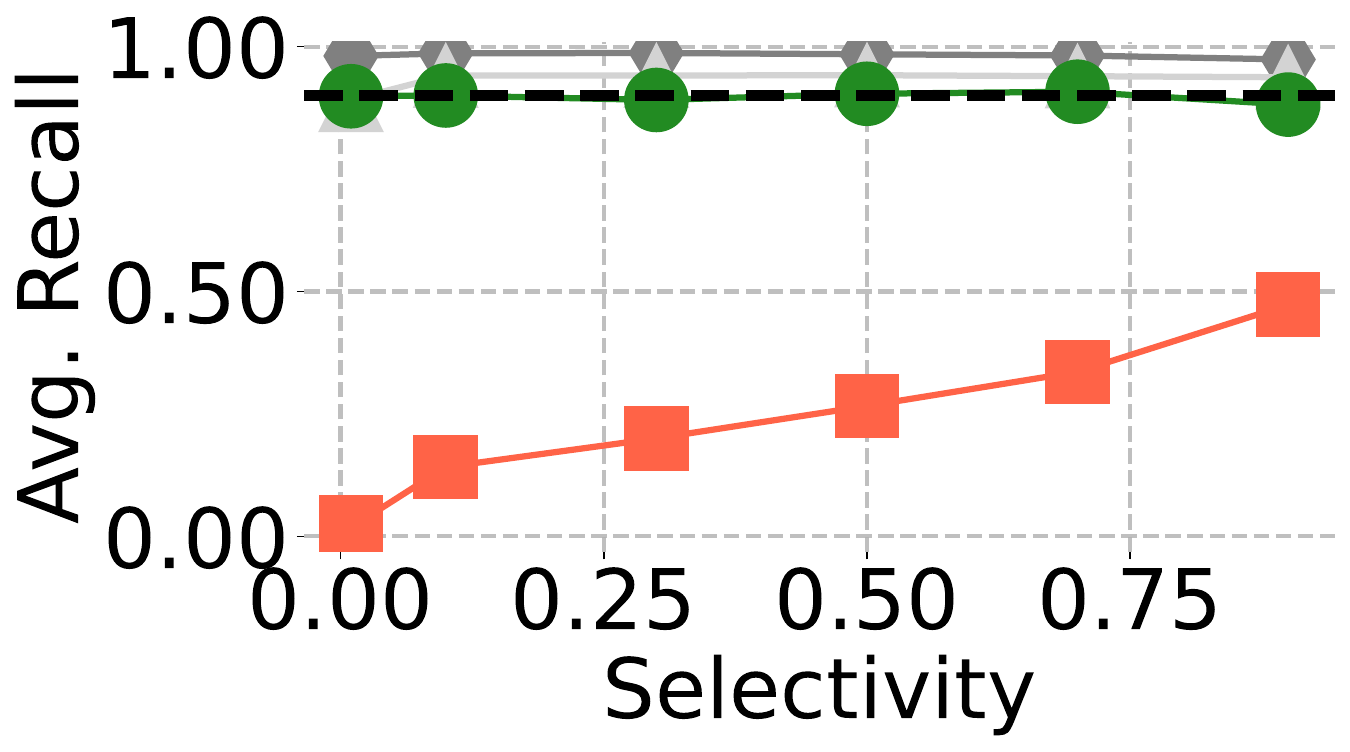}\hfill
        \includegraphics[width=0.19\textwidth]{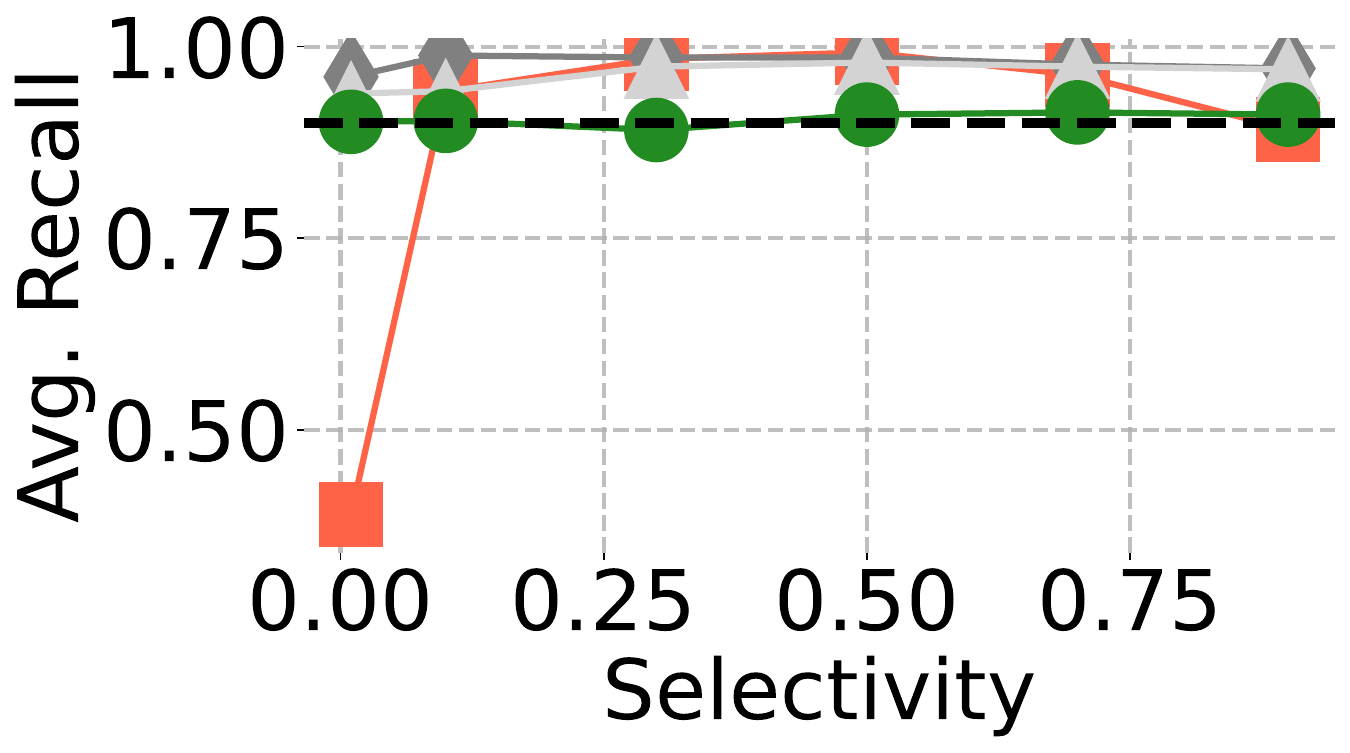}
        \vspace{-0.4cm}
        \caption{Sweeping: Achieved recalls for varying selectivity (Negative Correlation, $k=100$, $R_t=0.9$).}
        \label{fig:recall-vs-selectivity-rt0.9-k100-sweeping-neg-corr}
    \end{minipage}

    \begin{minipage}[t]{\textwidth}
        \includegraphics[width=0.19\textwidth]{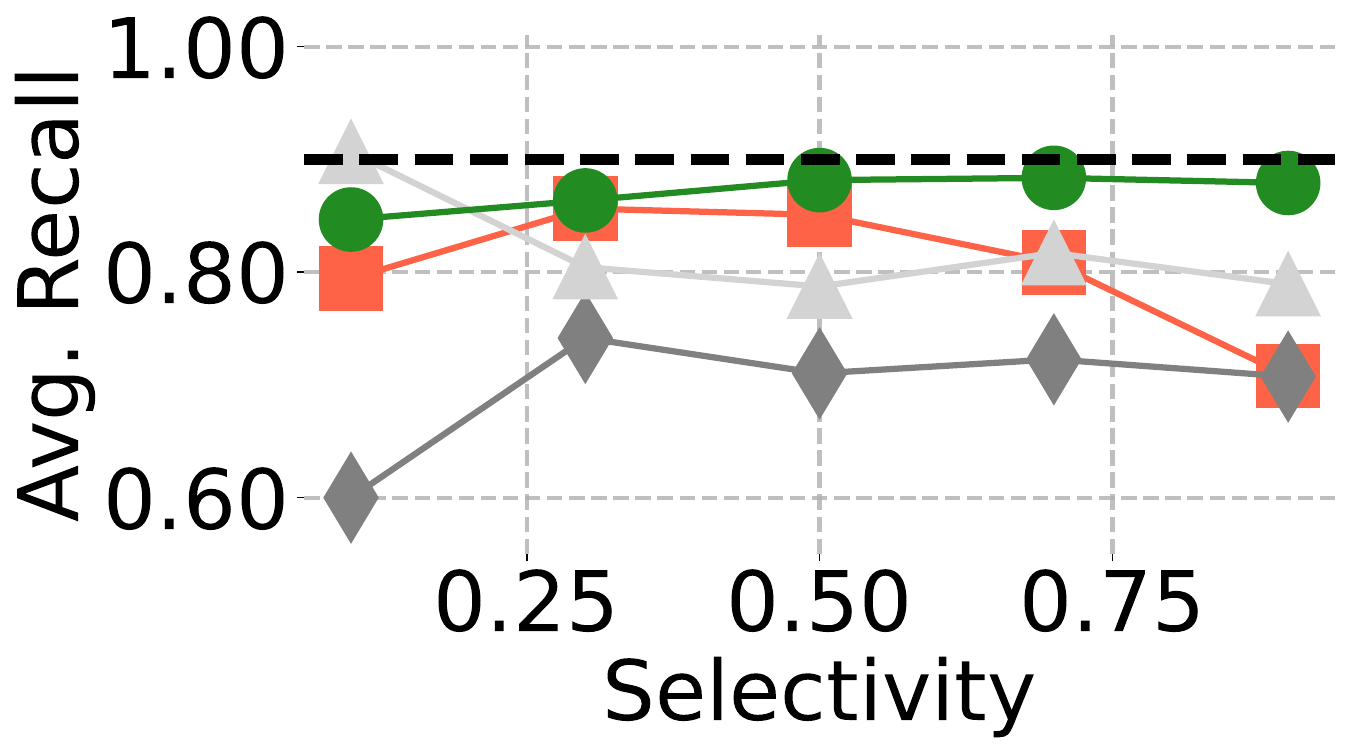}\hfill
        \includegraphics[width=0.19\textwidth]{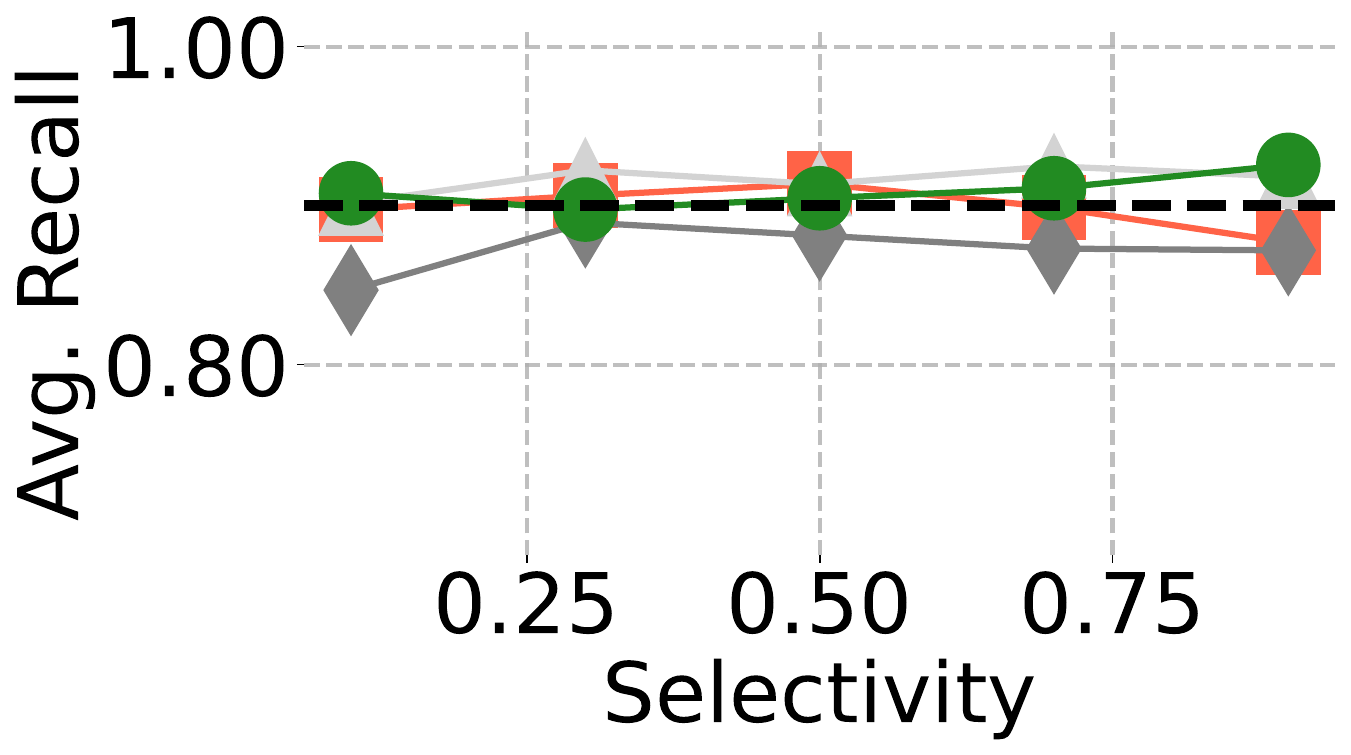}\hfill
        \includegraphics[width=0.19\textwidth]{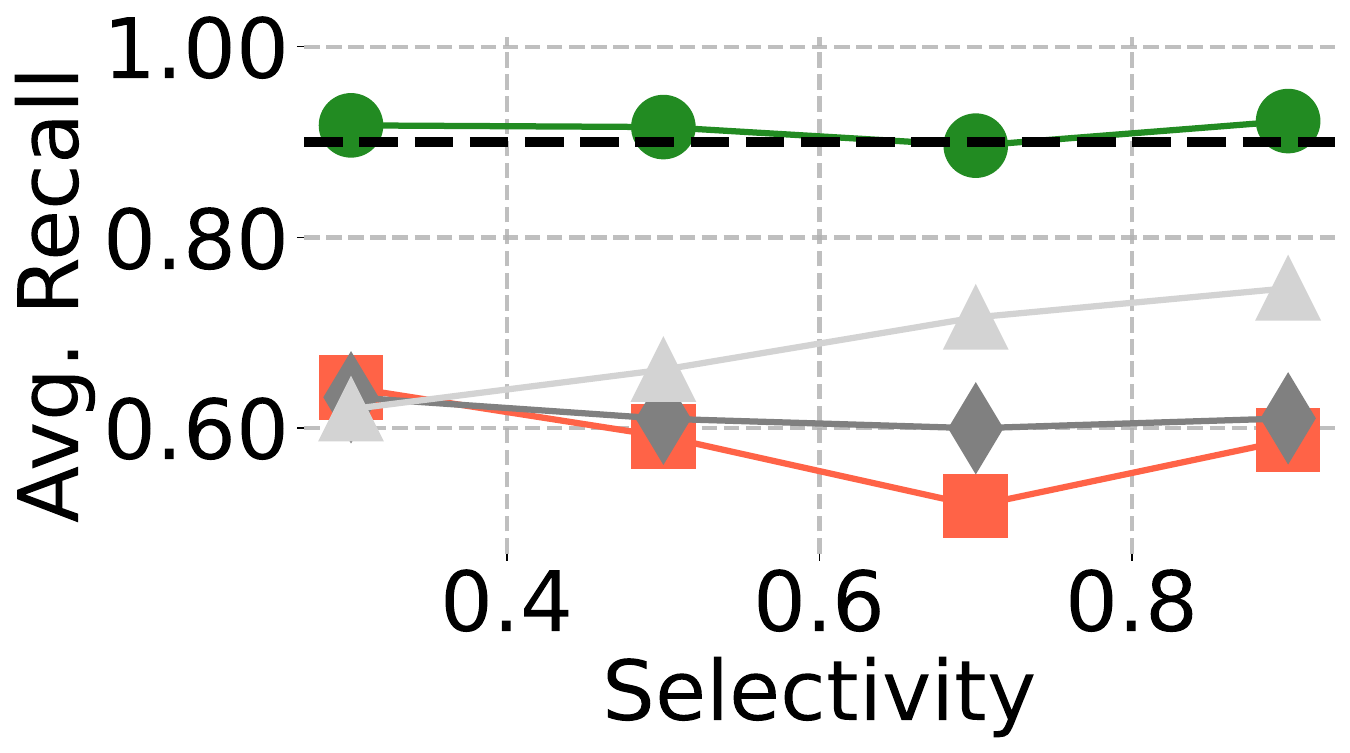}\hfill
        \includegraphics[width=0.19\textwidth]{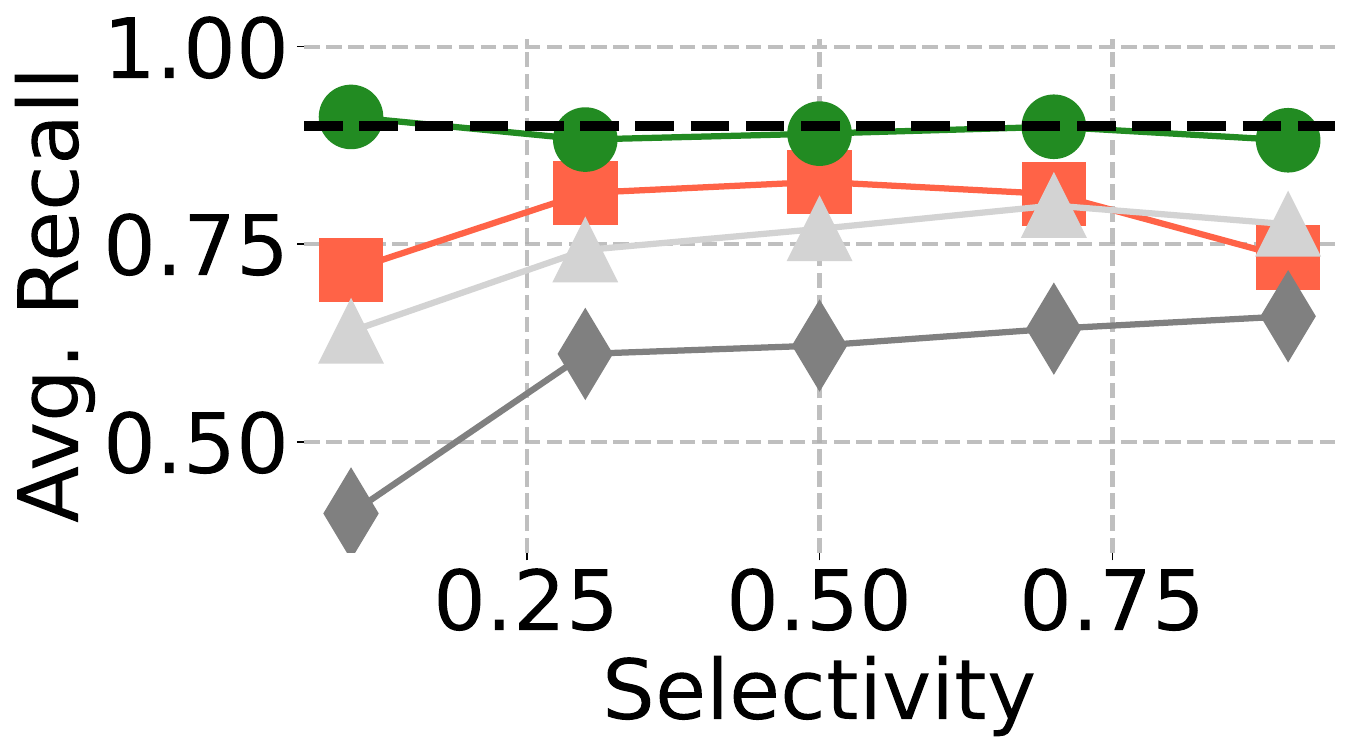}\hfill
        \includegraphics[width=0.19\textwidth]{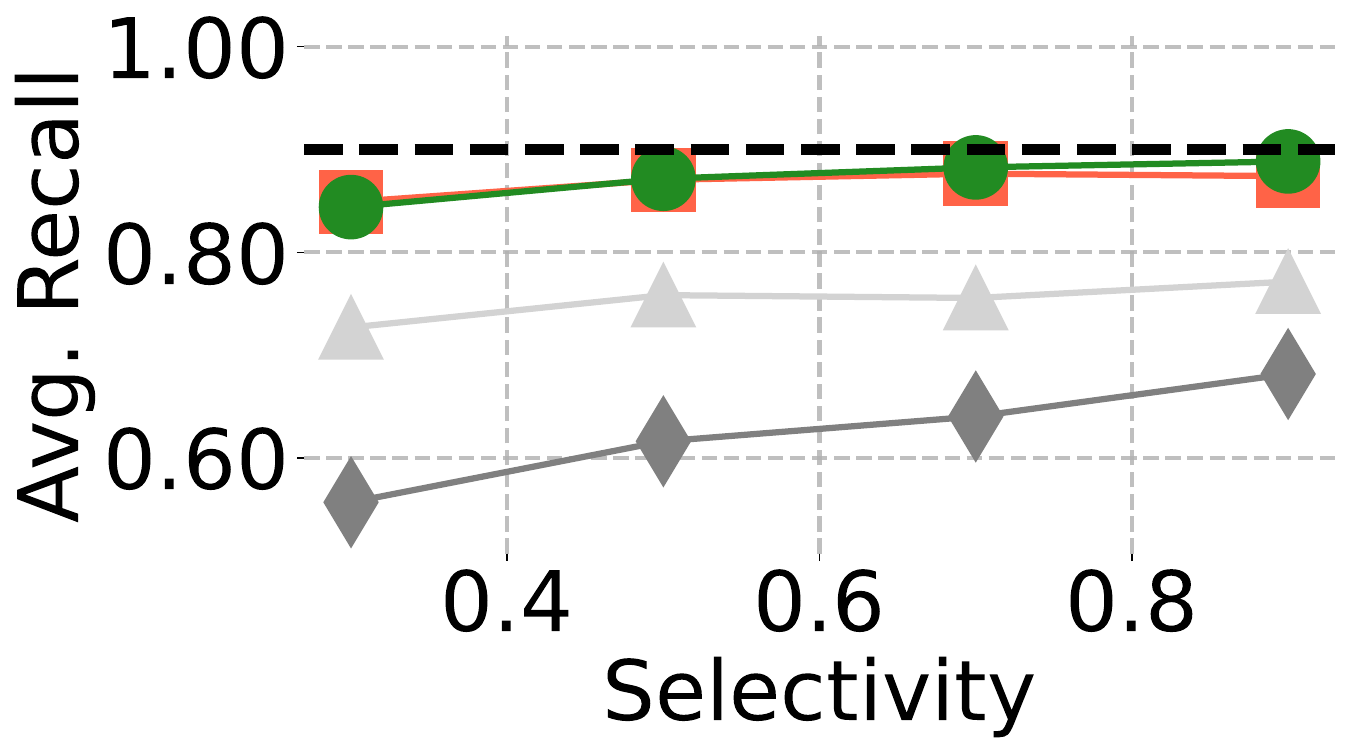}
        \vspace{-0.4cm}
        \caption{ACORN: Achieved recalls for varying selectivity (Negative Correlation, $k=100$, $R_t=0.9$).}
        \label{fig:recall-vs-selectivity-rt0.9-k100-acorn-neg-corr}
    \end{minipage}

    
    \begin{minipage}[t]{\textwidth}
        \includegraphics[width=0.19\textwidth]{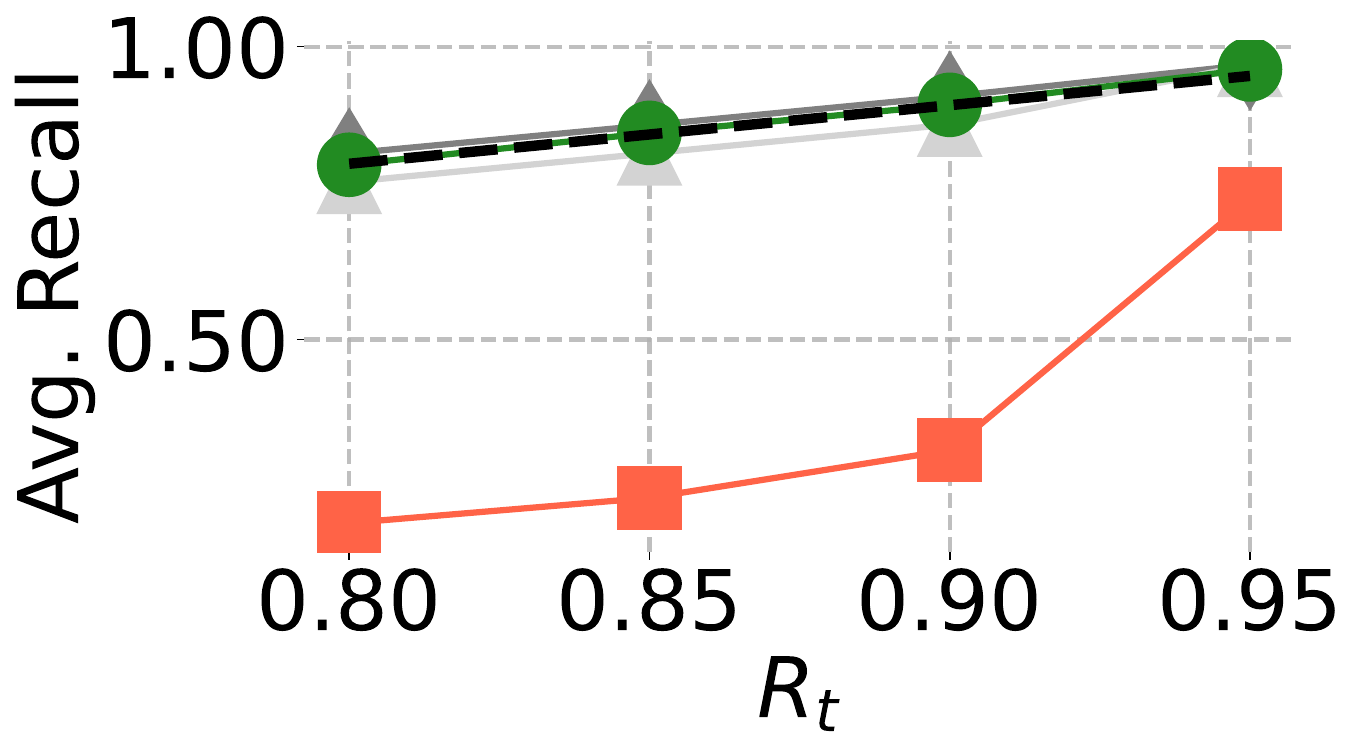}\hfill
        \includegraphics[width=0.19\textwidth]{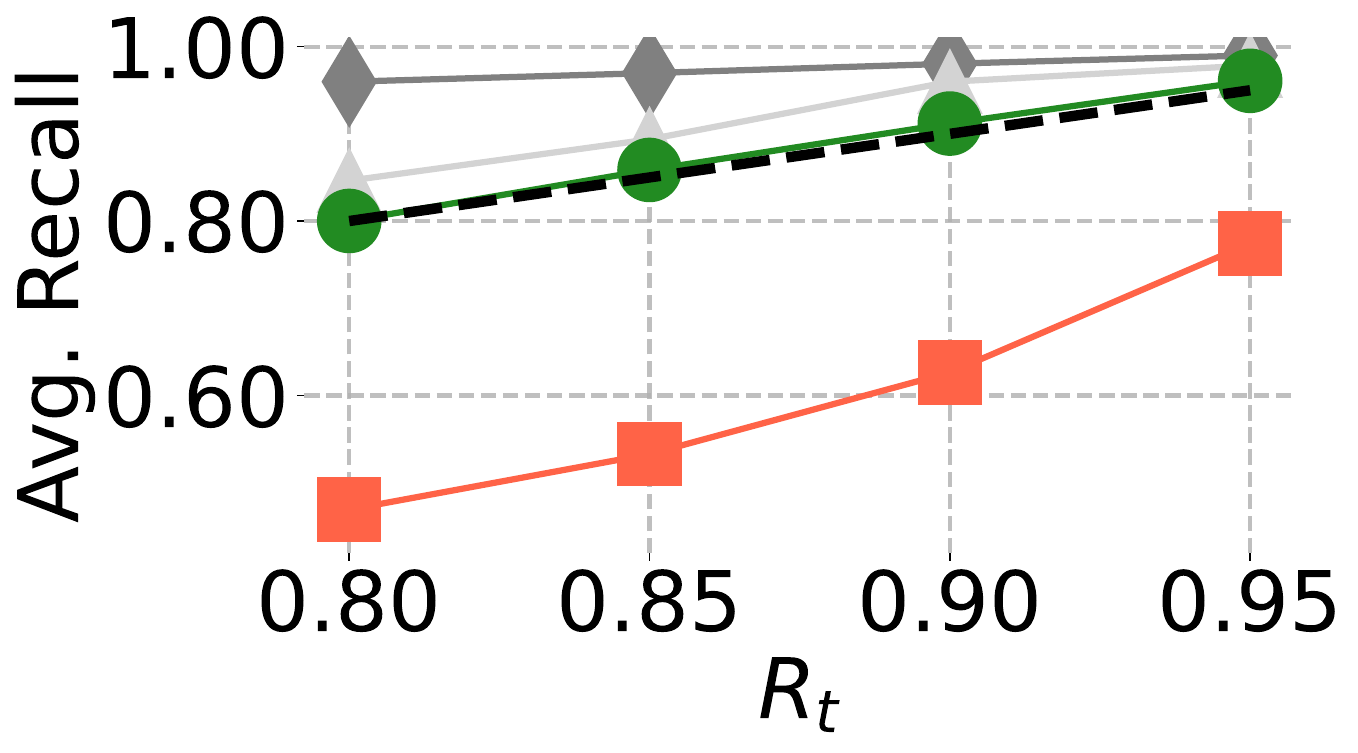}\hfill
        \includegraphics[width=0.19\textwidth]{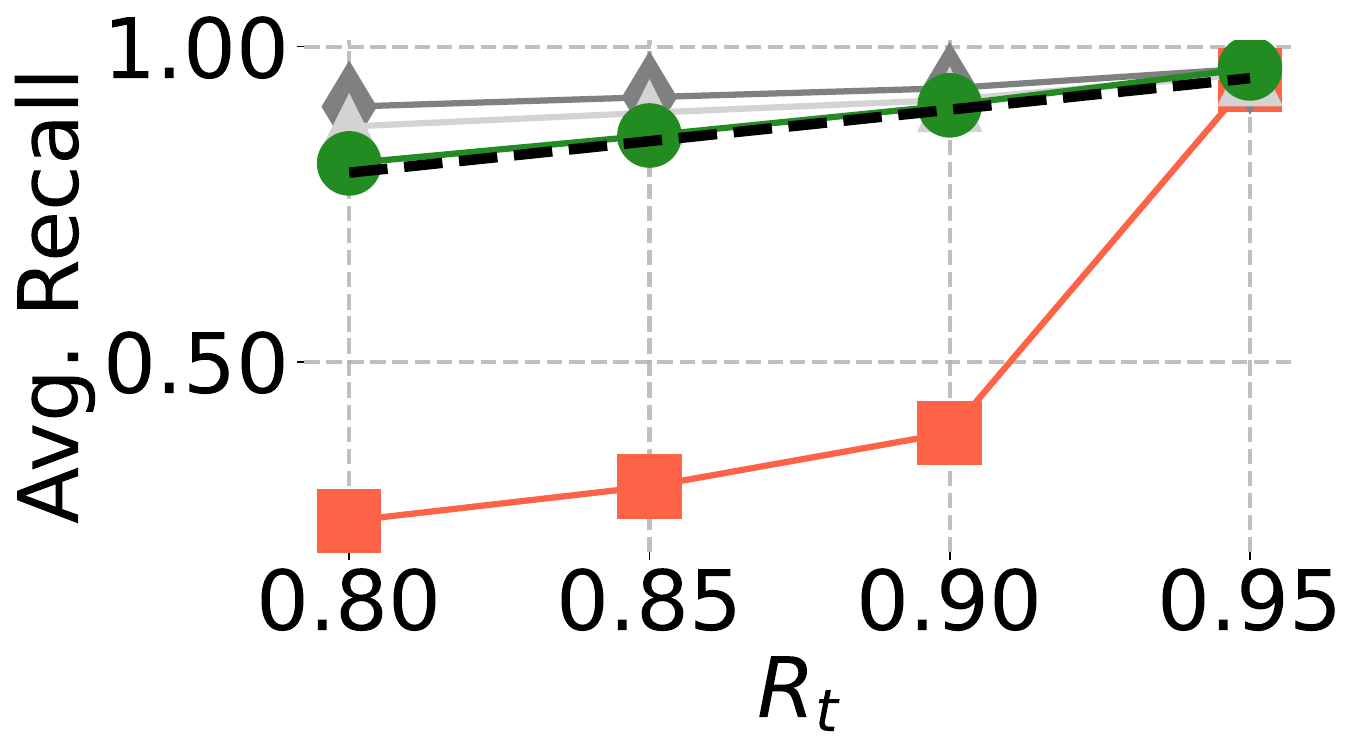}\hfill
        \includegraphics[width=0.19\textwidth]{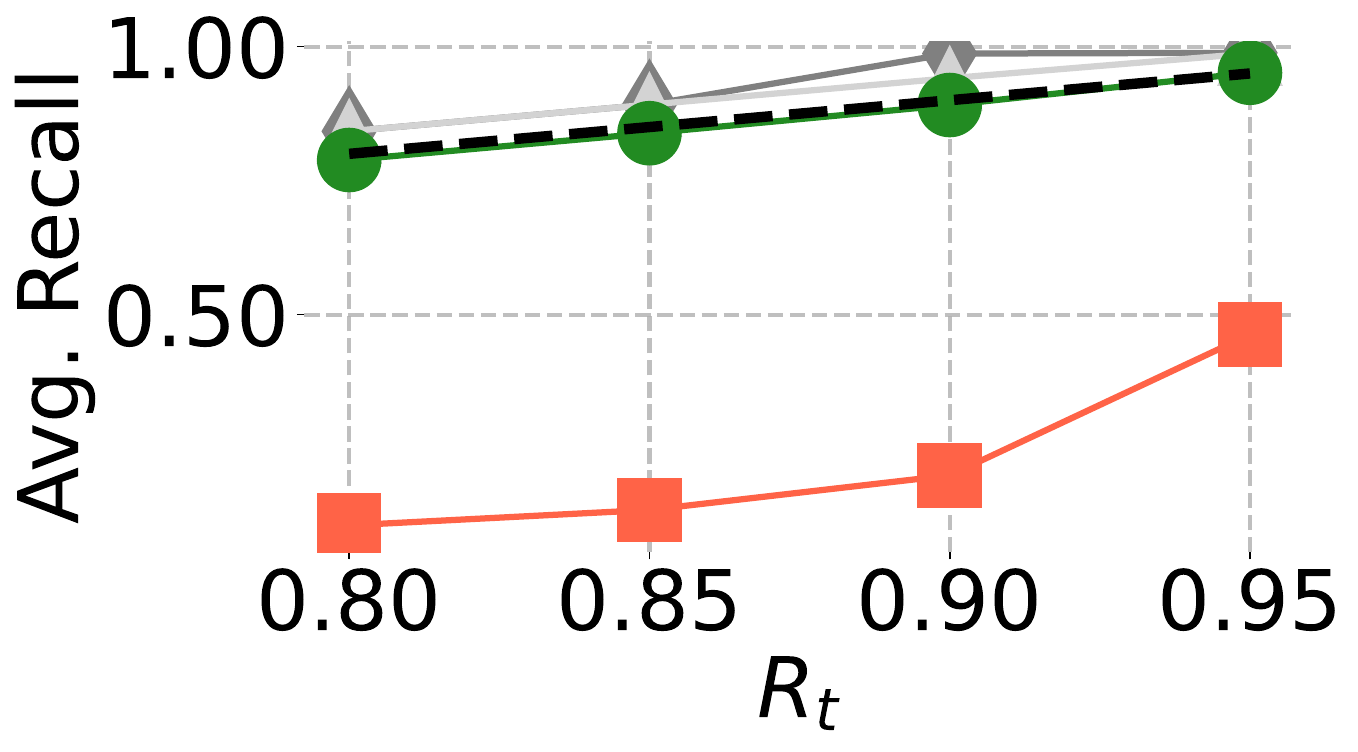}\hfill
        \includegraphics[width=0.19\textwidth]{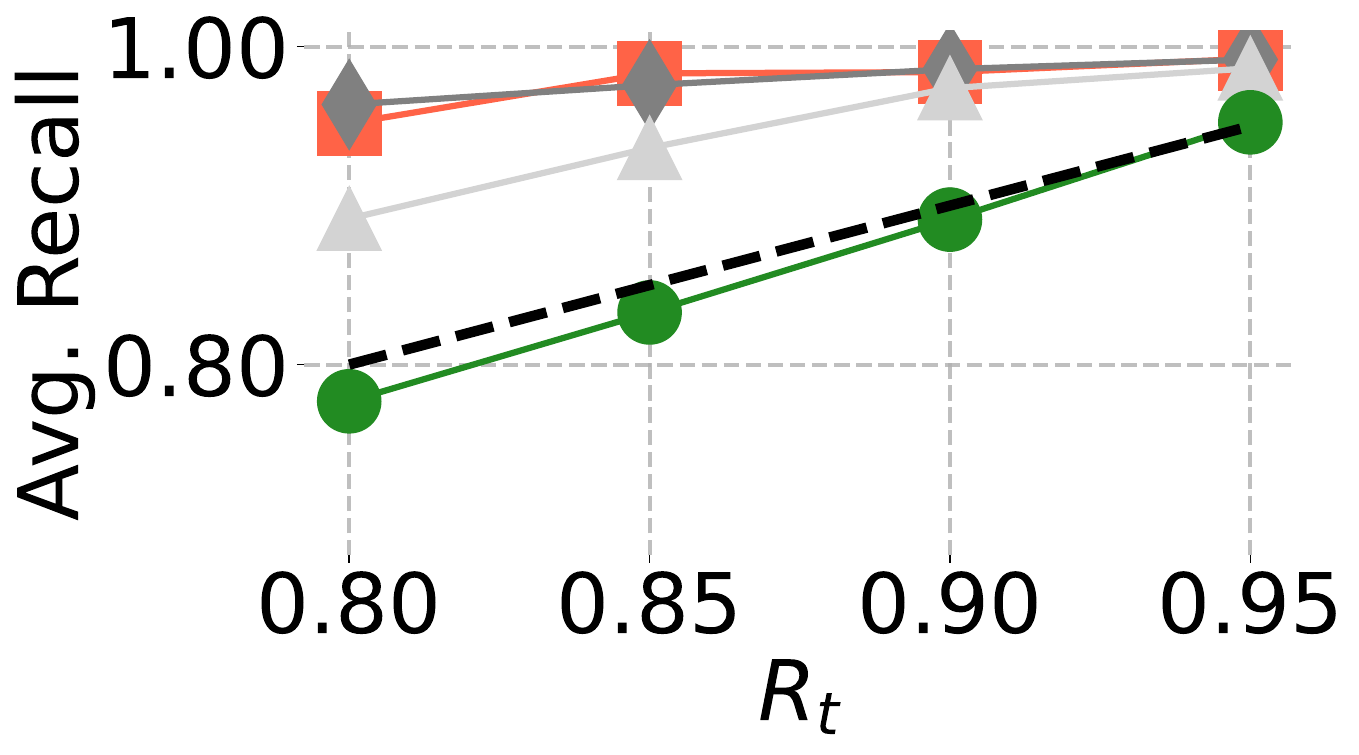}
        \vspace{-0.4cm}
        \caption{Sweeping: Achieved recalls for varying recall target (Negative Correlation, $k=100$, $sel=0.3$).}
        \label{fig:recall-vs-rt-k100-sel0.3-sweeping-neg-corr}
    \end{minipage}

    \begin{minipage}[t]{\textwidth}
        \includegraphics[width=0.19\textwidth]{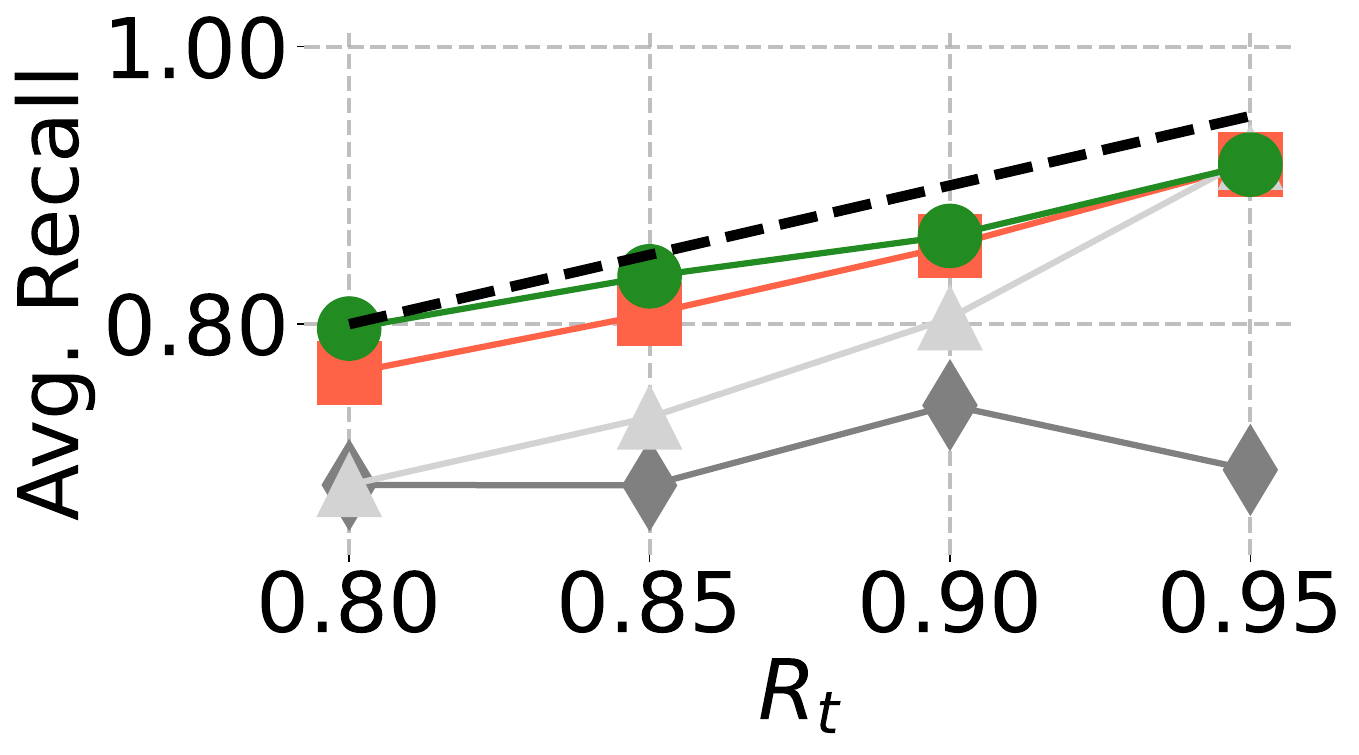}\hfill
        \includegraphics[width=0.19\textwidth]{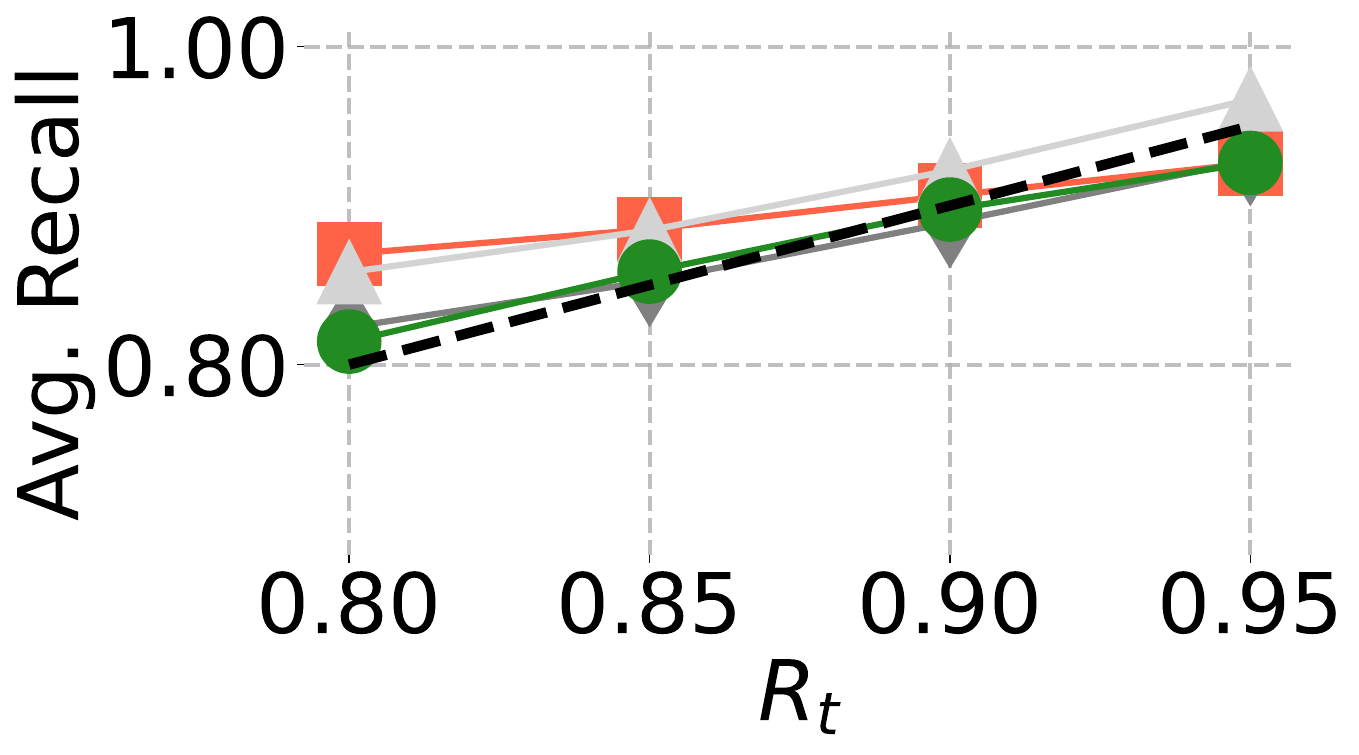}\hfill
        \includegraphics[width=0.19\textwidth]{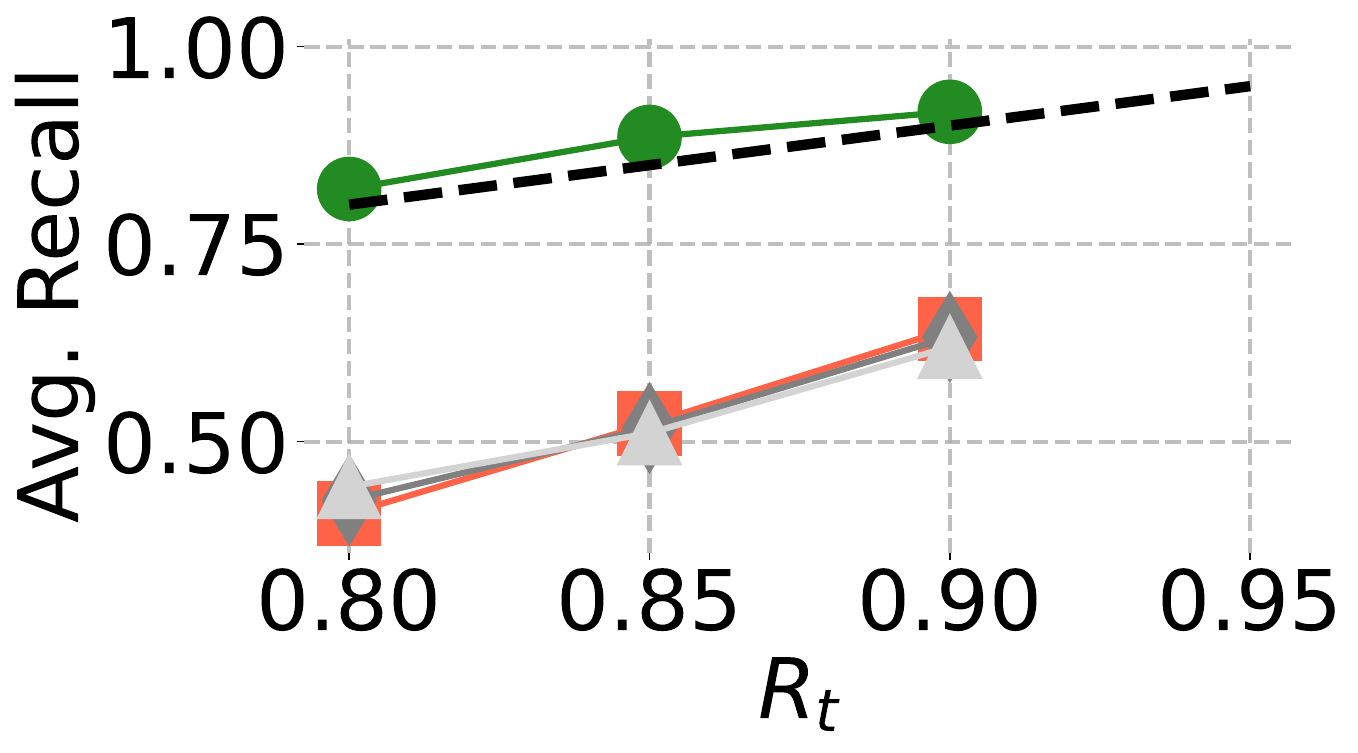}\hfill
        \includegraphics[width=0.19\textwidth]{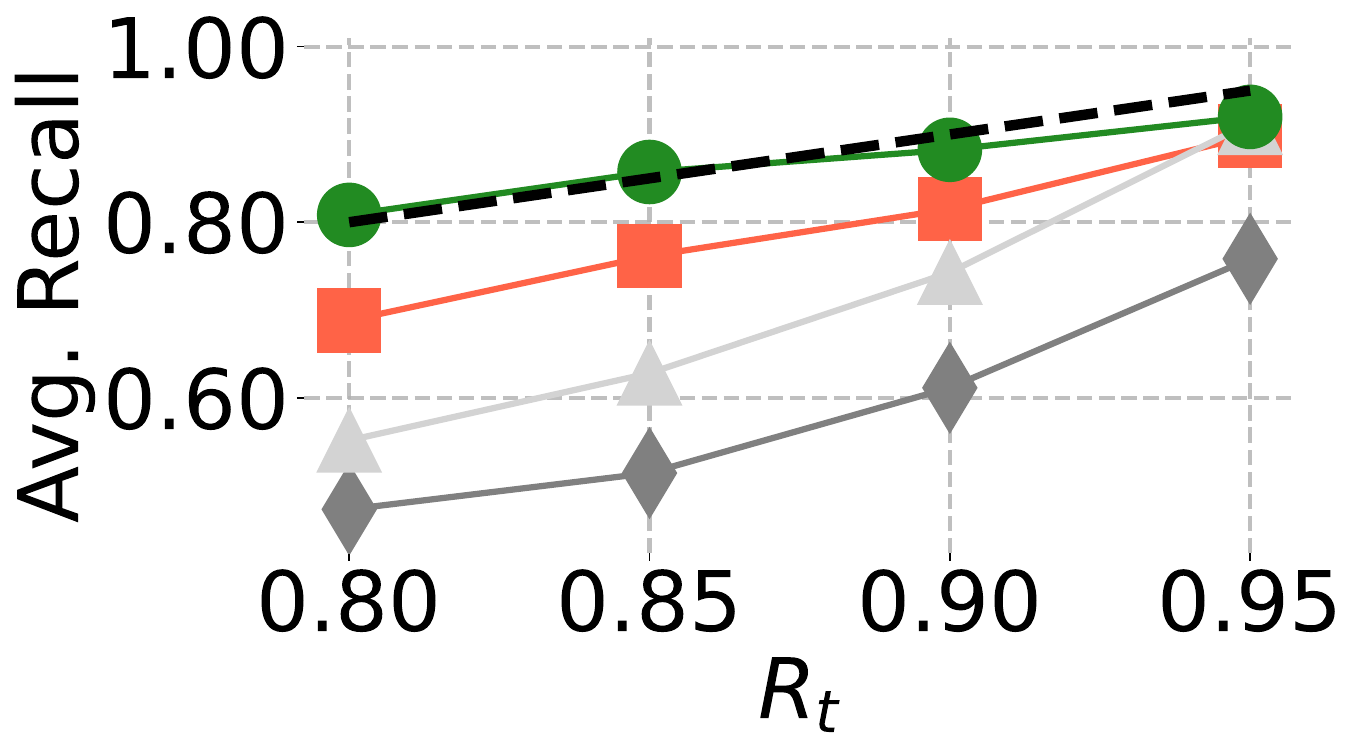}\hfill
        \includegraphics[width=0.19\textwidth]{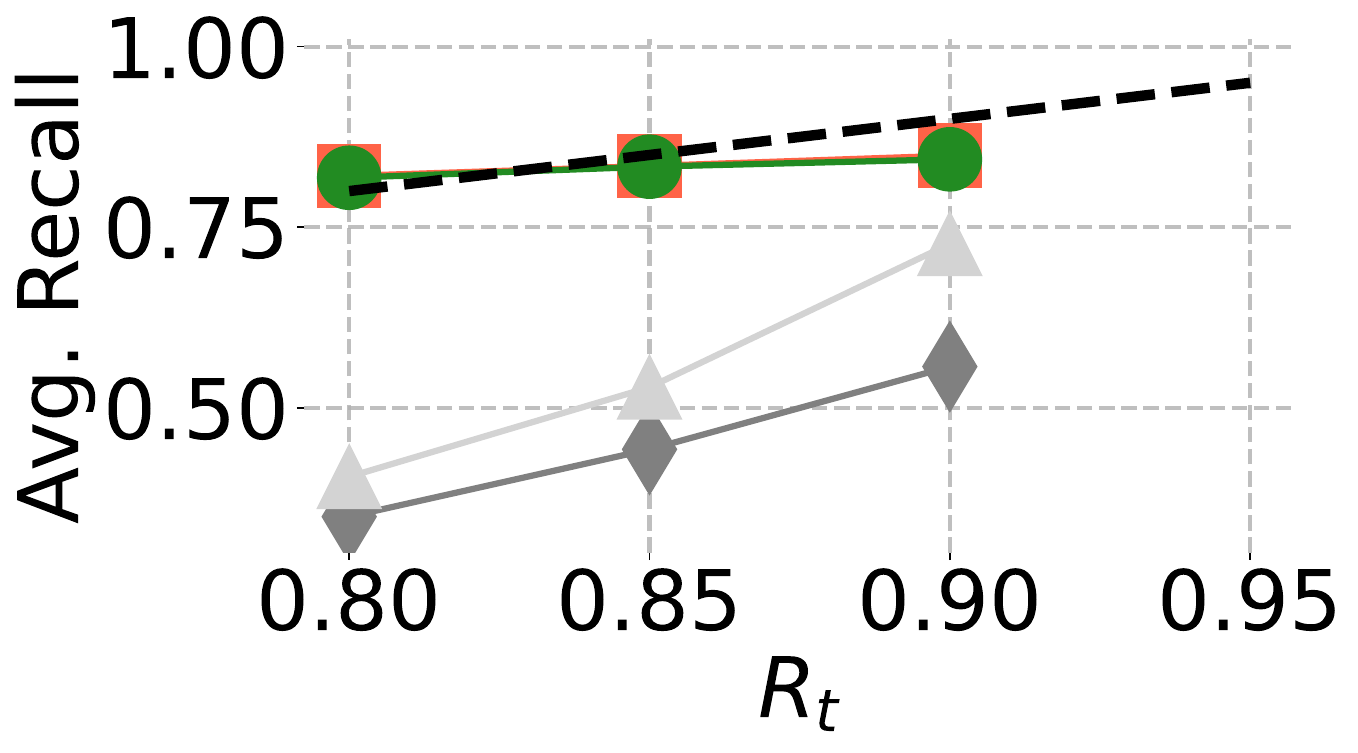}
        \vspace{-0.4cm}
        \caption{ACORN: Achieved recalls for varying recall target (Negative Correlation, $k=100$, $sel=0.3$).}
        \label{fig:recall-vs-rt-k100-sel0.3-acorn-neg-corr}
    \end{minipage}

    
    \begin{minipage}[t]{\textwidth}
        \includegraphics[width=0.19\textwidth]{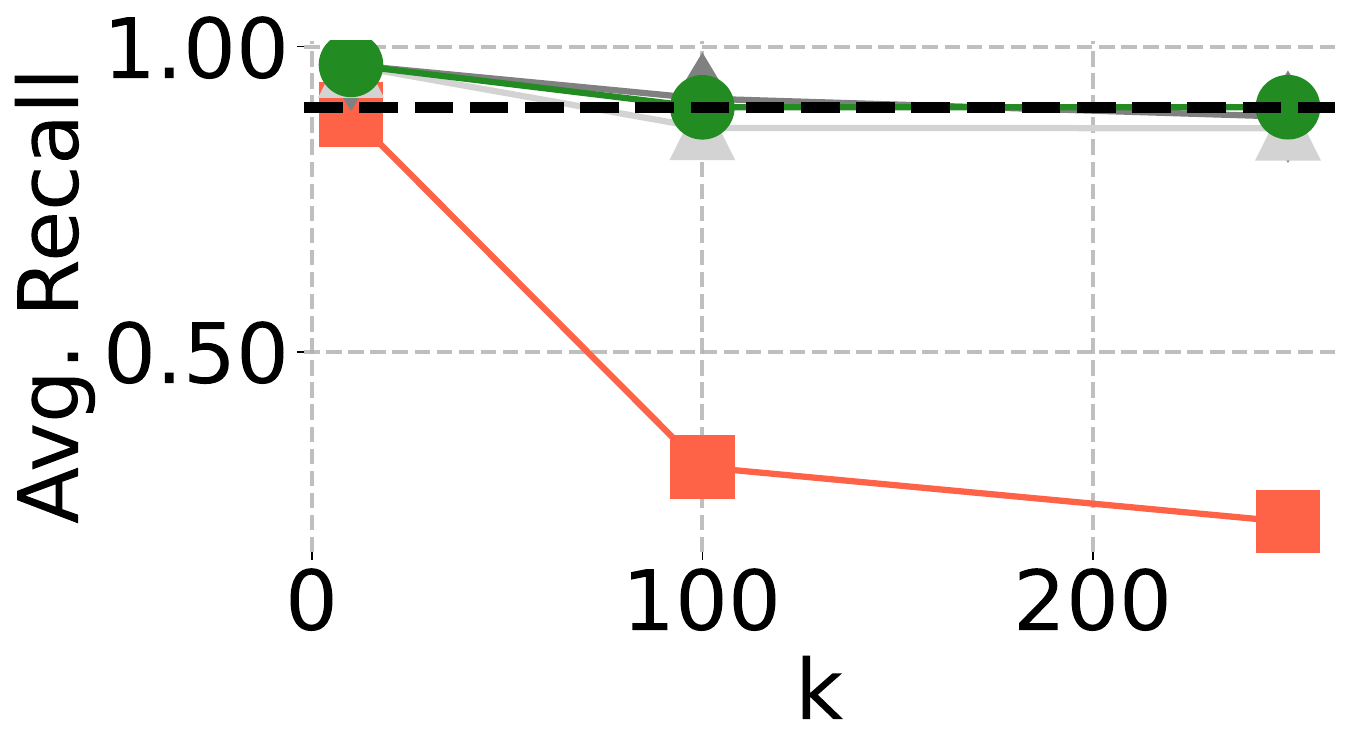}\hfill
        \includegraphics[width=0.19\textwidth]{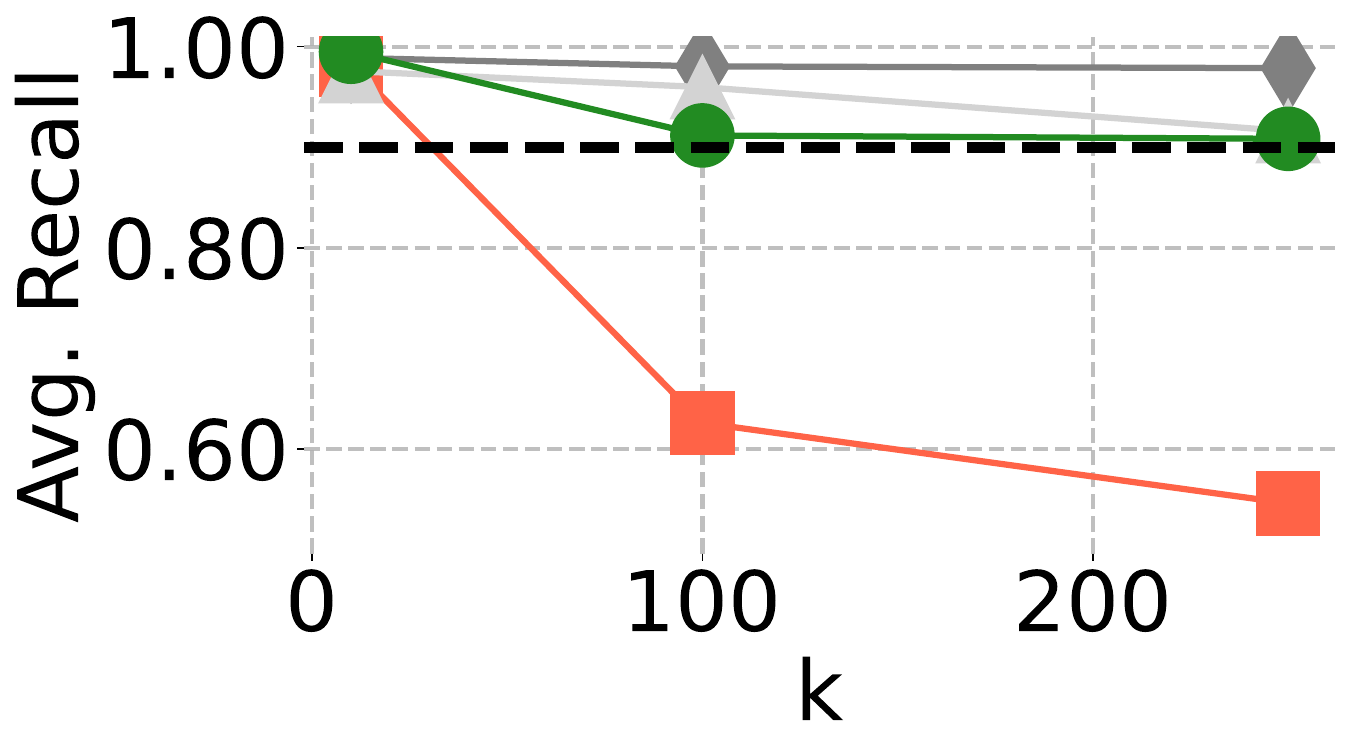}\hfill
        \includegraphics[width=0.19\textwidth]{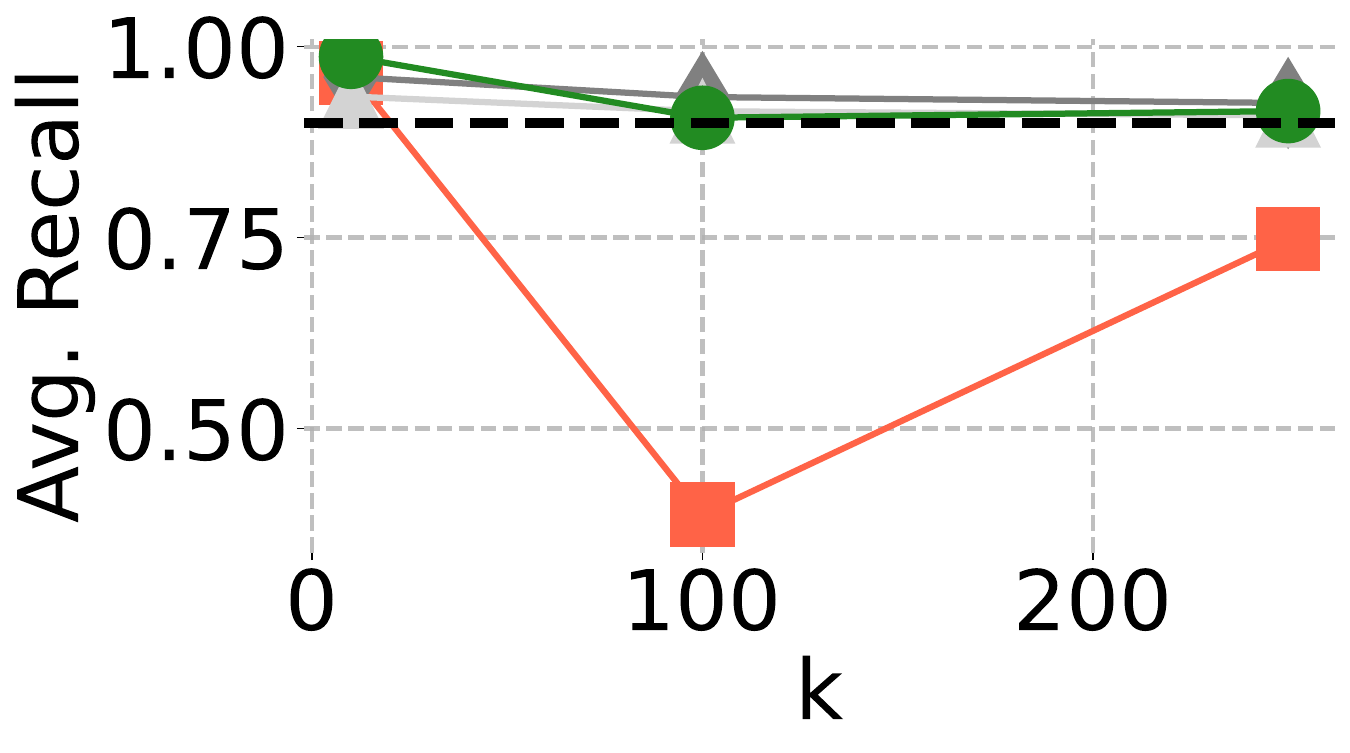}\hfill
        \includegraphics[width=0.19\textwidth]{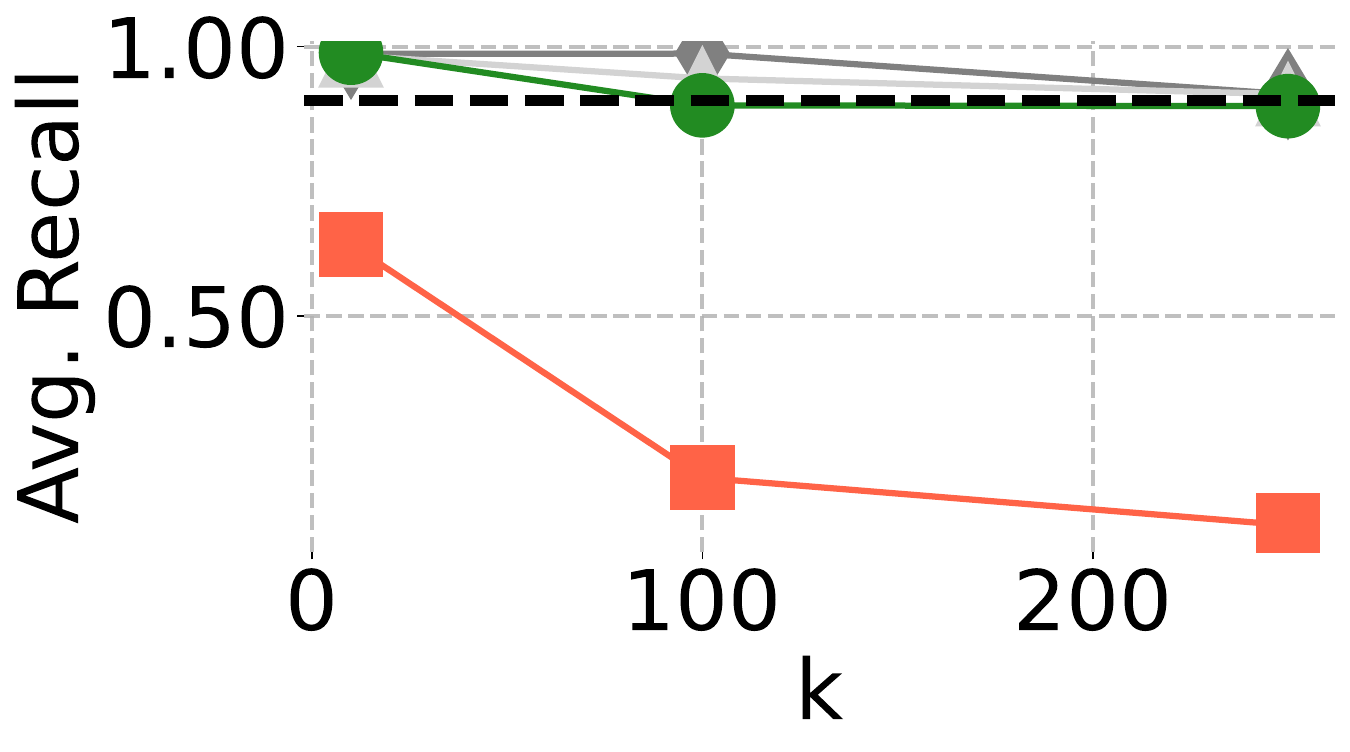}\hfill
        \includegraphics[width=0.19\textwidth]{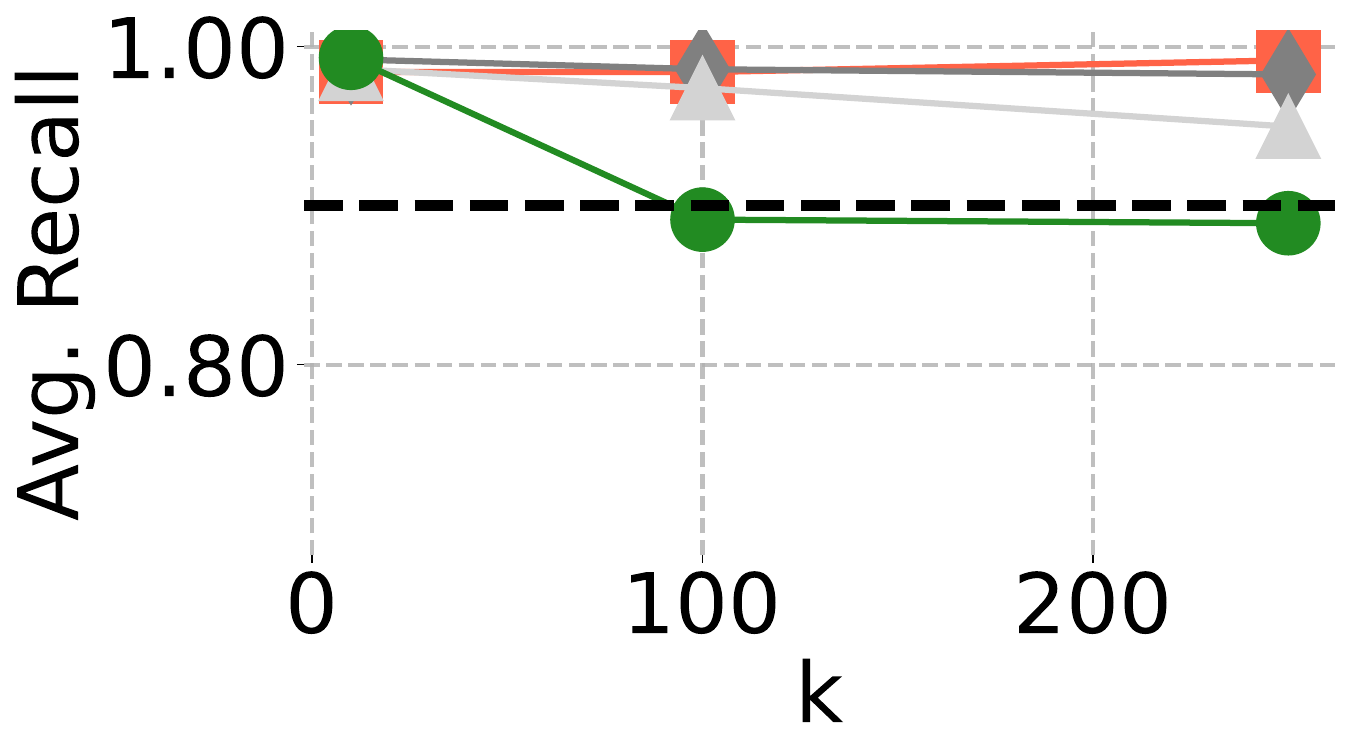}
        \vspace{-0.4cm}
        \caption{Sweeping: Achieved recalls for varying value of $k$ (Negative Correlation, $R_t=0.9$, $sel=0.3$).}
        \label{fig:recall-vs-k-rt0.9-sel0.3-sweeping-neg-corr}
    \end{minipage}
    
    \begin{minipage}[t]{\textwidth}
        \includegraphics[width=0.19\textwidth]{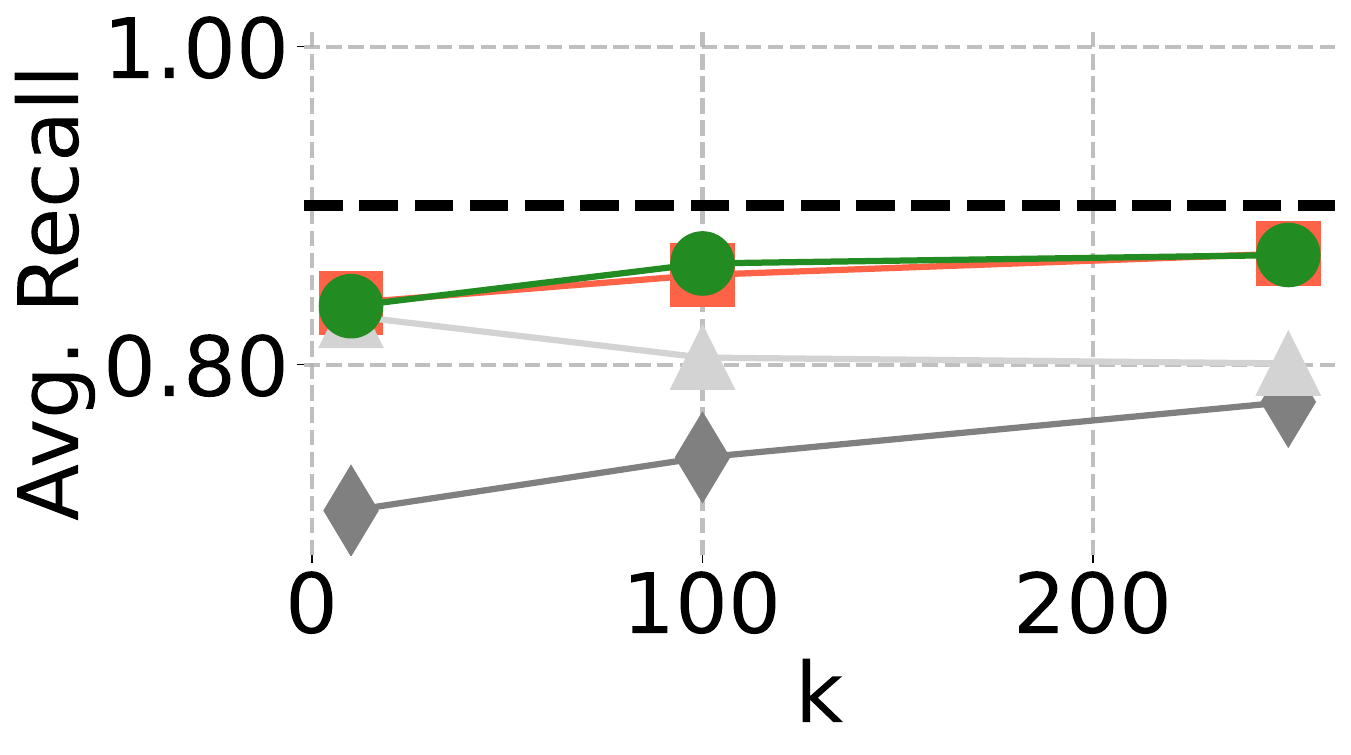}\hfill
        \includegraphics[width=0.19\textwidth]{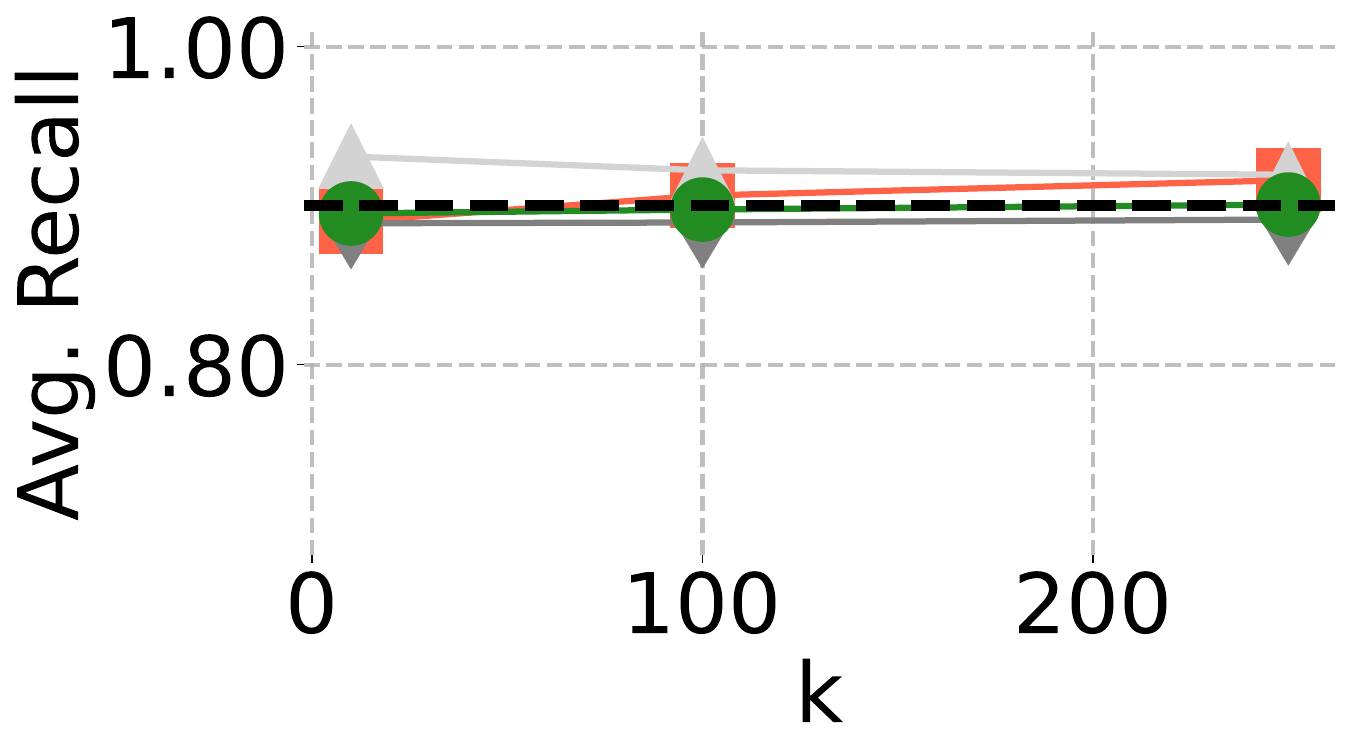}\hfill
        \includegraphics[width=0.19\textwidth]{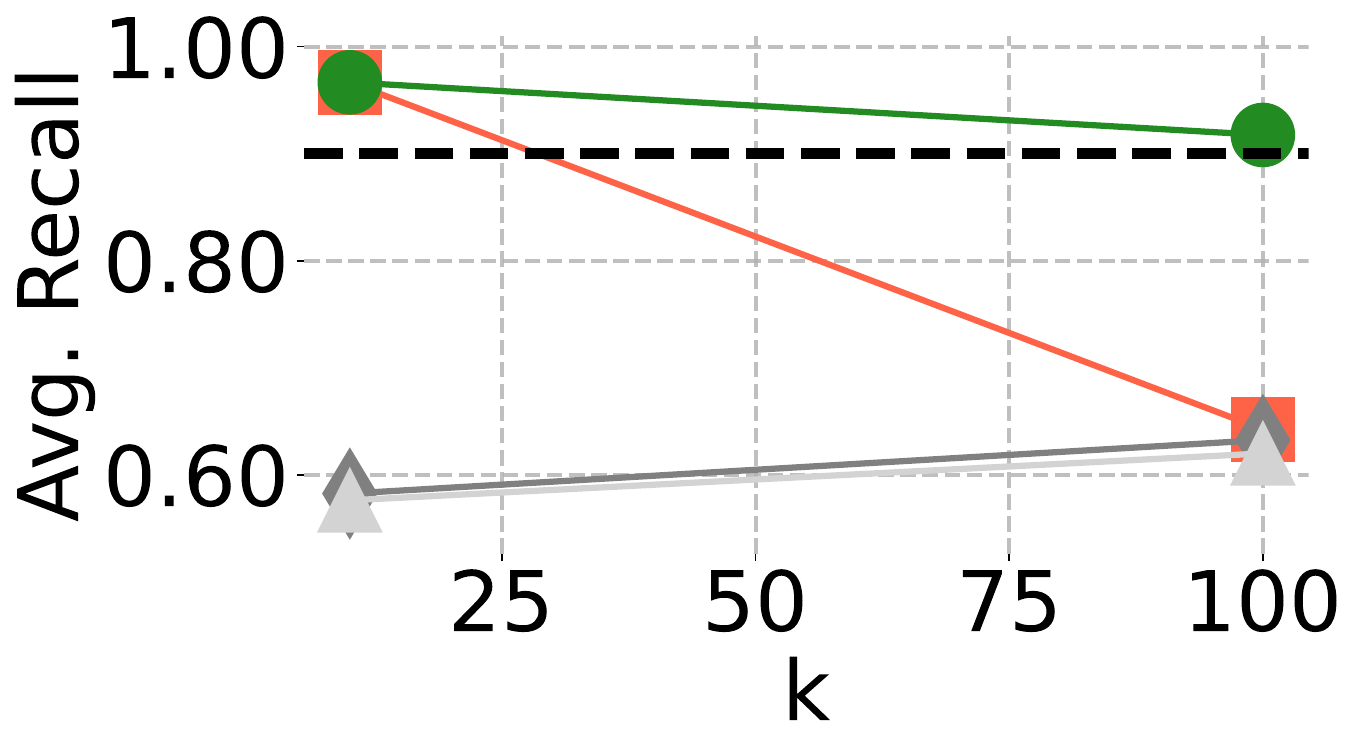}\hfill
        \includegraphics[width=0.19\textwidth]{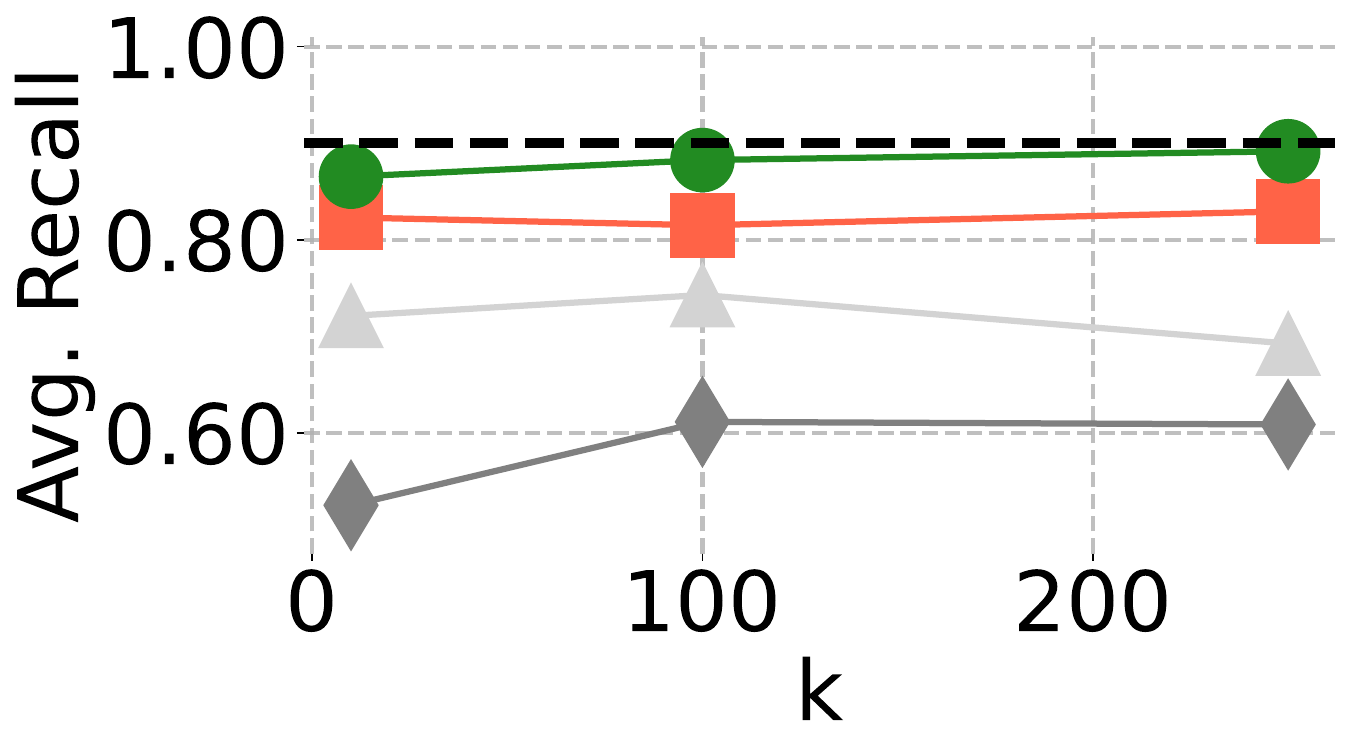}\hfill
        \includegraphics[width=0.19\textwidth]{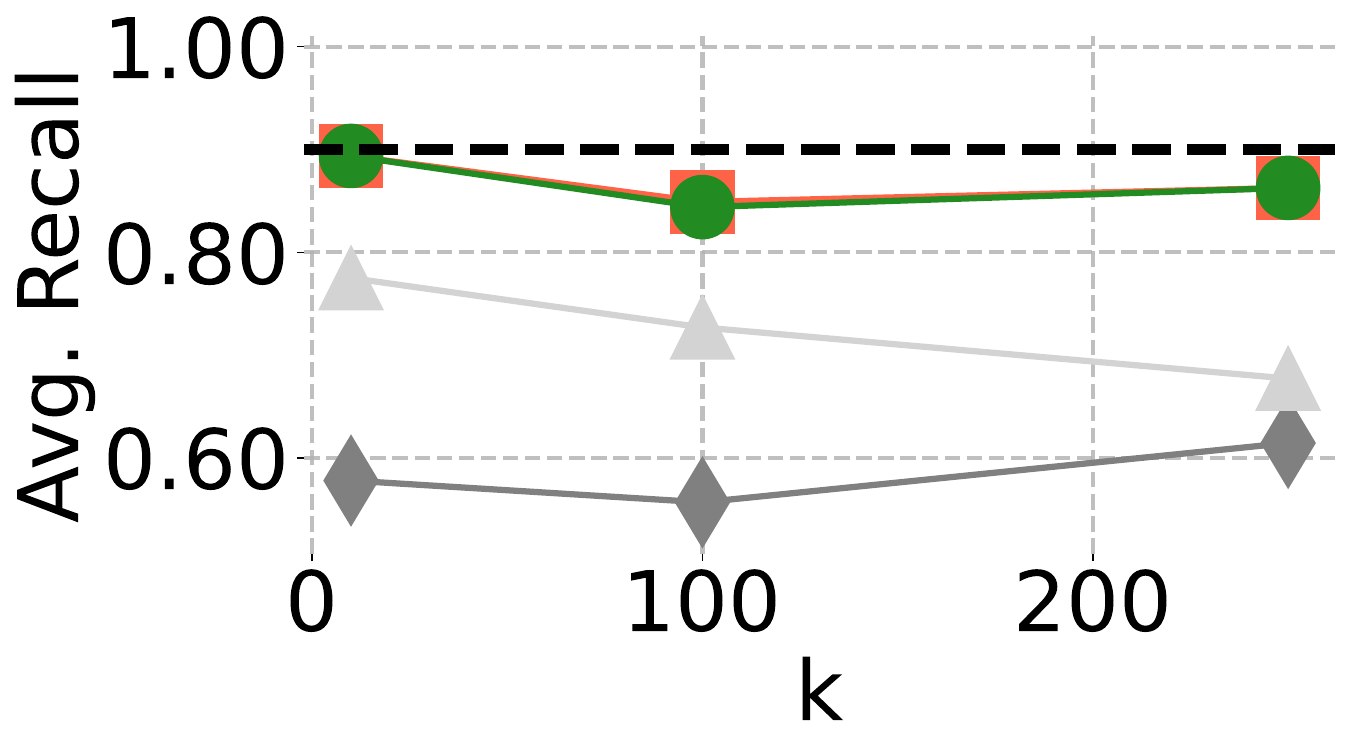}
        \vspace{-0.4cm}
        \caption{ACORN: Achieved recalls for varying value of $k$ (Negative Correlation, $R_t=0.9$, $sel=0.3$).}
        \label{fig:recall-vs-k-rt0.9-sel0.3-acorn-neg-corr}
    \end{minipage}
\end{figure*}

We now proceed to compare how VADER compares to other baselines, starting by comparing the results of different approaches for varying selectivities.
Throughout this experiment, we set $R_t=0.9$, and $k=100$, and we test all methods for all datasets, selectivities, and correlations.
The results are shown in Figure~\ref{fig:recall-vs-selectivity-rt0.9-k100-sweeping-no-corr} for Sweeping and Figure~\ref{fig:recall-vs-selectivity-rt0.9-k100-acorn-no-corr} for ACORN for no correlation.
The results for positive and negative correlation for Sweeping are in Figure~\ref{fig:recall-vs-selectivity-rt0.9-k100-sweeping-pos-corr} and Figure~\ref{fig:recall-vs-selectivity-rt0.9-k100-sweeping-neg-corr}, while for ACORN in Figure~\ref{fig:recall-vs-selectivity-rt0.9-k100-acorn-pos-corr} and Figure~\ref{fig:recall-vs-selectivity-rt0.9-k100-acorn-neg-corr}.
Note that the black dashed line corresponds to $R_t=0.9$, and thus, the closer a method is to the line, the better result quality it achieves.

Our results show that VADER is the approach with the best results.
From the results, and across all correlations, VADER is the only approach that consistently maintains recall values close to the target across all configurations, with an average deviation of 0.04.
This represents the best performance among all methods, outperforming REM-Filtered by 12\%, REM-Unfiltered by 25\%, and DARTH by 81\%.
Notably, the second-best method, REM-Filtered, is specifically tuned for each selectivity using the same FVS training queries as VADER, yet VADER still achieves superior accuracy.
In addition to improved result quality, VADER is 32\% faster than REM-Filtered and 52\% faster than REM-Unfiltered.
VADER is not faster than DARTH in this configuration, as DARTH fails to meet the target recall in most cases.
Furthermore, both REM variants tend to overshoot, performing more distance computations than necessary.
Finally, when we examine queries that fall below the target recall, VADER and REM-Unfiltered have the same average deviation, while VADER is 20\% better than REM-Filtered and 84\% better than DARTH.
Regarding ACORN, VADER again achieves the best performance, this time by an even larger margin.
The results show that VADER consistently achieves recall values closest to the target, with an average deviation of 0.04.
This corresponds to improvements of 28\% over REM-Filtered, 44\% over REM-Unfiltered, and 17\% over DARTH.
While VADER may incur slightly higher latency in ACORN due to its consistent effort to meet the recall target, competing methods often fail to do so, it maintains stable and high-quality results across all selectivities.
When focusing on queries that fall below the target recall, VADER again demonstrates superior performance, improving by 53\% over REM-Filtered, 72\% over REM-Unfiltered, and 60\% over DARTH.

\subsubsection{Performance for Varying Recall Targets.}
Fixing $sel=0.3$ and varying the recall target $R_t$ at $k=100$, Figures~\ref{fig:recall-vs-rt-k100-sel0.3-sweeping-no-corr} and~\ref{fig:recall-vs-rt-k100-sel0.3-acorn-no-corr} show the no correlation results for Sweeping and ACORN.
The results for positive and negative correlation with Sweeping are presented in Figures~\ref{fig:recall-vs-rt-k100-sel0.3-sweeping-pos-corr} and~\ref{fig:recall-vs-rt-k100-sel0.3-sweeping-neg-corr} for Sweeping and in Figures~\ref{fig:recall-vs-rt-k100-sel0.3-acorn-pos-corr} and~\ref{fig:recall-vs-rt-k100-sel0.3-acorn-neg-corr} for ACORN.
For Sweeping, and across all correlations, VADER is again closest to the ideal target, indicated by the dashed line.
It outperforms REM-Filtered by 5\%, REM-Unfiltered by 20\%, and DARTH by 75\%, while being 29\%, 48\%, and 47\% faster respectively.
On queries that fall below the target it trails REM-Unfiltered (average error 0.05 against 0.03, a 35\% gap) but still outperforms REM-Filtered by 12\% and DARTH by 80\%.
For ACORN, VADER outperforms REM-Filtered by 56\%, REM-Unfiltered by 65\%, and DARTH by 50\% in proximity to the target, and by 59\%, 70\%, and 54\% respectively on queries below it; it is slightly slower only because the other methods terminate prematurely without reaching the target.

\subsubsection{Performance for Varying Values of $k$.}
Finally, we vary $k$, with the no correlation results shown in Figure~\ref{fig:recall-vs-k-rt0.9-sel0.3-sweeping-no-corr} for Sweeping and Figure~\ref{fig:recall-vs-k-rt0.9-sel0.3-acorn-no-corr} for ACORN.
The results for positive and negative correlation with Sweeping are presented in Figures~\ref{fig:recall-vs-k-rt0.9-sel0.3-sweeping-pos-corr} and~\ref{fig:recall-vs-k-rt0.9-sel0.3-sweeping-neg-corr} for Sweeping and in Figures~\ref{fig:recall-vs-k-rt0.9-sel0.3-acorn-pos-corr} and~\ref{fig:recall-vs-k-rt0.9-sel0.3-acorn-neg-corr} for ACORN.

For Sweeping, VADER, DARTH, and REM-Unfiltered deviate from the target by similar amounts, with VADER 9\% better than REM-Unfiltered; REM-Unfiltered is marginally faster than VADER, which is in turn 11\% faster than DARTH, while REM-Filtered matches their latency but reaches the target less often.
For ACORN, VADER is again the most stable, improving by 45\% over REM-Filtered, 58\% over REM-Unfiltered, and 28\% over DARTH, and by 79\%, 83\%, and 74\% respectively on queries that fall below the target.

\subsubsection{Results on Real Structured Metadata}
\label{sec:experiments:caselaw}
\begin{figure*}
    \begin{minipage}[t]{0.99\textwidth}
        \centering
        \begin{subfigure}[t]{0.19\textwidth}
            \centering
            \includegraphics[width=\textwidth]{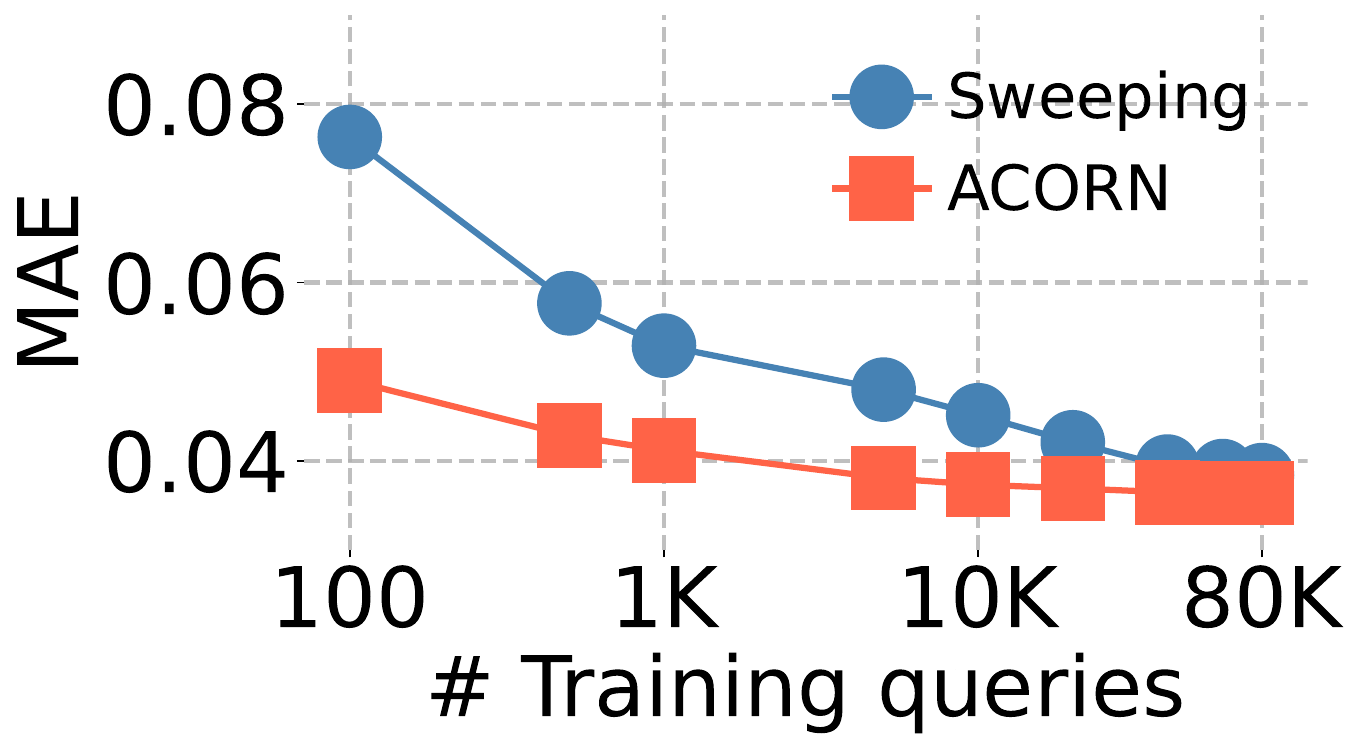}
            \caption{MAE for Varying training queries}
        \end{subfigure}
        \hfill
        \begin{subfigure}[t]{0.19\textwidth}
            \centering
            \includegraphics[width=\textwidth]{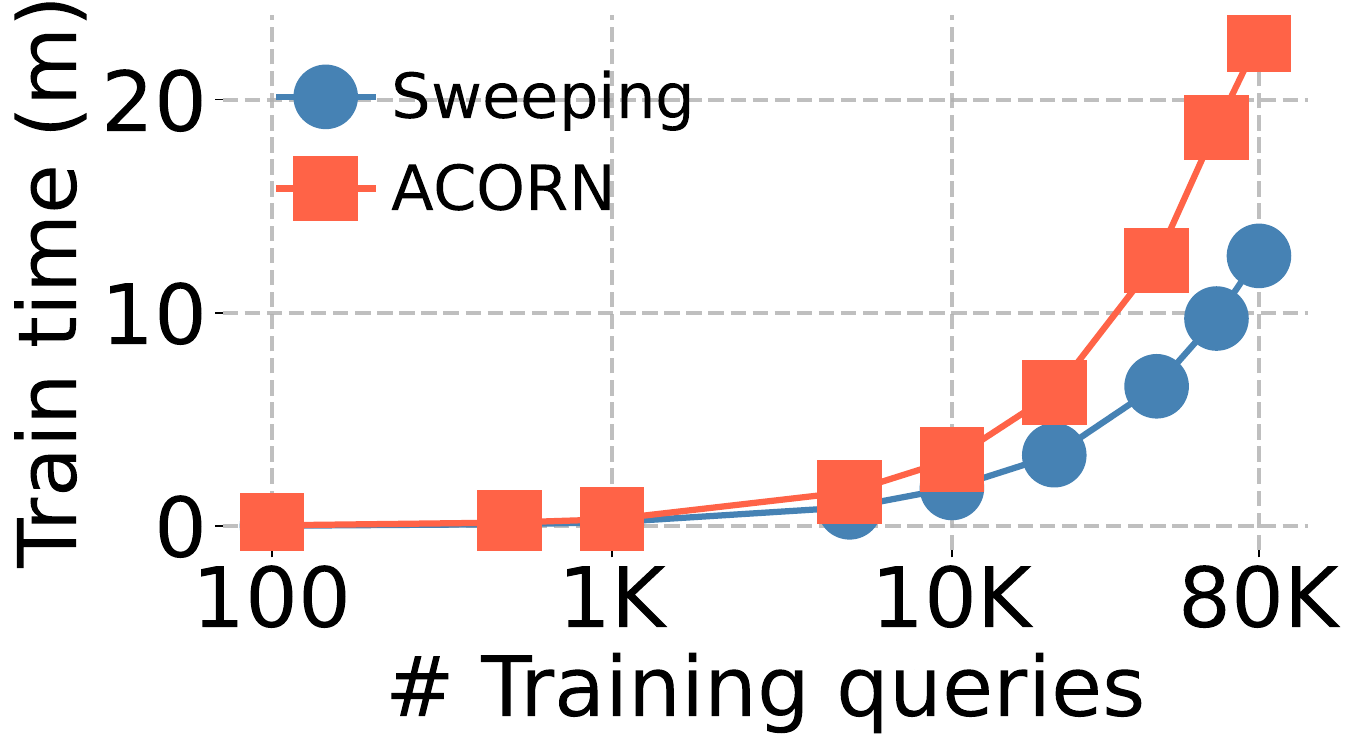}
            \caption{Training time for varying training queries}
        \end{subfigure}
        \hfill
        \begin{subfigure}[t]{0.19\textwidth}
            \centering
            \includegraphics[width=\textwidth]{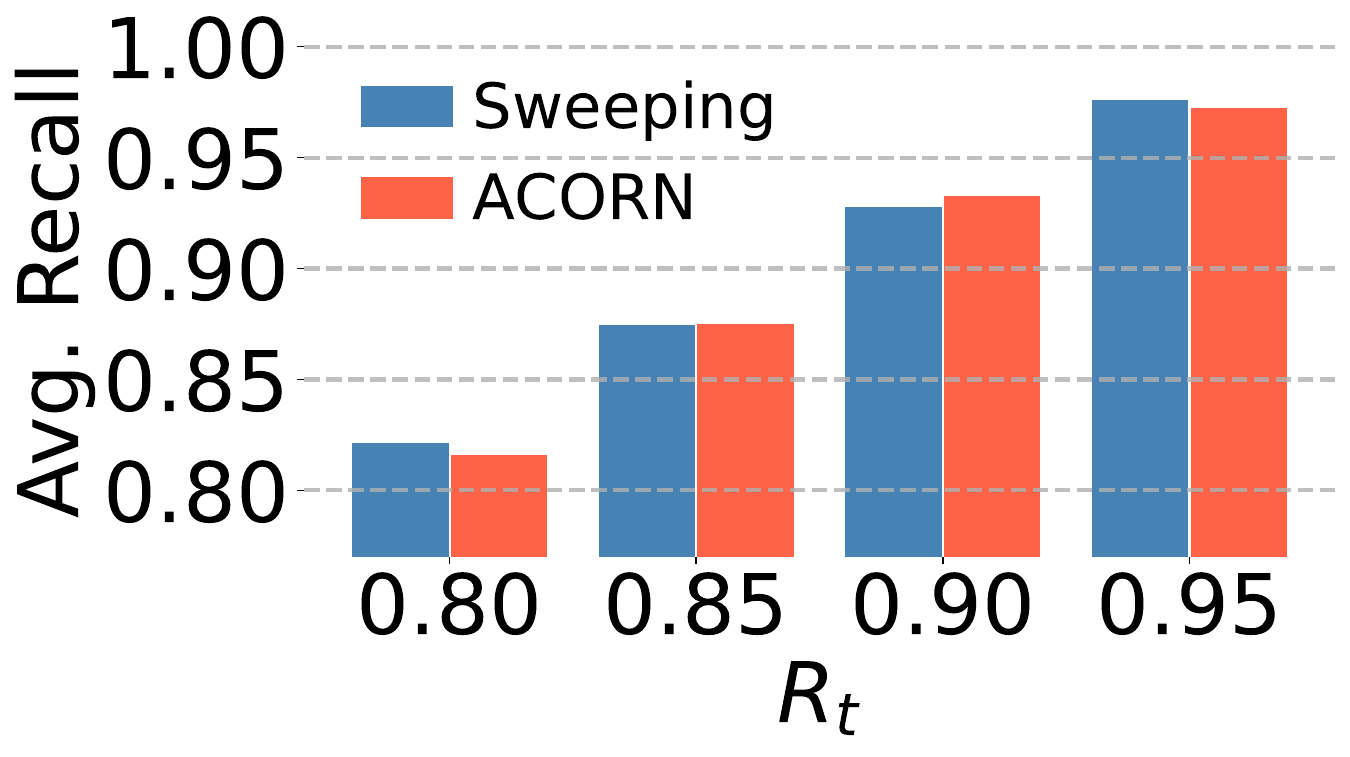}
            \caption{Varying $R_t$}
        \end{subfigure}
        \hfill
        \begin{subfigure}[t]{0.19\textwidth}
            \centering
            \includegraphics[width=\textwidth]{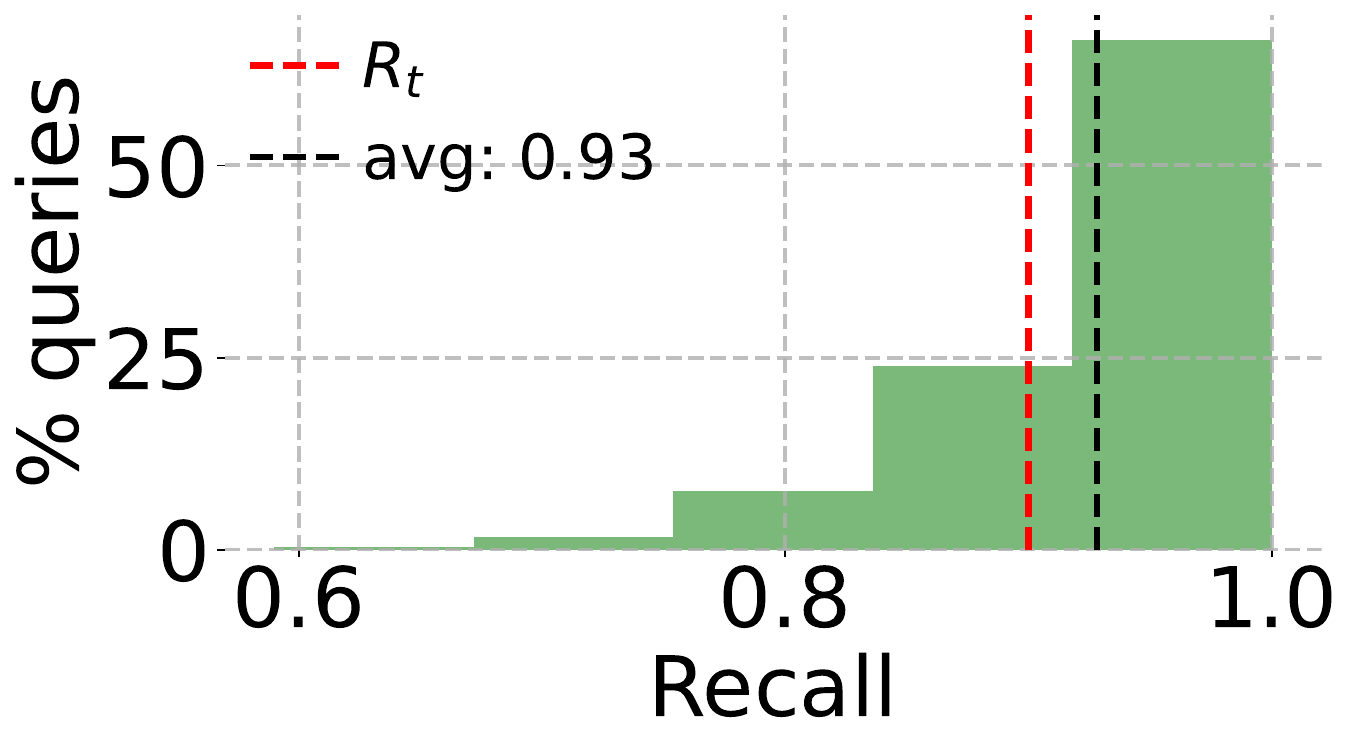}
            \caption{Recall distr. (Sweeping, $R_t=0.9$)}
        \end{subfigure}
        \hfill
        \begin{subfigure}[t]{0.19\textwidth}
            \centering
            \includegraphics[width=\textwidth]{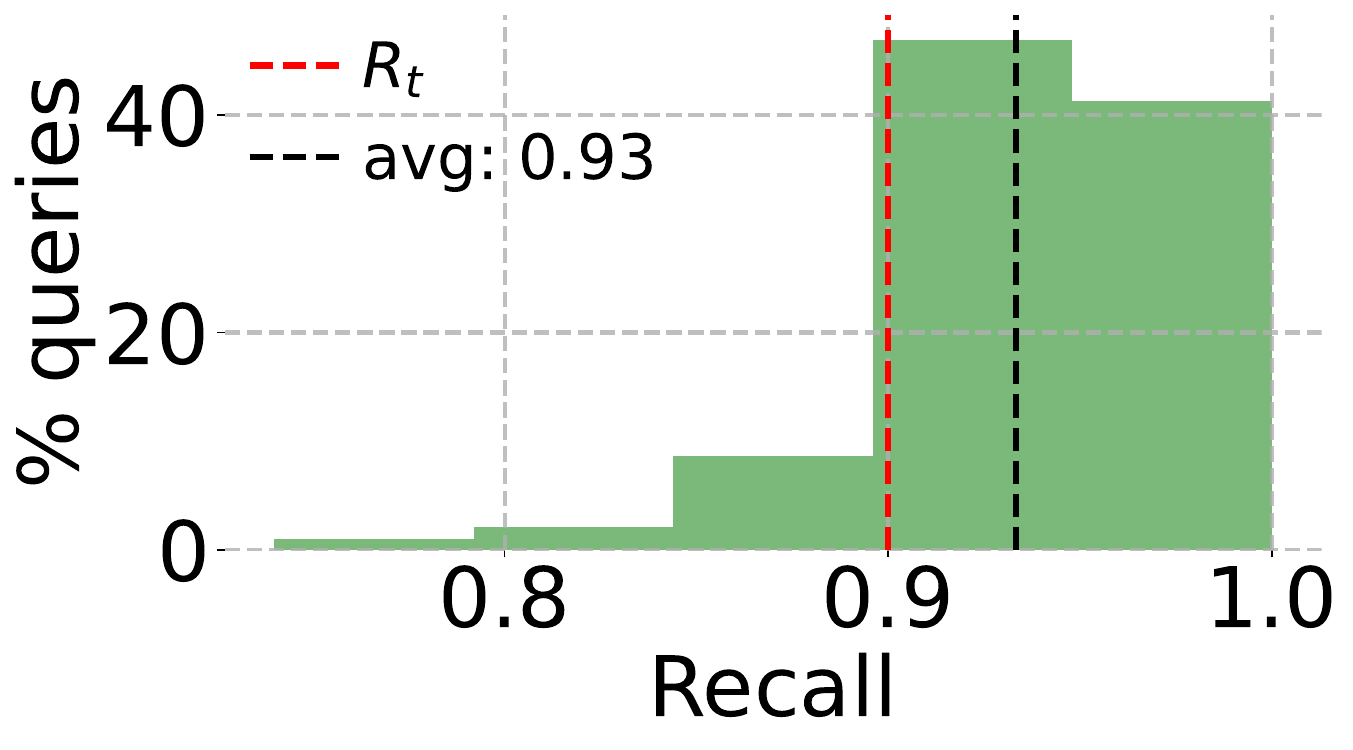}
            \caption{Recall distr. (ACORN, $R_t=0.9$)}
        \end{subfigure}
    \vspace{-0.3cm}
    \caption{VADER for the CASELAW7M dataset ($k=100$).}
    \label{fig:vader-caselaw-k100}
    \end{minipage}
    \vspace{-0.3cm}
\end{figure*}

We now evaluate VADER on the CASELAW7M~\cite{annbench2025caselaw} dataset, which contains filtered vector search queries over real structured metadata, allowing us to assess VADER without relying on a synthetic predicate generation process.
CASELAW7M consists of 7.4M vectors together with their associated structured metadata, and approximately 89K filtered queries defined over the real metadata, with an average predicate selectivity of 0.05.
From the provided workload, we reserve a subset of queries for training and use an additional 5K queries for evaluation.
To study the effect of training set size, we train VADER using between 100 and 80K training queries, with the results presented in Figure~\ref{fig:vader-caselaw-k100}(a).

VADER achieves recall prediction accuracy comparable to that observed on the previous datasets.
Specifically, for Sweeping, the MAE decreases from 0.076 with 100 training queries to 0.038 with 80K queries, while for ACORN it decreases from 0.049 to 0.036 over the same range.
Performance stabilizes at approximately 10K training queries, achieving $MAE=0.045$ for Sweeping and $MAE=0.037$ for ACORN.
The required training time, shown in Figure~\ref{fig:vader-caselaw-k100}(b), including both training data generation and GBDT model training, remains low, ranging from 0.02 to 13 minutes for Sweeping and from 0.04 to 23 minutes for ACORN.
Using 10K training queries, the total training time is only 1.8 minutes for Sweeping and 3.2 minutes for ACORN.
Although larger training sets provide marginal improvements in recall prediction accuracy, we adopt 10K training queries for the remainder of the evaluation in order to maintain consistency with 
the rest of our experiments.
Our training setup combines 500 query vectors with three correlation settings (no, positive, and negative correlation) and seven selectivities (0.01, 0.1, 0.2, 0.5, 0.7, 0.9, and 1.0), resulting in approximately 10.5K filtered training queries.
Therefore, using 10K CASELAW7M training queries provides a comparable amount of training data.
The CASELAW7M predictor is deployed using the same VADER configuration as in the previous experiments, including the proposed asymmetric loss function.

We also evaluate VADER across varying recall targets on CASELAW7M, in Figure~\ref{fig:vader-caselaw-k100}(c), for both Sweeping and ACORN.
In both algorithms, the achieved recall among all queries of the workload closely matches the corresponding target, confirming that the approach generalizes effectively to real-world filtered vector search workloads.
Regarding individual query recall distributions, and across all recall targets, Sweeping achieves an $RQUT=0.27$, while the deviation of queries from the recall target, both with recalls above and below the target, is approximately 0.07.
ACORN results achieve an $RQUT=0.17$, with queries deviating from the recall target by approximately 0.05 across all recall targets.
The individual query recall distributions for both Sweeping and ACORN are presented in Figures~\ref{fig:vader-caselaw-k100}(c) and (d) respectively, for $R_t=0.9$.

\begin{figure*}
    \begin{minipage}[t]{0.99\textwidth}
        \centering
        \begin{adjustbox}{max width=0.5\textwidth}
            \includegraphics{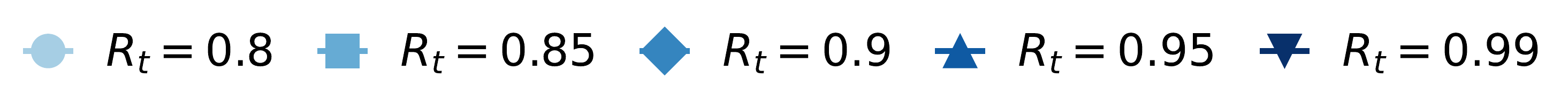}
        \end{adjustbox}
        
        \begin{subfigure}[t]{0.19\textwidth}
            \centering
            \includegraphics[width=\textwidth]{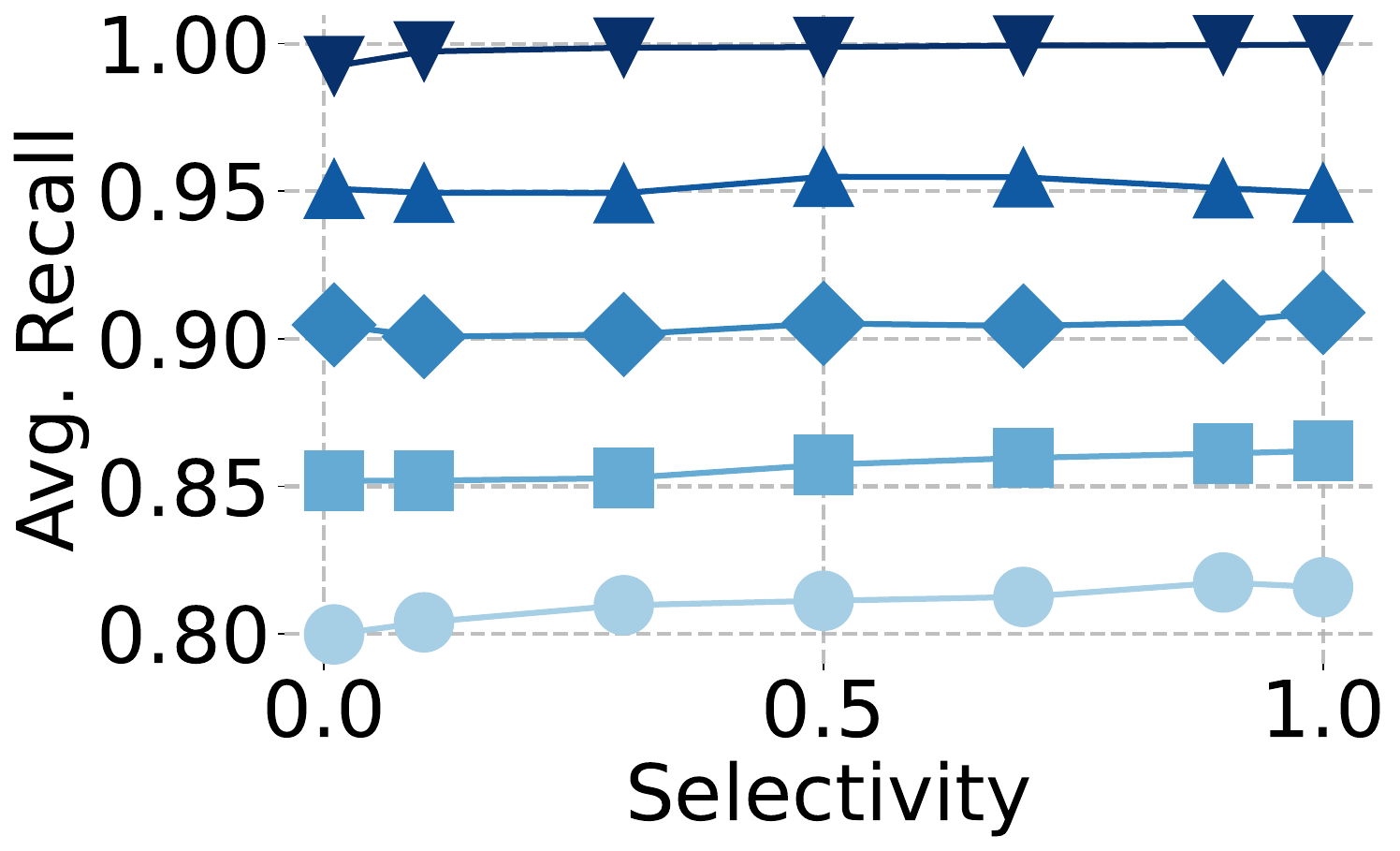}
            \caption{SIFT1M.}
        \end{subfigure}
        \hfill
        \begin{subfigure}[t]{0.19\textwidth}
            \centering
            \includegraphics[width=\textwidth]{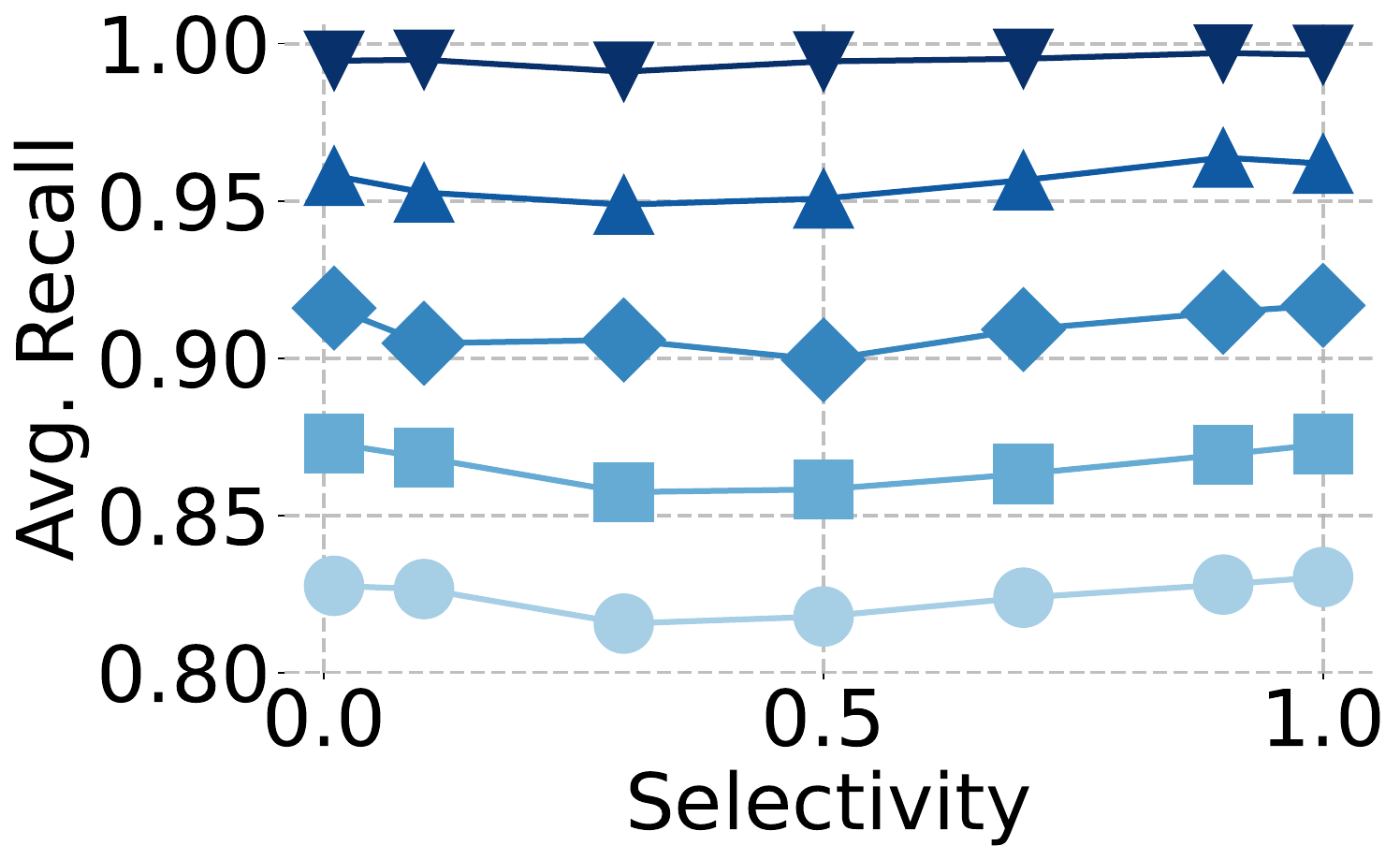}
            \caption{GIST1M.}
        \end{subfigure}
        \hfill
        \begin{subfigure}[t]{0.19\textwidth}
            \centering
            \includegraphics[width=\textwidth]{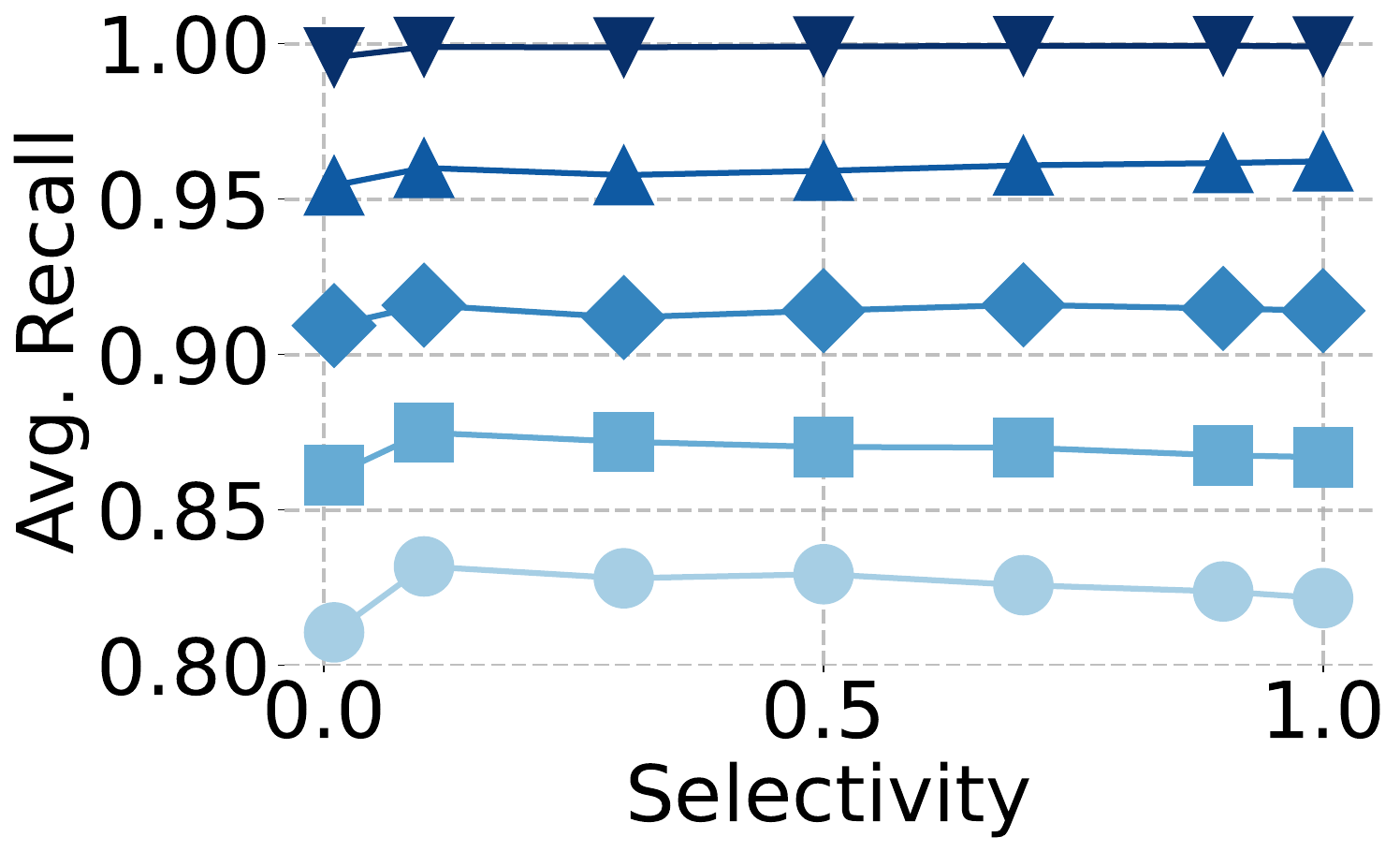}
            \caption{GLOVE1M.}
        \end{subfigure}
        \hfill
        \begin{subfigure}[t]{0.19\textwidth}
            \centering
            \includegraphics[width=\textwidth]{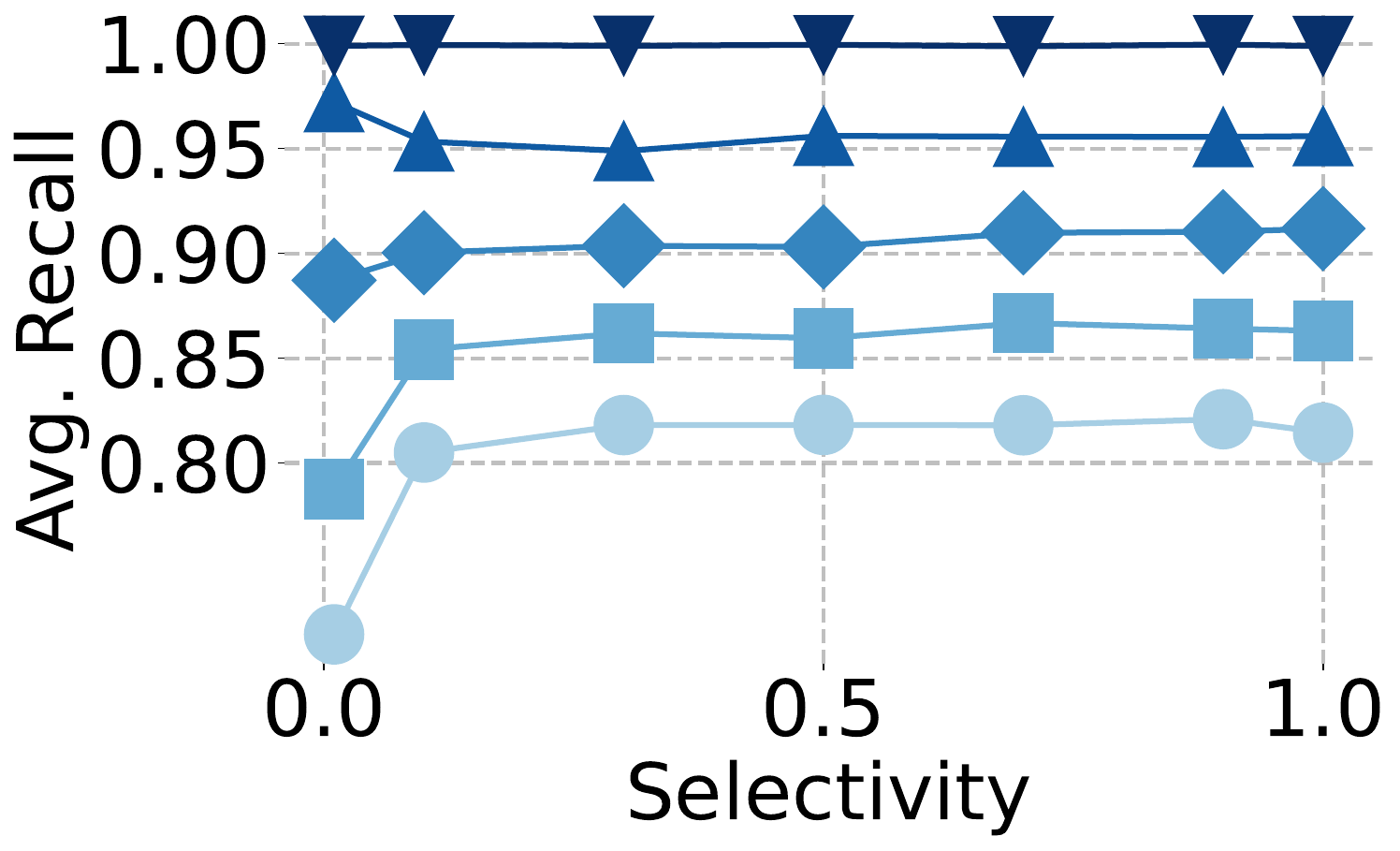}
            \caption{DEEP10M.}
        \end{subfigure}
        \hfill
        \begin{subfigure}[t]{0.19\textwidth}
            \centering
            \includegraphics[width=\textwidth]{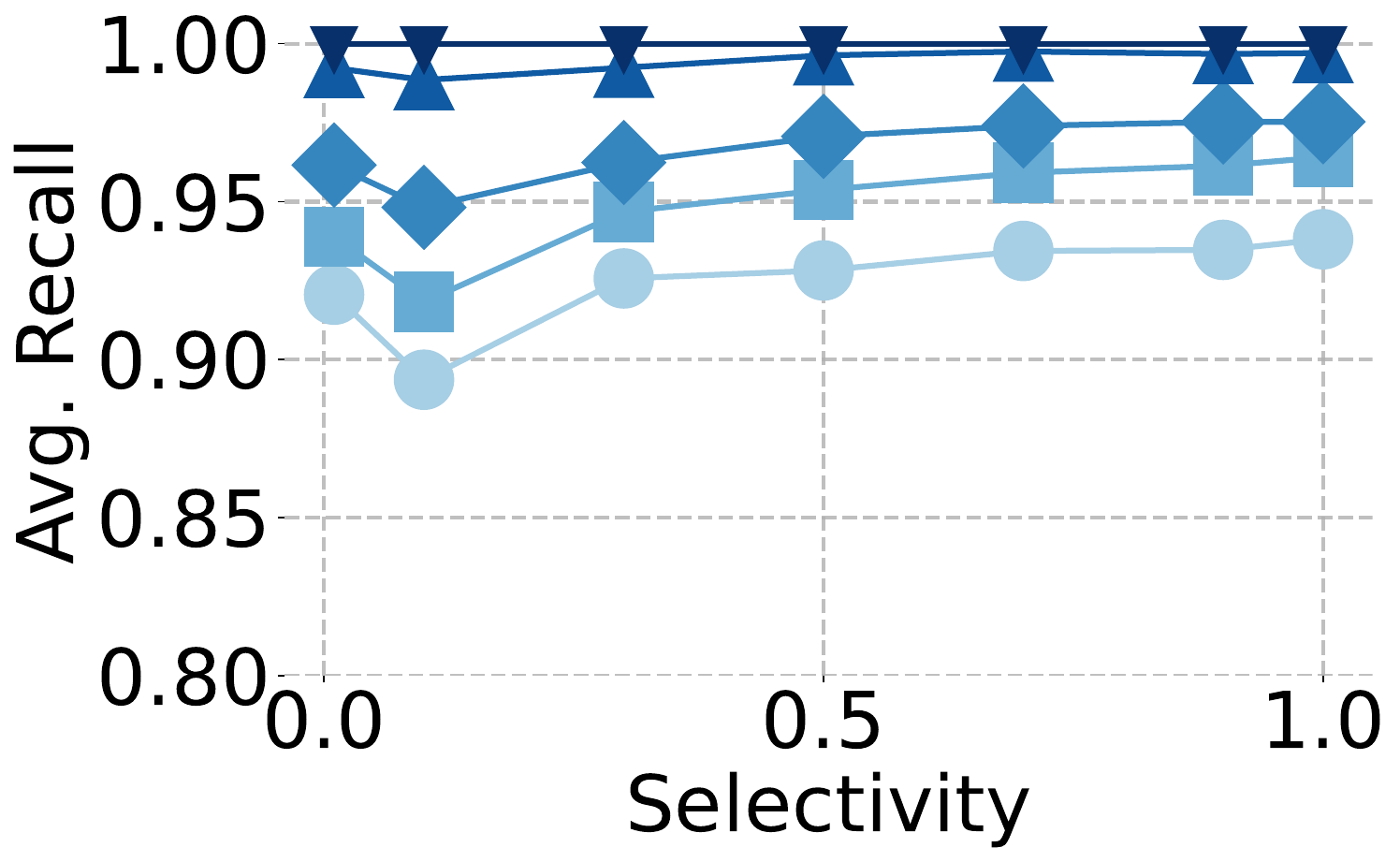}
            \caption{T2I10M.}
        \end{subfigure}

    \vspace{-0.3cm}
    \caption{VADER for Filtered-ScaNN with varying $Rt$ ($k=100$, No correlation).}
    \label{fig:vader-scann-k100-recall-targets}
    \end{minipage}

    \begin{minipage}[t]{0.99\textwidth}
        \centering

        \begin{subfigure}[t]{0.19\textwidth}
            \centering
            \includegraphics[width=\textwidth]{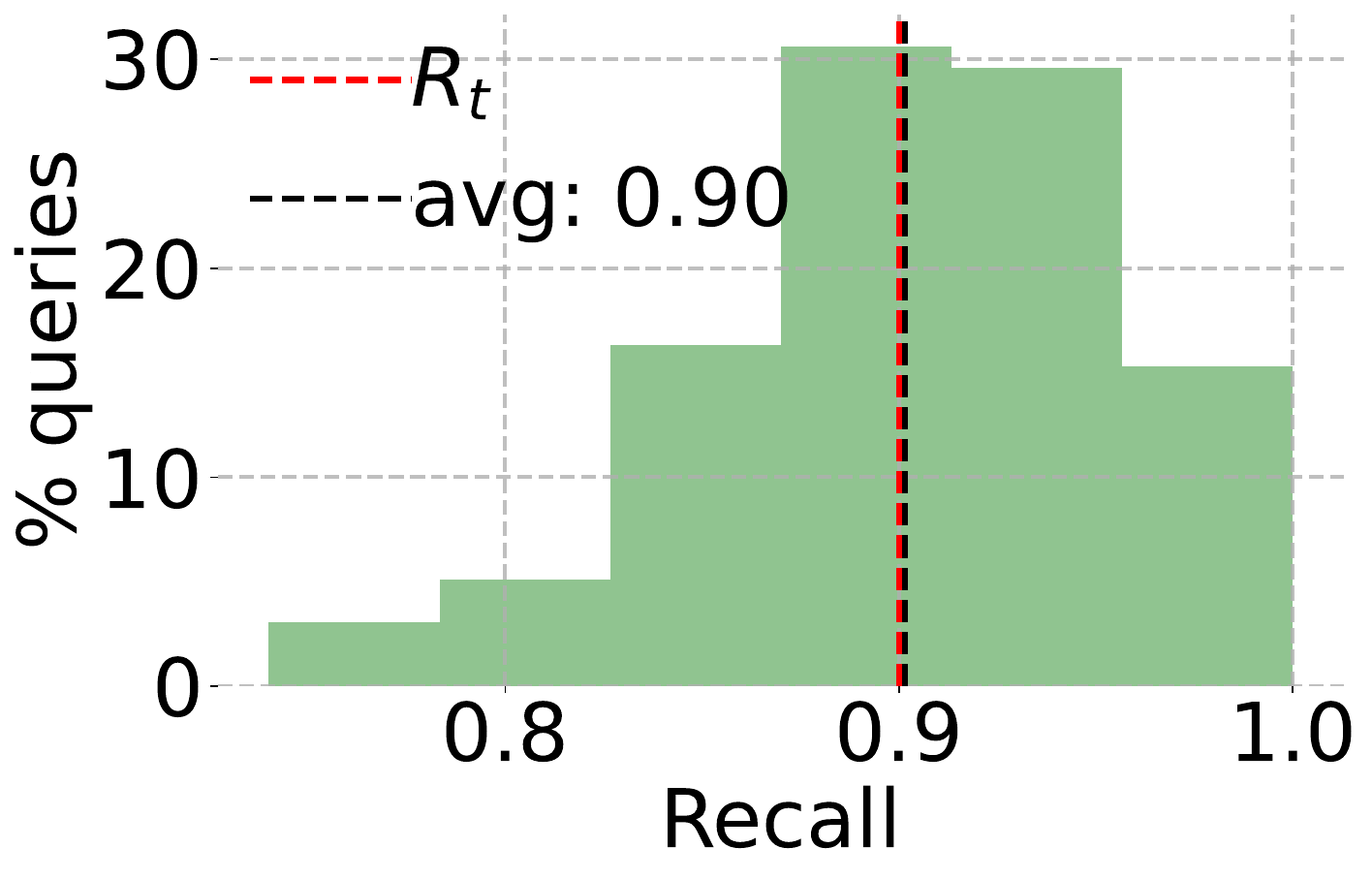}
            \caption{SIFT1M}
        \end{subfigure}
        \hfill
        \begin{subfigure}[t]{0.19\textwidth}
            \centering
            \includegraphics[width=\textwidth]{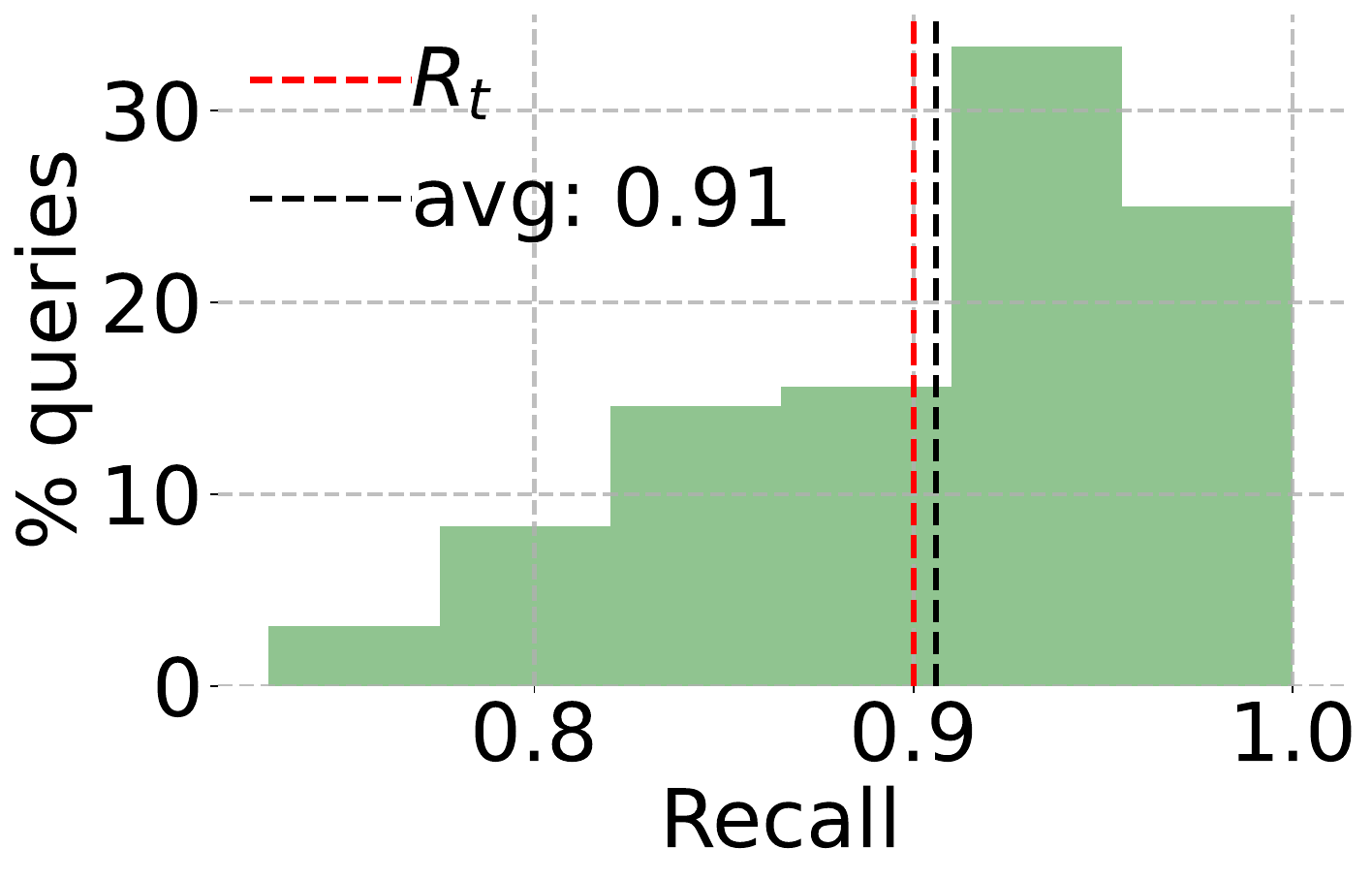}
            \caption{GIST1M}
        \end{subfigure}
        \hfill
        \begin{subfigure}[t]{0.19\textwidth}
            \centering
            \includegraphics[width=\textwidth]{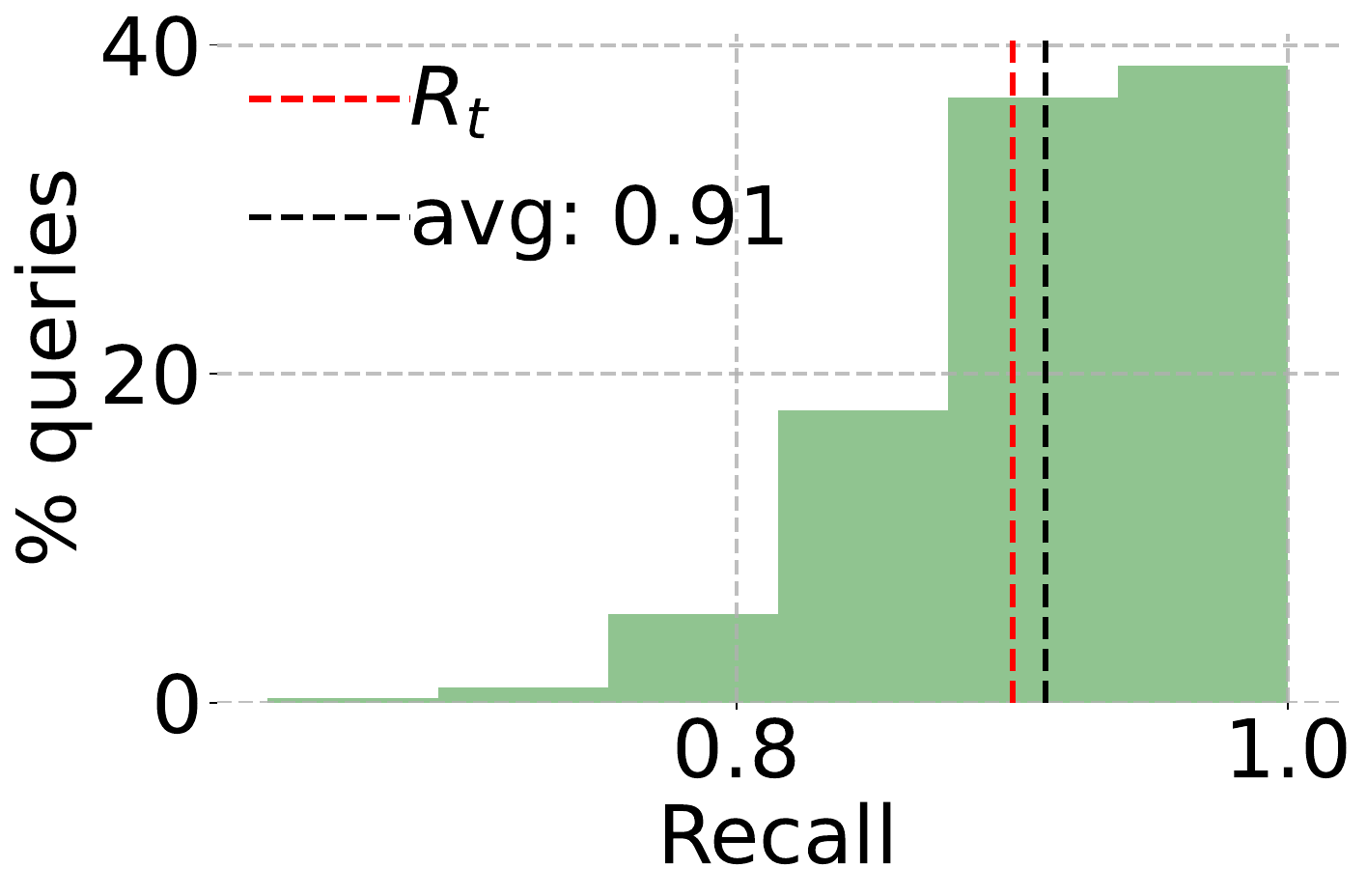}
            \caption{GLOVE1M}
        \end{subfigure}
        \hfill
        \begin{subfigure}[t]{0.19\textwidth}
            \centering
            \includegraphics[width=\textwidth]{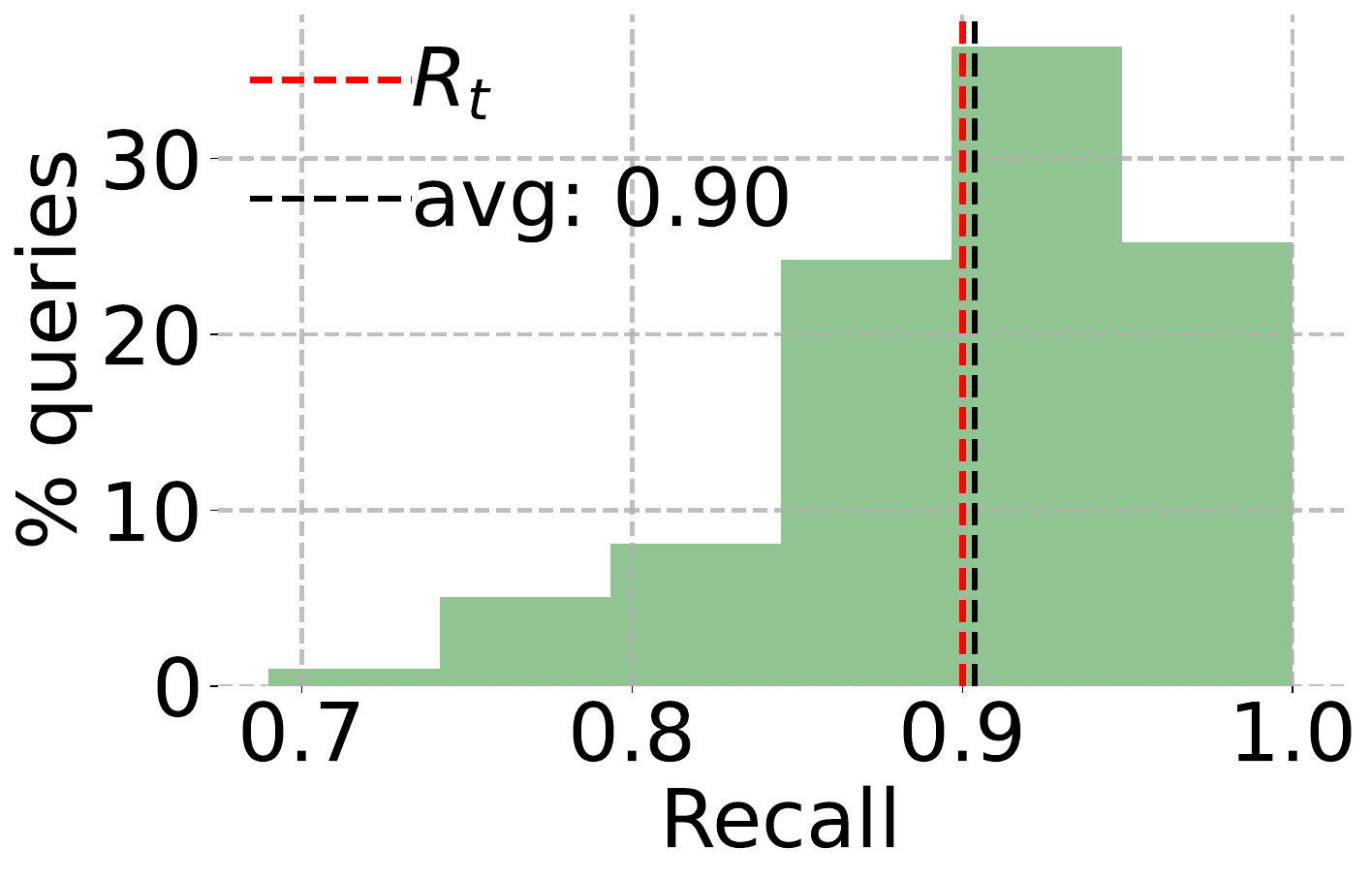}
            \caption{DEEP10M}
        \end{subfigure}
        \hfill
        \begin{subfigure}[t]{0.19\textwidth}
            \centering
            \includegraphics[width=\textwidth]{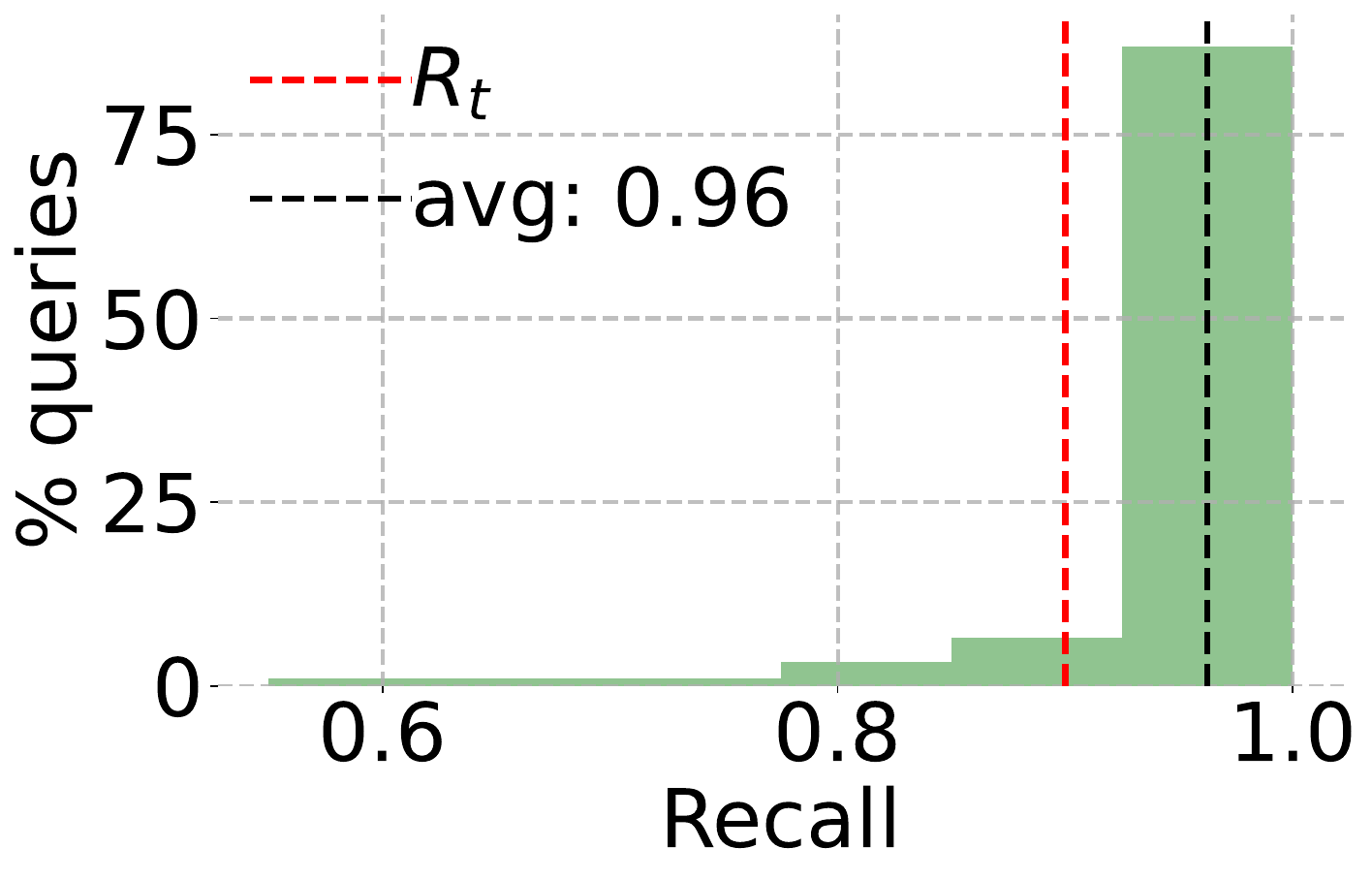}
            \caption{T2I10M}
        \end{subfigure}
    \vspace{-0.3cm}
    \caption{Recall distr. of VADER for Filtered-ScaNN ($k=100$, No correlation).}
    \label{fig:vader-scann-k100-distributions}
    \end{minipage}
\end{figure*}

\subsubsection{VADER for Filtered-ScaNN}
\label{sec:experiments:filtered-scann}

We now evaluate the VADER integration with Filtered-ScaNN~\cite{filtered_scann_alloydb_scann_2025}, the IVF-based~\cite{douze2024faiss} filter-agnostic FVS algorithm built on ScaNN~\cite{guo2020scann}, described in Section~\ref{sec:vader:integration-to-fvs}. 
The training procedure is the same as in the $k$-NN graph case.
For the experimental evaluation, we constructed one ScaNN index per dataset using the official ScaNN parameter selection heuristics~\cite{googlecloud_scann_parameters_2026}.
This yields 2K partitions for GIST1M, 10K for SIFT1M, 11K for GLOVE1M, and 100K for both DEEP10M and T2I10M, all with single-level IVF clustering.
Generating the training data and training the predictor requires approximately 4 minutes for the 1M-scale datasets and 20 minutes for DEEP10M and T2I10M.

Figure~\ref{fig:vader-scann-k100-recall-targets} reports the achieved recall results for varying selectivities and recall targets across all datasets at $k=100$ with no correlation; other configurations are similar.
In all cases, VADER manages to achieve a recall very close to the recall target. 
The only exception is the T2I10M dataset, where VADER achieves recall values above the recall target, though, not very close to it. 
This minor behavior glitch is attributed to the OOD nature of the dataset, which causes the following behavior. 
Since the VADER recall prediction is done after processing entire leafs, it may happen that the achieved recall before processing the leaf is much lower than the target, while the recall after processing the leaf surpasses the target. 
Since VADER does not terminate the search while processing the vectors of a leaf, we experience some recall overshooting in this particular dataset, however, VADER still achieves the recall targets.

Overall, across all of our datasets and recall targets, VADER always achieves recall very close to the target, achieving on average $RQUT=0.16$, while the deviation of individual query recalls from the target (including the queries above and below the target recall) is only 0.05, similar to the results of the k-NN graphs, with the individual query recall distributions presented in Figure~\ref{fig:vader-scann-k100-distributions} for all datasets.


  
  
  
  


\section{Conclusions}
\label{sec:conclusions}
We presented VADER, the first declarative recall approach for filtered approximate vector search.
VADER employs a filter-aware recall predictor that adaptively estimates per-query recall during execution and terminates the search once the specified target is reached.
Extensive experiments across diverse datasets and workloads show that VADER outperforms existing baselines and represents the state of the art for declarative recall in FVS.

\begin{acks}
Work supported by EU Horizon project DataGEMS ($101188416$). 
Manos Chatzakis is supported with a PhD  Scholarship from the Onassis Foundation.
\end{acks}

\bibliographystyle{ACM-Reference-Format}
\bibliography{references}

\end{document}